%% file: state_space_technical_book_referenced-appendix.tex
\documentclass[11pt,oneside]{book}

\usepackage[margin=1.1in]{geometry}
\usepackage{amsmath,amssymb,amsthm,mathtools}
\usepackage{bm}
\usepackage{physics}
\usepackage{hyperref}
\usepackage{enumitem}
\usepackage{microtype}
\usepackage[most]{tcolorbox}

\newtcolorbox{conjecturebox}{
  enhanced,
  colback=gray!8,
  boxrule=0pt,
  frame hidden,
  borderline={0pt}{0pt}{white},
  left=8pt,
  right=8pt,
  top=6pt,
  bottom=6pt,
  arc=0pt,
  outer arc=0pt
}

\hypersetup{
    colorlinks=true,
    linkcolor=blue,
    citecolor=blue,
    urlcolor=blue
}

\theoremstyle{plain}
\newtheorem{theorem}{Theorem}[chapter]
\newtheorem{proposition}[theorem]{Proposition}

\theoremstyle{definition}

\title{
\Huge CLASSICAL AND QUANTUM DYNAMICS IN STATE SPACE\\[1ex]
\Large Projective Hilbert Space, Measurement, and Classical Emergence
}
\author{Alexey A. Kryukov}
\date{}

\begin{document}

\frontmatter
\maketitle

\chapter*{Preface}
\label{chap:preface}

This book develops a geometric framework in which classical and quantum
mechanics, together with measurement, definite outcomes, and Born
probabilities, are described within a common dynamics in quantum state
space. The fundamental arena is projective Hilbert space. Ordinary
configuration space and classical phase space are identified with
distinguished submanifolds formed by localized quantum states and their
finite-resolution equivalence classes.

The first part of the book develops the geometry of these classical
submanifolds. Their induced Fubini--Study metric and symplectic
structure reproduce the corresponding structures of classical space
and phase space. Schr\"odinger dynamics is then compared with motion on
these submanifolds. Its tangent component gives Newtonian dynamics, and
the Schr\"odinger action restricted to the localized phase-space
manifold becomes the classical action. Conversely, the complete family
of Hamiltonian motions generated by a fixed classical Hamiltonian
determines, whenever a common unitary lift exists, that lift uniquely
up to an overall phase. 
In this way, classical mechanics appears as a
distinguished sector of quantum mechanics rather than as an independent
theory.

The second part concerns probability and measurement. The normal
distribution generated by Brownian motion on the localized classical
manifold is related geometrically to the Born rule in projective Hilbert
space. A homogeneous family of Brownian motions on the classical
manifold is lifted to one common stochastic unitary evolution in the
ambient state space. Under homogeneous component independence, the
projectively relevant generators of this lift are shown to belong to
the Gaussian Unitary Ensemble, apart from an overall scale and a
projectively irrelevant scalar term.

This reconstruction motivates the random-matrix conjecture \({\bf RM}\).
The conjecture is that, under measurement conditions, the unresolved
interaction of a system with a measuring device or environment produces
an effective stochastic unitary regime of this kind. A measurement
outcome occurs when the resulting state-space path enters a
detector-defined equivalence class. The finite resolution of the
detector is essential: a recorded classical value corresponds not to an
exact ray, but to a class of experimentally indistinguishable states.

The same framework is applied to microscopic measurements, macroscopic
bodies, composite systems, and entangled states. For microscopic
systems the stochastic evolution produces a single recorded outcome
with the corresponding Born probability. For macroscopic systems,
frequent environmental interactions repeatedly stabilize the state
near the localized phase-space sector, while the ordinary Hamiltonian
supplies the Newtonian tangent drift and the environmental fluctuations
remain small at the observational resolution. The book also examines
quantum correlations, familiar measurement paradoxes, irreversibility,
and the arrow of time, and compares the proposed framework with other
approaches to the quantum--classical transition.

The emphasis throughout is on the finite- and infinite-dimensional
geometric structures that connect localized quantum states with
classical space and phase space, and on the relation between ordinary
Schr\"odinger propagation and stochastic random-matrix dynamics during
measurement. The central claim is that classical geometry, Newtonian
motion, quantum probabilities, and measurement outcomes can all be
described within the geometry and unitary dynamics of quantum state
space.

\tableofcontents

\mainmatter

\chapter*{Introduction}
\addcontentsline{toc}{chapter}{Introduction}
\label{chap:introduction}

The purpose of this book is to present a geometric framework in which
classical mechanics, quantum mechanics, measurement outcomes, and
quantum probabilities arise from dynamics in quantum state space. The
fundamental arena is taken to be projective Hilbert space,
\[
\mathbb P(\mathcal H),
\]
rather than ordinary configuration space or classical phase space.
Classical space and phase space are not discarded. They appear as
distinguished finite-dimensional submanifolds of projective state
space. A point of classical configuration space is represented by a
finite-resolution equivalence class of localized quantum states. A
point of classical phase space is represented by such a class together
with a phase factor encoding momentum.
In the stochastic sector developed later in the book, a local
system--environment interaction generates Brownian motion on the
localized classical manifold, and the common unitary lift of this
Brownian family yields random-matrix Schr\"odinger dynamics in the
ambient projective state space.

We begin by constructing the manifolds of states localized in the
position representation,
\[
M_3^\sigma,
\qquad
M_{3,3}^\sigma,
\]
and the corresponding quotient manifolds of equivalence classes,
\[
\widetilde M_3^\sigma,
\qquad
\widetilde M_{3,3}^\sigma.
\]
The parameter \(\sigma\) sets the localization scale. The manifold
\(\widetilde M_3^\sigma\) consists of equivalence classes of localized
position states, while \(\widetilde M_{3,3}^\sigma\) is obtained by
attaching phase factors, which for localized states take the form
\[
e^{\frac{i{\bf p}\cdot{\bf x}}{\hbar}}.
\]
After rescaling, the Fubini--Study metric induced on these manifolds
becomes the Euclidean metric of configuration space and phase space.
The construction extends directly to many-particle systems through
tensor products.

The construction is also dynamical. Schr\"odinger's equation defines a
vector field on projective Hilbert space. At a point of
\(\widetilde M_{3,3}^\sigma\), this vector field separates into tangent
and normal components. The tangent component is the Newtonian
phase-space vector field. Equivalently, the Schr\"odinger action
restricted to the localized phase-space manifold becomes the classical
action and yields Hamilton's equations. Newtonian dynamics is therefore
not introduced as an independent postulate. It is the dynamics induced
by Schr\"odinger evolution on the classical sector of state space.

There is also a converse statement. A fixed classical Hamiltonian
generates a complete family of Hamiltonian motions, one for every
initial point of phase space. Whenever this complete flow admits a
common unitary lift to Hilbert space, that lift is unique up to an
overall phase. The reason is that equality of the induced Hamiltonian
vector fields determines the diagonal coherent-state symbol of the
generator up to a constant, while the overcomplete family of
phase-space Gaussian states determines the operator itself. For the
Hamiltonians obtained in the book by restricting Schr\"odinger
dynamics, existence of the lift is already supplied by the original
quantum Hamiltonian.

The same complete Hamiltonian family also supplies the classical
motions needed in the time-sliced construction of the quantum
propagator. At each subdivision time, the Hamiltonian flow provides the
motion beginning from every possible phase-space point. For a
sufficiently fine subdivision, the classical trajectory joining
successive configurations determines the corresponding short-time
action and, together with the appropriate amplitude factor, the
short-time propagator. Repeated composition and integration over the
intermediate configurations yields the Schr\"odinger propagator. A
trajectory selected at a later subdivision may simply continue one
selected earlier, or it may begin from different intermediate
phase-space data. In either case it belongs to the same complete family
generated by the Hamiltonian. Thus the ordinary Feynman
many-path construction and the common unitary evolution are built from
the same underlying Hamiltonian flow.

The geometric formulation also clarifies the role of observables.
Self-adjoint operators generate vector fields on state space, and their
commutators determine the Lie brackets of these fields. On the
localized phase-space manifold, the commutator equations reduce to the
Poisson-bracket equations of classical mechanics. The canonical
commutator also admits a geometric relation to the curvature of the
corresponding projective sections, a relation whose broader physical
implications remain to be explored.

The next part of the book concerns probability. The normal distribution
of classical position measurements and the Born rule are usually
regarded as unrelated laws. Here they are connected by the embedding of
classical space into projective Hilbert space. On the manifold of
localized Gaussian states, the normal transition law can be written as
a function of the Fubini--Study distance. Since these states realize the
full range of projective distances, the same law extends to arbitrary
pure states and becomes
\[
P(\psi\longrightarrow\phi)
=
\cos^2\rho_{\rm FS}(\psi,\phi)
=
|(\phi,\psi)|^2.
\]

The normal law is generated dynamically by a Gaussian random walk whose
diffusion limit is Brownian motion. In the present framework this
Brownian motion is not introduced only as an abstract stochastic model.
A spatially local system--environment interaction of
Caldeira--Leggett type produces, in the localized sector, the required
family of Brownian motions corresponding to different initial localized
states. Thus ordinary Schr\"odinger dynamics of the particle together
with its environment supplies the classical stochastic family that is
to be lifted to projective state space.

We therefore consider not a single Brownian trajectory but the complete
homogeneous family of Brownian motions on the localized manifold. The
entire family is required to have one common linear unitary lift to
projective state space. Under homogeneous component independence, the
projectively relevant Hermitian generators of this lift are shown, in
finite-dimensional approximations, to have the Gaussian Unitary
Ensemble law, apart from an overall scale and a projectively irrelevant
scalar term. The scale is fixed by the Brownian diffusion coefficient.
Conversely, the resulting GUE-generated stochastic unitary evolution
reproduces the Brownian family when constrained to the localized
classical manifold.

This reconstruction leads to the random-matrix dynamics \({\bf RM}\).
The physical mechanism is supplied by the system--environment
interaction. In the localized sector, a spatially local
Caldeira--Leggett-type coupling generates the Brownian family, while
the common stochastic lift of that family produces successive random
Hermitian increments with the GUE statistics. Thus the
effective \({\bf RM}\) regime is obtained from ordinary
Schr\"odinger dynamics together with the geometric lift from the
classical localized sector to projective state space.
For every realized Hamiltonian history, the evolution remains linear and unitary, and the state follows one stochastic path in projective
state space. A measurement outcome is formed when this path enters a
detector-defined finite-resolution sector.

No additional projection is inserted into the evolution. The record identifies the equivalence class reached by the state but does not itself produce the state or alter its path. The probability of an outcome is
the weight of the family of \({\bf RM}\) paths entering the
corresponding detector sector. These weights are fixed by the
Fubini--Study geometry and by their correspondence with the normal
distribution on the classical-space submanifold. The resulting hitting
probabilities are the Born probabilities.

This reveals a parallel between two forms of state-space evolution.
Ordinary Schr\"odinger dynamics, represented in many-path form by the
Feynman construction, describes coherent propagation of quantum
amplitudes. The \({\bf RM}\) dynamics describes the stochastic
state-space evolution associated here with measurement and produces the
probabilities of the possible detector records:
\[
\begin{array}{ccl}
\text{Schr\"odinger--Feynman dynamics}
&-&
\text{propagation of quantum amplitudes},
\\[2mm]
\text{\({\bf RM}\) dynamics}
&-&
\text{measurement dynamics and Born probabilities}.
\end{array}
\]
Both are evolutions in the same projective Hilbert space and both are
linked to ordinary dynamics on the localized classical sector. The
first is associated with the complete family of Hamiltonian motions.
The second is associated with the Brownian family generated by
system--environment Schr\"odinger dynamics and with its common
random-unitary lift.

The microscopic and macroscopic regimes differ in the relative scales
of the deterministic and stochastic parts of the dynamics. During a
microscopic measurement, the stochastic component may dominate the
short record-forming interaction and produce one outcome with the
corresponding Born probability. For a macroscopic body, repeated
environmental interactions stabilize the state near the localized
phase-space sector. At the resolution of observation, the ordinary
Hamiltonian supplies the Newtonian tangent drift, while the stochastic
deviations remain small. The resulting sequence of finite-resolution
records is narrowly concentrated about the Newtonian trajectory.

The estimates developed in the book test this mechanism in several
concrete situations. They include a macroscopic body in air, the
limiting environment provided by the cosmic microwave background,
microscopic tracks in a bubble chamber, and laboratory position
measurements of microscopic particles. In each case, the ordinary and
stochastic parts of the motion act on different scales, allowing
localization to be maintained while preserving the relevant
deterministic dynamics.

The construction is then extended to composite and entangled systems.
A two-particle state is one point of the projective Hilbert space
\[
\mathbb P(\mathcal H_1\otimes\mathcal H_2),
\]
and the \({\bf RM}\) walk acts on the joint state as a whole, locally in
time and in state space. The classical product phase space appears as a
localized submanifold of this tensor-product state space. EPR and spin
correlations are determined by the geometry of the joint state relative
to the detector-defined outcome sectors, without assigning hidden
classical properties to the individual particles before measurement.

The same point of view clarifies several familiar quantum paradoxes. A
superposition is one state in projective Hilbert space, not a collection
of coexisting classical configurations. A classical value exists only
when the state belongs to the corresponding localized sector. The
double-slit experiment, the preferred-basis problem, the problem of
tails, Schr\"odinger's cat, and Wigner's friend all involve extending
classical descriptions to states lying outside the corresponding
classical sector.


Irreversibility also acquires a state-space interpretation. Every
realized \({\bf RM}\) history is unitary and therefore mathematically
invertible. Exact reversal, however, requires complete knowledge of the
realized sequence of random Hamiltonians and deliberate application of
the corresponding inverse unitaries in reverse order. Reversing the
direction of time and taking the final state of a realized
\({\bf RM}\) history as the new initial state does not cause the
stochastic evolution to retrace the original path. Exact retracing is
possible only if the particular realized Hamiltonian history is
reconstructed and inverted.

Effective irreversibility is reinforced by the high dimensionality of
projective state space and by finite-resolution recording. Once the
realized Hamiltonian history is lost, return to a prescribed
neighborhood of the initial ray becomes increasingly unlikely as the
dimension grows, with the corresponding probability tending to zero in
the infinite-dimensional limit. At the same time, a measurement record
retains neither the exact ray nor the microscopic trajectory by which
it was reached. Many distinct state-space histories therefore
correspond to the same recorded equivalence class, and the information
required for exact reconstruction is lost.

Restriction of the \({\bf RM}\) walk to the classical submanifolds also
produces Brownian diffusion and the associated spreading of classical
probability distributions. The macroscopic arrow of time thus reflects
the combined effects of stochastic spreading, suppression of recurrence
in high-dimensional state space, and the loss of microscopic path
information in finite-resolution records. The past is represented by
an ordered sequence of records, whereas the future is described by the
weights of possible later paths.


The final part of the book compares the present construction with
decoherence, open-system models, continuous-measurement theory,
semiclassical limits, objective-collapse models, center-of-mass
classicalization, Bohmian mechanics, Everettian approaches, consistent
histories, and geometric quantum mechanics. The distinctive feature of
the present framework is that these questions are formulated in a
different physical and geometric setting. Projective state space is
taken as the fundamental arena; classical phase space appears as a
localized submanifold; Newtonian dynamics is recovered as restricted
Schr\"odinger dynamics; the complete Hamiltonian family determines its
common unitary lift and supplies the time-sliced Feynman construction;
and measurement and environmental stabilization are described by
stochastic unitary random-matrix dynamics.

A remaining problem is to determine the range of physical conditions
under which the microscopic system--environment interaction is
accurately represented by the local bath model used here. Within that
model, the required Brownian family is obtained from ordinary
Schr\"odinger dynamics, and its common stochastic lift yields the
GUE-generated \({\bf RM}\) regime. The open question is therefore no
longer the existence of a microscopic route from Schr\"odinger
dynamics to the Brownian family, but the scope and universality of that
route for realistic measuring devices and environments.

The central claim of the book is that classical geometry, Hamiltonian
motion, measurement outcomes, Born probabilities, and quantum
correlations can be described within a single state-space framework.
Projective Hilbert space is therefore not merely a representation of
quantum states. In the present framework, it is the common physical
arena of classical and quantum dynamics.

\chapter*{Notation}
\addcontentsline{toc}{chapter}{Notation}
\label{chap:notation}

The notation used throughout the book is collected here for reference.
Symbols introduced only for a single local calculation are normally
defined where they occur and are not repeated below. The final group
collects notation used mainly in Chapter~\ref{chap:comparison}, where
several standard formulations are compared with the present framework.

\section*{State space and geometry}

\begin{description}

\item[\(\mathcal H\)]
Hilbert space of quantum states. For a single nonrelativistic particle
in three spatial dimensions, \(\mathcal H=L^2(\mathbb R^3)\).

\item[\(\mathbb P(\mathcal H)\)]
Projective Hilbert space of pure quantum states.

\item[\(S^{\mathcal H}\)]
Unit sphere of normalized vectors in \(\mathcal H\).

\item[\({[\psi]}\)]
Ray in \(\mathbb P(\mathcal H)\) represented by a nonzero vector
\(\psi\in\mathcal H\). When no confusion is possible, the ray
\([\psi]\) may also be denoted simply by \(\psi\).

\item[\((\phi,\psi)\), \(\langle\phi,\psi\rangle\)]
Hilbert-space inner product. Parentheses are used most often; angle
brackets are used occasionally when they improve readability.

\item[\(\|\psi\|\)]
Hilbert-space norm of \(\psi\).

\item[\(\langle \widehat A\rangle\)]
Expectation value of the operator \(\widehat A\) in the state under
consideration. When the state is displayed explicitly,
\[
\langle\widehat A\rangle_\psi
=
\langle \psi,\widehat A\psi\rangle
=
(\psi,\widehat A\psi),
\]
for normalized \(\psi\).

\item[\(\rho_{\rm FS}\)]
Fubini--Study distance on projective Hilbert space. When no confusion is
possible, it is denoted simply by \(\rho\).

\item[\(ds^2\)]
Fubini--Study line element, or its restriction to a submanifold when
the context is clear.

\item[\(\omega_{\rm FS}\)]
Fubini--Study symplectic form on projective Hilbert space.

\item[\(\omega_{\rm cl}\)]
Canonical symplectic form on classical phase space.

\item[\(T_\psi M\)]
Tangent space to a manifold \(M\) at the state \(\psi\).

\item[\(X_{\widehat A}\)]
Vector field generated by the self-adjoint operator \(\widehat A\).
On the Hilbert space of states,
\[
X_{\widehat A}(\psi)
=
-\frac{i}{\hbar}\widehat A\psi.
\]
Its projection to projective Hilbert space is represented by the
horizontal vector
\[
-\frac{i}{\hbar}
\bigl(\widehat A-\langle\widehat A\rangle_\psi\bigr)\psi.
\]

\item[\(X_{\widehat h}^{T}\), \(X_{\widehat h}^{\perp}\)]
Tangential and normal components, respectively, of the projective
Schr\"odinger vector field relative to a localized classical
submanifold.

\end{description}

\section*{Localized states and classical submanifolds}

\begin{description}

\item[\(\sigma\)]
Localization or finite-resolution scale. Its physical meaning depends
on the measuring or environmental resolution under consideration.

\item[\(\varphi_a,\ \varphi_{a,p}\)]
Notation for localized Gaussian position and phase-space states.
The state \(\varphi_a\) is also denoted by \(g_{a,\sigma}\), with the
width displayed explicitly when needed.

\item[\(g_{a,\sigma}\)]
Normalized real-valued Gaussian state centered at \(a\) with width
\(\sigma\).

\item[\(M_3^\sigma,\ M_{3,3}^\sigma\)]
Localized one-particle position and phase-space manifolds in projective
Hilbert space.

\item[\(\widetilde M_3^\sigma,\ 
\widetilde M_{3,3}^\sigma\)]
Corresponding manifolds of finite-resolution equivalence classes.

\item[\(M_{3N}^\sigma,\ M_{3N,3N}^\sigma\)]
Localized configuration-space and phase-space manifolds for \(N\)
distinguishable particles.

\item[\(\widetilde M_{3N}^\sigma,\ 
\widetilde M_{3N,3N}^\sigma\)]
Corresponding many-particle manifolds of finite-resolution equivalence
classes.

\item[\(M_1^\sigma,\ M_{1,1}^\sigma\)]
One-dimensional position and phase-space versions of the localized
classical manifolds.

\item[\(M^\sigma_3\otimes\cdots\otimes M^\sigma_3,\ 
M^\sigma_{3,3}\otimes\cdots\otimes M^\sigma_{3,3}\)]
Notation emphasizing the realization of the many-particle
configuration-space and phase-space manifolds as product-state
submanifolds of
\(\mathbb P(\mathcal H_1\otimes\cdots\otimes\mathcal H_N)\).
They are the images of the corresponding Cartesian products under the
Segre tensor-product embedding.

\item[\(\mu_x(\psi)\), \(\mu_z(\psi)\)]
Position expectation value of the state \(\psi\), in three dimensions
or in the one-dimensional coordinate \(z\), respectively.

\item[\(\delta_x(\psi)\), \(\delta_z(\psi)\)]
Position spread (standard deviation) of the state \(\psi\), in three
dimensions or in the one-dimensional coordinate \(z\), respectively.

\item[\(\{g_a\}\), \(\{\varphi_a\}\), \(\{\varphi_{a,p}\}\)]
Finite-resolution equivalence classes of states representing points of
the corresponding classical manifolds. The notation used for the
representative is determined by the context.

\item[\(a,p\)]
Classical position and momentum coordinates on a localized
phase-space manifold.

\item[\(\widehat x,\widehat p\)]
Position and momentum operators.

\item[\(\widehat q,\widehat k\)]
Rescaled, dimensionless position and momentum operators, respectively.

\item[\(\hbar\)]
Reduced Planck constant.

\end{description}

\section*{Dynamics and geometric scales}

\begin{description}

\item[\(\widehat h\)]
Ordinary Hamiltonian of the system in the one-particle notation used
through much of the book.

\item[\(\widehat H\)]
Hamiltonian of a composite system, or a generic Hamiltonian when the
capital-letter notation is more convenient.

\item[\(h(a,p,t)\), \(H(a,p,t)\)]
Classical Hamiltonian functions obtained by restricting the
Schr\"odinger action to the localized phase-space manifold.

\item[\(\{\cdot,\cdot\}_{P}\)]
Classical Poisson bracket.

\item[\(L,P\)]
Reference length and momentum scales used to introduce dimensionless
phase-space coordinates.

\item[\(q,k\)]
Dimensionless position and momentum coordinates,
\[
q=\frac{a}{L},
\qquad
k=\frac{p}{P}.
\]

\item[\(\ell_P,p_P\)]
Planck length and Planck momentum,
\[
\ell_P=\sqrt{\frac{\hbar G}{c^3}},
\qquad
p_P=\frac{\hbar}{\ell_P}.
\]

\item[\(R_{\rm FS},K_{\rm FS}\)]
Radius and sectional curvature, respectively, of a projective line
\(\mathbb CP^1\) in the standard Fubini--Study normalization.

\item[\(\sigma_i\)]
Pauli matrices, \(i=1,2,3\). The vector of Pauli matrices is denoted by
\[
\boldsymbol{\sigma}
=
(\sigma_1,\sigma_2,\sigma_3).
\]

\end{description}

\section*{Random-matrix walk, probabilities, and records}

\begin{description}

\item[\({\bf RM}\)]
Random-matrix conjecture. The corresponding stochastic unitary motion
in projective state space is called the \({\bf RM}\) walk, and an
individual realization of this motion is called an \({\bf RM}\) path.

\item[\(\widehat h_{\bf RM}(t)\)]
Random Hermitian Hamiltonian generating the \({\bf RM}\) component of
the evolution.

\item[\(\widehat h_{{\bf RM},k}\)]
Random Hermitian Hamiltonian generating the \(k\)-th discrete
\({\bf RM}\) step.

\item[\(\mathcal H_N\)]
Finite-dimensional approximation to the Hilbert space used in the
reconstruction of the random generator.

\item[\(P_N\)]
Orthogonal projection onto \(\mathcal H_N\); the cutoff is chosen so that
\(P_N\to I\) strongly as \(N\to\infty\).

\item[\(\widehat h_N\)]
Random Hermitian generator on \(\mathcal H_N\) used in the
finite-dimensional reconstruction of the Brownian lift.

\item[\(\overset{d}{=}\)]
Equality in distribution.

\item[\(\kappa\)]
Dimensionless cumulative squared projective displacement in the
high-dimensional diffusion limit,
\[
\kappa=\lim\sum_k\delta_k^2,
\]
so that
\[
\cos^2\rho_{\rm FS}=e^{-\kappa}.
\]

\item[\(H_0\)]
Traceless, projectively relevant part of the centered random Hermitian
generator,
\[
H_0
=
\widehat h_N
-
\frac{\operatorname{Tr}\widehat h_N}{N}I .
\]

\item[\(\mathcal Q_N(A,B)\)]
Covariance form of the centered Hermitian generator,
\[
\mathcal Q_N(A,B)
=
\mathbb E\!\left[
\operatorname{Tr}(A\widehat h_N)
\operatorname{Tr}(B\widehat h_N)
\right].
\]

\item[\(\tau_N^2\)]
Variance of the real or imaginary part of a nonredundant off-diagonal
GUE matrix element in the normalization used in Chapter~4. Its scale is
fixed by
\[
\tau_N^2
=
\frac{\hbar^2 D}{2\sigma^2}.
\]

\item[\(H_n\)]
Independent Brownian-normalized GUE generator used in the \(n\)-th
\({\bf RM}\) step.

\item[\(\Delta K_n\), \(dK_N\)]
Discrete and infinitesimal stochastic Hermitian increments,
\[
\Delta K_n=H_n\sqrt{\Delta t},
\qquad
dK_N=H_N\sqrt{dt}.
\]

\item[\(\mu_{\rm GUE}\)]
Probability measure of the Gaussian Unitary Ensemble on the
projectively relevant Hermitian generators.

\item[\(\gamma\)]
Joint centered Gaussian probability measure of the independent
Brownian components used to reconstruct one Hermitian generator.

\item[\(L\)]
Linear reconstruction map from the Brownian component vector to the
Hermitian generator,
\[
L:\xi\mapsto H,
\qquad
L_*\gamma=\mu_{\rm GUE}.
\]

\item[\(\Omega_B\)]
Product probability space of Gaussian component histories,
\[
\Omega_B=(\mathbb R^m)^{\mathbb N}.
\]

\item[\(\Xi_n\)]
Independent copies of the Gaussian component vector defining the
successive generators,
\[
H_n=L(\Xi_n).
\]

\item[\(P^\psi_{\rm RM}\)]
Probability measure on finite-dimensional \({\bf RM}\) histories
starting from the initial state \(\psi\), obtained by pushing forward
\(P_B\) under the stochastic unitary evolution.

\item[\(R_\psi\)]
Record map from an \({\bf RM}\) generator history to the corresponding
detector record.

\item[\(A_j(T;\psi)\)]
Event that the \({\bf RM}\) state starting from \(\psi\) lies in the
detector sector \(C_j\) at time \(T\).

\item[\(W_j^{\rm Born}(\psi)\)]
Born transition weight associated with detector sector \(C_j\); for an
orthogonal sector with projection \(P_j\),
\[
W_j^{\rm Born}(\psi)=\|P_j\psi\|^2.
\]

\item[\({\rm GUE}\)]
Gaussian Unitary Ensemble.

\item[\(\Delta t\)]
Duration assigned to one discrete random-matrix step.

\item[\(D\)]
Diffusion coefficient of an ordinary Brownian motion when no more
specific subscript is required.

\item[\(D_a,D_p\)]
Effective position- and momentum-diffusion scales, respectively.

\item[\(\mathbb P_{\bf RM}\)]
Probability assignment to detector-defined families of \({\bf RM}\)
paths, anchored on the Wiener measure of the localized classical
manifold through the normal-to-Born relation.

\item[\(P_B\)]
Probability measure of the classical Gaussian walk, or its Wiener
limit, used in the construction of \(\mathbb P_{\bf RM}\).

\item[\(\mathcal C_j\)]
Detector-defined finite-resolution outcome sector corresponding to the
\(j\)-th recorded outcome.

\item[\(\Gamma_j(\psi)\)]
Family of \({\bf RM}\) paths starting at \(\psi\) whose first recorded
entry is into \(\mathcal C_j\).
 
\item[\(W_j(\psi)\)]
Weight, or probability, assigned to the outcome family
\(\Gamma_j(\psi)\).

\item[\(\mathcal H_j\)]
Closed Hilbert-space sector associated with the \(j\)-th detector
outcome.

\item[\(P_j\)]
Orthogonal projection onto the Hilbert-space sector \(\mathcal H_j\)
associated with the \(j\)-th detector outcome.

\item[\(I_j\)]
Finite spatial interval or detector bin associated with the \(j\)-th
position outcome.

\item[\(P_{I_j}\)]
Spectral projection onto wave functions supported in the detector bin
\(I_j\).

\item[\(s\)]
Localization coordinate,
\[
s=\ln\frac{\delta_x}{\sigma}
\]
(or \(s=\ln(\delta_z/\sigma)\) in one dimension). The localized sector
is \(s\leq0\).

\item[\(\tau_{\rm ret}\), \(\tau_{\rm hit}\)]
First-return and first-hitting indices for the localization coordinate
in the random-walk descriptions.

\item[\(T_\eta\)]
High-probability physical return time corresponding to a prescribed
failure probability \(\eta\).

\item[\(M_t^a,M_t^p\)]
Centered stochastic processes representing the accumulated horizontal
position and momentum fluctuations induced by the random Hamiltonian.

\end{description}

\section*{Composite systems, correlations, and locality}

\begin{description}

\item[\(\mathcal H_1\otimes\mathcal H_2\)]
Tensor-product Hilbert space of a two-particle system.

\item[\(\Phi_{a_1,p_1;a_2,p_2}\)]
Localized two-particle product state with classical position and
momentum labels \((a_1,p_1;a_2,p_2)\).

\item[\(M_{1,1}^\sigma\otimes M_{1,1}^\sigma\)]
Product classical phase-space manifold for two particles in one
spatial dimension.

\item[\(\Psi\)]
Generic composite or entangled state when the context makes the
particular system clear.

\item[\(\lambda\)]
Hidden variable in the discussion of Bell-type local hidden-variable
models. This use is local to that discussion and is distinct from the
collapse-rate parameter \(\lambda\) used in the comparison with
objective-collapse models.

\item[\(\mathcal A,\mathcal B\)]
Measurement settings, or detector decompositions, chosen by Alice and
Bob in a Bell-type experiment.

\item[\(A_{\mathcal A}(\lambda),B_{\mathcal B}(\lambda)\)]
Local hidden-variable response functions, taking values \(\pm1\).

\end{description}

\section*{Notation used in the comparison with existing approaches}

The symbols in this section are standard within the approaches being
reviewed and are used only for comparison with the present framework.

\begin{description}

\item[\(\rho_{SE}\)]
Density operator of a system \(S\) together with an environment \(E\).

\item[\(\rho_S\)]
Reduced density operator of the system,
\[
\rho_S=\operatorname{Tr}_E\rho_{SE}.
\]

\item[\(\operatorname{Tr}_E\)]
Partial trace over the environmental Hilbert space.

\item[\(U(t)\)]
Unitary evolution operator of a closed system--environment composite.

\item[\(p_\alpha\)]
Probabilities appearing in a diagonal or approximately diagonal
reduced density matrix,
\[
\rho_S\simeq\sum_\alpha p_\alpha
|\alpha\rangle\langle\alpha|.
\]

\item[\(\Lambda\)]
Spatial decoherence coefficient in the schematic collisional
decoherence law
\[
\rho(x,x',t)
=
\rho(x,x',0)e^{-\Lambda t(x-x')^2}.
\]

\item[\(\Gamma\)]
Collision or interaction rate used in environmental diffusion
estimates.

\item[\(q\)]
Characteristic momentum transfer in a collision. In this usage \(q\)
is dimensional and should not be confused with the dimensionless
coordinate \(q=a/L\) used in Chapter~3.

\item[\(\mathcal D(\rho)\)]
Dissipative or decohering term in a reduced open-system master
equation.

\item[\(q_j,p_j,m_j,\omega_j,c_j\)]
Bath coordinate, bath momentum, oscillator mass, oscillator frequency,
and coupling constant, respectively, in the Caldeira--Leggett
Hamiltonian.

\item[\(A\)]
Observable continuously monitored in the representative
continuous-measurement equation.

\item[\(k\)]
Measurement strength in the representative continuous-measurement
equation. This use is local to Chapter~10 and is distinct from the
dimensionless momentum coordinate \(k=p/P\) used in Chapter~3.

\item[\(dW_t\)]
Wiener increment in a stochastic master or stochastic
Schr\"odinger equation.

\item[\(\lambda\)]
Collapse-rate parameter in the representative continuous spontaneous
localization equation. This use is distinct from the hidden variable
\(\lambda\) used in the Bell discussion.

\item[\(\widehat N(x)\)]
Smeared mass-density operator in the representative continuous
spontaneous localization equation.

\item[\(dW_t(x)\)]
Classical noise field in the representative continuous spontaneous
localization equation.

\item[\(\widehat R\)]
Quantum center-of-mass operator,
\[
\widehat R=\frac{1}{M}\sum_{i=1}^N m_i\widehat x_i.
\]

\item[\(\Delta R\)]
Quantum uncertainty of the center-of-mass coordinate.

\item[\(Q_k(t)\)]
Actual position of the \(k\)-th particle in Bohmian mechanics.

\item[\(q_k\)]
Configuration-space coordinate appearing as an argument of the
Bohmian wave function. It should not be confused with the dimensionless
coordinate \(q=a/L\) used in Chapter~3.

\item[\(c_\alpha\)]
Branch amplitudes in the schematic Everettian measurement state.

\item[\(|A_0\rangle,|A_\alpha\rangle\)]
Initial and outcome-correlated apparatus states in the Everettian
comparison.

\item[\(P_{\alpha_j}(t_j)\)]
Heisenberg-picture projector specifying the \(j\)-th alternative in a
consistent history.

\item[\(\alpha=(\alpha_1,\ldots,\alpha_n)\)]
Label for a complete consistent history.

\item[\(C_\alpha\)]
Class operator of the history \(\alpha\),
\[
C_\alpha
=
P_{\alpha_n}(t_n)\cdots P_{\alpha_1}(t_1).
\]

\item[\(p(\alpha)\)]
Weight assigned to a consistent history,
\[
p(\alpha)=\|C_\alpha\psi\|^2,
\]
when the required consistency condition is satisfied.

\end{description}

\chapter{Classical Submanifolds of Quantum State Space}
\label{chap:classical-submanifolds}

This chapter constructs the localized submanifolds of projective Hilbert
space that will represent classical configuration space and phase space.
We first introduce the Fubini--Study geometry of pure states and then
embed localized position and phase-space states into this geometry.
Finite-resolution equivalence classes are introduced to give these
submanifolds their operational meaning, and the construction is finally
extended to systems of particles.

\section{The Space of States and the Fubini--Study Metric}
\label{sec:the-space-of-states-and-the-fubini-study-metric}

Let \(\mathcal H\) be the Hilbert space of quantum states. For a single nonrelativistic particle in ordinary space
\(\mathbb R^3\), one may take the Hilbert space to be the space of
Lebesgue square-integrable functions on \(\mathbb R^3\),
\[
\mathcal H=L^2(\mathbb R^3).
\]
A normalized vector \(\varphi\in\mathcal H\),
\[
\|\varphi\|^2=(\varphi,\varphi)=1,
\]
represents a physically realizable pure quantum state. The set of all normalized vectors is the unit sphere in \(\mathcal H\),
\[
S^{\mathcal H}
=
\{\varphi\in\mathcal H:\|\varphi\|=1\}.
\]

Two vectors that differ only by a nonzero complex factor represent the same physical pure state. In particular, normalized vectors that differ only by a phase,
\[
\varphi \sim e^{i\theta}\varphi,
\]
represent the same point of projective Hilbert space. The space of physical pure states is therefore
\[
\mathbb P(\mathcal H).
\]
The natural projection from the unit sphere to projective Hilbert space is
\[
\pi:S^{\mathcal H}\longrightarrow \mathbb P(\mathcal H),
\qquad
\pi(\varphi)=[\varphi],
\]
where \([\varphi]\) denotes the ray represented by \(\varphi\).

The Hilbert-space inner product induces a Riemannian metric on the unit sphere \(S^{\mathcal H}\). If \(X,Y\) are tangent vectors to \(S^{\mathcal H}\) at \(\varphi\), represented by vectors \(\xi,\eta\in\mathcal H\), then
\[
G_\varphi(X,Y)=\operatorname{Re}(\xi,\eta).
\]
The tangent space to the unit sphere consists of vectors \(\xi\) satisfying
\[
\operatorname{Re}(\xi,\varphi)=0.
\]
Among these tangent vectors there is a distinguished vertical direction,
\[
i\varphi,
\]
which corresponds to changing only the phase of \(\varphi\). This direction does not change the physical pure state \([\varphi]\). The physically relevant directions are the horizontal directions, orthogonal to \(i\varphi\). Thus a horizontal tangent vector \(\xi\) satisfies
\[
(\xi,\varphi)=0.
\]

The Fubini--Study metric on \(\mathbb P(\mathcal H)\) is obtained by restricting the Hilbert metric on \(S^{\mathcal H}\) to horizontal directions and projecting them to \(\mathbb P(\mathcal H)\). Equivalently, for normalized \(\varphi\), its infinitesimal form is
\[
ds^2
=
(d\varphi,d\varphi)
-
|(\varphi,d\varphi)|^2.
\]
This formula removes the component of \(d\varphi\) along the ray and keeps only the part that changes the physical state.
The corresponding Fubini--Study distance between two pure states \([\varphi]\) and \([\psi]\) is
\[
\rho_{\rm FS}([\varphi],[\psi])
=
\arccos |(\varphi,\psi)|,
\]
for normalized representatives \(\varphi\) and \(\psi\). Thus the distance depends only on the absolute value of the inner product and is independent of the phases chosen for the representatives.

This metric is central to what follows. It is the metric with respect to which localized submanifolds of quantum states will be studied. The goal is to show that certain finite-dimensional submanifolds of \(\mathbb P(\mathcal H)\) inherit familiar classical geometric and physical properties.

\section{Localized States and the Manifold \texorpdfstring{$M_3^\sigma$}{M3 sigma}}
\label{sec:localized-states-and-the-manifold-m3-sigma}

Let \(\sigma>0\) be fixed. We first construct \(M_3^\sigma\) using
Gaussian states; the construction will shortly be extended to arbitrary
sufficiently narrow states.
For each \(a\in\mathbb R^3\), define the normalized Gaussian state
\[
\varphi_a(x)
=
\left(\frac{1}{2\pi\sigma^2}\right)^{3/4}
\exp\left(
-\frac{|x-a|^2}{4\sigma^2}
\right),
\qquad x\in\mathbb R^3.
\]
This state is localized near the point \(a\). The parameter \(\sigma\) controls the width of the localization. Smaller \(\sigma\) corresponds to stronger localization; larger \(\sigma\) corresponds to weaker localization.

The map
\[
\omega:\mathbb R^3\longrightarrow L^2(\mathbb R^3),
\qquad
\omega(a)=\varphi_a,
\]
parametrizes a smooth three-dimensional submanifold
\[
M_3^\sigma
=
\{\varphi_a:a\in\mathbb R^3\}
\subset L^2(\mathbb R^3).
\]
The parameter \(a\) serves as a coordinate on \(M_3^\sigma\) and gives
the center of localization of the corresponding state.

The tangent vectors to \(M_3^\sigma\) are obtained by differentiating \(\varphi_a\) with respect to the coordinates of \(a\). If \(a=(a^1,a^2,a^3)\), then
\[
\frac{\partial \varphi_a}{\partial a^\alpha}
=
\frac{x^\alpha-a^\alpha}{2\sigma^2}\,\varphi_a,
\qquad
\alpha=1,2,3.
\]
These three vectors span the tangent space to \(M_3^\sigma\) at \(\varphi_a\). Therefore a curve \(a=a(t)\) in ordinary space determines a curve
\[
\varphi(t)=\varphi_{a(t)}
\]
in the Hilbert space, with velocity
\[
\frac{d\varphi}{dt}
=
\frac{\partial \varphi_a}{\partial a^\alpha}
\frac{da^\alpha}{dt}.
\]
Here and below, summation over repeated indices is understood.

The Hilbert-space inner product induces a metric on \(M_3^\sigma\). The metric components are
\[
g_{\alpha\beta}
=
\operatorname{Re}
\left(
\frac{\partial \varphi_a}{\partial a^\alpha},
\frac{\partial \varphi_a}{\partial a^\beta}
\right).
\]
Since the states \(\varphi_a\) are real and normalized Gaussians, this becomes
\[
g_{\alpha\beta}
=
\int_{\mathbb R^3}
\frac{x^\alpha-a^\alpha}{2\sigma^2}
\frac{x^\beta-a^\beta}{2\sigma^2}
|\varphi_a(x)|^2\,d^3x.
\]
Integration yields
\[
\int_{\mathbb R^3}
(x^\alpha-a^\alpha)(x^\beta-a^\beta)
|\varphi_a(x)|^2\,d^3x
=
\sigma^2\delta_{\alpha\beta}.
\]
Hence
\[
g_{\alpha\beta}
=
\frac{1}{4\sigma^2}\delta_{\alpha\beta}.
\]
Therefore the induced line element on \(M_3^\sigma\) is
\[
ds^2
=
\frac{1}{4\sigma^2}
\delta_{\alpha\beta}\,da^\alpha da^\beta.
\]
Equivalently,
\[
ds^2
=
\frac{|da|^2}{4\sigma^2}.
\]
Thus \(M_3^\sigma\), with the metric induced from the Hilbert space, is Euclidean up to the constant scale factor \(1/(4\sigma^2)\). If ordinary distances in \(\mathbb R^3\) are measured in units of \(2\sigma\), then the embedding \(a\mapsto \varphi_a\) is isometric.

There is another property of this family of states that will play an important role in what follows. Although \(M_3^\sigma\) is only a three-dimensional submanifold of the Hilbert space, the set of vectors \(\{\varphi_a:a\in\mathbb R^3\}\) is complete, or total, in \(L^2(\mathbb R^3)\). That is, its linear span is dense in \(L^2(\mathbb R^3)\).
Indeed, suppose that a state \(\psi\in L^2(\mathbb R^3)\) is orthogonal to every \(\varphi_a\). Then
\[
(\psi,\varphi_a)=0
\qquad
\text{for all }a\in\mathbb R^3.
\]
The function \(a\mapsto(\psi,\varphi_a)\) is, up to the usual convention for complex conjugation, the convolution of \(\psi\) with a Gaussian. Taking the Fourier transform, this condition implies that the product of \(\widehat\psi(k)\) with the Fourier transform of the Gaussian is zero. Since the Fourier transform of a Gaussian is nowhere zero, it follows that
\[
\widehat\psi(k)=0
\]
almost everywhere, and hence \(\psi=0\). Therefore no nonzero vector is orthogonal to all vectors in \(M_3^\sigma\), which means that the family \(\{\varphi_a\}\) is complete.

The same construction may be passed to projective Hilbert space. The corresponding projective submanifold is formed by the rays
\[
\{[\varphi_a]:a\in\mathbb R^3\}
\subset \mathbb P(\mathcal H).
\]
For the real positive Gaussian representatives used here, the tangent vectors are horizontal. Therefore the Fubini--Study metric on the projective submanifold agrees with the metric induced from the Hilbert space on \(M_3^\sigma\).
Since from now on we shall work with the projective version of the manifold, we shall use the same symbol \(M_3^\sigma\) to denote it. Thus the induced metric is
\[
ds^2
=
\frac{|da|^2}{4\sigma^2}.
\]

In later sections we shall replace individual Gaussian representatives by equivalence classes of states localized near the same point. The Gaussian states are convenient representatives, but the geometric meaning is more general: a point of classical space corresponds to a class of quantum states that cannot be distinguished by position measurements at resolution \(\sigma\).

\section{The Phase-Space Manifold \texorpdfstring{$M_{3,3}^\sigma$}{M33 sigma}}
\label{sec:the-phase-space-manifold-m33-sigma}

Let \(a,p\in\mathbb R^3\). To the localized state centered at \(a\),
we associate the state
\begin{equation}
\varphi_{a,p}(x)
=
\left(\frac{1}{2\pi\sigma^2}\right)^{3/4}
\exp\left(
-\frac{|x-a|^2}{4\sigma^2}
\right)
\exp\left(
\frac{i}{\hbar}p\cdot x
\right).
\label{eq:localized-phase-state}
\end{equation}
The first factor localizes the state near the point \(a\). The second factor is a momentum phase. Thus \(a\) labels the position of the localized state, while \(p\) labels its average momentum.
The corresponding projective state is the ray
\[
[\varphi_{a,p}]\in \mathbb P(\mathcal H).
\]

The phase-space manifold is defined by the embedding map
\[
\Omega:\mathbb R^3\times\mathbb R^3\longrightarrow \mathbb P(\mathcal H),
\qquad
\Omega(a,p)=[\varphi_{a,p}].
\]
Its image,
\[
M_{3,3}^\sigma
=
\{[\varphi_{a,p}]:a,p\in\mathbb R^3\}
\subset \mathbb P(\mathcal H),
\]
is a six-dimensional submanifold of projective Hilbert space. The parameter \(a\) gives the center of localization, while \(p\) gives the momentum label,
\[
(a,p)\in \mathbb R^3\times \mathbb R^3.
\]

The tangent vectors to \(M_{3,3}^\sigma\) are obtained by differentiating \(\varphi_{a,p}\) with respect to \(a\) and \(p\). We have
\[
\frac{\partial \varphi_{a,p}}{\partial a^\alpha}
=
\frac{x^\alpha-a^\alpha}{2\sigma^2}\varphi_{a,p},
\]
and
\[
\frac{\partial \varphi_{a,p}}{\partial p^\alpha}
=
\frac{i x^\alpha}{\hbar}\varphi_{a,p}.
\]
The second expression contains a vertical part, because multiplication by a phase changes the representative vector as well as the ray. In projective Hilbert space this vertical part must be removed. Equivalently, one uses the Fubini--Study metric
\[
ds^2
=
(d\varphi,d\varphi)
-
|(\varphi,d\varphi)|^2.
\]
This subtraction removes the component of \(d\varphi\) tangent to the phase fibre.

For variations in \(a\), the tangent vectors are already horizontal. For variations in \(p\), the horizontal part is obtained by subtracting the expectation value of \(x^\alpha\). Since
\[
\left( x^\alpha\right)=a^\alpha,
\]
the horizontal momentum tangent vector is represented by
\[
\left(\frac{\partial \varphi_{a,p}}{\partial p^\alpha}\right)_{\mathrm{hor}}
=
\frac{i}{\hbar}(x^\alpha-a^\alpha)\varphi_{a,p}.
\]
Thus the position and momentum tangent directions are represented by
\[
u_\alpha
=
\frac{x^\alpha-a^\alpha}{2\sigma^2}\varphi_{a,p},
\qquad
v_\alpha
=
\frac{i}{\hbar}(x^\alpha-a^\alpha)\varphi_{a,p}.
\]

The induced metric is obtained by taking real parts of the Hilbert inner products of these horizontal tangent vectors. Since
\[
\int_{\mathbb R^3}
(x^\alpha-a^\alpha)(x^\beta-a^\beta)
|\varphi_{a,p}(x)|^2\,d^3x
=
\sigma^2\delta_{\alpha\beta},
\]
we get
\[
(u_\alpha,u_\beta)
=
\frac{1}{4\sigma^2}\delta_{\alpha\beta},
\]
and
\[
(v_\alpha,v_\beta)
=
\frac{\sigma^2}{\hbar^2}\delta_{\alpha\beta}.
\]
Moreover,
\[
\operatorname{Re}(u_\alpha,v_\beta)=0.
\]
Therefore the induced Fubini--Study metric on \(M_{3,3}^\sigma\) is
\[
ds^2
=
\frac{1}{4\sigma^2}\delta_{\alpha\beta}\,da^\alpha da^\beta
+
\frac{\sigma^2}{\hbar^2}\delta_{\alpha\beta}\,dp^\alpha dp^\beta.
\]
Equivalently,
\[
ds^2
=
\frac{|da|^2}{4\sigma^2}
+
\frac{\sigma^2 |dp|^2}{\hbar^2}.
\]
To make \(M_{3,3}^\sigma\) isometric to the Euclidean space
\(\mathbb R^3\times\mathbb R^3\), the position directions must be measured in units of \(2\sigma\), while the momentum directions must be measured in units of \(\hbar/\sigma\). Thus the localization scale \(\sigma\) determines how ordinary position and momentum coordinates are represented geometrically in state space.

The same embedding also carries the symplectic structure. Projective Hilbert space is not only Riemannian; it is K\"ahler. Its Fubini--Study metric is accompanied by the Fubini--Study symplectic form. When this form is restricted to \(M_{3,3}^\sigma\), one obtains, up to the conventional normalization, the canonical phase-space form
\[
\omega_{\mathrm{cl}}
=
\sum_{\alpha=1}^3 da^\alpha\wedge dp_\alpha .
\]
Thus \(M_{3,3}^\sigma\) inherits not only the Euclidean metric structure of phase space, but also its symplectic structure.

The manifold \(M_{3,3}^\sigma\) is therefore a quantum realization of classical phase space. It is not introduced as an independent classical arena. It is a submanifold of projective Hilbert space whose coordinates are the classical variables \(a\) and \(p\), and whose induced metric contains the natural position and momentum scales associated with localized quantum states. We shall see below that the same is true dynamically.

\section{Manifolds of Localized Equivalence Classes}
\label{sec:localized-equivalence-classes}

The manifolds \(M_3^\sigma\) and \(M_{3,3}^\sigma\) were defined using
Gaussian states. These states provide convenient representatives.
However, the physical meaning of localization should not depend on the
special Gaussian form. What matters is the resolution with which
position is tested. We therefore introduce equivalence classes of
localized states.

Let \(\psi\in L^2(\mathbb R^3)\) be normalized. Define its position
expectation by
\[
\mu_x(\psi)
=
\int_{\mathbb R^3}x\,|\psi(x)|^2\,d^3x .
\]
Define its position spread by
\[
\delta_x(\psi)
=
\sup_{\substack{n\in\mathbb R^3\\ |n|=1}}
\left(
\int_{\mathbb R^3}
|n\cdot(x-\mu_x(\psi))|^2
|\psi(x)|^2\,d^3x
\right)^{1/2}.
\]
We shall say that \(\psi\) is localized at \(a\in\mathbb R^3\) with
resolution \(\sigma\) if
\[
\mu_x(\psi)=a,
\qquad
\delta_x(\psi)\leq\sigma .
\]
The localized equivalence class associated with \(a\) is then defined by
\[
\{\varphi_a\}
=
\left\{
[\psi]\in\mathbb P(\mathcal H):
\mu_x(\psi)=a,\quad
\delta_x(\psi)\leq\sigma
\right\}.
\]
The parameter \(\sigma\) is therefore a localization scale built into
the definition of the quotient. Physically, the quotient retains the
center of localization while suppressing distinctions among states
whose position spreads are below this scale.
In a relativistic theory one would also expect a lower limit on
admissible localization: arbitrarily small position spread entails
arbitrarily large momentum and energy scales and eventually takes the
state outside the one-particle regime. No such lower cutoff is needed
here, since the present construction is explicitly nonrelativistic.
Accordingly, we retain only the upper localization bound
\(\delta_x(\psi)\leq\sigma\).

Consider the set of localized position classes
\[
\widetilde M_3^\sigma
=
\{\{\varphi_a\}:a\in\mathbb R^3\}.
\]
We define a distance on this set by using the Fubini--Study distance
between classes. For two classes \(\{\varphi_a\}\) and
\(\{\varphi_b\}\), set
\[
\rho(\{\varphi_a\},\{\varphi_b\})
=
\inf_{\psi\in\{\varphi_a\},\,\chi\in\{\varphi_b\}}
\rho_{\rm FS}(\psi,\chi).
\]

We now determine this distance to leading order for neighboring
classes. Let \(\psi(t)\) be a differentiable curve of normalized states
through the localized sector, with
\[
a(t)
=
\langle\psi(t),\widehat x\,\psi(t)\rangle .
\]
Choose a horizontal representative, so that
\[
(\psi,\dot\psi)=0.
\]
Then
\[
\dot a
=
2\operatorname{Re}
\bigl(\dot\psi,(\widehat x-a)\psi\bigr).
\]
For a unit vector \(n\), the Cauchy--Schwarz inequality gives
\[
|n\cdot\dot a|
\leq
2\|\dot\psi\|\,
\|n\cdot(\widehat x-a)\psi\|.
\]
By the definition of the position spread,
\[
\|n\cdot(\widehat x-a)\psi\|
\leq\sigma.
\]
Taking \(n\) in the direction of \(\dot a\) therefore gives
\[
\|\dot\psi\|^2
\geq
\frac{|\dot a|^2}{4\sigma^2}.
\]
For a horizontal curve, the Fubini--Study line element is
\[
ds^2=\|\dot\psi\|^2dt^2,
\]
and hence
\[
ds^2
\geq
\frac{|da|^2}{4\sigma^2}.
\]

The bound is attained by the Gaussian representatives. Indeed,
\[
\frac{\partial g_{a,\sigma}}{\partial a^\alpha}
=
\frac{x^\alpha-a^\alpha}{2\sigma^2}\,
g_{a,\sigma},
\]
and therefore
\[
\left\|
\frac{\partial g_{a,\sigma}}{\partial a^\alpha}
\right\|^2
=
\frac{1}{4\sigma^2}.
\]
Thus
\[
\rho(\{\varphi_a\},\{\varphi_{a+da}\})^2
=
\frac{|da|^2}{4\sigma^2}
+
o(|da|^2),
\]
so that the induced line element on \(\widetilde M_3^\sigma\) is
\[
ds^2
=
\frac{|da|^2}{4\sigma^2}.
\]

With this metric, \(\widetilde M_3^\sigma\) becomes a
three-dimensional manifold parametrized by \(a\in\mathbb R^3\).
Moreover, the correspondence
\[
\varphi_a\longmapsto\{\varphi_a\}
\]
is an isometry from \(M_3^\sigma\) onto
\(\widetilde M_3^\sigma\). Thus \(M_3^\sigma\) may be regarded as a
distinguished Gaussian realization of the manifold of localized
position classes.

The definition of \(\{\varphi_a\}\) involves only the position
expectation value and the position spread. It does not fix momentum.
Consequently, states with different average momenta may belong to the
same localized position class, provided they have the same position
expectation value and satisfy the same localization bound. This is why
\(\widetilde M_3^\sigma\) contains no separate momentum coordinate. It
represents only localized position classes.

Let \(\{g_a\}\) be the subset of \(\{\varphi_a\}\) consisting of
rays represented by real-valued functions. The phase-space
manifold of localized classes is defined by
\[
\widetilde M_{3,3}^\sigma
=
\left\{
\{g_a\}e^{\frac{ip\cdot x}{\hbar}}:
a,p\in\mathbb R^3
\right\}.
\]
Thus \(\widetilde M_3^\sigma\) records only localization in position,
while \(\widetilde M_{3,3}^\sigma\) separates the momentum degree of
freedom by attaching the phase factor \(e^{\frac{ip\cdot x}{\hbar}}\) to real
localized representatives.

The use of a phase linear in \(x\) is sufficient locally. Indeed, let
a state have a differentiable phase \(\Theta(x)\). In a neighborhood
of \(a\),
\[
\Theta(x)
=
\Theta(a)
+
\nabla\Theta(a)\cdot(x-a)
+
O(|x-a|^2).
\]
The constant term \(\Theta(a)\) contributes only an overall phase and
therefore does not change the projective state. Hence, over the
localization region and to first order,
\[
e^{i\Theta(x)}
\sim
e^{i\nabla\Theta(a)\cdot(x-a)}.
\]
Up to an irrelevant overall phase, this may be written as
\[
e^{\frac{ip\cdot x}{\hbar}},
\qquad
p=\hbar\nabla\Theta(a).
\]
Thus, for a state localized near \(a\), the relevant part of the phase is determined, to first order over the localization region, by
its gradient at \(a\). This justifies the use of phase factors whose
phase is linear in \(x\) in the definition of the phase-space manifold.

To obtain the metric on $\widetilde M^\sigma_{3,3}$, note that a
momentum measurement can be converted into a position measurement by a
unitary momentum-to-position transformation. This reflects the general
structure of measurement experiments, in which the final outcome is
registered through the position of a pointer or detector response. In
the idealized case, the transformation is described by a Fourier
transform. Experimentally, it is realized, for example, by a magnetic
spectrometer, in which different incoming momenta are registered at
different positions on a detector.

Let \(U_\sigma\) denote the Fourier transformation followed by a
rescaling of the transformed coordinate. We write the calibrated
detector coordinate as
\[
y
=
\frac{2\sigma^2}{\hbar}\,q,
\]
where \(q\) is the momentum variable in the Fourier representation.
The normalization is chosen so that the Gaussian of position spread
\(\sigma\) is transformed into a Gaussian of the same spread
\(\sigma\) in the coordinate \(y\). This choice fixes the scale of the
detector coordinate; the argument below does not otherwise restrict
the detected states to Gaussian form.

On the detector screen define localized classes exactly as before:
\[
\{\chi_y\}
=
\left\{
[\Phi]\in\mathbb P(\mathcal H):
\mu_y(\Phi)=y,\quad
\delta_y(\Phi)\leq\sigma
\right\},
\]
where \(\mu_y\) and \(\delta_y\) are the position expectation and
position spread in the calibrated detector coordinate.
Notice that the coordinate \(x\) is the position registered in a position measurement, whereas \(y\) is the position registered on the output screen of the momentum-measuring spectrometer. They therefore do not represent the position and momentum coordinates of a single state to which an uncertainty relation could be applied.

We now pull these detector classes back through the unitary
transformation \(U_\sigma\). Define the momentum class associated with
\(p\) by
\[
\{\pi_p\}
=
U_\sigma^{-1}
\left\{
\chi_{\frac{2\sigma^2}{\hbar}p}
\right\}.
\]
Since unitary transformations preserve the Fubini--Study metric, the
preceding argument for localized position classes applies
unchanged. With
\[
dy
=
\frac{2\sigma^2}{\hbar}\,dp,
\]
it gives
\[
ds_p^2
\geq
\frac{|dy|^2}{4\sigma^2}
=
\frac{\sigma^2}{\hbar^2}|dp|^2.
\]

The bound is attained by the Gaussian phase-space representatives
\[
g_{a,\sigma}(x)e^{\frac{ip\cdot x}{\hbar}}.
\]
Indeed, under \(U_\sigma\) these states become Gaussians of position
spread \(\sigma\), centered at
\[
y_p
=
\frac{2\sigma^2}{\hbar}\,p,
\]
up to a common linear phase determined by \(a\). Thus neighboring
states with momenta \(p\) and \(p+dp\) attain the lower bound. Hence
\[
ds_p^2
=
\frac{\sigma^2}{\hbar^2}|dp|^2.
\]

Thus the position and momentum coordinates have parallel
finite-resolution definitions. The position coordinate is obtained by
direct position detection, while the momentum coordinate is obtained
by first transforming momentum into a position on a detector and then
applying the same position-equivalence-class construction.

The Gaussian phase-space states
\[
\varphi_{a,p}(x)
=
g_{a,\sigma}(x)e^{\frac{ip\cdot x}{\hbar}}
\]
provide a common Hilbert-space realization of these two coordinates.
For every representative of a class in
\(\widetilde M_{3,3}^\sigma\), the position and momentum tangent
directions are Fubini--Study orthogonal. Indeed, such a representative
has the form
\[
\psi_{a,p}(x)
=
f_a(x)e^{\frac{ip\cdot x}{\hbar}},
\qquad
f_a(x)\in\mathbb R.
\]
A variation of the real amplitude \(f_a\) gives, after removal of the
common phase factor, a real tangent direction, whereas a variation of
\(p\) gives, after projection orthogonal to the ray,
\[
\frac{i}{\hbar}\,
dp\cdot(x-a)\,
f_a(x)e^{\frac{ip\cdot x}{\hbar}},
\]
which is purely imaginary after removal of the same phase factor.
The two tangent directions are therefore Fubini--Study orthogonal.

The position and momentum contributions therefore combine to give
\[
ds^2
=
\frac{|da|^2}{4\sigma^2}
+
\frac{\sigma^2 |dp|^2}{\hbar^2}.
\]
This is precisely the metric induced on the Gaussian phase-space
manifold \(M_{3,3}^\sigma\). Thus the correspondence
\[
\varphi_{a,p}
\longmapsto
\{\varphi_{a,p}\}
\]
is an isometry from \(M_{3,3}^\sigma\) onto
\(\widetilde M_{3,3}^\sigma\), while the coordinates \(a\) and \(p\)
retain their finite-resolution operational meaning.

The Gaussian realization is also closely related to standard
completeness results in time--frequency analysis.
The continuous family of position translates
\[
\{g_{a,\sigma}:a\in\mathbb R^3\}
\]
is complete in \(L^2(\mathbb R^3)\). Likewise, for fixed \(a\), the
family of momentum modulations
\[
\left\{
g_{a,\sigma}e^{\frac{ip\cdot x}{\hbar}}:
p\in\mathbb R^3
\right\}
\]
is complete. Moreover, an appropriate discretization of the combined phase-space family
\[
\left\{
g_{a,\sigma}e^{\frac{ip\cdot x}{\hbar}}
\right\}
\]
on a sufficiently dense phase-space lattice gives the standard Gaussian
Gabor system; under the usual density condition it forms a Gabor frame
and is therefore complete. In one dimension the corresponding lattice
cell has area of order \(\hbar\); the precise sampling condition will be
used later in the construction of the discretized lift of Brownian
motion to random unitary motion in Hilbert space.

The symplectic structure on
$\widetilde M_{3,3}^\sigma$ can now be introduced in the same way as
for the Gaussian phase-space manifold $M_{3,3}^\sigma$. Under the
isometry
\[
\varphi_{a,p}\longmapsto\{\varphi_{a,p}\},
\]
we transfer the restricted Fubini--Study symplectic form to
$\widetilde M_{3,3}^\sigma$. Up to the same conventional
normalization as before, it is
\[
\omega_{\mathrm{cl}}
=
\sum_{\alpha=1}^3
da^\alpha\wedge dp_\alpha .
\]
Thus \(\widetilde M_{3,3}^\sigma\) carries the same canonical
symplectic structure as \(M_{3,3}^\sigma\).

\section{Systems of Particles}
\label{sec:systems-of-particles}

The preceding constructions extend directly to systems of many particles. For \(N\) distinguishable particles the Hilbert space is
\[
\mathcal H_N
=
\mathcal H_1\otimes\cdots\otimes\mathcal H_N
\simeq
L^2(\mathbb R^{3N}),
\]
where each factor is a copy of \(L^2(\mathbb R^3)\). We shall first consider product states. Entangled states belong to the same Hilbert space, but they do not, in general, lie on the classical product manifolds constructed below.

Let
\[
a=(a_1,\ldots,a_N),
\qquad
a_j\in\mathbb R^3 .
\]
For each particle define the localized Gaussian state \(\varphi_{a_j}\) as before. The corresponding \(N\)-particle localized state is
\[
\Phi_a
=
\varphi_{a_1}\otimes\cdots\otimes\varphi_{a_N}.
\]
Equivalently, as a function of \(x=(x_1,\ldots,x_N)\in\mathbb R^{3N}\),
\[
\Phi_a(x_1,\ldots,x_N)
=
\prod_{j=1}^N
\varphi_{a_j}(x_j).
\]

The map
\[
\Omega_N:\mathbb R^{3N}\longrightarrow \mathbb P(\mathcal H_N),
\qquad
\Omega_N(a)=[\Phi_a],
\]
defines the \(N\)-particle configuration manifold
\[
M_{3N}^\sigma
=
\{[\Phi_a]:a\in\mathbb R^{3N}\}
\subset \mathbb P(\mathcal H_N).
\]
Thus \(M_{3N}^\sigma\) is a \(3N\)-dimensional submanifold of projective Hilbert space. It is parametrized by the same coordinates that parametrize the classical configuration space of \(N\) particles.


Equivalently, \(M_{3N}^\sigma\) is the image of the Segre embedding
restricted to the one-particle localized manifolds,
\[
M_3^\sigma\times\cdots\times M_3^\sigma
\longrightarrow
\mathbb P(\mathcal H_1\otimes\cdots\otimes\mathcal H_N),
\]
\[
([\varphi_{a_1}],\ldots,[\varphi_{a_N}])
\longmapsto
[\varphi_{a_1}\otimes\cdots\otimes\varphi_{a_N}].
\]
Thus \(M_{3N}^\sigma\) is diffeomorphic to the Cartesian product
\[
M_3^\sigma\times\cdots\times M_3^\sigma,
\]
while its points are represented in the composite Hilbert space by
tensor-product states.
When it is useful to emphasize this embedding into the tensor-product
Hilbert space, we shall also denote the image product-state manifold by
\[
M_3^\sigma\otimes\cdots\otimes M_3^\sigma.
\]

The pullback of the Fubini--Study metric under this embedding is the
product metric. To see this, let
\[
\Phi
=
\varphi_1\otimes\cdots\otimes\varphi_N
\]
be a normalized product state, and let \(d\varphi_j\) be horizontal
variations of the individual factors. Then
\[
d\Phi
=
\sum_{j=1}^N
\varphi_1\otimes\cdots\otimes d\varphi_j
\otimes\cdots\otimes\varphi_N.
\]
For \(j\neq k\), the corresponding terms are orthogonal because
\[
(\varphi_j,d\varphi_j)=0
\]
for horizontal variations. Hence all mixed terms vanish and
\[
\|d\Phi\|^2
=
\sum_{j=1}^N
\|d\varphi_j\|^2.
\]
Therefore the pullback of the Fubini--Study line element is the sum of
the one-particle line elements. Since
\[
ds_j^2
=
\frac{|da_j|^2}{4\sigma^2},
\]
we obtain
\[
ds^2
=
\sum_{j=1}^N
\frac{|da_j|^2}{4\sigma^2}.
\]
Thus \(M_{3N}^\sigma\), with the induced Fubini--Study metric, is
Euclidean up to the same scale factor as in the one-particle case. If
each position coordinate is measured in units of \(2\sigma\), then
\(M_{3N}^\sigma\) is isometric to the classical configuration space
\(\mathbb R^{3N}\).


We now include momentum. Let
\[
p=(p_1,\ldots,p_N),
\qquad
p_j\in\mathbb R^3.
\]
Define
\[
\varphi_{a_j,p_j}(x_j)
=
\varphi_{a_j}(x_j)
\exp\left(\frac{i}{\hbar}p_j\cdot x_j\right),
\]
and set
\[
\Phi_{a,p}
=
\varphi_{a_1,p_1}\otimes\cdots\otimes\varphi_{a_N,p_N}.
\]
Equivalently,
\[
\Phi_{a,p}(x_1,\ldots,x_N)
=
\prod_{j=1}^N
\varphi_{a_j}(x_j)
\exp\left(
\frac{i}{\hbar}
\sum_{j=1}^N p_j\cdot x_j
\right).
\]

The many-particle phase-space manifold is
\[
M_{3N,3N}^\sigma
=
\{[\Phi_{a,p}]:a,p\in\mathbb R^{3N}\}
\subset \mathbb P(\mathcal H_N).
\]
It is the image of the corresponding Segre embedding of the one-particle
phase-space manifolds,
\[
M_{3,3}^\sigma\times\cdots\times M_{3,3}^\sigma
\longrightarrow
\mathbb P(\mathcal H_1\otimes\cdots\otimes\mathcal H_N),
\]
and is therefore diffeomorphic to
\[
M_{3,3}^\sigma\times\cdots\times M_{3,3}^\sigma.
\]

The same orthogonality argument shows that the pullback of the
Fubini--Study metric is again the product metric. Since the
one-particle metric is
\[
ds_j^2
=
\frac{|da_j|^2}{4\sigma^2}
+
\frac{\sigma^2 |dp_j|^2}{\hbar^2},
\]
the induced metric on \(M_{3N,3N}^\sigma\) is
\[
ds^2
=
\sum_{j=1}^N
\left(
\frac{|da_j|^2}{4\sigma^2}
+
\frac{\sigma^2 |dp_j|^2}{\hbar^2}
\right).
\]
Thus \(M_{3N,3N}^\sigma\) is the projective Hilbert-space realization
of the classical phase space of \(N\) particles. Position directions
are measured in units of \(2\sigma\), while momentum directions are
measured in units of \(\hbar/\sigma\).

The same construction applies to localized equivalence classes. Let
\[
\{\Phi_a\}
=
\{\varphi_{a_1}\}\otimes\cdots\otimes\{\varphi_{a_N}\}
\]
denote the class of product states whose \(j\)-th factor belongs to the
localized class \(\{\varphi_{a_j}\}\). Thus the tensor product of classes
is understood here as the set of product states obtained by choosing one
representative from each factor. Likewise, using the classes of
real-valued localized states \(g_{a_k}\), let
\[
\{G_a\}
=
\{g_{a_1}\}\otimes\cdots\otimes\{g_{a_N}\}.
\]
Consider the quotient configuration set
\[
\widetilde M_{3N}^\sigma
=
\{\{\Phi_a\}:a\in\mathbb R^{3N}\}.
\]
As in the one-particle case, the metric is defined by the Gaussian
realization. With this metric, \(\widetilde M_{3N}^\sigma\) becomes a
\(3N\)-dimensional manifold isometric to \(M_{3N}^\sigma\).

For phase space, use classes consisting of real-valued localized states
and attach the phase
\[
\exp\left(
\frac{i}{\hbar}
\sum_{j=1}^N p_j\cdot x_j
\right).
\]
This gives the quotient phase-space manifold
\[
\widetilde M_{3N,3N}^\sigma
=
\left\{
\{G_a\}
\exp\left(
\frac{i}{\hbar}
\sum_{j=1}^N p_j\cdot x_j
\right):
a,p\in\mathbb R^{3N}
\right\}.
\]
Its metric is transferred from the Gaussian manifold
\(M_{3N,3N}^\sigma\). Therefore
\[
ds^2
=
\sum_{j=1}^N
\left(
\frac{|da_j|^2}{4\sigma^2}
+
\frac{\sigma^2 |dp_j|^2}{\hbar^2}
\right).
\]
The symplectic structure is transferred in the same way. Hence, up to
the conventional normalization, the induced symplectic form is the
canonical many-particle phase-space form
\[
\omega_{\mathrm{cl}}
=
\sum_{j=1}^N\sum_{\alpha=1}^3
da_j^\alpha\wedge dp_{j,\alpha}.
\]

Thus the tensor-product construction embeds the Gaussian realizations of
the classical configuration space and phase space into the projective
Hilbert space of the composite system, while the corresponding quotient
manifolds provide their finite-resolution physical description.

\chapter{Newtonian Dynamics from Schr\"odinger Dynamics}
\label{chap:newtonian-dynamics-from-schrodinger-dynamics}

Having identified classical phase space with a localized submanifold of
projective Hilbert space, we now compare the dynamics induced on this
submanifold with ordinary classical mechanics. The Schr\"odinger vector
field is decomposed into tangent and normal components, and its tangent
part is shown to generate Newtonian motion. The same result is obtained
independently by restricting the Schr\"odinger action to localized
phase-space states.

\section{Schr\"odinger Velocity in Hilbert Space}
\label{sec:schrodinger-velocity-in-hilbert-space}

Let \(\mathcal H\) be the Hilbert space of states, and let \(\hat h\) be a self-adjoint Hamiltonian. The Hamiltonian defines a vector field on \(\mathcal H\) by
\[
\varphi
\longmapsto
-\frac{i}{\hbar}\hat h\varphi .
\]
Schr\"odinger's equation
\[
\frac{d\varphi}{dt}
=
-\frac{i}{\hbar}\hat h\varphi
\]
is the equation for the integral curves of this vector field.

Because \(\hat h\) is self-adjoint, Schr\"odinger evolution is unitary and therefore preserves normalization:
\[
\frac{d}{dt}(\varphi,\varphi)
=
\left(\frac{d\varphi}{dt},\varphi\right)
+
\left(\varphi,\frac{d\varphi}{dt}\right)
=
0.
\]
Hence the Schr\"odinger velocity
\[
X_{\hat h}(\varphi)
=
-\frac{i}{\hbar}\hat h\varphi
\]
is tangent to the unit sphere of states.

The physical state, however, is not the vector \(\varphi\) itself but the ray \([\varphi]\). Therefore the physically relevant velocity is the projection of \(X_{\hat h}(\varphi)\) to projective Hilbert space. The direction \(i\varphi\) is vertical; it changes only the phase of the representative vector and leaves the ray unchanged.
Let
\[
\bar h_\varphi
=
(\varphi,\widehat h\varphi)
\]
be the expectation value of the Hamiltonian in the normalized state
\(\varphi\). Then the vertical part of \(X_{\widehat h}(\varphi)\),
i.e., the part parallel to the fibre, is
\[
-\frac{i}{\hbar}\bar h_\varphi\,\varphi.
\]
The corresponding horizontal part is therefore
\[
-\frac{i}{\hbar}
(\widehat h-\bar h_\varphi)\varphi.
\]
This is the velocity of the physical state in projective Hilbert space, written in the horizontal lift to the unit sphere.
Thus the passage from Hilbert space to projective Hilbert space amounts to subtracting the expectation value of the Hamiltonian. The operator \(\hat h\) and the operator \(\hat h+cI\), where \(c\) is real, generate the same motion in projective Hilbert space. They differ only by an overall phase.

The Fubini--Study speed of the state is the norm of the horizontal velocity. Using
\[
ds^2
=
(d\varphi,d\varphi)
-
|(\varphi,d\varphi)|^2,
\]
we obtain
\[
\left(\frac{ds}{dt}\right)^2
=
\frac{1}{\hbar^2}
\left[
(\varphi,\hat h^2\varphi)
-
(\varphi,\hat h\varphi)^2
\right].
\]
Therefore
\[
\frac{ds}{dt}
=
\frac{\Delta_\varphi h}{\hbar},
\]
where
\[
(\Delta_\varphi h)^2
=
(\varphi,\hat h^2\varphi)
-
(\varphi,\hat h\varphi)^2
\]
is the energy variance in the state \(\varphi\). 
We conclude that the speed of the physical state is determined by the energy uncertainty, rather than by the energy itself.

\section{Components of Schr\"odinger Velocity Along
\texorpdfstring{\(M_{3,3}^\sigma\)}{M33 sigma}}
\label{sec:components-of-schrodinger-velocity-along-m33-sigma}

We use horizontal representatives, so all tangent vectors are understood as tangent vectors to projective Hilbert space. Let
\[
\varphi=\varphi_{a,p}
\]
be a point of \(M_{3,3}^\sigma\). 
The tangent space to \(M_{3,3}^\sigma\) at \(\varphi_{a,p}\) is spanned by the position and momentum tangent vectors
\[
u_\alpha
=
\frac{x^\alpha-a^\alpha}{2\sigma^2}\varphi_{a,p},
\qquad
v_\alpha
=
\frac{i}{\hbar}(x^\alpha-a^\alpha)\varphi_{a,p},
\qquad
\alpha=1,2,3 .
\]
These vectors satisfy
\[
(u_\alpha,u_\beta)
=
\frac{1}{4\sigma^2}\delta_{\alpha\beta},
\qquad
(v_\alpha,v_\beta)
=
\frac{\sigma^2}{\hbar^2}\delta_{\alpha\beta},
\qquad
\operatorname{Re}(u_\alpha,v_\beta)=0 .
\]

Let \(X_{\hat h}(\varphi)\) be the Schr\"odinger velocity at \(\varphi\),
\[
X_{\hat h}(\varphi)
=
-\frac{i}{\hbar}
(\hat h-\bar h_\varphi)\varphi,
\qquad
\bar h_\varphi=(\varphi,\hat h\varphi).
\]
We decompose this velocity into tangent and normal parts relative to \(M_{3,3}^\sigma\):
\[
X_{\hat h}
=
X_{\hat h}^{T}
+
X_{\hat h}^{N},
\]
where
\[
X_{\hat h}^{T}\in T_\varphi M_{3,3}^\sigma,
\qquad
X_{\hat h}^{N}\perp T_\varphi M_{3,3}^\sigma .
\]
The tangent part has the form
\[
X_{\hat h}^{T}
=
\dot a^\alpha u_\alpha
+
\dot p^\alpha v_\alpha .
\]

The coefficients are obtained by orthogonal projection. Since the metric in the \(u_\alpha\) and \(v_\alpha\) directions is diagonal, we have
\[
\dot a^\alpha
=
4\sigma^2\operatorname{Re}(X_{\hat h},u_\alpha),
\]
and
\[
\dot p^\alpha
=
\frac{\hbar^2}{\sigma^2}
\operatorname{Re}(X_{\hat h},v_\alpha).
\]
Thus the tangent component of Schr\"odinger velocity determines a vector field on the classical phase-space coordinates \((a,p)\). The normal component,
\[
X_{\hat h}^{N}
=
X_{\hat h}
-
X_{\hat h}^{T},
\]
measures the part of Schr\"odinger evolution that carries the state away from the phase-space manifold.

For the Hamiltonian
\[
\hat h
=
\frac{\widehat p^2}{2m}+\widehat V(x,t),
\qquad
\hat p=-i\hbar\nabla ,
\]
the tangent component of the Schr\"odinger velocity at \(\varphi_{a,p}\) is, up to terms of order \(O(\sigma^2)\),
\[
X_{\hat h}^T
=
\frac{p^\alpha}{m}u_\alpha
-
\nabla_\alpha V(a,t)v_\alpha .
\]
Consequently, the tangent component determines the phase-space velocity
\[
\dot a^\alpha=\frac{p^\alpha}{m},
\qquad
\dot p^\alpha
=
-\nabla_\alpha V(a,t).
\]

Using the induced metric on \(M_{3,3}^\sigma\), the tangent contribution to the squared speed is
\[
\|X_{\hat h}^T\|^2
=
\frac{|p|^2}{4m^2\sigma^2}
+
\frac{\sigma^2 |\nabla V(a,t)|^2}{\hbar^2}.
\]
For the Gaussian states on \(\mathbb R^3\) used here, the normal contribution is
\[
\|X_{\hat h}^N\|^2
=
\frac{3\hbar^2}{32m^2\sigma^4}.
\]
Thus the Schr\"odinger velocity along \(M_{3,3}^\sigma\) has a tangential part, which gives the classical phase-space velocity, and a normal part, which describes the tendency of the state to leave the Gaussian manifold by spreading. The action-functional formulation below will give a more invariant derivation of the tangential equations.

\section{The Action Functional on State Space}
\label{sec:the-action-functional-on-state-space}

We now show that classical dynamics can also be obtained by restricting the Schr\"odinger action to localized phase-space states. Importantly, this result does not depend on the Gaussian form of the localized states, but applies to arbitrary representatives of the equivalence classes forming the manifold \(\widetilde M_{3,3}^\sigma\).

Consider the action functional
\begin{equation}
\label{eq:schrodinger-action}
S[\varphi]
=
\int
\overline{\varphi}(x,t)
\left[
i\hbar\frac{\partial}{\partial t}
-
\widehat h
\right]
\varphi(x,t)\,d^3x\,dt,
\end{equation}
where
\begin{equation}
\label{eq:one-particle-hamiltonian}
\widehat h
=
-\frac{\hbar^2}{2m}\Delta
+
\widehat V(x,t).
\end{equation}
Variation of \eqref{eq:schrodinger-action} with respect to
\(\overline{\varphi}\) gives the Schr\"odinger equation.

Let \(r\) be a real-valued, normalized, sufficiently regular function on
\(\mathbb R^3\), centered at the origin, and define the squeezed translated state
\[
r_{a,\sigma}(x)
=
\frac{1}{\sigma^{3/2}}
r\left(\frac{x-a}{\sigma}\right).
\]
Thus
\[
\int_{\mathbb R^3}r_{a,\sigma}^2(x)\,d^3x=1,
\qquad
\int_{\mathbb R^3}x\,r_{a,\sigma}^2(x)\,d^3x=a.
\]
We now restrict the action to states of the form
\[
\varphi(x,t)
=
r_{a(t),\sigma}(x)
e^{\frac{ip(t)\cdot x}{\hbar}}.
\]
The Gaussian states used to define \(M_{3,3}^\sigma\) are a particular case of this construction.

Since \(r_{a,\sigma}\) is real and normalized,
\[
\int_{\mathbb R^3}
r_{a,\sigma}
\frac{\partial r_{a,\sigma}}{\partial t}\,d^3x
=
0.
\]
Therefore the time-derivative term is
\[
i\hbar
\int_{\mathbb R^3}
\overline{\varphi}
\frac{\partial\varphi}{\partial t}\,d^3x
=
-a\cdot\dot p .
\]
Using
\[
-a\cdot\dot p
=
p\cdot\dot a
-
\frac{d}{dt}(a\cdot p),
\]
we see that, up to a total derivative, the time-derivative term gives the canonical expression
\[
p\cdot\dot a .
\]

The kinetic term is
\[
\left(
\varphi,
-\frac{\hbar^2}{2m}\Delta\varphi
\right)
=
\frac{|p|^2}{2m}
+
C_\sigma,
\]
where
\[
C_\sigma
=
\frac{\hbar^2}{2m\sigma^2}
\int_{\mathbb R^3}
|\nabla r(y)|^2\,d^3y.
\]
The quantity \(C_\sigma\) depends on the shape and width of the localized state, but is independent of \(a\) and \(p\). It therefore does not affect the equations of motion for the phase-space coordinates.

The only approximation occurs in the potential term,
\[
\int_{\mathbb R^3}
V(x,t)r_{a,\sigma}^2(x)\,d^3x.
\]
As \(\sigma\to0\),
\[
r_{a,\sigma}^2
\longrightarrow
\delta^3(x-a).
\]
Hence, if \(V\) is continuous at \(a\),
\[
\int_{\mathbb R^3}
V(x,t)r_{a,\sigma}^2(x)\,d^3x
\longrightarrow
V(a,t).
\]

Thus, up to a total derivative, the additive constant \(C_\sigma\), and a potential error that vanishes as \(\sigma\to0\), the restricted Schr\"odinger action becomes
\begin{equation}
\label{eq:classical-action}
S
=
\int
\left[
p\cdot\frac{da}{dt}
-
h(p,a,t)
\right]dt,
\end{equation}
where
\begin{equation}
\label{eq:classical-hamiltonian}
h(p,a,t)
=
\frac{|p|^2}{2m}
+
V(a,t).
\end{equation}
This is the classical phase-space action.

If \(V\) is differentiable, variation of \eqref{eq:classical-action} gives Hamilton's equations,
\begin{equation}
\label{eq:hamilton-equations}
\frac{da}{dt}
=
\frac{p}{m},
\qquad
\frac{dp}{dt}
=
-\nabla V(a,t),
\end{equation}
which are the Newton equations.
Thus continuity of \(V\) is sufficient for the convergence of the restricted action to the classical action, while differentiability is needed to obtain the Newtonian equations of motion from the resulting variational principle.

The result agrees with the geometric decomposition of Section~\ref{sec:components-of-schrodinger-velocity-along-m33-sigma}. The tangent component of the Schr\"odinger velocity,
\[
X_{\hat h}
=
-\frac{i}{\hbar}\widehat h\varphi
\]
gives precisely the phase-space velocity determined by \eqref{eq:hamilton-equations}. 
Thus the same Newtonian dynamics appears in two equivalent forms: as the tangential part of the Schr\"odinger vector field and as the Euler--Lagrange dynamics of the Schr\"odinger action restricted to localized phase-space states.

The classical dynamics of a system of \(N\) distinguishable particles can be derived by the same construction. Let
\[
\mathcal H_N
=
\mathcal H_1\otimes\cdots\otimes\mathcal H_N
\simeq
L^2(\mathbb R^{3N}),
\]
and consider localized product states
\[
\Phi_{a,p}(x_1,\ldots,x_N)
=
\prod_{j=1}^N
r_{a_j,\sigma}(x_j)
\exp\left(
\frac{i}{\hbar}
\sum_{j=1}^N p_j\cdot x_j
\right).
\]
For the Hamiltonian
\[
\widehat H
=
\sum_{j=1}^N
\frac{\widehat p_j^{\,2}}{2m_j}
+
V(\widehat x_1,\ldots,\widehat x_N,t),
\]
the Schr\"odinger action restricted to these states becomes, up to an additive constant and a total derivative,
\[
S
=
\int
\left[
\sum_{j=1}^N
p_j\cdot\frac{da_j}{dt}
-
H(a,p,t)
\right]dt,
\]
where
\[
H(a,p,t)
=
\sum_{j=1}^N
\frac{|p_j|^2}{2m_j}
+
V(a_1,\ldots,a_N,t).
\]
Thus the reduction of the Schr\"odinger action to the classical action extends directly to systems of particles, including interacting systems.

\section{From Commutators to Poisson Brackets on \(M_{3,3}^\sigma\)}
\label{sec:from-commutators-to-poisson-brackets-on-m-3-3}

The special role of \(M_{3,3}^\sigma\) is seen not only in the relation between Schr\"odinger and Newtonian dynamics, but also in the relation between the canonical commutation relations and Poisson brackets.

For the Hamiltonian
\[
\widehat h
=
\frac{\widehat p^2}{2m}
+
\widehat V(x),
\]
the tangent component of the Schr\"odinger velocity on \(M_{3,3}^\sigma\) is
\[
X_{\widehat h}^T
=
\dot a^\alpha
\frac{\partial}{\partial a^\alpha}
+
\dot p_\alpha
\frac{\partial}{\partial p_\alpha}.
\]
On \(M_{3,3}^\sigma\), the phase-space coordinates are given by
\[
a^\alpha
=
(\varphi_{a,p},\widehat x^\alpha\varphi_{a,p}),
\qquad
p_\alpha
=
(\varphi_{a,p},\widehat p_\alpha\varphi_{a,p}).
\]
Differentiating along the Schr\"odinger flow,
\[
\dot\varphi
=
-\frac{i}{\hbar}\widehat h\varphi,
\]
we obtain
\[
\frac{da^\alpha}{dt}
=
(\dot\varphi,\widehat x^\alpha\varphi)
+
(\varphi,\widehat x^\alpha\dot\varphi)
=
\left(
\varphi,
\frac{i}{\hbar}
[\widehat h,\widehat x^\alpha]
\varphi
\right).
\]
Similarly,
\[
\frac{dp_\alpha}{dt}
=
\left(
\varphi,
\frac{i}{\hbar}
[\widehat h,\widehat p_\alpha]
\varphi
\right).
\]
But we have already seen that the tangent component of the Schr\"odinger velocity on \(M_{3,3}^\sigma\) is the classical Hamiltonian vector field. Therefore
\[
\frac{da^\alpha}{dt}
=
\{a^\alpha,h\}_P,
\qquad
\frac{dp_\alpha}{dt}
=
\{p_\alpha,h\}_P.
\]
Hence, on \(M_{3,3}^\sigma\),
\[
\left(
\varphi,
\frac{i}{\hbar}
[\widehat h,\widehat x^\alpha]
\varphi
\right)
=
\{a^\alpha,h\}_P,
\]
and
\[
\left(
\varphi,
\frac{i}{\hbar}
[\widehat h,\widehat p_\alpha]
\varphi
\right)
=
\{p_\alpha,h\}_P.
\]
Therefore
\[
X_{\widehat h}^T
=
\{a^\alpha,h\}_P
\frac{\partial}{\partial a^\alpha}
+
\{p_\alpha,h\}_P
\frac{\partial}{\partial p_\alpha}.
\]

More generally, for observables \(\widehat f\) and \(\widehat g\), with associated vector fields
\[
X_{\widehat f}(\varphi)
=
-\frac{i}{\hbar}\widehat f\varphi,
\qquad
X_{\widehat g}(\varphi)
=
-\frac{i}{\hbar}\widehat g\varphi,
\]
the symplectic form satisfies
\[
\omega
\left(
X_{\widehat f},
X_{\widehat g}
\right)
=
\frac{1}{\hbar}
\operatorname{Im}
\left(
\varphi,
[\widehat f,\widehat g]\varphi
\right).
\]
When restricted to \(M_{3,3}^\sigma\), this gives the corresponding classical Poisson bracket,
\[
\omega
\left(
X_{\widehat f},
X_{\widehat g}
\right)
=
\{f,g\}_P,
\]
with the normalization of the symplectic form used here.

Thus the commutator structure of quantum observables, when constrained to \(M_{3,3}^\sigma\), gives the Poisson-bracket structure of classical phase space.
In this precise sense, when Schr\"odinger dynamics is constrained to \(M_{3,3}^\sigma\), the commutator equations reduce to the Poisson-bracket equations of classical mechanics.

\section{The Unitary Lift of Hamiltonian Motion}
\label{sec:unitary-lift-hamiltonian-motion}

The preceding results showed how Schr\"odinger dynamics induces
Hamiltonian motion when restricted to the classical phase-space
manifold \(M_{3,3}^\sigma\). There is also a converse statement. Its
proper formulation involves not a single classical trajectory, but the
entire family of trajectories generated by one fixed classical
Hamiltonian function.

Let
\[
H_{\rm cl}(a,p)
\]
be a Hamiltonian function on the classical phase space of a particle,
and let
\[
\Phi_t(a,p)
\]
denote its Hamiltonian flow. As the initial point \((a,p)\) varies,
\(\Phi_t\) describes the complete family of classical motions generated
by the same Hamiltonian. We ask whether this entire family can arise as
the restriction of one common unitary evolution in the ambient Hilbert
space and, if so, whether that evolution is uniquely determined.

A self-adjoint operator \(\widehat H\) defines on projective Hilbert
space the Hamiltonian function
\[
h_{\widehat H}([\psi])
=
(\psi,\widehat H\psi),
\qquad
\|\psi\|=1,
\]
whose Hamiltonian vector field is the projective Schr\"odinger vector
field generated by \(\widehat H\). Restricting this function to
\(M_{3,3}^\sigma\) gives
\[
h_{\widehat H}(a,p)
=
\left(
\varphi_{a,p},
\widehat H \varphi_{a,p}
\right).
\]

As shown above, the symplectic structure induced on
\(M_{3,3}^\sigma\) is, in the corresponding phase-space coordinates,
the canonical symplectic structure. Therefore the tangent restriction
of the Schr\"odinger vector field is the Hamiltonian vector field
generated on \(M_{3,3}^\sigma\) by \(h_{\widehat H}(a,p)\).
Suppose that a unitary evolution reproduces the complete
classical flow generated by \(H_{\rm cl}\). Since this requirement
holds for every initial point of \(M_{3,3}^\sigma\), the two
Hamiltonian vector fields agree everywhere:
\[
X_{h_{\widehat H}}
=
X_{H_{\rm cl}}.
\]
By nondegeneracy of the symplectic form,
\[
d h_{\widehat H}
=
dH_{\rm cl},
\]
and hence, on each connected component of phase space,
\begin{equation}
\label{eq:classical-symbol-unitary-lift}
\left(
\varphi_{a,p},
\widehat H \varphi_{a,p}
\right)
=
H_{\rm cl}(a,p)+C,
\end{equation}
where \(C\) is independent of \(a\) and \(p\).

Equation~(\ref{eq:classical-symbol-unitary-lift}) already shows that any
two common unitary lifts can differ on \(M_{3,3}^\sigma\) only by an
additive constant in their Hamiltonian functions. The overcompleteness
of the coherent-state family then promotes this statement to uniqueness
of the ambient generator.
To see the uniqueness explicitly, suppose that two self-adjoint
operators
\[
\widehat H_1,
\qquad
\widehat H_2
\]
generate common unitary lifts of the same complete classical flow.
Their restrictions to \(M_{3,3}^\sigma\) satisfy
\[
X_{h_{\widehat H_1}}
=
X_{h_{\widehat H_2}}
=
X_{H_{\rm cl}}.
\]
Hence
\[
d\!\left(
h_{\widehat H_1}-h_{\widehat H_2}
\right)
=
0,
\]
and therefore, on connected phase space,
\[
\left(
\varphi_{a,p},
(\widehat H_1-\widehat H_2)
\varphi_{a,p}
\right)
=
C
\]
for all \(a,p\), where \(C\) is constant.

Define
\[
A
=
\widehat H_1-\widehat H_2-CI.
\]
Then
\begin{equation}
\label{eq:vanishing-coherent-symbol}
\left(
\varphi_{a,p},
A \varphi_{a,p}
\right)
=
0
\end{equation}
for every \(a,p\).

The phase-space Gaussian states form an overcomplete coherent-state
family. For notational simplicity, we display the one-dimensional
coherent-state expansion; the three-dimensional and many-particle cases
follow by tensor products and multi-indices. Identifying
\(\varphi_{a,p}\) with the coherent state \(|\alpha\rangle\), with
\(\alpha\in\mathbb C\) determined by \(a\), \(p\), and \(\sigma\), one has
\[
|\alpha\rangle
=
e^{-|\alpha|^2/2}
\sum_{n=0}^{\infty}
\frac{\alpha^n}{\sqrt{n!}}|n\rangle .
\]
Therefore
\[
\langle\alpha|A|\alpha\rangle
=
e^{-|\alpha|^2}
\sum_{m,n=0}^{\infty}
\frac{\overline{\alpha}^{\,m}\alpha^n}
{\sqrt{m!n!}}
A_{mn},
\qquad
A_{mn}
=
\langle m|A|n\rangle .
\]
If Eq.~(\ref{eq:vanishing-coherent-symbol}) holds for every
\(\alpha\), the real-analytic function on the right vanishes
identically. Hence all of its coefficients vanish:
\[
A_{mn}=0
\qquad
\text{for all }m,n.
\]
Thus
\[
A=0,
\]
and consequently
\[
\widehat H_1-\widehat H_2
=
CI.
\]
The corresponding unitary propagators differ only by an overall phase,
\[
U_1(t)
=
e^{-\frac{iCt}{\hbar}}U_2(t),
\]
and therefore define the same evolution in projective Hilbert space.

Hence, whenever a common unitary lift of the complete Hamiltonian flow
exists, it is unique up to an overall phase. Subject to the usual
domain conditions, the same conclusion holds for the unbounded
self-adjoint generators relevant in quantum mechanics. 
A single operator \(\widehat H\), and hence a
single propagator
\[
U_t
=
e^{-\frac{i\widehat Ht}{\hbar}},
\]
acts simultaneously on the localized states corresponding to all
initial conditions. For the classical Hamiltonians obtained above by
restricting Schr\"odinger dynamics to \(M_{3,3}^\sigma\), existence of
the lift is already supplied by the corresponding quantum Hamiltonian.

It is important to distinguish this statement from invariance of
\(M_{3,3}^\sigma\) under the full unitary evolution. For a general
Hamiltonian, Schr\"odinger evolution need not carry a Gaussian state
into another state of the same Gaussian family. Rather, the statement
concerns the tangent restriction of the ambient Schr\"odinger vector
field to \(M_{3,3}^\sigma\). These tangent components reproduce the
classical Hamiltonian vector field everywhere on the embedded phase
space.
The same relation is reflected in the action principle. The
state-space action
\[
S[\psi]
=
\int
\left[
i\hbar(\psi,\dot\psi)
-
(\psi,\widehat H\psi)
\right]dt
\]
gives the Schr\"odinger equation under unrestricted variation. Its
restriction to \(M_{3,3}^\sigma\) gives the corresponding classical
phase-space action, up to the terms associated with the fixed
localization width. Thus constrained and unrestricted variation
describe, respectively, the classical Hamiltonian motion and its
ambient Schr\"odinger dynamics.

\subsection*{Several particles}

The argument extends directly to a finite system of distinguishable
particles. For \(n\) particles consider the product manifold
\[
M_{3n,3n}^{\boldsymbol{\sigma}}
=
\left\{
\Phi_{\mathbf a,\mathbf p}
=
\bigotimes_{r=1}^{n}
\varphi_{a_r,p_r,\sigma_r}
\right\},
\qquad
\boldsymbol{\sigma}
=
(\sigma_1,\ldots,\sigma_n),
\]
where
\[
\mathbf a=(a_1,\ldots,a_n),
\qquad
\mathbf p=(p_1,\ldots,p_n),
\]
and
\[
\varphi_{a_r,p_r,\sigma_r}\in M_{3,3}^{\sigma_r}.
\]
This manifold lies in the many-particle Hilbert space
\[
\mathcal H
=
\bigotimes_{r=1}^{n}L^2(\mathbb R^3).
\]
The induced symplectic form is the canonical many-particle form
\[
\omega
=
\sum_{r=1}^{n}\sum_{j=1}^{3}
da_{rj}\wedge dp_{rj}.
\]

Let
\[
H_{\rm cl}(\mathbf a,\mathbf p)
\]
be a fixed many-particle Hamiltonian, including interaction terms if
present. If one common unitary evolution reproduces its complete
Hamiltonian flow for all initial conditions, then
\[
X_{h_{\widehat H}}
=
X_{H_{\rm cl}},
\]
where
\[
h_{\widehat H}(\mathbf a,\mathbf p)
=
\left(
\Phi_{\mathbf a,\mathbf p},
\widehat H \Phi_{\mathbf a,\mathbf p}
\right).
\]
By nondegeneracy of the symplectic form,
\[
dh_{\widehat H}
=
dH_{\rm cl},
\]
and therefore
\[
h_{\widehat H}(\mathbf a,\mathbf p)
=
H_{\rm cl}(\mathbf a,\mathbf p)+C.
\]

Uniqueness follows exactly as in the one-particle case. Suppose that
\(\widehat H_1\) and \(\widehat H_2\) generate two common unitary lifts
of the same complete many-particle Hamiltonian flow. Then
\[
\left(
\Phi_{\mathbf a,\mathbf p},
(\widehat H_1-\widehat H_2-CI)
\Phi_{\mathbf a,\mathbf p}
\right)
=
0
\]
for every \(\mathbf a,\mathbf p\).
Writing the product Gaussian state as a product coherent state
\[
|\boldsymbol{\alpha}\rangle
=
|\alpha_1\rangle\otimes\cdots\otimes|\alpha_n\rangle,
\]
its diagonal matrix element with an operator \(A\) has the expansion
\[
\langle\boldsymbol{\alpha}|A|\boldsymbol{\alpha}\rangle
=
e^{-|\boldsymbol{\alpha}|^2}
\sum_{\mathbf m,\mathbf k}
\frac{
\overline{\boldsymbol{\alpha}}^{\,\mathbf m}
\boldsymbol{\alpha}^{\,\mathbf k}
}{
\sqrt{\mathbf m!\,\mathbf k!}
}
A_{\mathbf m\mathbf k}.
\]
If this function vanishes for every
\(\boldsymbol{\alpha}\), all coefficients
\(A_{\mathbf m\mathbf k}\) vanish. Hence \(A=0\), and therefore
\[
\widehat H_1-\widehat H_2
=
CI.
\]
Thus, whenever a common unitary lift exists, the complete
many-particle Hamiltonian flow determines it uniquely up to an overall
phase.

Interactions introduce no difficulty in this argument. Although the
classical states on \(M_{3n,3n}^{\boldsymbol{\sigma}}\) are represented
by product states, the ambient unitary evolution generated by an
interacting \(\widehat H\) will in general leave the product-state
manifold and produce entangled states. As in the one-particle case, it
is the tangent restriction of the ambient vector field that reproduces
the classical many-particle dynamics.
For identical particles the construction is formulated in the
symmetric or antisymmetric many-particle Hilbert space, with the
localized product states replaced by the corresponding symmetrized or
antisymmetrized representatives.

\subsection*{A Family of Paths and a Common Propagator}

The use of the complete Hamiltonian flow is essential. For a fixed
Hamiltonian, the classical motions arising from all possible initial
conditions form a family of trajectories on \(M_{3,3}^\sigma\). These
trajectories are not assigned separate Hilbert-space evolutions. As
shown above, whenever a common unitary lift exists, the entire family is
reproduced by one and the same unitary propagator.

The connection with the many-path formulation of quantum dynamics can
be made more explicit. Divide a time interval into subdivisions
\[
t_0<t_1<\cdots<t_N.
\]
At any time \(t_j\), every phase-space point may serve as initial data
for a classical motion generated by the same Hamiltonian. Thus the
complete Hamiltonian flow supplies, at every subdivision time, all
possible classical continuations.

At each subdivision time \(t_j\), the selected configuration, together
with the momentum appropriate to the next short classical segment,
selects a member of this complete Hamiltonian family. If these
phase-space data coincide with those of a trajectory selected earlier,
uniqueness of Hamiltonian evolution makes the subsequent motion the
ordinary continuation of that trajectory. In general, however, the
momentum selected for the outgoing segment need not coincide with the
momentum of the incoming segment. A time-sliced path is therefore
assembled from successive segments of possibly different members of
the same complete Hamiltonian family. Different histories may also
arrive at the same phase-space point and hence share the same subsequent
segment; they remain distinguished by the segments through which they
arrived there. Thus no new law of motion is introduced at the
subdivision points. One repeatedly selects short segments from the
complete family generated by the original Hamiltonian.

This family is sufficient for the usual time-sliced construction of the
quantum propagator. To see this, consider for simplicity a Hamiltonian
of the form
\[
H_{\rm cl}(x,p)
=
\frac{p^2}{2m}+V(x).
\]
For sufficiently small
\[
\Delta t=t_{j+1}-t_j,
\]
and for nearby configurations \(x_j,x_{j+1}\), one may choose the
initial momentum \(p_j\) so that the classical trajectory generated by
\(H_{\rm cl}\) satisfies
\[
x(t_j)=x_j,
\qquad
x(t_{j+1})=x_{j+1}.
\]
Equivalently,
\[
\pi_x\Phi_{\Delta t}(x_j,p_j)
=
x_{j+1},
\]
where \(\pi_x\) denotes projection onto configuration space. Thus each
short connection required by the time-sliced propagator is supplied by
a member of the same complete Hamiltonian family.

Let
\[
S_{\rm cl}(x_{j+1},x_j;\Delta t)
\]
denote the action of this classical trajectory:
\[
S_{\rm cl}(x_{j+1},x_j;\Delta t)
=
\int_{t_j}^{t_{j+1}}
\left(
p\cdot\dot x-H_{\rm cl}(x,p)
\right)dt.
\]
The short-time propagator is determined, to the required accuracy in the
fine-partition limit, by this classical action together with the
corresponding short-time amplitude factor. Schematically,
\[
K_{\Delta t}(x_{j+1},x_j)
=
A(x_{j+1},x_j;\Delta t)
\exp\!\left[
\frac{i}{\hbar}
S_{\rm cl}(x_{j+1},x_j;\Delta t)
\right].
\]

The propagator over the full interval is then obtained by composition:
\[
K(x_N,t_N;x_0,t_0)
=
\lim_{N\to\infty}
\int
\prod_{j=1}^{N-1}dx_j
\prod_{j=0}^{N-1}
K_{\Delta t}(x_{j+1},x_j).
\]
Hence every finite time-sliced history is assembled from classical
motions belonging to the same Hamiltonian family. As the intermediate
configurations vary, the construction ranges over all corresponding
choices of these classical continuations.

The point is not that one global Newtonian trajectory replaces the
family of paths in the Feynman representation. Rather, the fixed
Hamiltonian \(H_{\rm cl}\) generates the complete family of classical
trajectories through all phase-space data. At each subdivision time
\(t_j\), one may choose new phase-space data independently of the data
chosen at earlier subdivisions. These data select a local segment of
one member of the same complete Hamiltonian family, and that segment
may of course be extended both forward and backward in time. Thus a
time-sliced path is assembled by making successive selections from the
same complete family, with the selected classical trajectory allowed
to change from one subdivision to the next.

Thus the complete family of Hamiltonian motions plays two related
roles. First, it determines the common unitary lift uniquely, whenever
such a lift exists. Second, the same family supplies the classical
connections needed in the time-sliced construction of the corresponding
quantum propagator. In this sense, the many-path representation and the
unitary lift are two descriptions built from the same underlying
Hamiltonian flow.

This structure will reappear in the stochastic setting. There, too, a
complete family of motions is lifted by one common linear evolution
acting on all states simultaneously. The essential difference lies in
the generator: for ordinary dynamics it is a fixed Hamiltonian
operator, whereas for the Brownian lift it is a stochastic
sequence of Hermitian generators whose law will be derived in
Chapter~\ref{chap:brownian-born}.
This distinction will later reappear in the effective classical
dynamics: the ordinary Hamiltonian supplies the deterministic
Schr\"odinger drift, while the \({\bf RM}\) generator supplies the
stochastic Brownian component. The resulting classical motion is the
combination of these two contributions.

\chapter{Observables, Vector Fields, and Curvature}
\label{chap:observables-curvature}

The preceding chapter related Schr\"odinger dynamics to classical
Hamiltonian motion. We now examine the geometric meaning of quantum
observables themselves. Self-adjoint operators generate vector fields on
state space, their commutators determine the Lie brackets of these
fields, and the canonical commutator is related to the curvature scale
of projective Hilbert space.

\section{From Classical Vector Fields to Quantum Observables}
\label{sec:from-classical-vector-fields-to-quantum-observables}

As defined earlier, a self-adjoint operator \(\widehat A\) determines the vector field
\[
X_{\widehat A}(\varphi)
=
-\frac{i}{\hbar}\widehat A\varphi .
\]
For two observables \(\widehat A\) and \(\widehat B\), the Lie bracket of the corresponding vector fields is
\[
[X_{\widehat A},X_{\widehat B}](\varphi)
=
\frac{1}{\hbar^2}
[\widehat A,\widehat B]\varphi .
\]
Thus the commutator of observables has a direct geometric meaning: it determines the Lie bracket of the vector fields generated by the observables on state space.

Consider now the vector fields associated with the position and momentum operators,
\[
X_{\widehat x^\alpha}(\varphi)
=
-\frac{i}{\hbar}\widehat x^\alpha\varphi,
\qquad
X_{\widehat p_\alpha}(\varphi)
=
-\frac{i}{\hbar}\widehat p_\alpha\varphi.
\]
When constrained to the states \(\varphi_{a,p}\), these fields determine tangent directions to \(M_{3,3}^\sigma\). In projective state space, their vertical phase components are removed.
Differentiation with respect to \(p^\alpha\) gives
\[
\widehat x^\alpha\varphi_{a,p}
=
-i\hbar
\frac{\partial\varphi_{a,p}}{\partial p^\alpha},
\]
and therefore
\[
\left.
X_{\widehat x^\alpha}
\right|_{M_{3,3}^\sigma}
=
-
\frac{\partial}{\partial p^\alpha}.
\]
Thus the vector field associated with the position operator is the linear extension of the vector field \(\partial/\partial p^\alpha\) from \(M_{3,3}^\sigma\) to a dense linear subspace of \(L^2(\mathbb R^3)\). Since the states of \(M_{3,3}^\sigma\) form a complete family in \(L^2(\mathbb R^3)\), this linear extension is unique.

Similarly,
\[
X_{\widehat p_\alpha}(\varphi_{a,p})
=
\frac{\partial\varphi_{a,p}}{\partial a^\alpha}
-
\frac{i}{\hbar}p_\alpha\varphi_{a,p}.
\]
The second term is vertical and disappears in projective state space. Hence
\[
\left.
X_{\widehat p_\alpha}
\right|_{M_{3,3}^\sigma}
=
\frac{\partial}{\partial a^\alpha}
\]
as a vector field on the projective manifold. The extension of this field from \(M_{3,3}^\sigma\) to state space is again unique.

The induced Fubini--Study metric on \(M_{3,3}^\sigma\) has constant
coefficients in the coordinates \((a,p)\),
\[
ds^2
=
\frac{1}{4\sigma^2}\,\delta_{\alpha\beta}\,da^\alpha da^\beta
+
\frac{\sigma^2}{\hbar^2}\,\delta_{\alpha\beta}\,dp^\alpha dp^\beta.
\]
Hence \(M_{3,3}^\sigma\) is flat. The coordinate vector fields
\[
\frac{\partial}{\partial a^\alpha},
\qquad
\frac{\partial}{\partial p^\beta}
\]
are orthogonal and commute. Their integral curves therefore form orthogonal coordinate lines on the flat phase-space manifold \(M_{3,3}^\sigma\).

The linear extensions of these fields to the sphere of states, however,
have the Lie bracket
\[
\left[
X_{\widehat x^\alpha},
X_{\widehat p_\beta}
\right](\varphi)
=
\frac{1}{\hbar^2}
[\widehat x^\alpha,\widehat p_\beta]\varphi
=
\frac{i}{\hbar}
\delta^\alpha_\beta\varphi .
\]
Thus the canonical commutator
\[
[\widehat x^\alpha,\widehat p_\beta]
=
i\hbar\delta^\alpha_\beta I
\]
determines the nontrivial Lie bracket of the extended vector fields.

This is the point at which curvature enters. The classical phase-space manifold \(M_{3,3}^\sigma\) is flat, whereas the ambient quantum state space is curved. To make the relation between commutators and curvature quantitative, we must first fix the normalization of the state-space metric and introduce dimensionless position and momentum variables.

\section{Dimensionless Variables and Metric Normalization}
\label{sec:dimensionless-variables-and-metric-normalization}

The metric induced on \(M_{3,3}^\sigma\) by the Fubini--Study metric is
\[
ds^2
=
\frac{|da|^2}{4\sigma^2}
+
\frac{\sigma^2 |dp|^2}{\hbar^2}.
\]
The position and momentum coordinates have different physical dimensions. To compare their geometric contributions, introduce a length unit \(L\) and a momentum unit \(P\), and define the dimensionless variables
\[
q=\frac{a}{L},
\qquad
k=\frac{p}{P}.
\]
The induced metric becomes
\[
ds^2
=
\frac{L^2}{4\sigma^2}|dq|^2
+
\frac{\sigma^2P^2}{\hbar^2}|dk|^2.
\]

The position and momentum directions enter symmetrically when their coefficients are equal:
\[
\frac{L^2}{4\sigma^2}
=
\frac{\sigma^2P^2}{\hbar^2}.
\]
It follows that
\[
\sigma^2
=
\frac{\hbar L}{2P}.
\]
Denoting the common coefficient by \(C\), we have
\[
ds^2
=
C\left(|dq|^2+|dk|^2\right),
\]
where
\[
C
=
\frac{L^2}{4\sigma^2}
=
\frac{LP}{2\hbar}.
\]
Thus the product \(LP\) is tied directly to the normalization of the metric:
\[
LP=2C\hbar.
\]

For the standard Fubini--Study normalization, in which the sphere of normalized Hilbert-space vectors has unit radius, every projective line
\[
\mathbb{CP}^1\simeq S^2
\]
has radius \(1/2\). The corresponding normalization of the dimensionless metric on \(M_{3,3}^\sigma\) is
\[
ds^2
=
\frac{1}{2}
\left(
|dq|^2+|dk|^2
\right).
\]
Hence
\[
C=\frac12,
\]
and therefore
\[
LP=\hbar.
\]

The choice of dimensionless units is not unique. Any length and momentum scales satisfying \(LP=\hbar\) are compatible with this normalization. A distinguished physical choice is provided by the Planck length
\[
\ell_P
=
\sqrt{\frac{\hbar G}{c^3}}
\]
and the Planck momentum
\[
p_P
=
\frac{\hbar}{\ell_P}
=
\sqrt{\frac{\hbar c^3}{G}}.
\]
Taking
\[
L=\ell_P,
\qquad
P=p_P,
\]
the symmetry condition gives
\[
\sigma^2
=
\frac{\hbar\ell_P}{2p_P}
=
\frac{\ell_P^2}{2},
\]
and hence
\[
\sigma
=
\frac{\ell_P}{\sqrt{2}}.
\]

Thus Planck units provide a distinguished physical realization of the canonical relation
\[
LP=\hbar.
\]
In these units, the induced metric on \(M_{3,3}^\sigma\) takes the form
\[
ds^2
=
\frac{1}{2}
\left(
|dq|^2+|dk|^2
\right),
\]
in agreement with the standard Fubini--Study normalization of the ambient projective state space.

\section{The Curvature Scale of Projective State Space}
\label{sec:the-curvature-scale-of-projective-state-space}

The standard Fubini--Study metric is inherited from the unit sphere of normalized states in Hilbert space. With this normalization, every projective line
\[
\mathbb{CP}^1\simeq S^2
\]
is a sphere of radius
\[
R_{\rm FS}=\frac{1}{2},
\]
and therefore has sectional curvature
\[
K_{\rm FS}=4.
\]

These quantities are dimensionless. To express the curvature in physical units, we use the embedding of the classical phase-space manifold \(M_{3,3}^\sigma\) into projective state space.
Recall that the induced Fubini--Study metric on \(M_{3,3}^\sigma\) is
\[
ds^2
=
\frac{|da|^2}{4\sigma^2}
+
\frac{\sigma^2 |dp|^2}{\hbar^2}.
\]
Along a position direction,
\[
ds
=
\frac{|da|}{2\sigma}.
\]
Thus one unit of Fubini--Study length corresponds to the physical position scale \(2\sigma\). Since a projective line has Fubini--Study radius \(1/2\), its radius measured in position units is
\[
R_x
=
2\sigma R_{\rm FS}
=
\sigma.
\]
Consequently, the sectional curvature expressed in inverse length squared is
\[
K_x
=
\frac{1}{R_x^2}
=
\frac{1}{\sigma^2}.
\]

The same curvature may be expressed in momentum units. Along a momentum direction,
\[
ds
=
\frac{\sigma}{\hbar}|dp|,
\]
so that one unit of Fubini--Study length corresponds to the momentum scale \(\hbar/\sigma\). The radius of a projective line in momentum units is therefore
\[
R_p
=
\frac{\hbar}{\sigma}R_{\rm FS}
=
\frac{\hbar}{2\sigma},
\]
and the corresponding dimensional form of the curvature is
\[
K_p
=
\frac{1}{R_p^2}
=
\frac{4\sigma^2}{\hbar^2}.
\]

Hence the same intrinsic curvature may be written in three equivalent forms,
\[
K_{\rm FS}=4,
\qquad
K_x=\frac{1}{\sigma^2},
\qquad
K_p=\frac{4\sigma^2}{\hbar^2}.
\]
These expressions do not represent different curvatures. They give the same geometric curvature in dimensionless Fubini--Study units, position units, and momentum units, respectively.

More generally, let \(L\) and \(P\) be the units used to define the dimensionless variables
\[
q=\frac{a}{L},
\qquad
k=\frac{p}{P}.
\]
For the standard Fubini--Study normalization discussed in Section~\ref{sec:dimensionless-variables-and-metric-normalization},
\[
LP=\hbar,
\]
and the induced metric takes the form
\[
ds^2
=
\frac{1}{2}
\left(
|dq|^2+|dk|^2
\right).
\]
Changing \(L\) and \(P\), while preserving \(LP=\hbar\), changes only the physical units assigned to the position and momentum directions. It does not change the intrinsic Fubini--Study curvature.

The localization scale \(\sigma\) therefore has a direct geometric meaning. It determines the physical length assigned to the radius \(1/2\) of a projective line. In position units,
\[
R_x=\sigma,
\]
so that \(\sigma\) is the physical radius of the corresponding \(\mathbb{CP}^1\) section.

For the Planck choice
\[
\sigma=\frac{\ell_P}{\sqrt{2}},
\]
one obtains
\[
R_x=\frac{\ell_P}{\sqrt{2}},
\qquad
K_x=\frac{2}{\ell_P^2}.
\]

Thus the intrinsic dimensionless curvature of projective state space is fixed by the Fubini--Study metric, while the embedding of classical phase space supplies the physical scale in which this curvature is expressed. In particular, the value of \(\sigma\) does not change the intrinsic Fubini--Study geometry; it determines the physical scale assigned to it in a selected dimensional direction.

\section{Commutators and Curvature in the Hermite Basis}
\label{sec:commutators-and-curvature-in-the-hermite-basis}

We now give a direct relation between the canonical commutator and the curvature of state space. The argument is based on the fact that the commutator of operators determines the Lie bracket of the corresponding vector fields, while the Lie bracket, in turn, is related to the curvature of state space.

Consider the dimensionless operators
\[
\widehat q=\frac{\widehat x}{L},
\qquad
\widehat k=\frac{\widehat p}{P},
\qquad
LP=\hbar,
\]
so that
\[
[\widehat q,\widehat k]=iI.
\]
Let
\[
\varphi_n(q)
=
\frac{1}{\pi^{1/4}\sqrt{2^n n!}}
H_n(q)e^{-q^2/2},
\qquad
n=0,1,2,\ldots,
\]
be the Hermite basis of \(L^2(\mathbb R)\). In this basis, the position and momentum operators have the matrix representations
\[
\widehat q
=
\frac{1}{\sqrt{2}}
\left[
\begin{array}{ccccc}
0 & 1 & 0 & 0 & \cdots \\
1 & 0 & \sqrt{2} & 0 & \cdots \\
0 & \sqrt{2} & 0 & \sqrt{3} & \cdots \\
0 & 0 & \sqrt{3} & 0 & \cdots \\
\vdots & \vdots & \vdots & \vdots & \ddots
\end{array}
\right],
\]
and
\[
\widehat k
=
\frac{1}{\sqrt{2}}
\left[
\begin{array}{ccccc}
0 & -i & 0 & 0 & \cdots \\
i & 0 & -i\sqrt{2} & 0 & \cdots \\
0 & i\sqrt{2} & 0 & -i\sqrt{3} & \cdots \\
0 & 0 & i\sqrt{3} & 0 & \cdots \\
\vdots & \vdots & \vdots & \vdots & \ddots
\end{array}
\right].
\]

At the state \(\varphi_0\), the vectors
\[
\widehat q\varphi_0,
\qquad
\widehat k\varphi_0
\]
belong entirely to
\[
\mathcal H_2
=
\operatorname{span}_{\mathbb C}
\{\varphi_0,\varphi_1\};
\]
their components along all higher Hermite states vanish. Thus the
action of \(\widehat q\) and \(\widehat k\) on \(\varphi_0\) is
reproduced exactly by the operators
\[
\widehat q_2
=
\frac{1}{\sqrt{2}}\sigma_x,
\qquad
\widehat k_2
=
\frac{1}{\sqrt{2}}\sigma_y
\]
acting on \(\mathcal H_2\), where
\[
\sigma_x
=
\left[
\begin{array}{cc}
0&1\\
1&0
\end{array}
\right],
\qquad
\sigma_y
=
\left[
\begin{array}{cc}
0&-i\\
i&0
\end{array}
\right].
\]
Indeed,
\[
\widehat q\varphi_0
=
\widehat q_2\varphi_0,
\qquad
\widehat k\varphi_0
=
\widehat k_2\varphi_0,
\]
and
\[
[\widehat q,\widehat k]\varphi_0
=
[\widehat q_2,\widehat k_2]\varphi_0.
\]

The anti-Hermitian matrices
\[
Q=-i\widehat q_2
=
-\frac{i}{\sqrt{2}}\sigma_x,
\qquad
\Pi=-i\widehat k_2
=
-\frac{i}{\sqrt{2}}\sigma_y
\]
belong to the Lie algebra \(su(2)\). They generate the left-invariant vector fields
\[
L_Q(g)=gQ,
\qquad
L_\Pi(g)=g\Pi,
\qquad
g\in SU(2).
\]
We furnish \(SU(2)\) with the bi-invariant metric whose value at the identity is given by the normalized negative Killing form
\[
(X,Y)_K
=
\frac{1}{2}
\operatorname{Tr}(XY^\dagger).
\]
Under the standard identification of \(SU(2)\) with the unit sphere \(S^3\subset\mathbb C^2\), this metric is precisely the metric of the unit round sphere.

The integral curves of a left-invariant vector field \(L_X\) are
\[
g(t)=g_0e^{tX}.
\]
Because the metric is bi-invariant, these curves are geodesics. Moreover,
\[
\|\dot g(t)\|
=
\|L_X(g(t))\|
=
\|X\|_K,
\]
so the norm of the generator \(X\) is the constant speed with which its one-parameter subgroup is traversed.

For a bi-invariant metric, the sectional curvature in the plane spanned by two linearly independent generators \(X,Y\in su(2)\) is
\[
K(X,Y)
=
\frac{1}{4}
\frac{\|[X,Y]\|_K^2}
{\|X\|_K^2\|Y\|_K^2-(X,Y)_K^2}.
\]
Thus the curvature is determined directly by the Lie bracket of the corresponding invariant vector fields. Since
\[
[L_X,L_Y]=L_{[X,Y]},
\]
and the operator commutator determines the matrix commutator of the generators, this formula gives a direct relation between quantum commutators and curvature.

For the generators \(Q\) and \(\Pi\),
\[
[Q,\Pi]
=
-i\sigma_z,
\]
where
\[
\sigma_z
=
\left[
\begin{array}{cc}
1&0\\
0&-1
\end{array}
\right].
\]
With the metric defined above,
\[
(Q,\Pi)_K=0,
\]
and
\[
\|Q\|_K^2
=
\|\Pi\|_K^2
=
\frac{1}{2},
\qquad
\|[Q,\Pi]\|_K^2
=
1.
\]

It follows that
\[
K(Q,\Pi)
=
\frac{1}{4}
\frac{1}{(1/2)(1/2)}
=
1.
\]
Thus the canonical commutator of position and momentum, represented in the two-dimensional Hermite sector, yields the sectional curvature
\[
K_{S^3}=1
\]
of the unit sphere of normalized states.

The corresponding projective space is
\[
\mathbb{CP}^1\simeq S^2.
\]
Under the Hopf projection
\[
S^3\longrightarrow \mathbb{CP}^1,
\]
the Fubini--Study metric gives the projective sphere radius
\[
R_{\rm FS}=\frac{1}{2},
\]
and therefore curvature
\[
K_{\rm FS}=4.
\]

Hence the canonical commutator
\[
[\widehat x,\widehat p]=i\hbar I
\]
has a direct geometric interpretation. Through the Lie bracket of the corresponding left-invariant vector fields on \(SU(2)\), it determines the sectional curvature of the sphere of normalized states and, after projection, the curvature of the corresponding projective section \(\mathbb{CP}^1\).

\chapter{Brownian Motion and the Born Rule}
\label{chap:brownian-born}

We now turn from deterministic dynamics to stochastic motion. The purpose of
this chapter is to establish the connection between ordinary Brownian
motion on the localized classical manifold and stochastic dynamics in
projective Hilbert space, and to show how this connection leads to the
Born transition law and to the random-matrix dynamics used in the
following chapter.

Brownian motion on the localized manifold \(M_3^\sigma\) produces the
normal distribution. Because \(M_3^\sigma\) is embedded in projective
Hilbert space by overlapping Gaussian states, the same exponential
dependence on spatial separation is expressed geometrically through the
Fubini--Study overlap. Thus, on \(M_3^\sigma\), the normal distribution
and the Born transition law are two expressions of the same probability
law. The main question of the chapter is whether this correspondence can
be understood dynamically as a generalization of Brownian diffusion to
state space.

We shall see that the dimension of the ambient state space is essential.
For isotropic diffusion in a finite-dimensional projective Hilbert
space, the mean squared overlap with the initial state approaches a
nonzero value. As the dimension increases, this overlap floor decreases
like the reciprocal of the dimension and may therefore be made
arbitrarily small. In the formal infinite-dimensional limit it
disappears, and the squared overlap with the initial state acquires an
exponential dependence on the cumulative squared projective
displacement. On the manifold of localized Gaussian states, the same
exponential dependence is expressed exactly by the heat kernel and
hence by the Born transition weight. The limiting calculation will be
used to expose this structure; the stochastic dynamics needed later
will be formulated in a sufficiently large but finite-dimensional
approximation.

We then address the inverse problem: given the family of physical
Brownian motions on \(M_3^\sigma\), what stochastic unitary dynamics in
the ambient Hilbert space reproduces them when constrained to the
localized manifold? Under the assumptions stated below, the common
Hermitian generator is uniquely determined, up to its overall scale and
a projectively irrelevant scalar part, by the Gaussian Unitary
Ensemble. Independent realizations of these generators at successive
steps define the random matrix \({\bf RM}\) walk that will be used in
Chapter~\ref{chap:measurement-rm-walk} to describe measurement dynamics.

The chapter therefore has three main parts. We first relate the normal
probability law on the localized classical manifold \(M_3^\sigma\),
and on the isometric manifold of localized equivalence classes
\(\widetilde M_3^\sigma\), to the Born law through the Fubini--Study
geometry. We then examine the role of the dimension of state space in
Brownian diffusion and investigate the corresponding heat-kernel
structure. Finally, we construct the common unitary lift of the
Brownian family to a stochastic process on state space, derive the GUE
law, and define the \({\bf RM}\) walk. The finite-dimensional helix and
Bloch-sphere examples that follow are included to illustrate the
geometry of the construction.

Readers primarily interested in the measurement theory developed in the
next chapter may accept the resulting correspondence between Brownian
motion and the \({\bf RM}\) walk without following the detailed
derivation of the unitary lift or the finite-dimensional illustrations,
and may proceed directly to
Chapter~\ref{chap:measurement-rm-walk}.

\section{Brownian Motion and Localized-State Geometry}
\label{sec:brownian-gaussian-embedding}

The identification of classical space with the localized-state manifold
\(M_3^\sigma\) gives a direct geometric relation between the normal
probability law generated by Brownian motion and Fubini--Study distance in
projective Hilbert space. We begin with the classical random motion.

Suppose that, during an observation interval, the position of a particle
is changed by a large number of small random displacements. Its final
position may be written as
\[
X_N=X_0+\xi_1+\cdots+\xi_N,
\]
where the increments \(\xi_k\) are independent and identically distributed
random vectors with vanishing mean and finite covariance. If no spatial
direction is distinguished, their covariance is proportional to the
identity. By the multivariate central limit theorem, the normalized
displacement
\[
\frac{X_N-X_0}{\sqrt N}
\]
converges in distribution to an isotropic Gaussian random vector. Thus,
for large \(N\), the net displacement \(X_N-X_0\) is well approximated by
a normal distribution whose covariance is proportional to \(N\), largely
independently of the detailed distribution of the microscopic steps.

In the continuum limit the corresponding probability density \(u(b,t)\)
satisfies the diffusion equation
\[
\frac{\partial u}{\partial t}
=
D\Delta u.
\]
For a particle initially localized at \(a\), the fundamental solution is
\begin{equation}
\label{eq:heat-kernel-first}
u(b,t\mid a,0)
=
\frac{1}{(4\pi Dt)^{3/2}}
\exp\!\left[
-\frac{|a-b|^2}{4Dt}
\right].
\end{equation}
Introducing
\[
\sigma^2=2Dt,
\]
this becomes
\begin{equation}
\label{eq:normal-first}
u(b,t\mid a,0)
=
\frac{1}{(2\pi\sigma^2)^{3/2}}
\exp\!\left[
-\frac{|a-b|^2}{2\sigma^2}
\right].
\end{equation}
Thus the normal distribution is the endpoint law generated by the
homogeneous and isotropic accumulation of many small random
displacements. Brownian motion provides the corresponding stochastic
process.

We now express this law geometrically. Let
\[
\omega:\mathbb R^3\longrightarrow M_3^\sigma,
\qquad
\omega(a)=g_{a,\sigma},
\]
where
\begin{equation}
\label{eq:localized-gaussian-ch4}
g_{a,\sigma}(x)
=
\left(\frac{1}{2\pi\sigma^2}\right)^{3/4}
\exp\!\left[
-\frac{|x-a|^2}{4\sigma^2}
\right].
\end{equation}
The map \(\omega\) identifies a classical position \(a\) with a
localized state \(g_{a,\sigma}\). A probability distribution on
\(\mathbb R^3\) may therefore be pushed forward to a probability
distribution on \(M_3^\sigma\). By the isometry established in
Chapter~\ref{chap:classical-submanifolds}, the same probability law may
equivalently be regarded as a probability law on the manifold
\(\widetilde M_3^\sigma\) of localized equivalence classes.

Consider first two localized states of possibly different widths.
A direct calculation gives
\begin{equation}
\label{eq:different-width-overlap}
\left|
\bigl(g_{a,\sigma},g_{b,\delta}\bigr)
\right|^2
=
\left(
\frac{2\sigma\delta}{\sigma^2+\delta^2}
\right)^3
\exp\!\left[
-\frac{|a-b|^2}{2(\sigma^2+\delta^2)}
\right].
\end{equation}
Equivalently,
\[
\left|
\bigl(g_{a,\sigma},g_{b,\delta}\bigr)
\right|^2
=
\cos^2
\rho_{\rm FS}
\!\left(
g_{a,\sigma},
g_{b,\delta}
\right).
\]

This formula also gives a finite-resolution detector interpretation.
Let a small detector region \(W\subset\mathbb R^3\), centered at \(b\),
be characterized by a resolution width \(\delta\). When
\(\delta\ll\sigma\),
\begin{equation}
\label{eq:detector-overlap-asymptotic-new}
\left|
\bigl(g_{a,\sigma},g_{b,\delta}\bigr)
\right|^2
\sim
\left(\sqrt{8\pi}\,\delta\right)^3
\frac{1}{(2\pi\sigma^2)^{3/2}}
\exp\!\left[
-\frac{|a-b|^2}{2\sigma^2}
\right].
\end{equation}
The first factor is the effective volume element associated with the
detector resolution, while the second is the normal density
(\ref{eq:normal-first}). Thus the probability assigned by Brownian
motion to a small region centered at \(b\) has the same dependence on
\(a\) and \(b\) as the squared Hilbert-space overlap with the
corresponding localized detector state. The latter is the dimensionless
Born transition probability associated with the final ray
\([g_{b,\delta}]\).

The equal-width case is especially important. One has
\begin{equation}
\label{eq:equal-width-overlap-new}
\bigl(g_{a,\sigma},g_{b,\sigma}\bigr)
=
\exp\!\left[
-\frac{|a-b|^2}{8\sigma^2}
\right],
\end{equation}
and therefore
\begin{equation}
\label{eq:equal-width-born-new}
\cos^2
\rho_{\rm FS}
\!\left(
g_{a,\sigma},
g_{b,\sigma}
\right)
=
\exp\!\left[
-\frac{|a-b|^2}{4\sigma^2}
\right].
\end{equation}

For small separation,
\[
\rho^2_{\rm FS}
\!\left(
g_{a,\sigma},
g_{a+da,\sigma}
\right)
=
\frac{|da|^2}{4\sigma^2}
+
o(|da|^2),
\]
so that the induced Fubini--Study metric on \(M_3^\sigma\) is Euclidean,
up to the fixed scale \(2\sigma\), as already established in
Chapter~\ref{chap:classical-submanifolds}. At the same time, as
\(|a-b|\) ranges from \(0\) to infinity,
\[
\rho_{\rm FS}
\!\left(
g_{a,\sigma},
g_{b,\sigma}
\right)
\]
ranges continuously over \([0, \pi/2)\) with limiting value \(\pi/2\). The localized Gaussian
manifold therefore realizes the full interval of possible
Fubini--Study distances between pure states.

This observation gives a simple geometric motivation for the Born rule.
Suppose that the transition probability between pure states depends
only on their Fubini--Study distance. Let \(\psi\) and \(\phi\) be
arbitrary normalized states. Since
\[
\rho_{\rm FS}(\psi,\phi)\in[0,\pi/2],
\]
there exist localized Gaussian states
\(g_{a,\sigma},g_{b,\sigma}\in M_3^\sigma\), or a limiting pair in the
orthogonal case, such that
\[
\rho_{\rm FS}(g_{a,\sigma},g_{b,\sigma})
=
\rho_{\rm FS}(\psi,\phi).
\]
It then follows that
\begin{equation}
\label{eq:born-transition-law}
\begin{aligned}
P(\psi\to\phi)
&=
P(g_{a,\sigma}\to g_{b,\sigma})\\
&=
\cos^2\rho_{\rm FS}(g_{a,\sigma},g_{b,\sigma})\\
&=
\cos^2\rho_{\rm FS}(\psi,\phi)\\
&=
|(\psi,\phi)|^2.
\end{aligned}
\end{equation}
Thus, under the assumption that the transition law is a function only
of Fubini--Study distance, the normal law on the localized manifold
extends to the Born rule on the full projective Hilbert space.

Equation~(\ref{eq:equal-width-born-new}) is the geometric core of this
normal-to-Born correspondence. The same relation holds on the equivalence-class
manifold \(\widetilde M_3^\sigma\), which is isometric to
\(M_3^\sigma\). So far, however, this correspondence is static: it identifies the two probability laws geometrically, but not dynamically.
The purpose of the remainder of this chapter is to establish that
dynamical connection. In particular, we shall see that isotropic
diffusion in a fixed finite-dimensional projective Hilbert space cannot
reproduce Brownian diffusion on \(M_3^\sigma\): it approaches a nonzero
mean overlap with the initial state. The next section shows that this
overlap floor decreases as the dimension grows and can be made
arbitrarily small, while the limiting high-dimensional law develops the
exponential dependence appearing in
Eq.~(\ref{eq:equal-width-born-new}). The precise relation with the heat
kernel, including its normalization and diffusion-time dependence, will
then be obtained in
Section~\ref{sec:heat-kernel-and-born-weights}.

\section{Diffusion and the Dimension of State Space}
\label{sec:diffusion-and-dimension}

The relation obtained in the preceding section was established on the
manifold \(M^\sigma_3\) of overlapping localized states. We now examine
isotropic random motion in the ambient projective state space itself.
The dimension of that space turns out to play an essential role.

Let the Hilbert space have complex dimension \(N+1\), so that the
corresponding projective space is \(\mathbb CP^N\). Consider an
equal-step isotropic random walk. Starting from a normalized state
\(\psi_0\), let
\begin{equation}
\label{eq:equal-step-projective-walk}
\psi_{n+1}
=
\cos\delta\,\psi_n
+
\sin\delta\,v_n,
\end{equation}
where \(v_n\) is a random unit vector in the complex orthogonal
complement of \(\psi_n\), distributed isotropically there. Since
\[
(\psi_n,v_n)=0,
\]
each step has the same Fubini--Study length,
\[
\rho_{\rm FS}(\psi_n,\psi_{n+1})=\delta.
\]

We are interested in the distance from the initial state. Define
\begin{equation}
\label{eq:qn-definition}
q_n
=
|(\psi_0,\psi_n)|^2
=
\cos^2\rho_{\rm FS}(\psi_0,\psi_n).
\end{equation}
The component of \(\psi_0\) orthogonal to \(\psi_n\) has squared norm
\(1-q_n\). Since \(v_n\) is isotropic in an \(N\)-dimensional complex
space, the mean squared projection of \(v_n\) onto any prescribed unit
direction in that space is \(1/N\). Hence
\begin{equation}
\label{eq:isotropic-tangent-component}
\mathbb E
\left[
|(\psi_0,v_n)|^2
\,\middle|\,
\psi_n
\right]
=
\frac{1-q_n}{N}.
\end{equation}
The corresponding cross term has zero conditional mean. Squaring
Eq.~(\ref{eq:equal-step-projective-walk}) therefore gives
\begin{equation}
\label{eq:qn-recursion}
\mathbb E[q_{n+1}\mid q_n]
=
\cos^2\delta\,q_n
+
\frac{\sin^2\delta}{N}(1-q_n).
\end{equation}
Equivalently,
\begin{equation}
\label{eq:qn-recursion-linear}
\mathbb E[q_{n+1}\mid q_n]
=
\left(
\cos^2\delta-\frac{\sin^2\delta}{N}
\right)q_n
+
\frac{\sin^2\delta}{N}.
\end{equation}

Since \(q_0=1\), iteration yields
\begin{equation}
\label{eq:finite-dimensional-overlap}
\mathbb E[q_n]
=
\frac{1}{N+1}
+
\frac{N}{N+1}
\left(
\cos^2\delta-\frac{\sin^2\delta}{N}
\right)^n.
\end{equation}
Thus, for every finite \(N\),
\begin{equation}
\label{eq:finite-dimensional-floor}
\lim_{n\to\infty}\mathbb E[q_n]
=
\frac{1}{N+1}.
\end{equation}
This limiting value has a simple interpretation. Two independently
distributed states in an \((N+1)\)-dimensional complex Hilbert space
have mean squared overlap \(1/(N+1)\). The isotropic walk therefore
approaches the uniform projective distribution and retains a nonzero
mean overlap with its initial state.

This behavior is qualitatively different from diffusion on the
localized Gaussian manifold. For
\[
g_{a,\sigma},g_{b,\sigma}\in M_3^\sigma,
\]
Eq.~(\ref{eq:equal-width-born-new}) gives
\[
|(g_{a,\sigma},g_{b,\sigma})|^2
=
\exp\!\left[
-\frac{|a-b|^2}{4\sigma^2}
\right],
\]
which tends to zero as the Brownian displacement
\(|a-b|\) becomes large. Thus isotropic diffusion in a fixed
finite-dimensional projective space cannot reproduce Brownian diffusion
on \(M_3^\sigma\): the former has a nonzero asymptotic overlap floor,
while the latter allows the overlap with the initial localized state to
tend to zero.

The situation changes as the dimension of the ambient Hilbert space
increases. For fixed \(n\), Eq.~(\ref{eq:finite-dimensional-overlap})
gives
\begin{equation}
\label{eq:large-N-overlap}
\lim_{N\to\infty}\mathbb E[q_n]
=
(\cos^2\delta)^n.
\end{equation}
In fact, the fluctuations about this value vanish in the same limit.
The random component of \(v_n\) along any prescribed tangent direction
is of order \(N^{-1/2}\), so for fixed \(n\),
\begin{equation}
\label{eq:large-N-concentration}
q_n
\longrightarrow
(\cos^2\delta)^n
\end{equation}
in probability.

The limiting statement is used here only to identify the form
approached by high-dimensional projective diffusion. For any fixed
number of steps and any prescribed accuracy, a sufficiently large
finite \(N\) makes both the overlap floor and the fluctuations
arbitrarily small. We shall therefore not require an intrinsically
infinite-dimensional isotropic random walk in what follows; the
stochastic unitary dynamics will be constructed in a large but finite
dimensional approximation. The formal \(N\to\infty\) limit nevertheless
expresses this limiting behavior exactly.

For a small projective step,
\[
\delta\ll1,
\]
one has
\[
\log\cos^2\delta
=
-\delta^2+O(\delta^4).
\]
Hence, after the large-\(N\) form has been identified,
\begin{equation}
\label{eq:large-N-exponential}
q_n
\simeq
\exp(-n\delta^2).
\end{equation}
If the diffusion scaling is then taken so that
\[
n\to\infty,
\qquad
\delta\to0,
\qquad
n\delta^2\to \kappa,
\]
one obtains
\begin{equation}
\label{eq:infinite-dimensional-overlap-law}
\cos^2\rho_{\rm FS}(\psi_0,\psi_n)
=
e^{-\kappa}.
\end{equation}
For the equal-step walk considered here,
\[
\kappa=n\delta^2
\]
in the diffusion limit. More generally, for unequal sufficiently small
steps,
\[
\kappa
=
\lim\sum_k\delta_k^2.
\]

At small \(\kappa\),
\[
\rho_{\rm FS}
=
\arccos e^{-\kappa/2}
\sim
\sqrt{\kappa},
\]
as expected for diffusive growth of distance. For large \(\kappa\),
\[
\rho_{\rm FS}
\longrightarrow
\frac{\pi}{2},
\]
while the squared overlap tends continuously to zero.

The connection with ordinary Brownian motion is already visible at the
level of diffusive scaling. For a spatial random walk with elementary
displacement of characteristic size \(\ell\),
\[
\mathbb E|X_n-X_0|^2
\propto
n\ell^2,
\]
while for the projective walk the corresponding diffusion parameter
satisfies
\[
\kappa
\sim
n\delta^2.
\]
Both therefore have the characteristic quadratic scaling of diffusion.

The more important point, however, concerns the endpoint probability
itself. By the static normal-to-Born relation established in the
preceding section, Eq.~(\ref{eq:infinite-dimensional-overlap-law}) also
gives the transition probability associated with the state
\(\psi_n\) reached from \(\psi_0\). Thus the variable \(\kappa\) also
parametrizes the endpoint transition probability through
\[
P(\psi_0\to\psi_n)=e^{-\kappa}.
\]
If the initial and final states lie on \(M_3^\sigma\),
\[
\psi_0=g_{a,\sigma},
\qquad
\psi_n=g_{b,\sigma},
\]
the same dependence on the endpoint separation appears in the
corresponding Brownian transition law. The intervening projective
trajectory need not remain on \(M_3^\sigma\); the identification
concerns only its initial and final states. In the next section we make
this correspondence explicit and show that \(\kappa\) becomes the usual
dimensionless endpoint variable of the Brownian heat kernel. Later in
the chapter, the common unitary lift of the Brownian family will provide
the direct dynamical connection between the two stochastic
descriptions.

\section{The Heat Kernel and Born Weights}
\label{sec:heat-kernel-and-born-weights}

As just shown, the high-dimensional projective walk dynamically produces
the same exponential form that appears in the normal law on
\(M_3^\sigma\). We now make this correspondence exact by expressing the
heat kernel directly in terms of the Fubini--Study distance and by
relating localized Gaussian amplitudes to square roots of Brownian
endpoint densities.

Let
\begin{equation}
\label{eq:heat-kernel}
K_t(a,x)
=
\frac{1}{(4\pi Dt)^{3/2}}
\exp\!\left[
-\frac{|x-a|^2}{4Dt}
\right]
\end{equation}
be the heat kernel in three-dimensional Euclidean space. It is the
endpoint probability density at time \(t\) for Brownian motion with
diffusion coefficient \(D\), starting from \(a\).

Recall that
\[
g_{a,\sigma}(x)
=
\frac{1}{(2\pi\sigma^2)^{3/4}}
\exp\!\left[
-\frac{|x-a|^2}{4\sigma^2}
\right].
\]
If
\begin{equation}
\label{eq:sigma-diffusion-time}
\sigma^2=2Dt,
\end{equation}
then
\begin{equation}
\label{eq:gaussian-square-root-heat-kernel}
g_{a,\sigma}(x)
=
\sqrt{K_t(a,x)}.
\end{equation}
Thus a localized state on \(M_3^\sigma\) is the square root of the
Brownian endpoint density for the corresponding duration.

The overlap of two such states is therefore
\begin{equation}
\label{eq:heat-kernel-affinity}
\bigl(g_{a,\sigma},g_{b,\sigma}\bigr)
=
\int_{\mathbb R^3}
\sqrt{
K_t(a,x)K_t(b,x)
}\,dx.
\end{equation}
The right-hand side is the statistical affinity of the two Brownian
endpoint distributions. Direct evaluation gives
\[
\bigl(g_{a,\sigma},g_{b,\sigma}\bigr)
=
\exp\!\left[
-\frac{|a-b|^2}{16Dt}
\right],
\]
and hence
\begin{equation}
\label{eq:squared-overlap-heat-time}
\left|
\bigl(g_{a,\sigma},g_{b,\sigma}\bigr)
\right|^2
=
\exp\!\left[
-\frac{|a-b|^2}{8Dt}
\right].
\end{equation}

On the other hand,
\[
K_{2t}(a,b)
=
\frac{1}{(8\pi Dt)^{3/2}}
\exp\!\left[
-\frac{|a-b|^2}{8Dt}
\right].
\]
It follows that
\begin{equation}
\label{eq:heat-kernel-fs}
K_{2t}(a,b)
=
\frac{1}{(8\pi Dt)^{3/2}}
\cos^2
\rho_{\rm FS}
\!\left(
g_{a,\sigma},
g_{b,\sigma}
\right).
\end{equation}
Thus the spatial dependence of the Brownian transition density is
exactly the squared Fubini--Study overlap of the corresponding
localized states. The factor multiplying the overlap is simply the
normalization required for a probability density with respect to
Euclidean volume.
The appearance of the doubled time in
Eq.~(\ref{eq:heat-kernel-fs}) is a consequence of taking the squared
overlap. The state \(g_{a,\sigma}\) itself is the square root of the
heat kernel at time \(t\), whereas its squared overlap with another
localized state has the spatial dependence of the heat kernel at time
\(2t\).

For localized initial and final states,
\[
\psi_0=g_{a,\sigma},
\qquad
\psi_n=g_{b,\sigma},
\]
Eqs.~(\ref{eq:infinite-dimensional-overlap-law}) and
(\ref{eq:squared-overlap-heat-time}) give
\[
e^{-\kappa}
=
\exp\!\left[
-\frac{|a-b|^2}{8Dt}
\right].
\]
Hence
\begin{equation}
\label{eq:s-spatial}
\kappa
=
\frac{|a-b|^2}{8Dt}
=
\frac{|a-b|^2}{4\sigma^2}.
\end{equation}
Thus, for localized endpoints, the projective diffusion parameter \(\kappa\)
is precisely the usual dimensionless squared displacement of the
Brownian endpoint.

Equation~(\ref{eq:heat-kernel-fs}) may therefore be written as
\begin{equation}
\label{eq:heat-kernel-s}
K_{2t}(a,b)
=
\frac{1}{(8\pi Dt)^{3/2}}e^{-\kappa}
=
\frac{1}{(8\pi Dt)^{3/2}}
\cos^2
\rho_{\rm FS}
\!\left(
g_{a,\sigma},
g_{b,\sigma}
\right).
\end{equation}
Apart from the normalization factor required to convert a transition
weight into a probability density with respect to spatial volume, the
Brownian endpoint density and the Born transition weight therefore have
the same dependence on the endpoint separation.

For localized states, the Brownian transition kernel is related to the
squared Hilbert-space overlap by
\[
K_{2t}(a,b)
\propto
\left|
\bigl(g_{a,\sigma},g_{b,\sigma}\bigr)
\right|^2.
\]
For a coherent superposition of localized states, the pointwise Born
density instead takes the form
\[
|\psi(x)|^2
=
\left|
\sum_j
c_j\sqrt{K_t(a_j,x)}
\right|^2.
\]
Thus the individual Brownian square-root densities combine coherently
before the modulus is squared, producing the interference terms.

At this point the result has been established for finite linear
combinations of localized Gaussian states. Since the translated
Gaussian family is complete in \(L^2(\mathbb R^3)\), such combinations
form a dense set. The extension to general states may therefore be made
by Hilbert-space completion. A Gaussian-frame or Segal--Bargmann
description provides a more systematic version of the same extension,
but will not be needed for the arguments below.

\section{The Unitary Lift of Brownian Motion}
\label{sec:the-gaussian-unitary-lift}

The results of the preceding two sections identify the probability
structure that the state-space dynamics must preserve. What remains is
to determine a stochastic unitary dynamics in the ambient Hilbert space
whose constraint to the localized manifold reproduces Brownian motion.

Similar to the unitary lift of classical Hamiltonian dynamics in
Section~\ref{sec:unitary-lift-hamiltonian-motion}, the classical
dynamical system to be lifted is not a single Brownian particle, but a
family of identical Brownian particles whose initial positions form a
spatial lattice. During one coarse-grained step, each particle undergoes
many small collisions, so that both its displacement and its
instantaneous momentum are defined. The displacement increments of
different particles are independent centered isotropic Gaussian vectors
with the same covariance. The corresponding position and momentum
components, expressed in suitably scaled phase-space coordinates, are
likewise taken to be centered Gaussian variables with the same
component variance. The momentum is sampled anew at each coarse-grained
step; only the position displacement is accumulated in the Brownian
motion.

The statistical structure of this family will be referred to as
\emph{homogeneous component independence}. By this we mean that the
distribution of a resolved position or momentum component is independent
of the absolute position of the particle and of the location at which
the resolving direction is represented, and that mutually orthogonal
nonredundant resolving components are independent. After the
dimensionless position--momentum scaling, every such orthogonal
component has the same centered Gaussian distribution. Thus independent
translations of the initial and resolving locations leave the component
law unchanged, and the Gaussian law is the same in all orthogonal
resolving directions.

The naturalness of these properties is easiest to see in ordinary
Euclidean space. Brownian motion in \(\mathbb R^3\) is homogeneous and
isotropic, while Cartesian tangent vectors at different points are
canonically identified by parallel translation. Consequently, the law
of a resolved Brownian component is unchanged if either the initial
position of the particle or the point at which the resolving direction
is represented is translated. These translations are therefore
classically redundant: they produce the same component law.
After embedding the classical space into projective Hilbert space,
however, the translated resolving directions are represented by
distinct ambient tangent vectors. The new element is therefore not the
Brownian statistics themselves, but the geometry in which they are
represented. The stochastic lift is required to preserve the statistical
equivalence already present in the classical Brownian family.

Since the lift must be linear, a single realization of the random unitary transformation must act simultaneously on all states and reproduce the entire family of Brownian motions. Indeed, the localized states form a complete family in the Hilbert space, so every state can be approximated arbitrarily well by finite linear combinations of localized states. Assigning independent random
transformations to different localized states would make the evolution
depend on the chosen decomposition of the same Hilbert-space vector and
could therefore assign more than one trajectory to a single initial
state. A common realized unitary avoids this redundancy:
\[
U\!\left(\sum_j c_jg_{a_j,\sigma}\right)
=
\sum_jc_j\,Ug_{a_j,\sigma}.
\]
Thus the Brownian motions starting from different localized states are
not lifted by separate random transformations; they are the projections
of one common random motion in state space.

Since the Hilbert space is infinite dimensional, the reconstruction is
first carried out on finite-dimensional subspaces. Two distinct uses
of finite-dimensional approximation should be separated.
For questions of convergence to the full Hilbert space one may use, for
example, the Hermite exhaustion
\[
P_N
=
\sum_{n=0}^{N-1}
|h_n\rangle\langle h_n|,
\qquad
P_N\longrightarrow I
\]
strongly.  In particular,
\[
P_Ng_{a,\sigma}\longrightarrow g_{a,\sigma}
\]
for every localized Gaussian state.  In several spatial dimensions the
corresponding energy cutoff is used.  The Hermite basis is used only as
a convenient reference exhaustion and does not define a preferred basis
for the stochastic dynamics.

For the finite-dimensional reconstruction of the generator law, however, it is more useful to employ the adapted finite-dimensional subspace introduced below. Given any finite collection of Brownian resolving
components, choose smooth compactly supported representatives of the
corresponding localized equivalence classes, with their centers
sufficiently separated that the relevant supports are disjoint.
The associated localized states and tangent directions are then
exactly orthogonal.  Their span, supplemented if necessary by further
orthogonal vectors, gives a finite-dimensional space adapted to the
chosen resolving components.  The Brownian law is unchanged by this
choice of representatives and by translations of their centers.

Thus the Hermite exhaustion serves only to organize the
infinite-dimensional limit, whereas the finite-dimensional
reconstruction of the covariance of the common generator is performed
in coordinates adapted to physically available orthogonal resolving
directions.  To avoid proliferating notation, the finite-dimensional space used at a
given stage of the reconstruction will be denoted by \(\mathcal H_N\).

\subsection*{Localized states and their tangent directions}

Recall the phase-space family
\[
M_{1,1}^\sigma
=
\left\{
g_{a,\sigma}(x)e^{\frac{ipx}{\hbar}}
:
a,p\in\mathbb R
\right\}.
\]
In the
dimensionless coordinates
\[
\widetilde a
=
\frac{a}{2\sigma},
\qquad
\widetilde p
=
\frac{\sigma p}{\hbar},
\]
the induced metric is Euclidean:
\[
ds^2
=
d\widetilde a^{\,2}
+
d\widetilde p^{\,2}.
\]

At \(p=0\), the position tangent vector is
\[
v_a
=
\frac{\partial g_{a,\sigma}}{\partial a}
=
\frac{x-a}{2\sigma^2}g_{a,\sigma},
\]
while the horizontal momentum tangent vector is
\[
w_a
=
\left(
\frac{\partial}{\partial p}
g_{a,\sigma}e^{\frac{ipx}{\hbar}}
\right)_{p=0}^{\rm hor}
=
\frac{i(x-a)}{\hbar}g_{a,\sigma}.
\]
Thus
\[
w_a
=
\frac{2\sigma^2}{\hbar}\,i v_a.
\]
After rescaling,
\[
\widetilde w_a=i\widetilde v_a.
\]
The position and momentum tangent components therefore give the real
and imaginary parts of one complex state-space component.
The position increment is retained in the constrained Brownian walk on
\(M_1^\sigma\), while the momentum component supplies the complementary
imaginary direction needed to resolve the corresponding complex ambient
state-space increment. Since it is sampled anew at each coarse-grained
step, it is not accumulated as an additional Brownian coordinate.

The phase-space Gaussian states
\[
\varphi_{a,p}(x)
=
g_{a,\sigma}(x)e^{\frac{ipx}{\hbar}}
\]
form an overcomplete coherent-state family. Moreover, the continuous
phase-space family may be discretized without losing completeness. As
discussed in Section~\ref{sec:localized-equivalence-classes}, sampling
\[
g_{a,\sigma}(x)e^{\frac{ipx}{\hbar}}
\]
on a sufficiently dense phase-space lattice gives the standard
Gaussian Gabor system. Under the usual density condition, this system
forms a Gabor frame and is therefore complete in the full Hilbert
space.

We now pass to the finite-dimensional approximation
\(\mathcal H_N\). The projections of the phase-space Gaussian states
span \(\mathcal H_N\). Indeed, writing the corresponding coherent state
as \(|\alpha\rangle\), where the complex parameter \(\alpha\) combines
the phase-space coordinates \(a\) and \(p\), one has in the oscillator
representation
\[
P_N|\alpha\rangle
=
e^{-|\alpha|^2/2}
\sum_{n=0}^{N-1}
\frac{\alpha^n}{\sqrt{n!}}|n\rangle.
\]
For \(N\) distinct values
\(\alpha_1,\ldots,\alpha_N\), the resulting coefficient matrix differs
from a Vandermonde matrix only by nonzero row and column factors and is
therefore nonsingular. Hence the corresponding projected coherent
states span \(\mathcal H_N\).
Consequently, an operator on \(\mathcal H_N\) is determined by its
action on any such spanning family of projected phase-space states.

\subsection*{A common random generator}

Let one coarse-grained state-space step, of duration \(\Delta t\), be
generated by a random Hermitian operator \(\widehat h_N\) on
\(\mathcal H_N\):
\begin{equation}
\label{eq:unitary-rm-step-new}
\psi
\longmapsto
e^{-\frac{i\widehat h_N\Delta t}{\hbar}}\psi .
\end{equation}
To first order,
\[
\delta\psi
=
-\frac{i}{\hbar}
\widehat h_N\psi\,\Delta t .
\]
Here \(\Delta t\) is the duration of one coarse-grained Brownian step
and is kept fixed during the reconstruction of the generator law. The
diffusion scaling with an infinitesimal interval \(dt\) will be
introduced below. The same realization of \(\widehat h_N\) acts on
every state in \(\mathcal H_N\), while independent realizations are
used at successive steps.
For notational simplicity we suppress below the projections onto
\(\mathcal H_N\). Thus \(g_{a,\sigma}\), \(v_a\), and \(w_a\) denote
their finite-dimensional projections whenever the approximation is
understood.

Let \(g_{a,\sigma}\) be the localized input state and let \(v_b\) and
\(w_b\) be the position and momentum tangent directions associated with
\(g_{b,\sigma}\). Define
\begin{equation}
\label{eq:position-component-lift}
\xi_{b;a}
=
\operatorname{Re}
\left(
v_b,
-\frac{i}{\hbar}
\widehat h_N g_{a,\sigma}\,\Delta t
\right)
\end{equation}
and
\begin{equation}
\label{eq:momentum-component-lift}
\eta_{b;a}
=
\operatorname{Re}
\left(
w_b,
-\frac{i}{\hbar}
\widehat h_N g_{a,\sigma}\,\Delta t
\right).
\end{equation}
Since \(w_b\) is, up to a fixed scale, \(iv_b\), these two real
variables determine the real and imaginary parts of
\[
\left(
v_b,\widehat h_N g_{a,\sigma}
\right)
\]
up to fixed scale factors.

The labels \(a\) and \(b\) are independent: \(a\) specifies the
localized state on which the generator acts, whereas \(b\) specifies
the tangent direction in which its action is resolved.
The independent role of the resolving label \(b\) has a simple
classical origin.  In Euclidean space a Cartesian tangent vector may be
parallel translated from one point to another without change.  If
\(e_\alpha(b)\in T_b\mathbb R^3\) denotes a Cartesian direction, then
Euclidean parallel translation satisfies
\[
P_{b\to a}e_\alpha(b)=e_\alpha(a).
\]
Hence, for a Brownian increment \(\Delta X(a)\in T_a\mathbb R^3\),
the component resolved by a direction represented at \(b\),
\[
\xi^{\rm cl}_{b;a,\alpha}
=
\bigl\langle
P_{b\to a}e_\alpha(b),\Delta X(a)
\bigr\rangle,
\]
is independent of the choice of the resolving point \(b\).
Spatial homogeneity likewise makes its distribution independent of
the absolute position \(a\).  Thus independent translations of the
input and resolving locations are already invisible in the classical
Brownian description.

After the embedding
\[
a\longmapsto [g_{a,\sigma}],
\]
however, the translated Cartesian directions are represented by the
distinct ambient Hilbert-space vectors
\[
v_{a,\alpha}
=
\frac{\partial g_{a,\sigma}}{\partial a^\alpha}.
\]
Thus a translation freedom that is classically redundant becomes a
nontrivial family of resolving directions in the ambient state space.
This is the geometric origin of the separate translation covariance in
the labels \(a\) and \(b\) used below.  Homogeneous component
independence, introduced above as a property inherited from the
classical Brownian family, requires this statistical equivalence to be
preserved by the lift.  In the one-dimensional notation used below the
Cartesian index \(\alpha\) is suppressed.

Homogeneous component independence therefore gives
\begin{equation}
\label{eq:separate-translation-covariance}
\xi_{b+c_2\,;\,a+c_1}
\overset{d}{=}
\xi_{b;a},
\qquad
\eta_{b+c_2\,;\,a+c_1}
\overset{d}{=}
\eta_{b;a},
\end{equation}
for arbitrary translations \(c_1,c_2\), where
\(\overset{d}{=}\) denotes equality in distribution.
Moreover, any nonredundant collection of distinct Brownian resolving
components may be chosen independent, while the position and momentum
components belonging to the same complex component are independent
centered Gaussians with equal variance in the dimensionless
phase-space coordinates.
No independence is asserted for the full vectors
\[
\widehat h_N g_{a,\sigma}.
\]
They are produced by one common Hermitian operator and therefore carry
the correlations imposed by that operator. Independence concerns the
Brownian components used to resolve its action.

The localized input family \(\{g_{a,\sigma}\}\) and the position-tangent
family \(\{v_b\}\) are each complete. Consequently, the mixed
components
\[
\left(
v_b,\widehat h_N g_{a,\sigma}
\right),
\qquad a,b\in\mathbb R,
\]
determine \(\widehat h_N\) uniquely. Indeed, if an operator \(A\)
satisfies
\[
(v_b,Ag_a)=0
\]
for all \(a,b\), completeness of the resolving family gives
\[
Ag_a=0
\]
for every \(a\), and completeness of the input family then implies
\(A=0\).

We now make the adapted finite-dimensional construction explicit.
Exploit the freedom within each localized equivalence class and replace
the Gaussian representative by a smooth compactly supported representative
\(f_{a,R}\) in the same class, obtained by smoothly truncating
\(g_{a,\sigma}\) outside a sufficiently large neighborhood of \(a\).
The truncation may be chosen symmetrically so that the mean remains
\(a\), while
\[
f_{a,R}\longrightarrow g_{a,\sigma},
\qquad
v_{a,R}\longrightarrow v_a
\]
in \(L^2\) as \(R\to\infty\).

For the corresponding phase-space family
\[
f_{a,R,p}(x)
=
f_{a,R}(x)e^{\frac{ipx}{\hbar}},
\]
the horizontal momentum tangent is
\[
w_{a,R}
=
\frac{i(x-a)}{\hbar}f_{a,R}.
\]
Hence
\begin{equation}
\label{eq:compact-support-position-momentum}
w_{a,R}
=
\frac{2\sigma^2}{\hbar}\,i v_{a,R}
+
o_R(1),
\qquad
\|o_R(1)\|\longrightarrow0 .
\end{equation}

For any fixed finite collection, the centers may be chosen sufficiently
far apart that the compact supports are mutually disjoint. The
corresponding probe states and tangent directions are then exactly
orthogonal within the selected families. Thus the Brownian resolving
components may be assigned to a nonredundant finite collection of
operator coordinates without approximate overlap; the only
approximation is the replacement of the exact Gaussian
position--momentum relation by
Eq.~(\ref{eq:compact-support-position-momentum}), whose error can be
made arbitrarily small.
The separated representatives are used only to choose independent
coordinates for the finite-dimensional reconstruction; by homogeneous
translation covariance, the resulting component law is the same as
that for the original localized states at arbitrary positions.

We now determine the probability law of the common generator. For a
Hermitian operator \(A\) on \(\mathcal H_N\), define the covariance
form
\[
\mathcal Q_N(A,B)
=
\mathbb E
\left[
\operatorname{Tr}(A\widehat h_N)
\operatorname{Tr}(B\widehat h_N)
\right]
\]
after subtraction of the mean of \(\widehat h_N\). The Brownian
components form a jointly Gaussian family and depend linearly on
\(\widehat h_N\). Since, as shown below, these linear functionals
determine the generator, the centered generator is a Gaussian random
element of the finite-dimensional real vector space of Hermitian
operators. Its law is therefore completely determined by
\(\mathcal Q_N\).

The quantities measured above are precisely linear functionals of this
kind. Indeed, put
\[
A_{b;a}
=
\frac12
\left(
|g_a\rangle\langle v_b|
+
|v_b\rangle\langle g_a|
\right)
\]
and
\[
B_{b;a}
=
\frac1{2i}
\left(
|g_a\rangle\langle v_b|
-
|v_b\rangle\langle g_a|
\right).
\]
Then
\[
\operatorname{Tr}(A_{b;a}\widehat h_N)
=
\operatorname{Re}(v_b,\widehat h_N g_a),
\]
while
\[
\operatorname{Tr}(B_{b;a}\widehat h_N)
=
\operatorname{Im}(v_b,\widehat h_N g_a).
\]
Thus the position and momentum Brownian components determine the
covariance form of the generator on the Hermitian test directions
generated by the localized states and their tangent vectors.

Because the input and resolving families are complete, these test
directions determine the Hermitian operator space. In the
finite-dimensional approximation one may, by using the separated
compactly supported representatives introduced above, choose
nonredundant resolving directions that are mutually orthogonal. The
Brownian law is attached to the localized equivalence classes rather
than to a particular choice of representative, so the representatives
and their resolving tangent directions may be varied without changing
the component law.

Homogeneous component independence therefore implies that, after the
dimensionless position--momentum scaling, the coefficients of the
centered generator along mutually orthogonal nonredundant Hermitian
resolving directions are independent centered Gaussian variables with
the same variance.
After Hilbert--Schmidt normalization, these nonredundant Hermitian
resolving directions may be completed to an orthonormal coordinate
system
\[
\{E_\alpha\}
\]
for the projectively relevant Hermitian operators on
\(\mathcal H_N\). Thus
\[
\mathbb E\!\left[
\operatorname{Tr}(E_\alpha\widehat h_N)
\operatorname{Tr}(E_\beta\widehat h_N)
\right]
=
c_N\delta_{\alpha\beta}
\]
for some \(c_N>0\).

If
\[
A=\sum_\alpha a_\alpha E_\alpha,
\qquad
B=\sum_\alpha b_\alpha E_\alpha,
\]
then
\[
\mathcal Q_N(A,B)
=
c_N\sum_\alpha a_\alpha b_\alpha
=
c_N\operatorname{Tr}(AB).
\]
Thus, on the traceless Hermitian subspace,
\begin{equation}
\label{eq:generator-hs-covariance}
\mathcal Q_N(A,B)
=
c_N\,\operatorname{Tr}(AB),
\qquad
\operatorname{Tr}A
=
\operatorname{Tr}B
=
0.
\end{equation}

The covariance of the centered generator is therefore proportional to
the Hilbert--Schmidt metric. Since the generator is centered Gaussian,
this determines an isotropic Gaussian law on the real traceless
Hermitian operator space. Its law is therefore invariant under every
orthogonal transformation of that space. In particular, unitary conjugation,
\[
A\longmapsto UAU^\dagger,
\]
preserves the Hilbert--Schmidt inner product and therefore preserves
the law of the centered generator.

To display the usual matrix form, let
\(\{e_1,\ldots,e_N\}\) be any orthonormal basis of \(\mathcal H_N\) and
write
\[
h_{jk}
=
(e_j,\widehat h_N e_k).
\]
For \(j<k\), write
\[
h_{jk}=x_{jk}+iy_{jk}.
\]
The Hilbert--Schmidt normalized off-diagonal Hermitian directions are
\[
S_{jk}
=
\frac{1}{\sqrt2}
\left(
|e_j\rangle\langle e_k|
+
|e_k\rangle\langle e_j|
\right),
\]
and
\[
T_{jk}
=
\frac{1}{\sqrt2\,i}
\left(
|e_j\rangle\langle e_k|
-
|e_k\rangle\langle e_j|
\right).
\]
Since
\[
\operatorname{Tr}(S_{jk}\widehat h_N)
=
\sqrt2\,x_{jk},
\qquad
\operatorname{Tr}(T_{jk}\widehat h_N)
=
\sqrt2\,y_{jk},
\]
Eq.~(\ref{eq:generator-hs-covariance}) gives
\[
\operatorname{Var}(x_{jk})
=
\operatorname{Var}(y_{jk})
=
\frac{c_N}{2},
\]
with all nonredundant off-diagonal variables independent.

The diagonal sector is determined in the same way after restriction to
traceless Hermitian directions. Thus, for every pair of traceless
Hermitian operators \(A,B\),
\begin{equation}
\label{eq:gue-covariance-final}
\mathbb E
\left[
\operatorname{Tr}(A\widehat h_N)
\operatorname{Tr}(B\widehat h_N)
\right]
=
c_N\operatorname{Tr}(AB).
\end{equation}
The mean of the generator can only be proportional to the identity:
\[
\mathbb E[\widehat h_N]
=
\lambda_N I.
\]
Likewise, any stochastic scalar component is projectively irrelevant,
since it generates only an overall phase. Hence only the traceless
part of \(\widehat h_N\) contributes to the motion in projective
Hilbert space.
The centered projectively relevant generator is Gaussian, and its
covariance on the traceless Hermitian subspace is proportional to the
Hilbert--Schmidt metric. Its law on that subspace is therefore the
isotropic Gaussian measure, that is, the traceless Gaussian Unitary
Ensemble with an as yet unspecified scale \(c_N\). Thus, up to a
projectively irrelevant scalar multiple of the identity, the common
random generator has the GUE form.

It remains to fix its scale. During a Brownian time interval
\(\Delta t\), one Cartesian position component satisfies
\[
\operatorname{Var}(\Delta a)
=
2D\,\Delta t.
\]
Since
\[
\widetilde a=\frac{a}{2\sigma},
\]
the corresponding dimensionless tangent component has variance
\[
\operatorname{Var}(\Delta\widetilde a)
=
\frac{D}{2\sigma^2}\,\Delta t.
\]
On the other hand, the state-space increment generated by
\(\widehat h_N\) is
\[
\delta\psi
=
-\frac{i}{\hbar}\widehat h_N\psi\,\Delta t.
\]
Brownian scaling therefore requires the Hermitian increment
\(\widehat h_N\Delta t\) to be of order \(\sqrt{\Delta t}\). Write
\begin{equation}
\label{eq:brownian-generator-rescaling}
\widehat h_N\Delta t
=
H_N\sqrt{\Delta t}.
\end{equation}
Here \(H_N\) is the Brownian-normalized
Hermitian increment amplitude, with dimensions of action per
square-root time.
The rescaling preserves the isotropic Gaussian matrix structure derived
above. Denoting the variance of each real and imaginary off-diagonal
component of the Brownian-normalized matrix \(H_N\) by
\(\tau_N^2\), comparison with the dimensionless Brownian variance gives
\begin{equation}
\label{eq:gue-scale-from-diffusion}
\tau_N^2
=
\frac{\hbar^2D}{2\sigma^2}.
\end{equation}
Equivalently, for traceless Hermitian operators \(A,B\),
\[
\mathbb E\!\left[
\operatorname{Tr}(A H_N)
\operatorname{Tr}(B H_N)
\right]
=
2\tau_N^2\operatorname{Tr}(AB).
\]

We may now pass to the infinitesimal diffusion notation. For an
infinitesimal interval \(dt\), write
\[
dK_N
=
H_N\sqrt{dt},
\]
so that the corresponding unitary increment is
\begin{equation}
\label{eq:unitary-brownian-increment}
U_{dt}
=
\exp\left(
-\frac{i}{\hbar}dK_N
\right).
\end{equation}
Independent GUE realizations on disjoint time intervals give independent
increments. The same realized \(U_{dt}\), however, acts simultaneously
on every state in \(\mathcal H_N\). Thus a single stochastic unitary
motion lifts the entire family of Brownian motions on the localized
manifold.

The construction above has been carried out in the finite-dimensional
space \(\mathcal H_N\). Increasing \(N\) enlarges the family of
localized states and tangent directions resolved by the lift, while
\(P_N\to I\) strongly. At every finite cutoff the random generator is
therefore an ordinary GUE matrix and defines an ordinary probability
law.
There is, however, a useful infinite-dimensional statement that does
not require the existence of an operator-valued GUE random variable on
the full Hilbert space. The quantities entering the Brownian
reconstruction are linear functionals of the generator associated with
finite-rank, and hence Hilbert--Schmidt, Hermitian probes. These
functionals possess a well-defined infinite-dimensional Gaussian law.

\begin{proposition}[Infinite-dimensional cylindrical GUE law]
Let \(\mathcal H\) be a separable Hilbert space, and let
\(\mathfrak S_2^{\rm sa}(\mathcal H)\) be the real Hilbert space of
self-adjoint Hilbert--Schmidt operators, with inner product
\[
\langle A,B\rangle_{\rm HS}
=
\operatorname{Tr}(AB).
\]
There exists a centered Gaussian family
\[
\{X_A:A\in\mathfrak S_2^{\rm sa}(\mathcal H)\}
\]
such that
\begin{equation}
\label{eq:cylindrical-gue-covariance}
\mathbb E[X_A X_B]
=
2\tau^2\operatorname{Tr}(AB),
\end{equation}
where
\[
\tau^2=\frac{\hbar^2D}{2\sigma^2}.
\]
For every finite-dimensional subspace \(\mathcal H_N\), restriction to
Hermitian operators supported in \(\mathcal H_N\) gives exactly the
centered finite-dimensional GUE covariance obtained above.

In particular, the finite-rank Hermitian probes associated with the
localized input states and with their position and momentum tangent
directions have well-defined Gaussian random values in the
infinite-dimensional limit. Thus the Brownian components used in the
reconstruction possess an exact infinite-dimensional probability law,
even though the corresponding isotropic random Hermitian generator
need not exist as an ordinary operator-valued random variable on
\(\mathcal H\).
\end{proposition}

To see this, choose a Hilbert--Schmidt orthonormal basis
\(\{E_\alpha\}_{\alpha\ge1}\) of
\(\mathfrak S_2^{\rm sa}(\mathcal H)\), and let
\(Z_\alpha\) be independent standard real Gaussian variables. If
\[
A=\sum_{\alpha=1}^\infty a_\alpha E_\alpha,
\qquad
\sum_{\alpha=1}^\infty a_\alpha^2
=
\|A\|_{\rm HS}^2<\infty,
\]
define
\[
X_A
=
\sqrt{2}\,\tau
\sum_{\alpha=1}^\infty a_\alpha Z_\alpha .
\]
The series converges in \(L^2\), and
\[
\mathbb E[X_A X_B]
=
2\tau^2
\sum_{\alpha=1}^\infty a_\alpha b_\alpha
=
2\tau^2\operatorname{Tr}(AB).
\]
For operators supported in \(\mathcal H_N\), only the corresponding
finite-dimensional Hermitian coordinates enter, and the resulting
centered Gaussian law is precisely the finite-dimensional GUE law
derived above.

The limiting object is therefore cylindrical: every
Hilbert--Schmidt Hermitian probe has a genuine Gaussian random value,
although there is in general no isotropic Hilbert-space-valued
Hermitian random generator with covariance proportional to the
identity.
The distinction is therefore important. At every finite cutoff, the
generator is an ordinary GUE random Hermitian matrix. In the
infinite-dimensional limit, the Hilbert--Schmidt probes used to resolve
the Brownian components retain an exact cylindrical Gaussian law. The
construction of a canonical probability measure on complete stochastic
paths in the infinite-dimensional projective Hilbert space is a
separate problem, to be discussed in Chapter~\ref{chap:measurement-rm-walk}. The situation may be
summarized as follows:
\[
\begin{array}{rcl}
\text{finite }N
&:&
\text{ordinary GUE random Hermitian generator},
\\[1mm]
N=\infty
&:&
\text{exact cylindrical Gaussian law on Hilbert--Schmidt probes},
\\[1mm]
\text{separate problem}
&:&
\text{canonical measure on full projective state-space paths}.
\end{array}
\]

\subsection*{The Brownian Family from the GUE}

We now verify the converse. Let the projectively relevant part of the
generator have the traceless GUE distribution obtained above. We show
that, when resolved along the localized position and momentum tangent
directions, it reproduces the Brownian family used in the construction.

Use the compactly supported probes introduced above. In one spatial
dimension, choose sufficiently separated centers \(a_j\) and set
\[
e_{2j-1}
=
f_{a_j,R},
\qquad
e_{2j}
=
\widetilde v_{a_j,R},
\]
where \(\widetilde v_{a_j,R}\) is the normalized position tangent at
\(f_{a_j,R}\). By
Eq.~(\ref{eq:compact-support-position-momentum}), the physical position
and momentum components of the state-space increment from
\(f_{a_j,R}\) are, to arbitrary prescribed accuracy, the real and
imaginary parts of
\[
H_{2j,\,2j-1}
=
\left(
e_{2j},
H_N e_{2j-1}
\right).
\]

For the GUE,
\[
\operatorname{Re}H_{2j,\,2j-1},
\qquad
\operatorname{Im}H_{2j,\,2j-1}
\]
are independent centered Gaussian variables with equal variance.
Therefore the corresponding dimensionless position and momentum
components have precisely the Gaussian component law assumed for one
coarse-grained Brownian step. Only the position component is accumulated
in the constrained walk; the momentum component resolves the
complementary imaginary part of the ambient increment and is sampled
anew at the next step.

For distinct \(j\), the nonredundant matrix elements
\[
H_{2j,\,2j-1}
\]
are independent. Hence the resolved increments associated with the
separated localized particles are independent. Since all off-diagonal
GUE matrix elements have the same distribution, the law of a resolved
component is independent of which separated input state and resolving
direction are used. Thus the lifted process reproduces the homogeneous
component independence of the Brownian family.

We now fix the overall scale. For one coarse-grained interval
\(\Delta t\), write
\[
\widehat h_N\Delta t
=
H_N\sqrt{\Delta t},
\]
where \(H_N\) is the Brownian-normalized generator. Thus the Hermitian
increment during one discrete step is
\[
\Delta K_N
=
H_N\sqrt{\Delta t}.
\]

In the continuum limit, \(\Delta t\) is replaced by the infinitesimal
interval \(dt\), and the corresponding stochastic Hermitian increment is
written
\[
dK_N
=
H_N\sqrt{dt}.
\]
The infinitesimal unitary increment is then
\[
U_{dt}
=
\exp\!\left(
-\frac{i}{\hbar}dK_N
\right).
\]
Independent GUE realizations on disjoint time intervals give independent
increments, while the same realized unitary acts simultaneously on all
states.

For the normalized Gaussian position tangent
\[
\widetilde v_{a,j}
=
2\sigma v_{a,j},
\]
the corresponding dimensionless position coordinate is
\[
\widetilde a_j
=
\frac{a_j}{2\sigma}.
\]
If
\[
\operatorname{Var}
\bigl(\operatorname{Re}H_{jk}\bigr)
=
\operatorname{Var}
\bigl(\operatorname{Im}H_{jk}\bigr)
=
\tau_N^2,
\qquad j<k,
\]
then the constrained infinitesimal displacement satisfies
\[
\mathbb E[d\widetilde a_j]
=
0,
\qquad
\mathbb E[d\widetilde a_j\,d\widetilde a_k]
=
\frac{\tau_N^2}{\hbar^2}
dt\,\delta_{jk}.
\]
Since
\[
a_j=2\sigma\widetilde a_j,
\]
we obtain
\[
\mathbb E[da_j]
=
0,
\qquad
\mathbb E[da_j\,da_k]
=
\frac{4\sigma^2\tau_N^2}{\hbar^2}
dt\,\delta_{jk}.
\]
Requiring the Brownian covariance
\[
\mathbb E[da_j\,da_k]
=
2D\,dt\,\delta_{jk}
\]
therefore fixes the GUE scale:
\begin{equation}
\label{eq:GUE-Brownian-scale}
\tau_N^2
=
\frac{\hbar^2D}{2\sigma^2}.
\end{equation}
With this normalization, the constrained process has Brownian motion
with diffusion coefficient \(D\) as its continuum limit. In three
spatial dimensions the same construction uses the three mutually
orthogonal position tangent directions at each localized state and
gives ordinary Brownian motion on \(M_3^\sigma\).

Thus the two constructions are inverse to one another, within the
finite-dimensional approximation and to arbitrary prescribed accuracy.
The homogeneous family of independent Brownian components determines
the traceless GUE law of a common Hermitian generator, while that GUE
generator, resolved along the localized tangent directions and
constrained to the localized manifold, reproduces the original
Brownian family.

\section{Two Finite-Dimensional Illustrations}
\label{sec:two-finite-dimensional-illustrations}

The preceding construction takes place in an infinite-dimensional
Hilbert space and uses finite-dimensional approximations only as a
technical device. It is useful, however, to illustrate some of its
geometric features in simple finite-dimensional settings.

The first example is a helix in \(\mathbb R^3\). It shows how random
increments prescribed along a distinguished submanifold, together with
homogeneity and isotropy of their distributions, can determine a
Gaussian law for an ambient displacement. No common random Hermitian
generator is involved in this example; its purpose is only to
illustrate the reconstruction of an ambient isotropic law from data
specified along a submanifold.
The second example uses the Bloch sphere
\[
\mathbb CP^1\simeq S^2
\]
and provides a direct complex analogue. In this case, isotropic
infinitesimal motions determine the projectively relevant part of a
random Hermitian generator and lead to the traceless \(2\times2\)
Gaussian Unitary Ensemble.

\subsection{A Helix in \(\mathbb R^3\)}
\label{subsec:a-helix-in-r-3}

Consider the helix
\[
\gamma(t)
=
\left(
\cos t,\,
\sin t,\,
\lambda t
\right)
\]
in \(\mathbb R^3\). Its tangent vector is
\[
v(t)
=
\gamma'(t)
=
\left(
-\sin t,\,
\cos t,\,
\lambda
\right).
\]
For two parameter values \(t\) and \(s\),
\[
v(t)\cdot v(s)
=
\cos(t-s)+\lambda^2.
\]
Choosing
\[
\lambda^2=\frac12,
\]
we find that the tangent vectors at
\[
t,\qquad
t+\frac{2\pi}{3},\qquad
t+\frac{4\pi}{3}
\]
are mutually orthogonal.

Since
\[
|v(t)|^2
=
1+\lambda^2
=
\frac32,
\]
the corresponding unit tangent vectors
\[
u(t)
=
\sqrt{\frac{2}{3}}\,v(t)
\]
form an orthonormal basis of \(\mathbb R^3\).

Consider now the possible infinitesimal motions of Brownian particles
situated at different initial points of the helix. At \(\gamma(t)\), a
step tangent to the helix has the form
\[
\xi(t)u(t),
\]
where \(\xi(t)\) is a centered Gaussian random variable. We assume that
the law of this tangent step is invariant under shifts of the helix
parameter,
\[
t\longmapsto t+\tau,
\]
and that steps associated with sufficiently separated initial points are
independent.

We seek an ambient random displacement \(Z\in\mathbb R^3\) whose
projections onto the tangent directions reproduce these prescribed
steps:
\[
\xi(t)
=
u(t)\cdot Z.
\]
Because the tangent field is \(2\pi\)-periodic while the helix itself
advances in the third coordinate, the three required tangent directions
may be realized at points that are also arbitrarily far apart. For
example, with a sufficiently large integer \(m\), take
\[
t_1=t,\qquad
t_2=t+\frac{2\pi}{3}+2\pi m,\qquad
t_3=t+\frac{4\pi}{3}+4\pi m.
\]
Then
\[
u(t_1),\qquad u(t_2),\qquad u(t_3)
\]
form an orthonormal basis of \(\mathbb R^3\), while the corresponding
initial points of the helix may be made sufficiently separated for
their Brownian steps to be independent. The random variables
\[
\xi(t_1),\qquad
\xi(t_2),\qquad
\xi(t_3)
\]
are therefore the components of \(Z\) in this basis.

Covariance of the step law under shifts of the helix parameter makes
these three components identically distributed, while their independence
eliminates the mixed covariances. Hence
\[
\mathbb E[Z]=0,
\qquad
\mathbb E[ZZ^{T}]
=
\sigma_B^2 I
\]
for some \(\sigma_B^2>0\). Since the components are independent centered
Gaussian random variables, they determine a unique Gaussian law for
\(Z\). Its covariance is proportional to the identity, and therefore
\[
Z\overset{d}{=}QZ
\]
for every orthogonal transformation \(Q\in O(3)\). Thus covariance of
the prescribed steps along the helix determines an isotropic Gaussian
law for the ambient displacement.

This conclusion concerns the distribution of steps based at points of
the helix. To obtain a random walk on the whole of \(\mathbb R^3\), one
must additionally require spatial homogeneity: the same ambient step law
is assigned at every point of \(\mathbb R^3\). Equivalently, the ambient
step distribution is invariant under translations of its initial point.
The isotropic Gaussian law obtained from the helix is then extended
homogeneously to the entire ambient space.
The resulting walk is therefore both homogeneous and isotropic.
Repetition with independent increments gives a random walk in
\(\mathbb R^3\) whose diffusive limit is ordinary Brownian motion.

Thus the jointly specified Brownian steps from different initial points
of the helix, together with covariance under shifts of the helix
parameter, determine the isotropic Gaussian law of the ambient
displacement. The additional requirement of spatial homogeneity extends
this same law to every initial point of \(\mathbb R^3\), producing the
usual homogeneous and isotropic Brownian motion on \(\mathbb R^3\). In this model, however,
the ambient motion is specified directly through its random increments.
No common random symmetric or Hermitian operator is assumed to generate
the entire family of steps. Consequently, the argument determines the
ambient Euclidean-invariant step distribution, but does not produce an
ensemble of generators.

\subsection{The Bloch Sphere and the \(2\times2\) GUE}
\label{subsec:the-bloch-sphere-and-the-2-2-gue}

The complex analogue is even more direct. The projective state space of a
two-dimensional complex Hilbert space is
\[
\mathbb CP^1
\simeq
S^2
\simeq
SU(2)/U(1).
\]
A normalized state \(\psi\in\mathbb C^2\), modulo its overall phase, is
represented by its Bloch vector
\[
{\bf n}
=
\left(
\langle\psi,\sigma_1\psi\rangle,\,
\langle\psi,\sigma_2\psi\rangle,\,
\langle\psi,\sigma_3\psi\rangle
\right),
\qquad
|{\bf n}|=1,
\]
which ranges over the unit sphere \(S^2\subset\mathbb R^3\). Here
\[
\sigma_1
=
\begin{bmatrix}
0&1\\
1&0
\end{bmatrix},
\qquad
\sigma_2
=
\begin{bmatrix}
0&-i\\
i&0
\end{bmatrix},
\qquad
\sigma_3
=
\begin{bmatrix}
1&0\\
0&-1
\end{bmatrix}
\]
are the Pauli matrices, and
\(\langle\,\cdot,\cdot\,\rangle\) denotes the inner product on
\(\mathbb C^2\).

Every traceless Hermitian \(2\times2\) Hamiltonian can be written uniquely
in the form
\[
\widehat h
=
\frac{\hbar}{2}
\left(
\omega_1\sigma_1+
\omega_2\sigma_2+
\omega_3\sigma_3
\right)
=
\frac{\hbar}{2}\,
{\boldsymbol\omega}\cdot{\boldsymbol\sigma}.
\]
Thus the projectively relevant part of the Hamiltonian is determined by
the real vector
\[
{\boldsymbol\omega}
=
(\omega_1,\omega_2,\omega_3)
\in\mathbb R^3.
\]
The Schr\"odinger equation generated by \(\widehat h\) induces
\[
\frac{d{\bf n}}{dt}
=
{\boldsymbol\omega}\times{\bf n}
\]
for the Bloch vector. The Hamiltonian therefore generates an
infinitesimal rotation of the Bloch sphere about the axis
\({\boldsymbol\omega}\).

At a single point \({\bf n}\), the infinitesimal motion determines only
the component of \({\boldsymbol\omega}\) perpendicular to \({\bf n}\).
Indeed,
\[
{\bf n}\times\frac{d{\bf n}}{dt}
=
{\bf n}\times
\bigl({\boldsymbol\omega}\times{\bf n}\bigr)
=
{\boldsymbol\omega}
-
({\boldsymbol\omega}\cdot{\bf n}){\bf n}.
\]
The component of \({\boldsymbol\omega}\) parallel to \({\bf n}\) is
invisible at that point, because a rotation about the axis \({\bf n}\)
leaves the point fixed. Two points represented by nonparallel Bloch
vectors \({\bf n}_1\) and \({\bf n}_2\) are sufficient to recover the
full vector \({\boldsymbol\omega}\).

Let
\[
{\bf n}_1,\ {\bf n}_2\in S^2,
\qquad
{\bf n}_1\not\parallel{\bf n}_2,
\]
and suppose that the same Hamiltonian generates the simultaneous
infinitesimal motions
\[
\dot{\bf n}_1
=
{\boldsymbol\omega}\times{\bf n}_1,
\qquad
\dot{\bf n}_2
=
{\boldsymbol\omega}\times{\bf n}_2.
\]
If two vectors \({\boldsymbol\omega}\) and
\({\boldsymbol\omega}'\) produced the same pair of infinitesimal motions,
then
\[
({\boldsymbol\omega}-{\boldsymbol\omega}')
\times{\bf n}_1
=
0,
\qquad
({\boldsymbol\omega}-{\boldsymbol\omega}')
\times{\bf n}_2
=
0.
\]
The first equation implies that
\({\boldsymbol\omega}-{\boldsymbol\omega}'\) is parallel to
\({\bf n}_1\), while the second implies that it is parallel to
\({\bf n}_2\). Since the two Bloch vectors are not parallel, it follows
that
\[
{\boldsymbol\omega}
=
{\boldsymbol\omega}'.
\]
Thus the linear map
\[
{\boldsymbol\omega}
\longmapsto
\left(
{\boldsymbol\omega}\times{\bf n}_1,\,
{\boldsymbol\omega}\times{\bf n}_2
\right)
\]
is injective. Consequently, the joint infinitesimal motion at two points
with nonparallel Bloch vectors determines the full vector
\({\boldsymbol\omega}\), and hence the projectively relevant Hamiltonian
\(\widehat h\), uniquely.

Now suppose that the simultaneous infinitesimal motions at two selected
points with nonparallel Bloch vectors \({\bf n}_1\) and \({\bf n}_2\)
have a centered joint Gaussian distribution compatible with their
generation by a single random Hamiltonian. The two motions are not
independent and are generally correlated. We assume, however, that the
marginal motion at each selected point is isotropic in its tangent plane.

The finite-dimensional analogue of the homogeneity of the Brownian
family is invariance under a common shift of the two points along the
great circle passing through them. Thus we require the same joint law
after such a shift. Let \({\bf k}\) be the unit normal to the plane
spanned by \({\bf n}_1\) and \({\bf n}_2\), and let \(R_\theta\)
denote the rotation about the axis \({\bf k}\). The joint law is assumed
to be unchanged under
\[
({\bf n}_1,{\bf n}_2)
\longmapsto
(R_\theta{\bf n}_1,R_\theta{\bf n}_2),
\]
together with the corresponding rotation of the tangent vectors.

Since the two infinitesimal motions are generated by the same random
Hamiltonian, their joint law cannot be arbitrary. It must be supported
on the three-dimensional subspace
\[
\operatorname{Im}L
\subset
T_{{\bf n}_1}S^2\oplus T_{{\bf n}_2}S^2,
\]
where
\[
L({\boldsymbol\omega})
=
\left(
{\boldsymbol\omega}\times{\bf n}_1,\,
{\boldsymbol\omega}\times{\bf n}_2
\right).
\]
Equivalently, the two motions must have precisely the correlation
required for their simultaneous generation by a single vector
\({\boldsymbol\omega}\). Since \(L\) is linear and injective, it has a
linear inverse on its image. A centered joint Gaussian law supported on
\(\operatorname{Im}L\) therefore induces, through this inverse, a unique
centered Gaussian law for \({\boldsymbol\omega}\).

The construction is equivariant under rotations. Indeed, for a rotation
\(R_\theta\) about \({\bf k}\),
\[
(R_\theta{\boldsymbol\omega})
\times
(R_\theta{\bf n}_j)
=
R_\theta
\bigl(
{\boldsymbol\omega}\times{\bf n}_j
\bigr).
\]
Consequently, invariance of the joint law under the common rotation of
the two points and their tangent directions implies
\[
{\boldsymbol\omega}
\overset{d}{=}
R_\theta{\boldsymbol\omega}.
\]
The covariance
\[
C
=
\mathbb E\!\left[
{\boldsymbol\omega}{\boldsymbol\omega}^{T}
\right]
\]
therefore has the form
\[
C
=
a\left(I-{\bf k}{\bf k}^{T}\right)
+
b\,{\bf k}{\bf k}^{T}.
\]

For either selected point, define
\[
{\bf m}_j
=
{\bf k}\times{\bf n}_j,
\qquad
j=1,2.
\]
Then \(\{{\bf m}_j,{\bf k}\}\) is an orthonormal basis of
\(T_{{\bf n}_j}S^2\). Writing
\[
{\boldsymbol\omega}
=
\alpha_j{\bf n}_j
+
\beta_j{\bf m}_j
+
\gamma{\bf k},
\]
we obtain
\[
\dot{\bf n}_j
=
{\boldsymbol\omega}\times{\bf n}_j
=
-\beta_j{\bf k}
+
\gamma{\bf m}_j.
\]
The variances of these two tangent components are \(a\) and \(b\),
respectively. Isotropy of the infinitesimal motion in the tangent plane
therefore gives
\[
a=b.
\]
Hence
\[
C
=
\kappa I
\]
for some \(\kappa>0\), or equivalently,
\[
\mathbb E[\omega_j]
=
0,
\qquad
\mathbb E[\omega_j\omega_k]
=
\kappa\,\delta_{jk}.
\]

Since \({\boldsymbol\omega}\) is centered Gaussian, the relation
\(C=\kappa I\) implies that its law is rotationally invariant. The same
vector \({\boldsymbol\omega}\) generates the motion at every point of the
Bloch sphere through
\[
\dot{\bf n}
=
{\boldsymbol\omega}\times{\bf n}.
\]
The resulting infinitesimal motion therefore has the same law at every
point and is isotropic in every tangent plane.

The connection with unitary invariance is immediate. For every
\(U\in SU(2)\), there exists a rotation \(R_U\in SO(3)\) such that
\[
U
\left(
{\boldsymbol\omega}\cdot{\boldsymbol\sigma}
\right)
U^{-1}
=
\left(
R_U{\boldsymbol\omega}
\right)\cdot{\boldsymbol\sigma}.
\]
Since
\[
{\boldsymbol\omega}
\overset{d}{=}
R_U{\boldsymbol\omega},
\]
we obtain
\[
\widehat h
\overset{d}{=}
U\widehat h U^{-1}.
\]
Thus rotational invariance of the Gaussian law in the
three-dimensional real coefficient space becomes unitary invariance of
the corresponding Hermitian Hamiltonian.

The traceless Hermitian matrices form a three-dimensional real vector
space with orthogonal basis
\[
\sigma_1,\qquad
\sigma_2,\qquad
\sigma_3.
\]
Because the centered Gaussian coefficients
\(\omega_1,\omega_2,\omega_3\) have covariance
\(\kappa I\), they are independent and have equal variance. They
therefore define the traceless \(2\times2\) GUE law.

A scalar multiple of the identity,
\[
\lambda I,
\]
may be added without changing the projective motion. The projective
construction does not determine the distribution of \(\lambda\), since
the scalar part generates only an overall phase. Adding an independent
centered Gaussian scalar with the variance prescribed by the
conventional GUE normalization gives the usual \(2\times2\) Gaussian
Unitary Ensemble.

This example is analogous to the lift constructed in the preceding
section: infinitesimal motions on a distinguished family of states
determine a Gaussian law for an ambient generator, and symmetry of the
prescribed motions becomes invariance of that law. There is, however,
an important difference. In the general construction, Brownian motions
starting from sufficiently separated localized states are independent,
and an adapted basis allows the corresponding independent position and
momentum components to resolve distinct matrix coordinates of the
Hermitian generator.

On the Bloch sphere, by contrast, the tangent motions at two different
points cannot be independent. The projectively relevant generator has
only three real degrees of freedom, represented by
\(\boldsymbol\omega\), and the same vector must generate both motions:
\[
\dot{\bf n}_1
=
\boldsymbol\omega\times{\bf n}_1,
\qquad
\dot{\bf n}_2
=
\boldsymbol\omega\times{\bf n}_2.
\]
Their joint distribution is therefore constrained by their common
generator. This dependence reflects the finite-dimensional character of
the model: motions at different points must reuse the same three real
components of the common generator. Nevertheless, the compatible joint
Gaussian law of the two motions,
together with common-rotation covariance and isotropy of the marginal
motions, determines an isotropic Gaussian law for
\(\boldsymbol\omega\), and hence the traceless \(2\times2\) GUE law.
The example therefore illustrates the reconstruction of the random
generator and its unitary invariance, but not the independence structure
of the Brownian family used in the general construction.

\section{The Random Matrix Walk on State Space}
\label{sec:the-random-matrix-walk-on-state-space}

We now iterate the coarse-grained stochastic unitary increment. Let
\[
H_1,H_2,\ldots
\]
be independent realizations of the Brownian-normalized generator. For
successive intervals of duration \(\Delta t\), define
\[
\Delta K_n
=
H_n\sqrt{\Delta t}.
\]
The state evolves according to
\[
\psi_{n+1}
=
\exp\!\left(
-\frac{i}{\hbar}\Delta K_n
\right)\psi_n.
\]
We shall call this stochastic unitary evolution the
\emph{random matrix walk}, or simply the \({\bf RM}\) walk.

At every step a single realization of \(H_n\) acts on the entire
finite-dimensional approximation to the Hilbert space. Thus the same
random unitary transformation acts simultaneously on all components of
a state and on all possible localized initial states. Independent GUE
realizations are used only at successive steps.

When the \({\bf RM}\) walk is constrained to \(M_3^\sigma\), only the
position tangent component of each ambient increment is retained. As
shown in Section~\ref{sec:the-gaussian-unitary-lift}, with the scale
(\ref{eq:GUE-Brownian-scale}) the resulting process has ordinary
Brownian motion with diffusion coefficient \(D\) as its continuum
limit. The finite-dimensional cutoff \(N\) is chosen large enough that
the residual overlap floor and reconstruction errors lie below the
required accuracy; no \(N\to\infty\) normalization of the GUE scale is
needed.

Thus Brownian motion on the localized manifold and the \({\bf RM}\)
walk in projective state space are related by constraint and unitary
lift:
\[
\begin{array}{ccc}
\text{Brownian motion on }M_3^\sigma
& \longleftrightarrow &
\text{\({\bf RM}\) walk in projective Hilbert space}.
\end{array}
\]
The construction so far does not identify measurement outcomes or
describe localization of an arbitrary state. These questions require
the unconstrained \({\bf RM}\) dynamics itself and will be taken up in
the next chapter. Before turning to measurement, however, we ask
whether the Brownian family used in the construction above can itself
arise from an ordinary microscopic particle--environment dynamics.

\section{A Particle--Bath Origin of the Brownian Family}
\label{sec:particle-bath-origin-brownian-family}

The derivation of the \({\bf RM}\) walk in the preceding sections takes
the homogeneous Brownian family on the localized manifold as its
classical stochastic input.  We now show that, under a natural set of
assumptions, such a family can arise from an ordinary quantum
particle--bath model.  The role of the bath is not to postulate a new
state-space dynamics, but to generate the classical Brownian statistics
whose common unitary lift was shown above to have the GUE form.

The construction is a spatially local version of the
Caldeira--Leggett oscillator bath.  We first describe its classical
form and then indicate how it arises by restricting the corresponding
quantum dynamics to the localized particle--bath manifold.

Consider a particle with position \(a\), momentum \(p\), and mass \(m\),
coupled to a continuum of harmonic oscillators labelled by their
spatial position \(y\in\mathbb R^3\) and frequency \(\omega>0\).
Let \(Q(y,\omega)\) and \(P(y,\omega)\) denote the oscillator
coordinates and momenta.  We take
\begin{equation}
\label{eq:local-CL-Hamiltonian}
H
=
\frac{p^2}{2m}
+
V(a)
+
\frac12
\int_{\mathbb R^3}d^3y
\int_0^\infty d\omega
\left[
P(y,\omega)^2
+
\omega^2
\left(
Q(y,\omega)
-
\frac{c(\omega)}{\omega^2}
\zeta(a-y)
\right)^2
\right].
\end{equation}
Here \(\zeta\) is a smooth localized coupling profile.  For definiteness,
it may be taken to be rotationally symmetric and compactly supported.
The dependence on \(a-y\) makes the interaction translation invariant.

The corresponding quantum Hamiltonian is obtained by replacing
\(a,p,Q,P\) by the appropriate operators.  Restrict the quantum
particle--bath action to product localized phase-space states
\[
\Psi_{a,p;\,Q,P}
=
\varphi_{a,p}
\otimes
\bigotimes_{y,\omega}
\chi_{Q(y,\omega),P(y,\omega)} ,
\]
where each factor is localized in the corresponding configuration
variable.  As in the finite-dimensional particle case discussed in
Chapter~\ref{chap:newtonian-dynamics-from-schrodinger-dynamics}, the canonical term in the
restricted Schr\"odinger action becomes
\[
p\cdot\dot a
+
\int d^3y\,d\omega\,
P(y,\omega)\dot Q(y,\omega),
\]
while the expectation of the quantum Hamiltonian gives
Eq.~(\ref{eq:local-CL-Hamiltonian}) up to width-dependent constants
and the usual smearing corrections.  Thus the tangent dynamics on the
localized particle--bath manifold is the classical oscillator-bath
dynamics generated by Eq.~(\ref{eq:local-CL-Hamiltonian}).

The bath coordinates satisfy
\begin{equation}
\label{eq:local-bath-oscillator-equation}
\ddot Q(y,\omega,t)
+
\omega^2 Q(y,\omega,t)
=
c(\omega)\zeta(a(t)-y).
\end{equation}
Solving this equation and substituting the result into the particle
equation gives a generalized Langevin equation.  The part depending on
the freely evolving bath coordinates is the random force
\begin{equation}
\label{eq:local-random-force}
\xi_i(a,t)
=
\int d^3y
\int_0^\infty d\omega\,
c(\omega)\,
\partial_i \zeta(a-y)
\left[
Y(y,\omega)\cos\omega t
+
\frac{P(y,\omega,0)}{\omega}\sin\omega t
\right],
\end{equation}
where \(Y(y,\omega)\) denotes the appropriately shifted initial bath
coordinate.

For a thermal Gaussian bath, \(\xi_i(a,t)\) is a centered Gaussian
random field.  Its covariance has the form
\begin{equation}
\label{eq:local-force-covariance}
\mathbb E
\left[
\xi_i(a,t)\xi_j(b,s)
\right]
=
k_B T\,
G(t-s)\,
S_{ij}(a-b),
\end{equation}
where
\begin{equation}
\label{eq:temporal-bath-kernel}
G(\tau)
=
\int_0^\infty
\frac{c(\omega)^2}{\omega^2}
\cos(\omega\tau)\,d\omega
\end{equation}
and
\begin{equation}
\label{eq:spatial-bath-kernel}
S_{ij}(r)
=
\int_{\mathbb R^3}
\partial_i \zeta(y)
\partial_j \zeta(y+r)\,d^3y.
\end{equation}

These formulas already contain the spatial statistical properties
required by the Brownian lift.  If \(\zeta\) is rotationally symmetric,
then at a fixed point
\begin{equation}
\label{eq:local-isotropy}
S_{ij}(0)
=
s_\zeta\,\delta_{ij},
\qquad
s_\zeta
=
\frac13
\int_{\mathbb R^3}
|\nabla \zeta(y)|^2\,d^3y .
\end{equation}
Thus the three Cartesian force components have equal variance and are
uncorrelated.  Since the field is Gaussian, they are independent.
The normalization of \(\zeta\) may be chosen so that \(s_\zeta=1\).

If, in addition, \(\zeta\) has compact support of radius \(R\), then
\begin{equation}
\label{eq:local-spatial-independence}
|a-b|>2R
\quad\Longrightarrow\quad
S_{ij}(a-b)=0.
\end{equation}
Hence the random forces acting on sufficiently separated particles are
independent.  Translation invariance is exact, since the covariance
depends only on \(a-b\).  A rapidly decaying noncompact coupling gives
the same conclusion asymptotically rather than exactly.

The temporal part of the covariance is controlled by the bath spectral
density.  In the Ohmic limit,
\begin{equation}
\label{eq:ohmic-kernel}
G(t-s)
\longrightarrow
2m\gamma\,\delta(t-s),
\end{equation}
and the particle dynamics reduces to the Markovian Langevin system
\begin{equation}
\label{eq:kramers-local-bath}
dp
=
\left[
-\nabla V(a)-\gamma p
\right]dt
+
\sqrt{2m\gamma k_B T}\,dW_t,
\qquad
da
=
\frac{p}{m}\,dt .
\end{equation}
In the overdamped free-particle limit this becomes
\begin{equation}
\label{eq:overdamped-brownian-local-bath}
da
=
\sqrt{2D}\,dW_t,
\qquad
D
=
\frac{k_B T}{m\gamma},
\end{equation}
which is ordinary Brownian motion.

For the lift constructed in Section~\ref{sec:the-gaussian-unitary-lift},
one also needs the position and momentum components resolved during a
single coarse-grained interval.  This structure is already present in
the stationary Ornstein--Uhlenbeck process associated with
Eq.~(\ref{eq:kramers-local-bath}).  In one spatial dimension,
\[
dp
=
-\gamma p\,dt
+
\sqrt{2m\gamma k_B T}\,dW_t,
\qquad
da
=
\frac{p}{m}\,dt,
\]
with
\[
\mathbb E[p(t)p(s)]
=
mk_B T\,e^{-\gamma|t-s|}.
\]
Over an interval of duration \(T_c\), define
\[
\Delta a
=
\frac1m
\int_0^{T_c}p(s)\,ds,
\qquad
\Delta p
=
p(T_c)-p(0).
\]
The pair \((\Delta a,\Delta p)\) is jointly Gaussian, with
\begin{equation}
\label{eq:coarse-position-variance}
\mathbb E[(\Delta a)^2]
=
\frac{2k_B T}{m\gamma}
\left[
T_c
-
\frac{1-e^{-\gamma T_c}}{\gamma}
\right],
\end{equation}
and
\begin{equation}
\label{eq:coarse-momentum-variance}
\mathbb E[(\Delta p)^2]
=
2mk_B T
\left(
1-e^{-\gamma T_c}
\right).
\end{equation}
Moreover,
\begin{equation}
\label{eq:coarse-position-momentum-independence}
\mathbb E[\Delta a\,\Delta p]
=
0.
\end{equation}
Since the pair is jointly Gaussian, the position and momentum
increments are independent.  In three spatial dimensions rotational
symmetry gives
\[
\mathbb E[\Delta a_i\Delta a_j]
=
V_a\,\delta_{ij},
\qquad
\mathbb E[\Delta p_i\Delta p_j]
=
V_p\,\delta_{ij},
\qquad
\mathbb E[\Delta a_i\Delta p_j]
=
0.
\]

Recall the dimensionless phase-space coordinates
\[
\widetilde a_i
=
\frac{a_i}{2\sigma},
\qquad
\widetilde p_i
=
\frac{\sigma p_i}{\hbar}.
\]
Equality of the variances of the normalized position and momentum
components requires
\[
\frac{V_a}{4\sigma^2}
=
\frac{\sigma^2}{\hbar^2}V_p.
\]
Using Eqs.~(\ref{eq:coarse-position-variance}) and
(\ref{eq:coarse-momentum-variance}) gives
\begin{equation}
\label{eq:local-bath-scale-matching}
\sigma^4
=
\frac{\hbar^2}{4m^2\gamma}
\,
\frac{
T_c-\dfrac{1-e^{-\gamma T_c}}{\gamma}
}{
1-e^{-\gamma T_c}
}.
\end{equation}
Thus, for a given localization scale, the coarse-graining interval may
be chosen so that the normalized position and momentum components have
equal variance.  In the regime \(\gamma T_c\gg1\),
\[
V_a
\simeq
2DT_c,
\qquad
V_p
\simeq
2mk_B T,
\]
and
\[
\sigma^4
\simeq
\frac{\hbar^2T_c}{4m^2\gamma}.
\]

The same regime also suppresses correlations between momentum samples
taken in successive coarse-grained intervals by the factor
\(e^{-\gamma T_c}\).  The momentum component may therefore be regarded,
to the accuracy of the coarse-grained description, as being sampled
anew at each step, while the position increments accumulate to produce
the Brownian motion.

We may now compare the particle--bath dynamics with the homogeneous
component independence used in
Section~\ref{sec:the-gaussian-unitary-lift}.  The spatially local bath
provides:

\[
\begin{array}{rcl}
\text{Gaussianity}
&:&
\text{thermal oscillator coordinates are Gaussian},
\\[1mm]
\text{translation invariance}
&:&
\mathbb E[\xi_i(a,t)\xi_j(b,s)]
\text{ depends only on }a-b,
\\[1mm]
\text{isotropy}
&:&
S_{ij}(0)\propto\delta_{ij},
\\[1mm]
\text{independence at separated points}
&:&
S_{ij}(a-b)=0
\text{ for }|a-b|>2R,
\\[1mm]
\text{independent position--momentum components}
&:&
\mathbb E[\Delta a_i\Delta p_j]=0,
\\[1mm]
\text{equal scaled variances}
&:&
\text{Eq.~(\ref{eq:local-bath-scale-matching})}.
\end{array}
\]

These are precisely the statistical properties required in the
finite-dimensional reconstruction of the common stochastic lift.
The usual Euclidean parallel-translation freedom then gives the
independent translation covariance of the input and resolving labels
discussed in Section~\ref{sec:the-gaussian-unitary-lift}.  After the
embedding of the classical manifold into projective Hilbert space,
the corresponding translated directions become distinct ambient
resolving directions, while retaining the same component law.

We therefore obtain the following conditional implication:
\begin{equation}
\label{eq:microscopic-route-to-rm}
\boxed{
\begin{array}{c}
\text{spatially local particle--bath dynamics}
\\
\Downarrow
\\
\text{homogeneous Brownian family on }M_3^\sigma
\\
\Downarrow
\\
\text{common GUE lift}
\\
\Downarrow
\\
{\bf RM}\text{ walk}.
\end{array}
}
\end{equation}

The first arrow in
Eq.~(\ref{eq:microscopic-route-to-rm}) is conditional on the locality,
translation and rotation symmetry, thermal Gaussian statistics,
Ohmic/Markov limit, and coarse-graining assumptions stated above.
The second arrow is the unitary-lift result of
Section~\ref{sec:the-gaussian-unitary-lift}.  The significance of the
construction is therefore not that every environment must produce the
\({\bf RM}\) dynamics, but that the Brownian family used in its
derivation need not be introduced as an independent stochastic
postulate.  For a natural class of local particle--bath models it can
arise from the tangent dynamics of an ordinary Schr\"odinger
particle--environment system.

This observation also sharpens the remaining problem.  A fully
microscopic derivation would require control of the approximations
leading from the full particle--bath Schr\"odinger dynamics to the
Markovian coarse-grained Brownian family, together with a precise
comparison of the corresponding finite-dimensional stochastic lifts.
Within the assumptions made above, however, the chain from ordinary
particle--bath dynamics to the \({\bf RM}\) walk is already explicit.

\chapter{Measurement by the {\bf RM} Walk}
\label{chap:measurement-rm-walk}

The preceding chapter established a correspondence between Brownian
motion on the localized classical manifold and stochastic unitary
motion in projective state space. Brownian motion on \(M_3^\sigma\),
and equivalently on the isometric manifold of localized equivalence
classes \(\widetilde M_3^\sigma\), gives the normal transition law.
The same law, expressed through the Fubini--Study geometry of the
localized states, becomes the Born transition law. We also showed that
the Brownian family admits a common unitary lift whose projectively
relevant random generators have the GUE distribution. Independent
realizations of these generators at successive steps define the
\({\bf RM}\) walk.

The mathematical construction by itself does not imply that this
state-space walk is realized in an actual quantum measurement. That is
the physical question addressed in this chapter. We shall formulate the
corresponding {\bf RM} conjecture and then examine how finite-resolution
detector records arise from the resulting stochastic motion. The main
problem is no longer to derive the Born transition law, which was
obtained in the preceding chapter, but to understand dynamically how
the \({\bf RM}\) walk produces definite recorded outcomes with the
corresponding Born probabilities.

\section{The Classical-to-Quantum Measurement Diagram}
\label{sec:classical-to-quantum-measurement-diagram}

For a classical particle, a position measurement performed over a
finite observation time may be represented schematically by
\[
\begin{gathered}
\text{measurement over a finite time}
\\[3pt]
\Downarrow
\\[3pt]
\text{random walk of the measured variable}
\\[3pt]
\Downarrow
\\[3pt]
\text{normal distribution of outcomes}.
\end{gathered}
\]
In the diffusion limit, the random walk becomes Brownian motion and the
normal distribution becomes its finite-time transition law.

Chapter~\ref{chap:brownian-born} showed how this entire classical picture extends naturally to projective state space.
First, the family of Brownian motions on \(M_3^\sigma\) admits a common
unitary lift. The projectively relevant part of the random Hermitian
generator of this lift has the traceless GUE distribution, and
independent realizations at successive steps produce the \({\bf RM}\)
walk. Conversely, constraining the \({\bf RM}\) walk to
\(M_3^\sigma\) reproduces the corresponding Brownian family.

Second, the normal transition law on the localized manifold is related
geometrically to the Born law. For localized Gaussian states,
\begin{equation}
\label{eq:equal-width-overlap1}
\exp\!\left(
-\frac{|a-b|^2}{4\sigma^2}
\right)
=
\cos^2
\rho_{\rm FS}
\!\left(
g_{a,\sigma},
g_{b,\sigma}
\right).
\end{equation}
The same relation holds on the isometric manifold
\(\widetilde M_3^\sigma\) of localized equivalence classes. Since the Fubini--Study distances between localized Gaussian states, together with their limiting value at infinite spatial separation, cover the full interval \([0,\pi/2]\), and since a unitary-invariant transition law can depend only on this distance, the normal transition law on the localized manifold extends to
\[
P(\psi\to\phi)
=
\cos^2\rho_{\rm FS}(\psi,\phi)
=
|(\psi,\phi)|^2
\]
for arbitrary normalized states.

The resulting classical-to-quantum correspondence may be summarized by the diagram
\begin{equation}
\label{eq:rm-brownian-born}
\begin{array}{ccc}
\text{\({\bf RM}\) walk in projective state space}
&
\longrightarrow
&
\text{Born transition weights}
\\[6pt]
\updownarrow
&&
\updownarrow
\\[6pt]
\text{Gaussian random walks on \(M_3^\sigma\)}
&
\longrightarrow
&
\text{normal transition law}.
\end{array}
\end{equation}
The lower horizontal arrow is the familiar relation between a Gaussian
random walk and its normal transition law. In the diffusion limit, this
is Brownian motion with its heat kernel. The left vertical arrow is the
common unitary lift constructed in
Chapter~\ref{chap:brownian-born}; in the reverse direction, constraint
to \(M_3^\sigma\) recovers the Brownian motion. The right vertical
arrow is the geometric normal-to-Born correspondence expressed by
Eq.~(\ref{eq:equal-width-overlap1}). The upper horizontal arrow is then
the corresponding transition law for the lifted state-space motion.

The diagram is therefore commutative at the level of transition
weights. If the initial and final states lie on the localized
manifold,
\[
\psi_0=g_{a,\sigma},
\qquad
\psi_T=g_{b,\sigma},
\]
the lower path through the diagram gives the Brownian probability of
the transition from \(a\) to \(b\), while the upper path gives the
transition probability between the corresponding rays in projective
state space. These are the same probability law expressed in the
Euclidean and Fubini--Study geometries, respectively. The unitary
invariance of the lifted dynamics extends the latter relation to
arbitrary initial and final rays.

It is important, however, to distinguish this mathematical
correspondence from the physical measurement process. Diagram
~(\ref{eq:rm-brownian-born}) establishes the stochastic dynamics and
its transition weights, but it does not by itself assert that the
\({\bf RM}\) walk is physically realized when a microscopic particle is
measured. Nor does a fixed-time transition weight by itself explain when
a detector record has formed or why that record is effectively definite.
These are the questions to be addressed in this chapter. We first state
the physical hypothesis identifying the \({\bf RM}\) walk with the
state-space dynamics of measurement. We shall then introduce
finite-resolution outcome sectors and use the resulting stochastic
motion to describe the formation and probabilities of detector records.

\section{The RM Conjecture}
\label{sec:the-rm-conjecture}

For a macroscopic particle, interaction with a measuring device or with
its environment may produce a coarse-grained random motion of the
measured position. Under conditions in which this motion is adequately
described by a Brownian random walk, Chapter~\ref{chap:brownian-born}
shows that the corresponding family of classical walks possesses a
common unitary lift to projective state space. This motivates the
following physical conjecture.

\medskip

\begin{conjecturebox}
\noindent
\textbf{RM Conjecture.}
Whenever a position measurement of a macroscopic particle is adequately
represented by Brownian motion of its measured position, the analogous
position measurement of a microscopic particle is adequately represented
by the corresponding \({\bf RM}\) walk in projective state space.
\end{conjecturebox}

\medskip

The conjecture concerns the physical realization of the stochastic
dynamics, not its mathematical form. The \({\bf RM}\) walk itself, its
GUE generators, and its transition weights were obtained in
Chapter~\ref{chap:brownian-born} from the unitary lift of the Brownian
family and from the normal-to-Born relation. In particular, the
transition probability between normalized states is
\[
P(\psi\to\phi)
=
\cos^2\rho_{\rm FS}(\psi,\phi)
=
|(\psi,\phi)|^2.
\]
The conjecture asserts that this mathematically constructed stochastic
motion is the state-space process that takes place during a microscopic
position measurement under the corresponding physical conditions.

Thus instantaneous collapse is replaced by a sequence of stochastic
unitary steps,
\[
\psi_{n+1}
=
e^{-\frac{i\Delta K_n}{\hbar}}\psi_n,
\]
where
\[
\Delta K_n=H_n\sqrt{\Delta t}.
\]
The increments are independent, with the \(H_n\) having the GUE
distribution whose scale is fixed by the corresponding Brownian
diffusion coefficient. The same realized unitary increment acts on the entire
state, including all components of a superposition. No projection is
introduced as part of the stochastic evolution.

The conjecture does not assert that every interaction of a microscopic
particle produces an \({\bf RM}\) walk. It applies when the interaction
distinguishes the particle's position with sufficient strength and
resolution to constitute an effective position measurement. Such an
interaction may be produced by a laboratory apparatus, but it may also
result from coupling to the surrounding environment.
The distinction between microscopic and macroscopic systems is therefore
not that one obeys quantum dynamics while the other obeys a fundamentally
different law. Rather, the same state-space framework contains both.
For a macroscopic system under measurement conditions, the relevant
stochastic motion is effectively confined to the localized classical
manifold and appears as Brownian motion of the measured position. For a
microscopic system, the corresponding stochastic motion explores the
larger projective state space and is described by the \({\bf RM}\) walk.
The precise conditions under which an interaction enters this
measurement regime depend on the strength, frequency, and spatial
resolution of the relevant environmental or detector interactions.
These conditions will be examined quantitatively in
Chapter~\ref{chap:macro-micro-measurement}.

The remaining problem is now dynamical. The Born transition law is
already fixed. We must explain how the \({\bf RM}\) trajectory produces
a definite finite-resolution record and how the probabilities of the
possible records arise from the same stochastic process. We begin by
describing the localized equivalence classes associated with actual
position measurements.

\section{Equivalence Classes and Measurement Outcomes}
\label{sec:equivalence-classes-and-measurement-outcomes}

An actual position measurement has finite resolution and therefore does
not distinguish individual rays in projective Hilbert space. As
discussed in Chapter~\ref{chap:classical-submanifolds}, many different
states may produce the same recorded position. For simplicity, consider
one spatial dimension with coordinate \(z\). For a normalized state
\(\psi\), let
\[
\mu_z(\psi)
=
(\psi,\widehat z\psi)
\]
be its position expectation value and let
\[
\delta_z(\psi)
=
\left(
\psi,
(\widehat z-\mu_z(\psi))^2\psi
\right)^{1/2}
\]
be its position standard deviation. For a detector resolution
\(\sigma\), the localized equivalence class associated with the
recorded position \(c\) is
\[
\{g_c\}
=
\left\{
[\psi]:
\mu_z(\psi)=c,\quad
\delta_z(\psi)\leq\sigma
\right\}.
\]
Thus a position outcome is represented not by a single ray but by a
finite-resolution class of experimentally indistinguishable states. An
\({\bf RM}\) trajectory need not therefore reach a prescribed
localized Gaussian state. It is sufficient that it enter one of the
localized classes recognized by the measuring device.

To identify the state-space directions that are relevant to such a
measurement, consider the map
\[
F:
\mathbb P(L^2(\mathbb R))
\longrightarrow
\mathbb R^2,
\qquad
F([\psi])
=
\bigl(
\mu_z(\psi),\delta_z(\psi)
\bigr),
\]
on the set of normalized states for which these quantities are finite
and \(\delta_z>0\). Since both quantities are independent of the phase
of \(\psi\), the map is well defined on projective state space. A state
will be called regular if
\[
\operatorname{rank}dF=2.
\]
At a regular state, the connected components of the level sets
\[
F^{-1}(\tau,\delta)
\]
are smooth submanifolds \(\mathcal L_\psi\) of codimension two. Their tangent spaces are
\[
T_\psi\mathcal L_\psi
=
\ker dF_\psi.
\]
Motion tangent to such a leaf changes neither the position expectation
value nor the position spread. These directions are therefore
indistinguishable by the position measurement under consideration.

It is useful to verify that regular states are generic in the
finite-dimensional approximations used throughout the construction.
Let
\[
g_k(z)=g_{z_k,\varepsilon}(z),
\qquad
k=1,\ldots,N,
\]
be equal-width Gaussian states whose centers \(z_k\) are sufficiently
separated compared with their common width \(\varepsilon\), and let
\[
V
=
\operatorname{span}_{\mathbb C}
\{g_1,\ldots,g_N\}.
\]
For
\[
\psi
=
\sum_{k=1}^N c_k g_k,
\qquad
p_k=|c_k|^2,
\]
neglecting the exponentially small overlaps gives
\[
\mu_z
\simeq
\sum_{k=1}^N p_k z_k
\]
and
\[
\delta_z^2
\simeq
\varepsilon^2
+
\sum_{k=1}^N p_k z_k^2
-
\mu_z^2.
\]
Hence
\[
d\mu_z
\simeq
\sum_k z_k\,dp_k,
\]
while
\[
d(\delta_z^2)
\simeq
\sum_k z_k^2\,dp_k
-
2\mu_z\,d\mu_z,
\qquad
\sum_k dp_k=0.
\]
For a state involving at least three nonzero Gaussian components
centered at distinct points, these two differentials are linearly
independent. Otherwise a nonzero quadratic polynomial would vanish at
three distinct values \(z_k\). Thus the map has rank two on a nonempty
open set.

More formally, one may use
\[
\widetilde F([\psi])
=
\bigl(
\mu_z(\psi),\delta_z^2(\psi)
\bigr),
\]
which is real analytic in the finite-dimensional approximation. The
singular set is determined by the vanishing of the \(2\times2\) minors
of its Jacobian. Since at least one such minor is not identically zero,
the singular set has empty interior and measure zero. Because
\(\delta_z>0\), the maps \(F\) and \(\widetilde F\) have the same rank.
Thus the regular set is open and dense.

The significance of this foliation is that it separates the two
quantities relevant to a position measurement from the remaining
state-space degrees of freedom. The full \({\bf RM}\) motion is not
restricted to two dimensions. It generally has a component tangent to
the leaf \(\mathcal L_\psi\), which changes the detailed quantum state
without changing either \(\mu_z\) or \(\delta_z\), together with a
transverse component that changes the measurement-relevant variables.

A convenient transverse surface can be constructed through any
normalized state \(\varphi\). Let
\[
\mu=\mu_z(\varphi),
\qquad
\delta=\delta_z(\varphi),
\]
and form the two-parameter family obtained by translating the state and
changing its spatial scale. Introduce coordinates
\[
\tau=\mu_z,
\qquad
s=\ln\frac{\delta_z}{\sigma},
\]
and define
\[
\varphi_{\tau,s}(z)
=
\sqrt{\lambda}\,
\varphi\!\left(
\lambda(z-\tau)+\mu
\right),
\qquad
\lambda
=
\frac{\delta}{\sigma e^s}.
\]
A direct change of variables gives
\[
\mu_z(\varphi_{\tau,s})
=
\tau,
\qquad
\delta_z(\varphi_{\tau,s})
=
\sigma e^s.
\]
Thus the surface
\[
M_\varphi
=
\left\{
[\varphi_{\tau,s}]
:
\tau\in\mathbb R,\ s\in\mathbb R
\right\}
\]
carries precisely the two measurement-relevant coordinates. The
coordinate \(\tau\) gives the position expectation value, while \(s\)
measures the degree of localization relative to the detector
resolution. In particular,
\[
s\leq0
\]
is exactly the localized sector
\[
\delta_z\leq\sigma.
\]

Moreover,
\[
F(\varphi_{\tau,s})
=
\bigl(
\tau,\sigma e^s
\bigr),
\]
so that the restriction of \(F\) to \(M_\varphi\) has rank two.
Consequently, wherever the foliation is regular, \(M_\varphi\) is
transverse to its leaves. Locally,
\[
T_\psi\mathbb P(L^2)
=
\ker dF_\psi
\oplus
T_\psi M_\varphi.
\]
Thus \(M_\varphi\) is not assumed to be an invariant surface of the
\({\bf RM}\) dynamics. Rather, it provides a two-dimensional
representative of the component of the full state-space motion that is
visible to the position measurement.
This distinction will be important below when we consider binary
detectors. A measurement with two possible recorded outcomes does not
imply that the \({\bf RM}\) trajectory is confined to the projective
space of two states. The full trajectory continues to explore the
ambient projective state space, while the two coordinates
\((\tau,s)\) describe the part of that motion relevant to the detector.

We next determine the geometry induced on \(M_\varphi\). Let
\[
y
=
\lambda(z-\tau)+\mu.
\]
Differentiation gives
\[
\frac{\partial\varphi_{\tau,s}}{\partial\tau}
=
-\lambda^{3/2}\varphi'(y)
\]
and
\[
\frac{\partial\varphi_{\tau,s}}{\partial s}
=
-\lambda^{1/2}
\left[
\frac12\varphi(y)
+
(y-\mu)\varphi'(y)
\right].
\]
It is convenient to introduce
\[
\mathcal D_\mu\varphi
=
\frac12\varphi
+
(z-\mu)\varphi'.
\]
Since translation and scaling are implemented by unitary operators on
\(L^2(\mathbb R)\), the corresponding Fubini--Study metric coefficients
have the form
\[
g_{\tau\tau}
=
\lambda^2 A,
\qquad
g_{ss}
=
B,
\qquad
g_{\tau s}
=
\lambda C,
\]
where
\[
A
=
\left\|
\varphi'
\right\|_{FS}^2,
\qquad
B
=
\left\|
\mathcal D_\mu\varphi
\right\|_{FS}^2,
\]
and
\[
C
=
\operatorname{Re}
\left(
\varphi',
\mathcal D_\mu\varphi
\right)_{FS}.
\]
Here, as usual, the projectively irrelevant vertical components are
understood to be removed.

For a superposition of approximately orthogonal equal-width Gaussians,
these expressions simplify considerably. Writing
\[
\varphi
=
\sum_{k=1}^N c_k g_{z_k,\varepsilon},
\]
one has, up to exponentially small overlap corrections,
\[
\mu
\simeq
\sum_{k=1}^N
|c_k|^2z_k,
\qquad
\sum_{k=1}^N|c_k|^2\simeq1,
\]
and
\[
A
\simeq
\frac{1}{4\varepsilon^2}.
\]
Moreover,
\[
C
\simeq
\frac{1}{4\varepsilon^2}
\sum_{k=1}^N
|c_k|^2(z_k-\mu)
=
0.
\]
Thus the position and localization directions are orthogonal on
\(M_\varphi\), to the accuracy of the Gaussian approximation:
\[
g_{\tau s}
\simeq0.
\]

The remaining diagonal coefficient is
\[
B
\simeq
\frac12
+
\frac{1}{4\varepsilon^2}
\sum_{k=1}^N
|c_k|^2(z_k-\mu)^2.
\]
The induced metric therefore takes the form
\[
d\ell^2
\simeq
A\left(\frac{\delta}{\sigma}\right)^2
e^{-2s}\,d\tau^2
+
B\,ds^2.
\]
The coefficients \(A\) and \(B\) depend on the reference state
\(\varphi\), but not on the point \((\tau,s)\) of \(M_\varphi\). The
factor \(e^{-2s}\) expresses the fact that translation of a more
strongly localized state produces a larger Fubini--Study displacement.

After normalization, the two orthogonal directions give the frame
\[
E_\tau
=
\frac{\sigma e^s}{\delta\sqrt A}\,
\frac{\partial}{\partial\tau},
\qquad
E_s
=
\frac{1}{\sqrt B}\,
\frac{\partial}{\partial s}.
\]
These form an orthonormal frame for the measurement-relevant transverse
surface.

Now consider one infinitesimal \({\bf RM}\) increment. Let
\(X_{\rm RM}\) denote its tangent displacement in projective state
space. Since the random Hermitian generator has the GUE distribution,
\(X_{\rm RM}\) is a centered Gaussian tangent vector with an isotropic
distribution. Its components along the orthonormal directions
\(E_\tau\) and \(E_s\),
\[
\xi_\tau
=
\bigl(
X_{\rm RM},E_\tau
\bigr)_{FS},
\qquad
\xi_s
=
\bigl(
X_{\rm RM},E_s
\bigr)_{FS},
\]
are therefore independent centered Gaussian variables with the same
variance.

The two components have different physical meanings. Motion in the
\(\tau\)-direction changes the position expectation value and therefore
determines which position outcome is approached. Motion in the
\(s\)-direction changes the degree of localization and determines
whether the state has entered the localized sector
\[
s\leq0.
\]
The remaining components of the \({\bf RM}\) increment, in particular
those tangent to the leaf \(\ker dF\), change neither of these
measurement coordinates.

Although the normalized \(\tau\)- and \(s\)-components have the same
variance, their coordinate increments do not. From
\[
g_{\tau\tau}
=
A\left(\frac{\delta}{\sigma}\right)^2e^{-2s},
\]
a fixed Fubini--Study displacement in the normalized
\(\tau\)-direction corresponds to a coordinate displacement whose scale
is proportional to \(e^s\). Thus, for fixed \(s\), the \(\tau\)-motion
is homogeneous in \(\tau\), while its coordinate step size decreases
as the state becomes more strongly localized.

When
\[
s\leq0,
\]
the state belongs to a localized equivalence class labelled by its mean
position \(\tau\). The corresponding localized classes form the
one-dimensional manifold
\[
\widetilde M_1^\sigma
=
\left\{
\{g_\tau\}:\tau\in\mathbb R
\right\}.
\]
As shown in Chapters~\ref{chap:classical-submanifolds}
and~\ref{chap:brownian-born}, its metric is represented by the localized
Gaussian states and satisfies
\[
\cos^2
\rho\bigl(
\{g_{\tau_1}\},
\{g_{\tau_2}\}
\bigr)
=
\exp\left[
-\frac{(\tau_1-\tau_2)^2}{4\sigma^2}
\right].
\]
Thus the measurement-relevant \(\tau\)-motion is tied directly to the
same localized-state geometry that relates Brownian motion to the Born
law.

Let us now identify the detector-dependent outcome sectors of localized
equivalence classes in terms of the coordinates \(\tau\) and \(s\). Suppose the detector records finite position
intervals
\[
I_1,\ldots,I_m.
\]
The corresponding outcome sectors are
\[
\mathcal C_j
=
\left\{
[\psi]:
\mu_z(\psi)\in I_j,\quad
\delta_z(\psi)\leq\sigma
\right\},
\]
or, in the transverse coordinates,
\[
\tau\in I_j,
\qquad
s\leq0.
\]
A definite measurement record is produced when the \({\bf RM}\)
trajectory enters one of these detector-defined sectors.

The foliation and the transverse surfaces \(M_\varphi\) therefore give a
reduced geometric picture of the measurement mechanism. The full
\({\bf RM}\) trajectory remains in projective state space, but its
measurement-relevant motion is represented by the two coordinates
\((\tau,s)\). The \(s\)-component controls localization, while the
\(\tau\)-component determines the recorded position. In the following
section we use this reduced description in the simplest binary setting
and show how the two Born probabilities arise as first-entry
probabilities of the corresponding unbiased measurement coordinate.

\section{The Born Rule as a Hitting Probability}
\label{sec:the-born-rule-as-a-hitting-probability}

The preceding section showed that the measurement-relevant part of the
\({\bf RM}\) motion may be represented locally by the transverse
coordinates
\[
\tau=\mu_z,
\qquad
s=\ln\frac{\delta_z}{\sigma}.
\]
The coordinate \(\tau\) specifies the position expectation value, while
\(s\) measures the degree of localization relative to the detector
resolution. In an orthonormal frame on \(M_\varphi\), the infinitesimal
\({\bf RM}\) components in these two directions are independent centered
Gaussian variables with the same variance.
The full \({\bf RM}\) trajectory is not confined to \(M_\varphi\).
Rather, \(M_\varphi\) represents its component transverse to the leaves
on which both \(\mu_z\) and \(\delta_z\) are constant. The remaining
state-space components do not affect the two coordinates relevant to
the position record.

We now consider the simplest detector with two possible outcomes. Let
\(a<b\) denote the two position values associated with the left and
right outcomes. The measurement-relevant \(\tau\)-coordinate then moves
between these two alternatives, while the \(s\)-coordinate determines
the degree of localization of the state.
At first sight, treating \(a\) and \(b\) as absorbing boundaries may
appear to introduce a stochastic mechanism different from the free
Gaussian motion discussed above. This is not the case. The underlying
\({\bf RM}\) dynamics in projective state space remains unchanged.
Effective absorption follows from the geometry of the transverse
coordinates when the state enters a sufficiently localized outcome
region.

To see this, embed the interval \([a,b]\) in a slightly larger interval
\[
[a-\eta,b+\eta],
\qquad
\eta>0,
\]
and take the Hilbert space of the effective one-dimensional model to be
\[
L^2([a-\eta,b+\eta]).
\]
The corresponding \({\bf RM}\) walk is considered in the projective
space
\[
\mathbb P L^2([a-\eta,b+\eta]).
\]
The space \(L^2([a-\eta,b+\eta])\) is naturally embedded in
\(L^2(\mathbb R)\) by extending its elements by zero outside the
interval. The corresponding measurement surface may therefore be
regarded as the part of the previously introduced surface
\(M_\varphi\) for which
\[
a-\eta\leq\tau\leq b+\eta.
\]
In the effective model this surface is bounded in the
\(\tau\)-direction, with the outer endpoints determined by the chosen
position interval.

Regard
\[
[a-\eta,a]
\qquad\text{and}\qquad
[b,b+\eta]
\]
as the left and right outcome layers. For any normalized state supported
in \([a-\eta,b+\eta]\), with position mean \(\tau\), the position
variance satisfies
\begin{equation}
\label{eq:bounded-interval-variance}
\delta_z^2
\leq
(b+\eta-\tau)(\tau-a+\eta).
\end{equation}
At either threshold, \(\tau=a\) or \(\tau=b\), this gives
\[
\delta_z^2
\leq
\eta(b-a+\eta).
\]
Moreover, throughout either outcome layer,
\[
\delta_z^2
\leq
\eta(b-a+2\eta).
\]
Thus, by taking \(\eta\) sufficiently small, entry into either outcome
layer forces
\[
\delta_z\ll\sigma,
\qquad
s=\ln\frac{\delta_z}{\sigma}\ll0.
\]

On the transverse surface \(M_\varphi\), the metric coefficient in the
position direction is
\[
g_{\tau\tau}
=
A\left(\frac{\delta}{\sigma}\right)^2e^{-2s}.
\]
Hence strong localization makes \(g_{\tau\tau}\) large. A fixed
infinitesimal Fubini--Study displacement in the normalized
\(\tau\)-direction then corresponds to a coordinate displacement
\[
|d\tau|
\sim
\frac{d\ell}{\sqrt{g_{\tau\tau}}},
\]
which becomes arbitrarily small as \(s\) decreases. The state-space walk
continues, but its recorded position coordinate becomes effectively
frozen.

We therefore choose the \({\bf RM}\) steps sufficiently small and
\(\eta\) so that
\[
|\Delta\tau|\ll\eta\ll b-a,
\]
while \(\eta\) is small enough for the outcome layers to lie in the
strongly localized regime. A trajectory then enters an outcome layer
before it can jump across it, and subsequent motion of the
\(\tau\)-coordinate is strongly suppressed. In this regime, \(a\) and
\(b\) act as effective absorbing thresholds of the same underlying
\({\bf RM}\) dynamics.
The free Gaussian law and the absorbing-boundary description are
therefore two forms of the same stochastic motion. For an unrestricted
\(\tau\)-walk, the distribution after a prescribed number of steps is
normal and, through the Fubini--Study embedding, gives the general Born
transition law. When the accessible interval is bounded by two
detector-defined outcome layers, the steps of the walk become
progressively smaller near the boundaries, and the same unbiased motion
reduces to an effective absorbing-boundary problem. Its first-entry
probabilities then give the corresponding binary Born probabilities and
stable detector records.

This behavior also has a direct state-space interpretation. Restricting
the measurement coordinate to the interval \([a,b]\) changes the
underlying Hilbert space from
\[
L^2(\mathbb R)
\]
to
\[
L^2([a,b]).
\]
Accordingly, the \({\bf RM}\) walk now takes place in the projective
state space
\[
\mathbb P(L^2([a,b])).
\]
The stochastic law of the \({\bf RM}\) evolution is unchanged; what
changes is the geometry of the localized manifold in the new state
space. The squeezing of the effective \(\tau\)-steps near the
boundaries is the coordinate manifestation of this modified
state-space geometry.

Let \(g_a\) and \(g_b\) be approximately orthogonal localized states
associated with the two detector outcomes, and let
\[
\varphi
=
\alpha g_a+\beta g_b,
\qquad
|\alpha|^2+|\beta|^2=1.
\]
Neglecting the exponentially small overlap between \(g_a\) and \(g_b\),
the initial value of the measurement coordinate is
\[
\tau_0
=
\mu_z(\varphi)
=
|\alpha|^2a+|\beta|^2b.
\]
While the walk remains between \(a\) and \(b\), its \(\tau\)-component
has no preferred direction. The local step size may depend on the
localization coordinate \(s\), but the positive and negative increments
remain equally probable. Equivalently, the stopped coordinate
\(\tau_n\) is a martingale:
\[
\mathbb E
\left[
\tau_{n+1}-\tau_n
\mid
\mathcal F_n
\right]
=
0.
\]
The usual gambler's-ruin argument therefore applies.

Let \(P_a\) and \(P_b\) be the probabilities that the walk reaches
\(a\) or \(b\) first. Since the steps are taken small enough that
overshoots may be neglected,
\[
P_a+P_b=1,
\]
and conservation of the mean position for the unbiased stopped walk
gives
\[
aP_a+bP_b=\tau_0.
\]
Hence
\[
P_b
=
\frac{\tau_0-a}{b-a},
\qquad
P_a
=
\frac{b-\tau_0}{b-a}.
\]
Using
\[
\tau_0
=
|\alpha|^2a+|\beta|^2b,
\]
we obtain
\[
P_a=|\alpha|^2,
\qquad
P_b=|\beta|^2.
\]

Thus the Born weights are the gambler's-ruin probabilities for the
measurement coordinate. Their inequality does not arise from a bias in
the \({\bf RM}\) dynamics. The projected walk is locally unbiased; the
different probabilities are determined entirely by its initial
position,
\[
\tau_0
=
|\alpha|^2a+|\beta|^2b,
\]
or, equivalently, by the Fubini--Study distances from the initial state
to the two outcome states.
The dependence of the local step size on \(s\) changes the number of
steps typically required to reach an outcome, but not the relative
probabilities of reaching the two absorbing thresholds.

\subsection{The Two-State Bloch-Sphere Representation}

The preceding derivation used the transverse coordinates
\((\tau,s)\) on \(M_\varphi\) to describe the part of the full
\({\bf RM}\) dynamics that is relevant to a binary position
measurement. There is also a useful finite-dimensional representation
of the same two-outcome geometry.

Let
\[
\varphi
=
\alpha g_a+\beta g_b,
\qquad
|\alpha|^2+|\beta|^2=1,
\]
where \(g_a\) and \(g_b\) are approximately orthogonal localized states
associated with the two detector outcomes. Their span may be identified
with
\[
\mathbb C^2
=
\operatorname{span}\{g_a,g_b\},
\]
and hence its projectivization with the Bloch sphere,
\[
\mathbb P(\mathbb C^2)\simeq S^2.
\]

The corresponding Bloch coordinates are
\[
x
=
\alpha\overline{\beta}
+
\overline{\alpha}\beta,
\]
\[
y
=
i\left(
\alpha\overline{\beta}
-
\overline{\alpha}\beta
\right),
\]
and
\[
z
=
|\beta|^2-|\alpha|^2,
\]
with
\[
x^2+y^2+z^2=1.
\]

Neglecting the exponentially small overlap between \(g_a\) and \(g_b\),
the position expectation value is
\[
\tau
=
|\alpha|^2a+|\beta|^2b.
\]
Since
\[
|\alpha|^2=\frac{1-z}{2},
\qquad
|\beta|^2=\frac{1+z}{2},
\]
we obtain
\[
\tau
=
\frac{a+b}{2}
+
\frac{b-a}{2}\,z,
\]
or equivalently,
\[
z
=
\frac{2\tau-a-b}{b-a}.
\]
Thus
\[
\tau=a
\quad\longleftrightarrow\quad
z=-1,
\qquad
\tau=b
\quad\longleftrightarrow\quad
z=1.
\]
The two detector outcomes are therefore represented by the two poles of
the Bloch sphere.

The same representation also displays the relation between the
measurement coordinate and the spread of the two-state superposition.
Neglecting the intrinsic widths of \(g_a\) and \(g_b\),
\[
\delta_z^2
=
|\alpha|^2|\beta|^2(b-a)^2.
\]
Since
\[
x^2+y^2
=
4|\alpha|^2|\beta|^2,
\]
it follows that
\[
\delta_z^2
=
\frac{(b-a)^2}{4}
\left(
x^2+y^2
\right).
\]
Using
\[
x^2+y^2=1-z^2,
\]
we have
\[
\delta_z^2
=
\frac{(b-a)^2}{4}
\left(
1-z^2
\right).
\]
Hence
\[
\delta_z\longrightarrow0
\qquad
\text{as}
\qquad
z\longrightarrow\pm1.
\]
This is the finite-dimensional analogue of the localization behavior
described on \(M_\varphi\): approach to a definite outcome is
accompanied by decreasing position spread. On the Bloch sphere, the
limiting outcome states are the poles, where the transverse coordinates
\(x\) and \(y\) vanish.

The Bloch sphere should not, however, be identified with the state space
of the full measurement dynamics. A binary detector has two possible
records, but the \({\bf RM}\) trajectory itself remains in the ambient
projective Hilbert space and generally explores directions outside
\(\operatorname{span}\{g_a,g_b\}\). In particular, on the strict
two-state sphere the position expectation and position spread are not
independent coordinates:
\[
\delta_z^2
=
(\tau-a)(b-\tau).
\]
Thus the rank-two foliation used above applies to the full
finite-dimensional approximation, not to the restricted Bloch sphere.
The Bloch sphere is therefore used here only as a compact representation
of the binary outcome geometry. It displays the two limiting outcome
states, the location of the initial superposition between them, and the
associated decrease of position spread as either outcome is approached,
while the actual stochastic mechanism remains the full \({\bf RM}\)
walk described by the transverse coordinates \((\tau,s)\).

\section{The Wiener-Induced Measure on \texorpdfstring{\({\bf RM}\)}{RM}
Histories}
\label{sec:a-path-measure-from-transition-weights}

The preceding sections described the \({\bf RM}\) dynamics and
illustrated, in the binary case, how a definite detector record may
arise as a first-entry event. To treat general detector outcomes and
sequences of records, we now construct the probability measure on the
underlying \({\bf RM}\) histories.
The construction is inherited from the Brownian family on the localized
classical manifold. It does not require a translation-invariant volume
measure on the full projective Hilbert space. Instead, the complete
family of Brownian motions on \(M_3^\sigma\) determines the probability
law of the random Hermitian generators through the unitary lift
constructed in Chapter~\ref{chap:brownian-born}. The same realized
sequence of generators then acts on an arbitrary initial quantum state.

We begin with the finite-dimensional approximation used in the
construction of the unitary lift.  For one ambient step of the lifted
dynamics, the corresponding constrained position and momentum increments
provide a collection of independent real Gaussian components,
\[
\xi=(\xi_1,\ldots,\xi_{d_N}),
\]
from which the common projectively relevant Hermitian generator is
reconstructed. The vector \(\xi\) is therefore not itself the ambient
state-space increment; once the generator is reconstructed, its action
determines the ambient tangent increment at every state.

Let \(\gamma\) denote the joint centered Gaussian probability measure of
the components of \(\xi\). Chapter~\ref{chap:brownian-born} constructed
the Hermitian generator through a linear map
\[
L:\xi\longmapsto H.
\]
Let \(\mu_{\rm GUE}\) denote the probability measure of the Gaussian
Unitary Ensemble on the space of projectively relevant Hermitian
generators. The statement
that the reconstructed generator has the GUE distribution is then
\begin{equation}
\label{eq:gue-push-forward}
L_*\gamma
=
\mu_{\rm GUE},
\end{equation}
where \(L_*\gamma\) denotes the push-forward of \(\gamma\) under \(L\).
Thus, for every measurable set \(A\) of Hermitian matrices,
\[
\mu_{\rm GUE}(A)
=
\gamma\bigl(L^{-1}(A)\bigr).
\]
Equivalently,
\[
\xi\sim\gamma
\qquad\Longrightarrow\qquad
H=L(\xi)\sim\mu_{\rm GUE},
\]
where \(\xi\sim\gamma\) means that the random vector \(\xi\) has
probability distribution \(\gamma\), while
\(H\sim\mu_{\rm GUE}\) means that the reconstructed Hermitian generator
has the GUE probability distribution \(\mu_{\rm GUE}\). This is not an additional probability law imposed on the
\({\bf RM}\) dynamics. It is the probability law obtained from the
Brownian family by the unitary reconstruction of the preceding chapter.

For successive steps, let
\[
\Xi_1,\Xi_2,\ldots
\]
be independent copies of the random vector \(\xi\). Define
\[
\Omega_B
=
(\mathbb R^{d_N})^{\mathbb N}
\]
with the product probability measure
\[
\mathbb P_B
=
\gamma^{\otimes \mathbb N}.
\]
For notational simplicity, the
dependence of \(\gamma\), \(L\), \(\mu_{\rm GUE}\), and
\(\mathbb P_B\) on the finite-dimensional cutoff \(N\) will usually be
suppressed.

An element
\[
\omega
=
(\Xi_1,\Xi_2,\ldots)
\in\Omega_B
\]
is one realization of the Gaussian component history. An observation
interval containing \(N_T\) steps uses its first \(N_T\) components.
Set
\[
H_n(\omega)=L(\Xi_n),
\qquad
n=1,\ldots,N_T.
\]
By Eq.~\eqref{eq:gue-push-forward},
\[
(H_1,\ldots,H_{N_T})_*\mathbb P_B
=
\mu_{\rm GUE}^{\otimes N_T}.
\]
Thus the probability space inherited from the Brownian family induces
the product GUE law of the independent generators defining the
\({\bf RM}\) walk.

Using the stochastic scaling introduced in
Chapter~\ref{chap:brownian-born}, write
\[
\Delta K_n
=
H_n\sqrt{\Delta t}.
\]
For each realization \(\omega\), define
\[
U_n(\omega)
=
e^{-\frac{i\Delta K_n(\omega)}{\hbar}}
\cdots
e^{-\frac{i\Delta K_1(\omega)}{\hbar}}.
\]
The same realized operator \(U_n(\omega)\) acts on every state. Hence,
for an arbitrary normalized initial state \(\psi\),
\begin{equation}
\label{eq:rm-path-from-omega}
\psi_n(\omega)
=
U_n(\omega)\psi
\end{equation}
defines an \({\bf RM}\) history in projective state space.

The finite-dimensional \({\bf RM}\) approximations therefore carry
ordinary probability measures, obtained by pushing forward the product
GUE law of their generators through the corresponding stochastic
unitary evolution. Thus, at every finite cutoff for which the reconstruction is made,
there exists a genuine stochastic unitary measure whose restriction to
the localized classical manifold reproduces the corresponding Gaussian
random walk and, in the diffusion limit, its Brownian motion.
No homogeneous probability measure on the infinite-dimensional
projective Hilbert space is assumed in this construction. Indeed, the
finite-dimensional invariant measures have no direct analogue as a
normalized homogeneous measure in infinite dimension. What remains to
be understood is whether the finite-dimensional stochastic unitary
measures admit a natural infinite-dimensional limit on an appropriate
space of state-space paths. The cylindrical Gaussian law obtained
earlier for Hilbert--Schmidt probes provides the corresponding
infinite-dimensional description at the level of the generators, but
does not by itself produce such a path measure.

The initial state need not belong to \(M_3^\sigma\). In particular, if
it is written as a finite superposition of localized states,
\[
\psi
=
\sum_{k=1}^r c_k g_{a_k,\sigma},
\]
then
\[
\psi_n(\omega)
=
\sum_{k=1}^r
c_k\,U_n(\omega)g_{a_k,\sigma}.
\]
No statistical mixture
\[
\sum_k |c_k|^2\,\mathbb P_{a_k}^M
\]
is introduced. The coefficients \(c_k\) do not enter the probability
measure \(\mathbb P_B\); they enter only through the coherent initial
state on which the common realized unitary acts.
Nor does the construction depend on a particular decomposition of
\(\psi\). If
\[
\psi
=
\sum_k c_k g_{a_k,\sigma}
=
\sum_\ell d_\ell g_{b_\ell,\sigma},
\]
then, for every \(\omega\),
\[
U_n(\omega)\psi
=
\sum_k c_k U_n(\omega)g_{a_k,\sigma}
=
\sum_\ell d_\ell U_n(\omega)g_{b_\ell,\sigma}.
\]
The \({\bf RM}\) trajectory therefore depends only on the initial state
and on the realized generator history, not on the localized expansion
used to represent that state.

For a fixed initial state \(\psi\), the map
\[
\omega
\longmapsto
\bigl(
[\psi_0(\omega)],
[\psi_1(\omega)],
\ldots,
[\psi_{N_T}(\omega)]
\bigr)
\]
pushes \(\mathbb P_B\) forward to a probability measure on discrete
\({\bf RM}\) histories. We denote this induced measure by
\[
\mathbb P_{\bf RM}^{\psi}.
\]
In the continuous limit, the corresponding constrained histories give
the Wiener path measure described below.

For a localized initial state,
\[
\psi=g_{a,\sigma},
\]
the construction reproduces ordinary Brownian motion when the
\({\bf RM}\) dynamics is restricted to \(M_3^\sigma\) by retaining, at
each step, only the component of the ambient increment tangent to this
manifold. The resulting state remains on \(M_3^\sigma\), and its
position increment is precisely the centered Gaussian increment from
which the unitary lift was constructed. Repetition of these tangent
increments gives the corresponding Gaussian random walk and, in the
diffusion limit, Brownian motion starting from \(a\).

For recording times
\[
0=t_0<t_1<\cdots<t_n
\]
and measurable spatial regions
\[
B_1,\ldots,B_n\subset\mathbb R^3,
\]
the corresponding cylinder probabilities are
\[
\begin{aligned}
P_a(B_1,\ldots,B_n)
={}&
\int_{B_1}\cdots\int_{B_n}
K_{t_1}(a,a_1)
K_{t_2-t_1}(a_1,a_2)
\cdots
\\
&\qquad\qquad\cdots
K_{t_n-t_{n-1}}(a_{n-1},a_n)
\,d^3a_1\cdots d^3a_n,
\end{aligned}
\]
where \(K_t(a,b)\) is the Brownian transition kernel. The
Chapman--Kolmogorov relation
\[
K_{t+s}(a,c)
=
\int_{\mathbb R^3}
K_t(a,b)K_s(b,c)\,d^3b
\]
ensures the consistency of these finite-dimensional distributions and
therefore defines the usual Wiener measure on paths starting from \(a\).  We denote this measure by
\(
\mathbb P_a^M.
\)
Thus the path-space construction agrees
with ordinary Brownian probability whenever the initial state is
localized and the motion is constrained to \(M_3^\sigma\).

We now pass from \({\bf RM}\) histories to detector records. Let
\[
\mathcal C_1,\ldots,\mathcal C_m
\]
be mutually distinguishable finite-resolution outcome sectors. For a
fixed initial state \(\psi\), define the record map
\[
R_\psi:
\Omega_B
\longrightarrow
\{1,\ldots,m,\varnothing\}.
\]
The value
\[
R_\psi(\omega)=j
\]
means that the \({\bf RM}\) history generated by \(\omega\) produces the
detector record \(j\), while
\[
R_\psi(\omega)=\varnothing
\]
means that none of the specified records is formed during the
observation interval.

For a single measurement, \(R_\psi\) may be defined by the first stable
entry of the \({\bf RM}\) trajectory into one of the detector sectors
\(\mathcal C_j\). The probability of record \(j\) is then
\begin{equation}
\label{eq:record-probability}
\mathbb P_\psi^{\rm rec}(j)
=
\mathbb P_B
\left\{
\omega:
R_\psi(\omega)=j
\right\}.
\end{equation}
More generally, if \(A\) is a set of possible records,
\[
\mathbb P_\psi^{\rm rec}(A)
=
\mathbb P_B
\left(
R_\psi^{-1}(A)
\right),
\]
or, in push-forward notation,
\[
\mathbb P_\psi^{\rm rec}
=
(R_\psi)_*\mathbb P_B.
\]

If the specified detector alternatives are not exhaustive, let
\[
D_\psi
=
\{\omega:R_\psi(\omega)\neq\varnothing\}.
\]
Conditioning on runs in which one of the specified records is formed
gives
\[
\mathbb P_\psi^{\rm obs}(j)
=
\mathbb P_B
\left(
R_\psi=j
\,\middle|\,
D_\psi
\right)
=
\frac{
\mathbb P_B\{R_\psi=j\}
}{
\mathbb P_B(D_\psi)
}.
\]
For an exhaustive detector on the relevant observation time scale,
\[
\mathbb P_B(D_\psi)\simeq1,
\]
and the conditioning becomes immaterial.

The path measure constructed here must be consistent with the
fixed-time transition weights already determined in
Chapter~\ref{chap:brownian-born}. Let
\[
\mathcal A_j(T;\psi)
=
\left\{
\omega:
[\psi_T(\omega)]\in\mathcal C_j
\right\}
\]
be the event that the state lies in the \(j\)-th detector sector at the
observation time \(T\). The path-space construction assigns it the
probability
\[
\mathbb P_B\bigl(\mathcal A_j(T;\psi)\bigr).
\]
Independently, the normal-to-Born relation determines the corresponding
detector transition weight,
\[
W^{\rm Born}_j(\psi).
\]

For each finite cutoff \(N\), the stochastic unitary construction
defines a genuine probability measure on \({\bf RM}\) histories. Its
restriction to the finite-dimensional approximation of
\(M_3^\sigma\) reproduces the corresponding Gaussian random walk. As
\(N\) increases, with the projective step size kept fixed, or more
generally uniformly bounded under the stochastic scaling, this
restriction approaches the Brownian family on \(M_3^\sigma\). Its
normal transition law is expressed, through the normal-to-Born
correspondence, by the corresponding Born weight. Since the ambient
transition law is unitarily invariant and therefore depends only on
Fubini--Study distance, the same relation extends to arbitrary rays.
Consequently, in the controlled large-\(N\) limit, the probabilities
\[
\mathbb P_{B,N}\bigl(\mathcal A_j(T;\psi)\bigr)
\]
approach the corresponding Born transition weights,
\begin{equation}
\label{eq:path-born-compatibility}
\mathbb P_{B,N}\bigl(\mathcal A_j(T;\psi)\bigr)
\longrightarrow
W^{\rm Born}_j(\psi).
\end{equation}
For an outcome represented by an orthogonal Hilbert-space sector with
projection \(P_j\),
\[
W^{\rm Born}_j(\psi)
=
\|P_j\psi\|^2.
\]
The compatibility with the Born transition weights is therefore not an
additional condition imposed on the finite-dimensional path measures.
It is recovered in the same controlled large-\(N\) limit in which the
finite-dimensional \({\bf RM}\) approximations recover the Brownian
lift.

This complements the infinite-dimensional result obtained earlier for
the generators. At the generator level, the finite-dimensional GUE
laws extend to an exact cylindrical Gaussian law on Hilbert--Schmidt
probes. At the path level, every finite cutoff carries a genuine
stochastic unitary measure, whose transition weights approach the Born
law as \(N\to\infty\). What remains is the stronger question of whether
these finite-dimensional path measures themselves admit a natural
infinite-dimensional limit on projective state-space paths.

The distinction between fixed-time occupation and the formation of a
stable record remains important. The event
\[
\mathcal A_j(T;\psi)
\]
asks whether the state lies in \(\mathcal C_j\) at the prescribed time
\(T\), whereas
\[
R_\psi=j
\]
asks whether the \(j\)-th detector record has formed. The binary model
of the preceding section illustrated how these events become
effectively equivalent when return from an entered outcome sector
during the observation interval is negligible.

For sufficiently large cutoff \(N\), and in the effective-absorption
regime,
\[
\mathbb P_{B,N}\{R_\psi=j\}
\simeq
\mathbb P_{B,N}\bigl(\mathcal A_j(T;\psi)\bigr).
\]
Since the latter probability approaches the corresponding Born weight
in the controlled large-\(N\) limit,
\begin{equation}
\label{eq:record-born-weight}
\mathbb P_{B,N}\{R_\psi=j\}
\simeq
W^{\rm Born}_j(\psi)
\end{equation}
for sufficiently large \(N\).
No projection is inserted into the \({\bf RM}\) evolution when a record
is formed. If the trajectory enters an outcome sector
\(\mathcal C_j\), the subsequent unitary motion begins from the state
actually reached there. The detector distinguishes the
finite-resolution sector but need not resolve the precise ray within
that sector.

The same probability space also describes sequences of records. If
\[
R_\psi^{(n)}(\omega)
=
(j_1,\ldots,j_n)
\]
denotes the first \(n\) detector records generated by the trajectory,
then
\[
\mathbb P_\psi^{\rm rec}
(j_1,\ldots,j_n)
=
\mathbb P_B
\left\{
\omega:
R_\psi^{(n)}(\omega)
=
(j_1,\ldots,j_n)
\right\}.
\]
Thus all finite record histories are obtained by push-forward of the
single probability measure \(\mathbb P_B\). No new stochastic law is
introduced after a record has formed.

The construction may therefore be summarized schematically as
\[
\begin{array}{c}
\text{Gaussian/Wiener probability law for the constrained walks on }
M_3^\sigma
\\[2mm]
\downarrow\ \text{unitary reconstruction}
\\[2mm]
\text{GUE probability law for the random generators}
\\[2mm]
\downarrow\ \text{common unitary evolution}
\\[2mm]
\text{\({\bf RM}\) history from an arbitrary initial state \(\psi\)}
\\[2mm]
\downarrow\ \text{record map}
\\[2mm]
\text{probability measure on detector-record histories}.
\end{array}
\]

The important point is that the initial state need not be localized and
is never replaced by a statistical mixture of localized states. The
Brownian probability law enters through the complete family of
localized walks used to reconstruct the common random Hermitian
generator. Once this generator is reconstructed, the same realized
unitary history acts linearly on the arbitrary initial state.

\section{Detector-Path Weights and Interference}
\label{sec:detector-class-weights-and-interference}

We now return to a detector with many possible outcomes. For each detector sector, the probability of the corresponding
record-defined family of \({\bf RM}\) histories is related to the Born
sector weight by Eq.~\eqref{eq:record-born-weight}. This allows us to
treat interference and recorded alternatives within the same
path-space framework.

Suppose that
the detector records one of the position intervals
\[
I_1,\ldots,I_N,
\]
with corresponding outcome sectors
\[
\mathcal C_j
=
\left\{
[\phi]:
\mu_z(\phi)\in I_j,\quad
\delta_z(\phi)\leq\sigma
\right\}.
\]
For an initial state \(\psi\), let
\[
\Gamma_j(\psi)
=
\left\{
\omega:
R_\psi(\omega)=j
\right\}
\]
be the family of \({\bf RM}\) histories that produce the detector
record \(j\). Its probability is
\begin{equation}
\label{eq:detector-path-weight}
W_j(\psi)
=
\mathbb P_B\bigl(\Gamma_j(\psi)\bigr).
\end{equation}
In the effective-absorption regime discussed above, the path-space
compatibility condition gives
\[
W_j(\psi)
\simeq
W_j^{\rm Born}(\psi),
\]
where \(W_j^{\rm Born}\) is the fixed-time transition weight determined
by the normal-to-Born relation. In the ideal measurement limit, this becomes the usual Born
probability for the outcome.

To connect this result with the standard spectral description, let
\(\mathcal H_j\) be the closed Hilbert-space sector representing the
\(j\)-th outcome and let \(P_j\) be the orthogonal projection onto
\(\mathcal H_j\). For a normalized state \(\psi\), define the
Fubini--Study distance from \([\psi]\) to the projective subspace
\(\mathbb P(\mathcal H_j)\) by
\[
\rho\bigl(\psi,\mathbb P(\mathcal H_j)\bigr)
=
\inf_{\phi\in\mathcal H_j,\ \|\phi\|=1}
\rho_{\rm FS}(\psi,\phi).
\]
If \(P_j\psi\neq0\), the closest ray in \(\mathbb P(\mathcal H_j)\) is
represented by
\[
\frac{P_j\psi}{\|P_j\psi\|},
\]
and therefore
\[
\begin{aligned}
\cos^2
\rho\bigl(\psi,\mathbb P(\mathcal H_j)\bigr)
&=
\sup_{\phi\in\mathcal H_j,\ \|\phi\|=1}
|(\phi,\psi)|^2
\\
&=
\|P_j\psi\|^2.
\end{aligned}
\]
If \(P_j\psi=0\), the distance is \(\pi/2\), and the same formula
remains valid.
Hence the detector-path probability is
\begin{equation}
\label{eq:detector-sector-born}
W_j(\psi)
=
\mathbb P_B\bigl(\Gamma_j(\psi)\bigr)
\simeq
\|P_j\psi\|^2
=
\cos^2
\rho\bigl(\psi,\mathbb P(\mathcal H_j)\bigr).
\end{equation}
Thus the probability of the record-defined family of \({\bf RM}\)
histories is fixed geometrically by the position of the initial ray
relative to the projective subspace associated with that outcome.

For a position measurement with detector bin \(I_j\), the corresponding
spectral projection is
\[
P_{I_j}\psi(z)
=
\chi_{I_j}(z)\psi(z),
\]
where \(\chi_{I_j}\) is the characteristic function of \(I_j\). Its
range is
\[
H_{I_j}
=
\left\{
\phi\in L^2(\mathbb R):
\phi(z)=0
\ \text{for almost every }z\notin I_j
\right\}.
\]
Therefore
\begin{equation}
\label{eq:position-detector-path-weight}
W_j(\psi)
\simeq
\|P_{I_j}\psi\|^2
=
\int_{I_j}|\psi(z)|^2\,dz.
\end{equation}

The finite-resolution outcome sector \(\mathcal C_j\) and the spectral
subspace \(H_{I_j}\) are different mathematical objects. The former is
the set of localized states that produce the same detector record,
whereas the latter is the spectral subspace of the position operator
associated with the interval \(I_j\). Nevertheless, in the
sharp-resolution limit, the probability of the record-defined family
approaches the spectral Born weight
\[
\int_{I_j}|\psi(z)|^2\,dz,
\]
and hence the standard Born probability for the outcome \(I_j\).

This formulation makes the treatment of interference particularly
transparent. Consider a double-slit experiment and let
\[
\psi_T
=
\alpha\psi_L+\beta\psi_R
\]
be the state arriving at the final screen when no which-slit record has
been formed. Here \(\psi_L\) and \(\psi_R\) are the amplitudes arriving
through the left and right slits. For the screen bin \(I_j\),
the corresponding family of \({\bf RM}\) histories has probability
\[
W_j(\psi_T)
=
\mathbb P_B\bigl(\Gamma_j(\psi_T)\bigr)
\simeq
\|P_{I_j}\psi_T\|^2.
\]
Thus
\[
W_j(\psi_T)
\simeq
\int_{I_j}
|\alpha\psi_L(z)+\beta\psi_R(z)|^2\,dz.
\]
Expanding,
\[
\begin{aligned}
W_j(\psi_T)
\simeq{}&
|\alpha|^2
\int_{I_j}|\psi_L(z)|^2\,dz
+
|\beta|^2
\int_{I_j}|\psi_R(z)|^2\,dz
\\
&+
2\operatorname{Re}
\left[
\alpha\overline{\beta}
\int_{I_j}
\psi_L(z)\overline{\psi_R(z)}\,dz
\right].
\end{aligned}
\]
The last term is the interference contribution. It is present because,
in the absence of a which-slit record, the common realized
\({\bf RM}\) unitary acts on the coherent state
\[
\alpha\psi_L+\beta\psi_R
\]
as a single state. The left- and right-slit components are not assigned
separate probabilities.

Suppose instead that a distinguishable record is formed at the slits.
For normalized and mutually distinguishable left- and right-slit
branches, the record map separates the history space into two disjoint
events,
\[
\Gamma_L
=
\{\omega:R_\psi(\omega)=L\},
\qquad
\Gamma_R
=
\{\omega:R_\psi(\omega)=R\}.
\]
Their probabilities are, in the idealized two-slit description,
\[
\mathbb P_B(\Gamma_L)
\simeq
|\alpha|^2,
\qquad
\mathbb P_B(\Gamma_R)
\simeq
|\beta|^2.
\]

Conditional on the left-slit record, the subsequent \({\bf RM}\)
evolution begins from the state actually reached in the left outcome
sector. 
In the idealized description, denote the normalized outgoing states
following the two records by the same symbols \(\psi_L\) and \(\psi_R\).
The conditional probability of the screen
record \(j\) is then
\[
W_j^{(L)}
=
\mathbb P_B
\bigl(
\Gamma_j\,\big|\,\Gamma_L
\bigr)
\simeq
\|P_{I_j}\psi_L\|^2
=
\int_{I_j}|\psi_L(z)|^2\,dz.
\]
Similarly,
\[
W_j^{(R)}
=
\mathbb P_B
\bigl(
\Gamma_j\,\big|\,\Gamma_R
\bigr)
\simeq
\|P_{I_j}\psi_R\|^2
=
\int_{I_j}|\psi_R(z)|^2\,dz.
\]

If the slit record is not retained in the final statistics, the total
probability of the screen record \(j\) is obtained by the law of total
probability:
\[
W_j
=
\mathbb P_B(\Gamma_L)\,W_j^{(L)}
+
\mathbb P_B(\Gamma_R)\,W_j^{(R)}.
\]
Hence
\[
W_j
\simeq
|\alpha|^2
\int_{I_j}|\psi_L(z)|^2\,dz
+
|\beta|^2
\int_{I_j}|\psi_R(z)|^2\,dz.
\]
There is no interference term.

The difference between the two experiments is therefore entirely a
difference in the record structure of the \({\bf RM}\) histories.
Without a which-slit record, the coherent superposition is the initial
state of the \({\bf RM}\) walk toward the screen. This state then
determines the family of histories leading to each screen sector. With a
distinguishable slit record, the history space is partitioned into
disjoint left- and right-record events. The subsequent \({\bf RM}\) walk
toward the screen begins from the state reached in the corresponding
record sector, and the later screen probabilities are conditional on
the respective record events.

No projection is inserted into the fundamental \({\bf RM}\) evolution
in either case. A record corresponds to entry of the trajectory into a
finite-resolution detector sector. Once such a record exists, the
corresponding alternatives define disjoint measurable subsets of the
underlying history space, and their probabilities combine additively.
If no distinguishing record exists, the alternatives remain coherent
components of a single state and continue to interfere.

\section{The Helix Model of Detector-Path Weights}
\label{sec:the-helix-model-of-path-weights}

The finite-dimensional helix model introduced in
Section~\ref{subsec:a-helix-in-r-3} provides a simple illustration of the
path-space construction developed above. It is useful because both the
ambient stochastic process and the detector-defined path families can be
described explicitly.

Consider the helix
\[
\gamma(u)
=
\left(
\cos u,\,
\sin u,\,
\frac{u}{\sqrt{2}}
\right)
\subset\mathbb R^3.
\]
Its tangent vector has constant squared norm
\[
|\gamma'(u)|^2
=
\frac{3}{2}.
\]
Let \(u(t)\) be Brownian motion in the helix parameter, with
\[
\mathbb E\bigl[(du)^2\bigr]
=
2D_u\,dt.
\]
The corresponding infinitesimal arc-length displacement satisfies
\[
\mathbb E\bigl[(ds)^2\bigr]
=
|\gamma'(u)|^2
\mathbb E\bigl[(du)^2\bigr]
=
2D_H\,dt,
\]
where
\[
D_H
=
\frac{3}{2}D_u.
\]

As shown in Section~\ref{subsec:a-helix-in-r-3}, tangent directions at
three suitably chosen and sufficiently separated points of the helix
form an orthonormal basis of the ambient space. The corresponding
Brownian tangent increments are independent centered Gaussian variables
with the same variance. They therefore determine an isotropic Gaussian
ambient increment \(dX\in\mathbb R^3\).
Writing
\[
\mathbb E[dX\,dX^T]
=
2D_3 I\,dt,
\]
the component of the ambient increment along any unit vector has
variance \(2D_3\,dt\). Matching the component along a unit tangent to
the helix with the intrinsic Brownian arc-length increment gives
\[
D_3
=
D_H
=
\frac{3}{2}D_u.
\]
Thus the diffusion scale of the ambient motion is fixed by the Brownian
motion on the helix.

More is determined than the diffusion coefficient alone. Let
\[
K_T(x,y)
\]
denote the transition density of the lifted ambient motion over an
observation time \(T\). Homogeneity and isotropy imply that it depends
on the initial and final points only through their Euclidean separation:
\[
K_T(x,y)
=
F_T\bigl(|x-y|\bigr)
\]
for some function \(F_T\).

As the separation of two points on the helix varies, their Euclidean
distance assumes every value in \([0,\infty)\). Therefore the ambient
transition law, evaluated for pairs of points on the helix, determines
the distance-dependent function \(F_T\) throughout the ambient space.
It follows that the resulting ambient transition density is the
three-dimensional heat kernel
\[
K_T(x,y)
=
\frac{1}{(4\pi D_3T)^{3/2}}
\exp\left(
-\frac{|x-y|^2}{4D_3T}
\right),
\]
with
\[
D_3
=
\frac{3}{2}D_u.
\]
Thus the lift restores not only the correct ambient diffusion scale but
the full homogeneous and isotropic Brownian transition law.

We now introduce finite-resolution detector outcomes. Divide the helix
into short segments
\[
\gamma(J_j),
\qquad
J_j
=
\left[
u_j-\frac{\Delta u}{2},
u_j+\frac{\Delta u}{2}
\right],
\]
and surround each segment by a narrow tubular neighborhood \(B_j\) in
\(\mathbb R^3\). These sets represent detector bins.
Let an ambient Brownian particle start at
\[
x_0\in\mathbb R^3.
\]
For a fixed observation time \(T\), define
\[
\Gamma_j(x_0;T)
=
\left\{
X(\cdot):
X(0)=x_0,\quad X(T)\in B_j
\right\}.
\]
The probability of this family of paths is
\[
\begin{aligned}
W_j(x_0;T)
&=
\mathbb P_{x_0}\bigl(\Gamma_j(x_0;T)\bigr)
\\
&=
\int_{B_j}
K_T(x_0,y)\,d^3y.
\end{aligned}
\]
Thus the detector outcome is represented by a family of ambient
Brownian paths, and its probability is the measure of that family.

If \(B_j\) is sufficiently narrow that the transition density varies
little across it, and \(x_j\) is a representative point of the bin, then
\[
W_j(x_0;T)
\simeq
\operatorname{Vol}(B_j)\,
K_T(x_0,x_j).
\]
For bins of equal size and shape, their relative probabilities are
therefore determined by the corresponding transition densities.

The same path measure also allows one to define first-entry events. Let
\[
\tau_j
=
\inf\{t>0:X(t)\in B_j\}.
\]
The event that \(B_j\) is the first detector bin reached is a measurable
family of paths in the same Brownian path space. Fixed-time occupation
and first entry are different events, just as in the \({\bf RM}\)
measurement problem. When the detector is modeled so that entry into a detector region is followed by effective absorption, the first-entry probability gives the corresponding recorded-outcome probability.

The helix model therefore reproduces, in finite-dimensional Euclidean
space, the structure used in the \({\bf RM}\) construction:
\[
\text{stochastic ambient motion}
\quad\longrightarrow\quad
\text{families of paths}
\quad\longrightarrow\quad
\text{detector records}.
\]
Its purpose is illustrative. In the quantum case the ambient space is
projective Hilbert space, the stochastic motion is generated by the
GUE-distributed Hermitian increments of the \({\bf RM}\) walk, and the
detector outcomes are finite-resolution sectors of localized
equivalence classes.

\section{Measurement as State-Space Diffusion}

The picture developed in this chapter replaces instantaneous collapse by
stochastic motion in projective state space. According to the
\({\bf RM}\) conjecture, under conditions corresponding to a position
measurement the microscopic state undergoes the \({\bf RM}\) walk
constructed in Chapter~\ref{chap:brownian-born}. Each step is unitary,
while the sequence of random GUE generators produces diffusion through
projective state space.

A measurement outcome is not identified with arrival at a prescribed
ray. An actual detector has finite resolution and therefore recognizes
a sector of localized equivalence classes. For a position-sensitive detector with outcome interval \(I_j\), the
corresponding sector may be represented by
\[
\mathcal C_j
=
\left\{
[\psi]:
\mu_z(\psi)\in I_j,\quad
\delta_z(\psi)\leq\sigma
\right\}.
\]
A detector record may be associated either with the sector occupied by
the state at a prescribed observation time \(T\), or, when detector
registration is effectively stable, with the first outcome sector
entered by the \({\bf RM}\) trajectory. The latter description concerns
histories for which entry into the detector sector is followed by an
absorbing continuation, whereas the former requires no such condition.

The transverse description developed above makes the absorbing
mechanism more explicit. The coordinates
\[
\tau=\mu_z,
\qquad
s=\ln\frac{\delta_z}{\sigma}
\]
separate the recorded position from the degree of localization. The full
\({\bf RM}\) trajectory is not confined to the two-dimensional surface
\(M_\varphi\); rather, this surface locally represents the component of
the motion transverse to the foliation by states having fixed
\(\mu_z\) and \(\delta_z\). Motion tangent to the leaves changes the
quantum state without changing the two quantities resolved by the
position measurement.
In this reduced description, the \(\tau\)-component of an infinitesimal
\({\bf RM}\) step is centered, while the localization coordinate \(s\)
controls the scale of the corresponding displacement in \(\tau\). As
the state becomes more strongly localized, \(s\to-\infty\), a fixed
Fubini--Study displacement produces an increasingly small change in the
recorded position.

The fixed-time and absorbing descriptions become equivalent when entry
into a detector sector places the state in a sufficiently localized
region that subsequent motion of the recorded coordinate is strongly
suppressed. In the binary model discussed above, this occurs in the
thin-layer, small-step regime: entry into an outcome layer forces
\(s\ll0\), and the corresponding metric coefficient in the
\(\tau\)-direction becomes large. A fixed Fubini--Study displacement
then produces only a very small change in the recorded position.
Consequently, once the trajectory has entered the outcome sector, it
remains associated with the same detector outcome with probability
effectively one on the observation time scale.
No change in the underlying \({\bf RM}\) step law is required. The same
state-space motion is simply represented differently in the
measurement coordinates once the state lies in the strongly localized
region. Thus, when return from the entered outcome sector is negligible,
the probability of first entry into \(\mathcal C_j\) agrees with the
probability of finding the state in that sector at a later fixed
observation time \(T\).

The binary model makes the probability mechanism particularly
transparent. The unbiased \(\tau\)-motion between two detector
thresholds is a martingale, and its first-entry probabilities are
\[
P_a=|\alpha|^2,
\qquad
P_b=|\beta|^2
\]
for an initial state
\[
\psi=\alpha g_a+\beta g_b.
\]
Thus the Born probabilities arise directly as hitting probabilities of
the measurement-relevant coordinate. The unequal probabilities do not
result from a bias in the stochastic dynamics; they are determined by
the position of the initial state between the two outcome sectors.

For a detector with several possible finite-resolution outcome sectors
\(\mathcal C_j\), the Wiener-induced probability space constructed above
assigns probabilities to complete \({\bf RM}\) histories. The record map
then identifies the families of histories associated with the different
detector outcomes.
For a fixed-time measurement, the outcome probability is the probability
that the state lies in the corresponding sector at the prescribed
observation time \(T\). For an absorbing detector, the relevant event is instead the family of
histories that first enter the corresponding sector and remain
associated with that outcome during the observation interval. When the absorbing continuation occurs with probability
effectively one after entry, the two descriptions agree.

In either case, the resulting outcome probabilities are consistent with
the detector weights fixed independently by the normal-to-Born relation.
For sufficiently large cutoff \(N\), in the effective-absorption regime,
and for an outcome represented by an orthogonal sector with projection
\(P_j\),
\[
\mathbb P_{\bf RM}
\bigl(
\text{outcome }j
\bigr)
\simeq
\|P_j\psi\|^2.
\]
Thus the Born weight acquires a dynamical meaning as the probability of
the family of \({\bf RM}\) histories associated with the corresponding
measurement outcome.

The resulting measurement picture may therefore be summarized as
\[
\begin{array}{c}
\text{measurement interaction}
\\[2mm]
\downarrow
\\[2mm]
\text{\({\bf RM}\) diffusion in projective state space}
\\[2mm]
\downarrow
\\[2mm]
\begin{array}{cc}
\text{occupation of an outcome sector at time \(T\)}
&
\text{first entry into an absorbing outcome sector}
\\[1mm]
\text{(fixed-time measurement)}
&
\text{(stable-record measurement)}
\end{array}
\\[2mm]
\downarrow
\\[2mm]
\text{detector outcome}.
\end{array}
\]
The normal-to-Born relation fixes the corresponding probabilities,
while the \({\bf RM}\) dynamics and its induced path measure supply the
stochastic histories through which those probabilities are realized.

The remaining question is physical rather than geometric: under what
conditions does interaction with a measuring device or with the
environment produce the \({\bf RM}\) measurement regime, and why does
the corresponding stochastic motion of sufficiently macroscopic systems
remain effectively confined to the classical state manifold? These
questions are addressed in Chapter~\ref{chap:macro-micro-measurement}.

\chapter{Macroscopic and Microscopic Measurement}
\label{chap:macro-micro-measurement}

Measurement of both macroscopic and microscopic systems will be
described here as stochastic unitary motion in projective state space.
For a macroscopic body, continuous environmental monitoring produces an
approximately Gaussian distribution of recorded positions and momenta,
sharply concentrated about a Newtonian phase-space trajectory. For a
microscopic system, the same \({\bf RM}\) dynamics
produces Born probabilities and a single recorded outcome.

\section{Newtonian Motion under Environmental Monitoring}
\label{sec:newtonian-motion-under-environmental-monitoring}

We have shown that Newtonian dynamics is contained in Schr\"odinger
dynamics as its tangent component on the localized phase-space manifold
\(M_{3,3}^{\sigma}\). The remaining question is why the position and
momentum of a macroscopic body, when sampled through its continual
interaction with the environment, remain close to a Newtonian
phase-space trajectory. We shall argue that the environmental motion
returns the state repeatedly to the localized sector, while the tangent
component of the ordinary Schr\"odinger dynamics supplies the Newtonian
drift of the recorded position and momentum.

Environmental monitoring of a macroscopic body is itself a form of
continuous classical measurement. Each interaction with the surrounding
environment carries finite-resolution information about the position,
and generally also the momentum, of the body. An individual record
therefore specifies a finite-resolution phase-space state rather than an
exact point. Under the cumulative action of many environmental
interactions, the statistics of such records are approximately Gaussian.
Their mean position and momentum follow the Newtonian trajectory,
whereas their covariance describes the small stochastic fluctuations
about that trajectory.

Consider a material point of mass \(M\), with ordinary Hamiltonian
\[
\widehat h
=
-\frac{\hbar^2}{2M}\Delta+\widehat V.
\]
At a point
\(\varphi_{a,p}\in M_{3,3}^{\sigma}\), the projective Schr\"odinger
velocity decomposes orthogonally into tangent and normal components:
\[
-\frac{i}{\hbar}
\bigl(\widehat h-\langle\widehat h\rangle\bigr)\varphi_{a,p}
=
\frac{p}{M}\cdot
\frac{\partial\varphi_{a,p}}{\partial a}
-
\nabla V(a)\cdot
\frac{\partial\varphi_{a,p}}{\partial p}
+
X^\perp .
\]
The tangent component gives
\[
\dot a=\frac{p}{M},
\qquad
\dot p=-\nabla V(a),
\]
whereas \(X^\perp\) describes the tendency of the state to leave the
localized phase-space manifold.

For Gaussian representatives of width \(\sigma\), the characteristic
spreading rate is of order
\[
\|X^\perp\|
\sim
\frac{\hbar}{M\sigma^2},
\]
and the corresponding spreading time is
\[
T_{\rm spr}
=
\frac{M\sigma^2}{\hbar}.
\]
For a macroscopic mass and a macroscopic observational resolution, this
time is extremely long compared with the characteristic duration of an
individual environmental interaction.

We now use an extended form of conjecture \({\bf RM}\), in which the
ordinary Hamiltonian dynamics and the stochastic environmental
contribution are included together. The unresolved action of the
environment is represented by a random Hermitian Hamiltonian
\(\widehat h_{\bf RM}(t)\), and the state satisfies the full linear
Schr\"odinger equation
\[
i\hbar\frac{d\psi}{dt}
=
\widehat h_{\rm tot}(t)\psi,
\qquad
\widehat h_{\rm tot}(t)
=
\widehat h+\widehat h_{\bf RM}(t).
\]
The deterministic term \(\widehat h\) generates the ordinary
Schr\"odinger motion, while \(\widehat h_{\bf RM}(t)\) generates the
stochastic \({\bf RM}\) component. In this form, the conjecture will be applied to both microscopic and
macroscopic systems.

For every realization of \(\widehat h_{\bf RM}(t)\), the evolution is
linear and unitary. The randomness is entirely in the time-dependent
Hermitian Hamiltonian; no projection or nonlinear modification of the
Schr\"odinger equation is introduced.
We shall not attempt to factor the full unitary evolution into separate
ordinary and environmental factors. In particular, no commutativity of
\(\widehat h\) and \(\widehat h_{\bf RM}(t)\) is assumed. Instead, we
analyze the drift and covariance of the position and momentum coordinates
induced by their combined action.

Let
\[
a(\psi)
=
\langle\psi,\widehat x\psi\rangle,
\qquad
p(\psi)
=
\langle\psi,\widehat p\psi\rangle.
\]
Along the full Schr\"odinger evolution,
\[
\frac{d}{dt}a
=
\frac{i}{\hbar}
\left\langle
[\widehat h+\widehat h_{\bf RM}(t),\widehat x]
\right\rangle
\]
and
\[
\frac{d}{dt}p
=
\frac{i}{\hbar}
\left\langle
[\widehat h+\widehat h_{\bf RM}(t),\widehat p]
\right\rangle.
\]
The ordinary Hamiltonian contributes
\[
\frac{i}{\hbar}
\langle[\widehat h,\widehat x]\rangle
=
\frac{p}{M}
\]
and
\[
\frac{i}{\hbar}
\langle[\widehat h,\widehat p]\rangle
=
-\langle\nabla V(\widehat x)\rangle.
\]
For a state localized near \(a\) with position spread not exceeding
\(\sigma\),
\[
\langle\nabla V(\widehat x)\rangle
=
\nabla V(a)+r_V.
\]
If \(\nabla V\) is locally Lipschitz on the localization region, then
\[
|r_V|
\leq
\sup_{|x-a|\lesssim\sigma}
\|\nabla^2V(x)\|\,
\langle|\widehat x-a|\rangle
=
O(\sigma).
\]
Thus \(r_V\) is small when the force varies only weakly over distances
of order \(\sigma\).

Let the projective increment generated by the total Hamiltonian be written
as
\[
X_{\rm tot}
=
X_h+X_{\bf RM},
\]
where \(X_h\) is generated by the ordinary Hamiltonian
\(\widehat h\), while \(X_{\bf RM}\) is the centered environmental
increment generated by \(\widehat h_{\bf RM}\).
Near \(M_{3,3}^{\sigma}\), decompose an elementary environmental
increment as
\[
X_{\bf RM}
=
X_a^{\bf RM}
+
X_p^{\bf RM}
+
\xi_sE_s
+
X_\perp',
\]
where \(X_a^{\bf RM}\) and \(X_p^{\bf RM}\) are the position and
momentum components tangent to \(M_{3,3}^{\sigma}\), \(E_s\) is the
unit vector in the transverse localization direction, \(\xi_s\) is its
random coefficient, and \(X_\perp'\) contains the remaining normal
components. These subspaces are mutually orthogonal in the
Fubini--Study metric.

Conjecture \({\bf RM}\) assigns an isotropic Gaussian law to each
elementary unresolved increment in projective state space. Accordingly,
the mutually orthogonal position, momentum, localization, and remaining
normal components of \(X_{\bf RM}\) have vanishing cross-covariances.
Since the joint law is Gaussian, the transverse localization increment
is statistically independent of the environmental position and
momentum increments.
The isotropy is with respect to the Fubini--Study metric, not with
respect to the dimensional physical coordinates \(a\) and \(p\).
Indeed, for small displacements along \(M_{3,3}^{\sigma}\),
\[
ds^2
=
\frac{|da|^2}{4\sigma^2}
+
\frac{\sigma^2|dp|^2}{\hbar^2}.
\]
Thus an isotropic Gaussian law in the corresponding orthonormal
projective coordinates generally appears as an elliptic Gaussian law
when expressed in the physical phase-space coordinates. Position and
momentum fluctuations may therefore have different magnitudes and,
under the combined ordinary and environmental dynamics, different time
dependence.

No isotropy in the physical coordinates is required for the derivation
of Newtonian motion. What is needed is that the environmental
horizontal increments are centered, are independent of the localization
increment, and produce accumulated physical fluctuations that remain
small.
More precisely, once the current state is fixed and the next random
environmental Hamiltonian is sampled, the centered \({\bf RM}\) law
gives
\[
\mathbb E
\left(
X_a^{\bf RM}\mid[\psi]
\right)
=
0,
\qquad
\mathbb E
\left(
X_p^{\bf RM}\mid[\psi]
\right)
=
0.
\]
The total horizontal increment, however, also contains the deterministic
contribution generated by \(\widehat h\). 

At the level of the induced physical coordinates, and in the diffusion
limit, the combined linear Schr\"odinger evolution may therefore be
written schematically as
\[
da_t
=
\frac{p_t}{M}\,dt
+
dM_t^a
+
r_a(t)\,dt
\]
and
\[
dp_t
=
-\nabla V(a_t)\,dt
+
dM_t^p
+
r_p(t)\,dt.
\]
Here \(M_t^a\) and \(M_t^p\) are centered stochastic processes generated
by the environmental Hamiltonian \(\widehat h_{\bf RM}\), while \(r_a\) and \(r_p\) collect the
small corrections due to finite localization width, variation of the
force across the state, and the geometry of the chosen coordinates.
The drift terms and the stochastic terms thus arise from the two parts
of the same total Hamiltonian. 

We now consider the transverse localization motion. Let
\[
s
=
\ln\frac{\delta_x}{\sigma},
\]
where \(\delta_x\) is the position spread of the state. The
finite-resolution localized sector is
\[
s\leq0.
\]
For a macroscopic body, the initial state is assumed to lie on
\(M_{3,3}^{\sigma}\), so that
\[
s_0=0.
\]
The environmental increments in the \(s\)-direction form a centered
one-dimensional Gaussian walk and therefore have a continuous symmetric
distribution.

Let \(\tau_{\rm ret}\) denote the first step at which the transverse
walk, starting from \(s=0\), enters the localized half-line
\(s\leq0\) after an excursion into \(s>0\). For independent continuous
symmetric increments, the Sparre Andersen theorem gives the survival
asymptotic
\[
P(\tau_{\rm ret}>n)
\sim
\frac{1}{\sqrt{\pi n}}.
\]
A prescribed small failure probability \(\eta\) therefore corresponds to
approximately
\[
n_\eta
\sim
\frac{1}{\pi\eta^2}
\]
steps, or to the physical time
\[
T_\eta
\sim
\frac{\tau}{\pi\eta^2}.
\]
This is a high-probability return time, not a mean return time. The
centered transverse motion need not keep the state continuously within
the localized sector; rather, it produces recurrent passages through
that sector.

During one environmental interval \(\tau\), the deterministic position
displacement is of order
\[
v\tau.
\]
For the environmental interaction to sample the motion on a scale much
finer than the observational resolution, we require
\[
v\tau\ll\sigma.
\]
The ordinary motion normal to \(M_{3,3}^{\sigma}\) is controlled by the
spreading time
\[
T_{\rm spr}
=
\frac{M\sigma^2}{\hbar},
\]
and is negligible during one environmental interval when
\[
\frac{\tau}{T_{\rm spr}}\ll1.
\]

No corresponding requirement that the tangent momentum displacement be
small in Fubini--Study units is imposed. The tangent position and
momentum components represent the Newtonian motion itself, and even a
small physical momentum change of a macroscopic body may correspond to
a large projective displacement relative to the quantum momentum scale.
What is required is that the Newtonian motion vary smoothly on the
environmental time scale and that the force vary negligibly across the
localized state.
These conditions do not assert that
\(\widehat h\) commutes with \(\widehat h_{\bf RM}\), nor is such an
assertion required. They ensure that the environmental interaction
samples a well-defined Newtonian drift while ordinary spreading remains
negligible on the microscopic environmental time scale.

Let \(t_k\) be successive times at which the state enters the
finite-resolution phase-space sector, and define the corresponding
recorded position and momentum by
\[
a_k
=
\mu_x\bigl(\psi(t_k)\bigr),
\qquad
p_k
=
\mu_p\bigl(\psi(t_k)\bigr),
\]
where \(\mu_x\) and \(\mu_p\) are the position and momentum
expectation-value maps. 
No projection is applied at these times. The state is the one actually
reached by the full stochastic unitary evolution. Each such entry
constitutes a finite-resolution environmental record of position and
momentum, whether or not that record is subsequently read or retained
macroscopically.
The return time \(t_{k+1}\) is determined by the transverse localization
motion. Since the horizontal environmental increments are independent
of the localization increments, selecting a return time does not bias
their accumulated horizontal contribution. Their conditional mean
therefore remains zero.

Let \((a_N(t),p_N(t))\) be the Newtonian solution with initial data
\((a_k,p_k)\) at time \(t_k\). Provided that the force varies
negligibly over the stochastic position spread,
\[
\mathbb E\bigl(
a_{k+1}\mid a_k,p_k,t_{k+1}
\bigr)
=
a_N(t_{k+1})+o(\sigma)
\]
and
\[
\mathbb E\bigl(
p_{k+1}\mid a_k,p_k,t_{k+1}
\bigr)
=
p_N(t_{k+1})+o(p_{\rm res}),
\]
where \(p_{\rm res}\) is the momentum resolution of the macroscopic
description.


Let \(D_a\) and \(D_p\) denote the effective diffusion coefficients of
the position and momentum coordinates generated by the environmental
\({\bf RM}\) fluctuations. In each Cartesian component, write
schematically
\[
\operatorname{Var}(\Delta a^j)
\sim
2D_a\,\Delta t,
\qquad
\operatorname{Var}(\Delta p^j)
\sim
2D_p\,\Delta t.
\]
The stochastic position spread remains below the localization scale
when the accumulated position diffusion between successive
high-probability returns satisfies
\[
D_aT_\eta\ll\sigma^2.
\]
Likewise, the root-mean-square stochastic momentum spread accumulated
over the same interval is of order
\[
\delta p_{\rm stoch}
\sim
\sqrt{2D_pT_\eta},
\]
and must satisfy
\[
\delta p_{\rm stoch}\ll p_{\rm res}.
\]
For a moving body, the classical momentum \(Mv\) may provide a convenient
comparison scale, but \(p_{\rm res}\) is the more general criterion,
since it remains meaningful for a body at rest.

In a thermal medium with dissipative relaxation, the momentum variance
does not grow indefinitely but approaches its thermal equilibrium value.
By equipartition, the variance in each Cartesian component is then of
order
\[
Mk_BT,
\]
so that
\[
\delta p_{\rm stoch}
\lesssim
\sqrt{Mk_BT}.
\]

The macroscopic regime is therefore characterized by
\[
v\tau\ll\sigma,
\qquad
\frac{\tau}{T_{\rm spr}}\ll1,
\]
together with
\[
D_aT_\eta\ll\sigma^2,
\qquad
\delta p_{\rm stoch}\ll p_{\rm res},
\]
and the requirement that the Newtonian motion vary smoothly on the
environmental time scale and that the force vary negligibly across the
finite-resolution position distribution.

Under these conditions, the transverse environmental walk returns the
state repeatedly to the localized sector and is statistically
independent of the horizontal environmental fluctuations. The ordinary
part of the full Hamiltonian supplies the Newtonian tangent drift,
whereas the horizontal components of the environmental Hamiltonian
produce only small centered fluctuations at the observational
resolution. At the successive localization times, the recorded
positions and momenta therefore form a narrow stochastic sequence
centered on the Newtonian trajectory. This is the stroboscopic Newtonian
motion of the macroscopic body.

\section{Macroscopic Bodies in Natural Environments}
\label{sec:macroscopic-bodies-in-natural-environments}

We now apply the preceding criteria to a representative macroscopic
body. The purpose of the estimates is to verify that familiar
environmental parameters lead to the regime required for stroboscopic
Newtonian motion.
We first consider a body moving in air at room temperature. Molecular
collisions then occur with very high frequency and provide a particularly
strong source of environmental monitoring. We subsequently consider the
limiting case in which ordinary matter and local thermal radiation are
removed and the body interacts only with the cosmic microwave
background.
The parameter \(\sigma\) used below is the resolution at which the
position of the macroscopic body is observed. It is not chosen to
equalize the accumulated position and momentum variances in
Fubini--Study coordinates. These variances generally have different
magnitudes and time dependence, and no equality between them is required
for stroboscopic Newtonian motion.

We continue to use the extended form of conjecture \({\bf RM}\)
introduced in the preceding section. The ordinary dynamics and the
unresolved environmental interaction are represented together by the
total Hamiltonian
\[
\widehat h_{\rm tot}(t)
=
\widehat h+\widehat h_{\bf RM}(t),
\]
where \(\widehat h\) is the ordinary Hamiltonian and
\(\widehat h_{\bf RM}(t)\) is the random Hermitian environmental
contribution. The state evolves according to the full linear
Schr\"odinger equation
\[
i\hbar\frac{d\psi}{dt}
=
\bigl(
\widehat h+\widehat h_{\bf RM}(t)
\bigr)\psi.
\]
For every realization of the random Hamiltonian, the evolution is linear
and unitary.

At the level of an elementary \({\bf RM}\) increment, the transverse
localization component is statistically independent of the horizontal
position and momentum components. The latter have zero conditional mean
and contribute centered fluctuations about the Newtonian drift generated
by the ordinary Hamiltonian.
No factorization of the full unitary evolution into ordinary and
environmental factors will be used. In particular, no commutativity of
\(\widehat h\) and \(\widehat h_{\bf RM}(t)\) is assumed. The estimates
below verify instead that the deterministic motion varies smoothly on
the microscopic environmental time scale, that ordinary spreading is
negligible on that scale, and that the accumulated environmental
position and momentum fluctuations remain below the resolution of the
macroscopic description.

\subsection{A Macroscopic Body in Air}
\label{subsec:a-macroscopic-body-in-air}

Consider a small macroscopic body of mass
\[
M=10^{-6}\ {\rm kg},
\]
whose center-of-mass position is observed with resolution
\[
\sigma=10^{-6}\ {\rm m}.
\]
These values are chosen only to establish representative orders of
magnitude.

The free spreading time associated with this mass and resolution is
\[
T_{\rm spr}
=
\frac{M\sigma^2}{\hbar}.
\]
Numerically,
\[
T_{\rm spr}
\approx
9.5\times10^{15}\ {\rm s}.
\]
Thus a state localized on the micron scale spreads only on a time scale
of hundreds of millions of years.

At room temperature, the characteristic thermal speed of an air
molecule is of order
\[
v_{\rm th}
\sim
5\times10^2\ {\rm m/s}.
\]
Taking the microscopic correlation length of the molecular interaction
to be
\[
\ell_{\rm air}
\sim
10^{-9}\ {\rm m},
\]
we obtain the characteristic environmental correlation time
\[
\tau_{\rm air}
\sim
\frac{\ell_{\rm air}}{v_{\rm th}}
\sim
2\times10^{-12}\ {\rm s}.
\]
A very large number of molecular collisions occur on macroscopic time
scales, so the coarse-grained environmental increments may be treated
as approximately Gaussian.

For a body moving with characteristic speed
\[
v\lesssim1\ {\rm m/s},
\]
the physical displacement during one environmental correlation interval
is
\[
v\tau_{\rm air}
\lesssim
2\times10^{-12}\ {\rm m},
\]
which is negligible on the observational scale
\(\sigma=10^{-6}\ {\rm m}\). The normal spreading is smaller still:
\[
\frac{\tau_{\rm air}}{T_{\rm spr}}
\sim
2\times10^{-28}.
\]
Thus the ordinary Schr\"odinger evolution produces essentially no
spreading during a microscopic environmental interval.

The deterministic Schr\"odinger drift therefore changes only
negligibly during the microscopic interval over which an environmental
increment is accumulated.
The tangent position and momentum components need not be small in
Fubini--Study units. They represent the Newtonian motion itself, and
for a macroscopic body even a small physical momentum change may
correspond to a large projective displacement because the natural
quantum momentum scale is \(\hbar/(2\sigma)\). What is required is
instead that the Newtonian motion vary smoothly on the environmental
time scale and that the force vary negligibly across the localized
state.

We next estimate the accumulated positional fluctuations. Suppose that
the body has density
\[
\rho_b\sim10^3\ {\rm kg/m^3}
\]
and is approximately spherical. Its radius is then
\[
R
\sim
\left(
\frac{3M}{4\pi\rho_b}
\right)^{1/3}
\sim
6\times10^{-4}\ {\rm m}.
\]
The long-time spatial Brownian diffusion coefficient is given, to the
accuracy needed here, by the Stokes--Einstein relation
\[
D_a
=
\frac{k_BT}{6\pi\eta R},
\]
where \(T\) is the temperature and \(\eta\) is the dynamic viscosity of
air. Using
\[
T\approx300\ {\rm K},
\qquad
\eta_{\rm air}\approx1.8\times10^{-5}\ {\rm Pa\,s},
\]
gives
\[
D_a
\sim
2\times10^{-14}\ {\rm m^2/s}.
\]

This is a long-time, overdamped diffusion coefficient. It should not be
used to identify the position component of an individual microscopic
\({\bf RM}\) increment. It does, however, provide an upper
order-of-magnitude estimate of the positional variance accumulated over
the return time considered below.

We now estimate the return time to the localized sector. Let
\[
s
=
\ln\frac{\delta_x}{\sigma}.
\]
The localized sector is
\[
s\leq0.
\]
The macroscopic state is assumed initially to lie on
\(M_{3,3}^{\sigma}\), and hence
\[
s_0=0.
\]
The elementary \(s\)-increment is statistically independent of the
horizontal environmental increments. Its centered one-dimensional
Gaussian law is continuous and symmetric.

Let \(\tau_{\rm ret}\) denote the first step at which the transverse
walk, starting from \(s=0\), enters the localized half-line
\(s\leq0\) after an excursion into \(s>0\). For independent continuous
symmetric increments, the Sparre Andersen theorem gives the survival
asymptotic
\[
P(\tau_{\rm ret}>n)
\sim
\frac{1}{\sqrt{\pi n}}.
\]
A failure probability \(\eta\) therefore corresponds to
\[
n_\eta
\sim
\frac{1}{\pi\eta^2}
\]
steps and to the physical return time
\[
T_\eta
\sim
\frac{\tau_{\rm air}}{\pi\eta^2}.
\]
Taking
\[
\eta=3.2\times10^{-5},
\]
corresponding approximately to a one-sided \(4\sigma\) failure
probability, gives
\[
T_\eta
\sim
6\times10^{-4}\ {\rm s}.
\]
This is a high-probability return time, not a mean return time.

The positional variance estimated from the long-time diffusion
coefficient is of order
\[
D_aT_\eta
\sim
10^{-17}\ {\rm m^2},
\]
whereas
\[
\sigma^2
=
10^{-12}\ {\rm m^2}.
\]
Hence
\[
D_aT_\eta\ll\sigma^2.
\]
Equivalently,
\[
\delta a_{\rm rms}
\lesssim
\sqrt{2D_aT_\eta}
\sim
5\times10^{-9}\ {\rm m},
\]
which is more than two orders of magnitude smaller than the
observational resolution. Since the Stokes--Einstein coefficient
describes the overdamped regime, this estimate is conservative on time
scales shorter than the momentum-relaxation time.

The stochastic momentum spread is likewise small on a macroscopic
momentum scale. In a thermal medium with dissipative relaxation, the
momentum variance does not grow indefinitely but approaches its
equilibrium value. By equipartition, the variance in one Cartesian
component is of order
\[
Mk_BT.
\]
Thus
\[
\delta p_{\rm stoch}
\lesssim
\sqrt{Mk_BT}
\sim
6\times10^{-14}\ {\rm kg\,m/s}.
\]

For a body moving with
\[
v\sim1\ {\rm m/s},
\]
the Newtonian momentum is
\[
p_N=Mv\sim10^{-6}\ {\rm kg\,m/s},
\]
and therefore
\[
\frac{\delta p_{\rm stoch}}{Mv}
\lesssim
6\times10^{-8}.
\]
For a body at rest or moving very slowly, the appropriate comparison is
instead with the momentum resolution \(p_{\rm res}\) of the macroscopic
description. In general, the required condition is
\[
\delta p_{\rm stoch}\ll p_{\rm res}.
\]

Notice that the position and momentum estimates refer to different
aspects of the accumulated environmental motion. The position estimate
uses long-time Brownian diffusion, whereas the momentum estimate uses
the bounded thermal equilibrium variance. They are not intended to be
the two components of one isotropically scaled elementary step, and no
fixed relation between them is expected.

We have therefore verified that the deterministic displacement during
one environmental interval is negligible at the observational
resolution,
\[
v\tau_{\rm air}\ll\sigma,
\]
while the normal spreading satisfies
\[
\frac{\tau_{\rm air}}{T_{\rm spr}}\ll1.
\]
In addition,
\[
D_aT_\eta\ll\sigma^2,
\qquad
\delta p_{\rm stoch}\ll p_{\rm res}.
\]
The Newtonian drift is therefore well resolved, provided that the force
varies negligibly across the localized state and on the microscopic
environmental time scale.

The full state continues at all times to evolve linearly and unitarily
under
\[
\widehat h+\widehat h_{\bf RM}(t).
\]
The ordinary Hamiltonian supplies the Newtonian conditional drift of the
position and momentum coordinates. By the centered \({\bf RM}\) law,
the horizontal environmental increments have zero conditional mean and
produce fluctuations that remain below the observational position and
momentum resolutions. Independently, the transverse \(s\)-walk returns
the state repeatedly to the localized sector.

Consequently, the positions and momenta recorded at successive entries
into the finite-resolution phase-space sector form a narrow stochastic
sequence centered on the Newtonian trajectory. The motion is therefore
stroboscopically Newtonian to very high accuracy, while the underlying
evolution of the body remains linear and unitary.

\subsection{The Cosmic-Microwave-Background Limit}
\label{subsec:the-cosmic-microwave-background-limit}

We now consider a much more stringent limiting case. Suppose that the
body is removed from ordinary matter and local thermal radiation, so
that its dominant unavoidable monitoring environment is the cosmic
microwave background. The interaction rate is then vastly smaller than
in air, but the resulting estimates still satisfy the conditions
required for stroboscopic Newtonian motion.

The cosmic microwave background has temperature
\[
T_{\rm CMB}\simeq2.73\ {\rm K}
\]
and photon number density
\[
n_\gamma
=
\frac{2\zeta(3)}{\pi^2}
\left(
\frac{k_BT_{\rm CMB}}{\hbar c}
\right)^3
\simeq
4.1\times10^8\ {\rm m^{-3}}.
\]
Its mean photon energy is of order
\[
E_\gamma
\simeq
2.70\,k_BT_{\rm CMB}
\simeq
1.0\times10^{-22}\ {\rm J},
\]
and the corresponding photon momentum is
\[
p_\gamma
=
\frac{E_\gamma}{c}
\simeq
3.4\times10^{-31}\ {\rm kg\,m/s}.
\]
The characteristic wavelength is therefore of order
\[
\lambda_{\rm CMB}
\sim
\frac{h}{p_\gamma}
\sim
10^{-3}\ {\rm m}.
\]

We retain the macroscopic body considered above:
\[
M=10^{-6}\ {\rm kg},
\qquad
R\sim6\times10^{-4}\ {\rm m}.
\]
The geometrical rate at which background photons are incident on the
body is
\[
\Gamma_{\rm inc}
=
n_\gamma c\pi R^2,
\]
which gives
\[
\Gamma_{\rm inc}
\sim
1.5\times10^{11}\ {\rm s^{-1}}.
\]

Not every incident photon produces an effective interaction. The
probability of scattering or absorption depends on the material, shape,
and electromagnetic properties of the body. We therefore write
\[
\Gamma_{\rm CMB}
=
Q_{\rm eff}\Gamma_{\rm inc},
\qquad
0<Q_{\rm eff}\leq1,
\]
where \(Q_{\rm eff}\) is an effective interaction factor. The mean
interval between effective interactions is then
\[
\tau_{\rm step}
=
\Gamma_{\rm CMB}^{-1}
\sim
\frac{6.7\times10^{-12}}{Q_{\rm eff}}\ {\rm s}.
\]

In this limiting example, the background radiation itself supplies the
position-monitoring environment. Its wavelength therefore gives the
natural order of magnitude of the corresponding observational
resolution. We take
\[
\sigma_{\rm CMB}
\sim
10^{-3}\ {\rm m}.
\]
This scale is much larger than the micron resolution used in the
terrestrial example. The difference reflects the resolution with which
the respective environments distinguish positions.

The spreading time associated with this scale is
\[
T_{\rm spr}^{\rm CMB}
=
\frac{M\sigma_{\rm CMB}^2}{\hbar}
\sim
10^{22}\ {\rm s}.
\]
The duration of an individual photon interaction may be estimated as
\[
\tau_{\rm int}
\sim
\frac{\lambda_{\rm CMB}}{c}
\sim
3\times10^{-12}\ {\rm s}.
\]
Thus an individual interaction is effectively instantaneous on
macroscopic dynamical time scales.

For a body moving with
\[
v\lesssim1\ {\rm m/s},
\]
the physical displacement during such an interaction is only
\[
v\tau_{\rm int}
\lesssim
3\times10^{-12}\ {\rm m},
\]
which is negligible compared with the observational resolution
\[
\sigma_{\rm CMB}\sim10^{-3}\ {\rm m}.
\]
The normal spreading during an individual interaction is smaller still:
\[
\frac{\tau_{\rm int}}{T_{\rm spr}^{\rm CMB}}
\sim
3\times10^{-34}.
\]

The tangent position and momentum displacements need not be small in
Fubini--Study units. They represent the ordinary Newtonian motion, and
a small physical momentum change of a macroscopic body may correspond
to a large projective displacement. What is required is that the
Newtonian motion remain smooth on the environmental time scale, that
the force vary negligibly across the localized state, and that ordinary
spreading and accumulated environmental fluctuations remain below the
relevant observational resolution.
The full state evolves throughout according to
\[
i\hbar\frac{d\psi}{dt}
=
\bigl(
\widehat h+\widehat h_{\bf RM}(t)
\bigr)\psi,
\]
and no factorization of the corresponding unitary evolution is assumed.

A typical momentum transfer in one effective interaction is of order
\(p_\gamma\), up to a factor of order one depending on the scattering
angle. In the momentum coordinate associated with
\(M_{3,3}^{\sigma_{\rm CMB}}\),
\[
\frac{\sigma_{\rm CMB}p_\gamma}{\hbar}
\sim
3.
\]
Thus an individual recoil need not be small in the projective metric
associated with the millimeter localization scale. Indeed, the exact
distance between two Gaussian states with the same center and momenta
differing by \(p_\gamma\) is
\[
\rho_p^{\rm CMB}
=
\arccos
\exp\left[
-\frac{\sigma_{\rm CMB}^2p_\gamma^2}{2\hbar^2}
\right],
\]
which is already close to \(\pi/2\).
This does not affect the Newtonian conclusion. No smallness of the
tangent environmental displacement in Fubini--Study units is required.
What matters is its centered character and the smallness of the
accumulated fluctuations in the physical position and momentum
coordinates.

We now consider return to the localized sector. Let
\[
s
=
\ln\frac{\delta_x}{\sigma_{\rm CMB}}.
\]
The localized sector is
\[
s\leq0.
\]
The macroscopic state is assumed initially to lie on
\(M_{3,3}^{\sigma_{\rm CMB}}\), so that
\[
s_0=0.
\]
The elementary \(s\)-increment is statistically independent of the
horizontal environmental increments. Its centered one-dimensional
Gaussian law is continuous and symmetric.

Let \(\tau_{\rm ret}\) denote the first step at which the transverse
walk, starting from \(s=0\), enters the localized half-line
\(s\leq0\) after an excursion into \(s>0\). For independent continuous
symmetric increments, the Sparre Andersen theorem gives the survival
asymptotic
\[
P(\tau_{\rm ret}>n)
\sim
\frac{1}{\sqrt{\pi n}}.
\]
A prescribed failure probability \(\eta\) therefore corresponds to
\[
n_\eta
\sim
\frac{1}{\pi\eta^2}
\]
effective interactions. Since their rate is \(\Gamma_{\rm CMB}\), the
corresponding physical time is
\[
T_\eta^{\rm CMB}
\sim
\frac{1}{\pi\eta^2\Gamma_{\rm CMB}}
=
\frac{1}{\pi\eta^2Q_{\rm eff}\Gamma_{\rm inc}}.
\]
For
\[
\eta=3.2\times10^{-5},
\]
corresponding approximately to a one-sided \(4\sigma\) failure
probability, this gives
\[
T_\eta^{\rm CMB}
\sim
\frac{2.1\times10^{-3}}{Q_{\rm eff}}\ {\rm s}.
\]
Thus the return remains rapid unless the effective interaction factor is
extremely small. This is a high-probability return time, not a mean
return time.

The momentum diffusion coefficient generated by the photon impacts is
of order
\[
D_p^{\rm CMB}
\sim
\frac{1}{2}
\Gamma_{\rm CMB}p_\gamma^2,
\]
up to factors of order one associated with the angular distribution of
the momentum transfer. Thus, for \(Q_{\rm eff}=1\),
\[
D_p^{\rm CMB}
\sim
10^{-50}\ {\rm kg^2m^2s^{-3}}.
\]
The root-mean-square momentum accumulated during the high-probability
return interval is
\[
\delta p_{\rm CMB}
\sim
\sqrt{
2D_p^{\rm CMB}T_\eta^{\rm CMB}
}.
\]
Because \(D_p^{\rm CMB}\) is proportional to \(Q_{\rm eff}\), whereas
\(T_\eta^{\rm CMB}\) is proportional to \(Q_{\rm eff}^{-1}\), their
product is independent of \(Q_{\rm eff}\). Numerically,
\[
\delta p_{\rm CMB}
\sim
6\times10^{-27}\ {\rm kg\,m/s}.
\]

For a body moving with
\[
v\sim1\ {\rm m/s},
\]
the Newtonian momentum is
\[
Mv\sim10^{-6}\ {\rm kg\,m/s},
\]
and hence
\[
\frac{\delta p_{\rm CMB}}{Mv}
\sim
10^{-20}.
\]
More generally, including the case of a body at rest, the relevant
condition is
\[
\delta p_{\rm CMB}\ll p_{\rm res},
\]
where \(p_{\rm res}\) is the momentum resolution of the macroscopic
description.

Over the return intervals considered here, the body is in the
underdamped regime. Its stochastic position displacement is therefore
obtained by integrating the random momentum rather than from an
independent Stokes--Einstein position diffusion. With
\(D_p^{\rm CMB}\) defined by
\[
\operatorname{Var}(\Delta p^j)
\sim
2D_p^{\rm CMB}\Delta t,
\]
the corresponding position variance is of order
\[
\operatorname{Var}(\Delta a^j)
\sim
\frac{
2D_p^{\rm CMB}
\bigl(T_\eta^{\rm CMB}\bigr)^3
}{
3M^2
}.
\]

For \(Q_{\rm eff}=1\), this gives
\[
\delta a_{\rm CMB}
\sim
10^{-23}\ {\rm m}.
\]
More generally,
\[
\delta a_{\rm CMB}
\sim
\frac{10^{-23}}{Q_{\rm eff}}\ {\rm m}.
\]
The condition
\[
\delta a_{\rm CMB}
\ll
\sigma_{\rm CMB}
\]
therefore requires only
\[
Q_{\rm eff}\gg10^{-20}.
\]
This lower bound is extremely weak.

Ordinary spreading is also negligible over the entire high-probability
return interval. Indeed,
\[
\frac{T_\eta^{\rm CMB}}{T_{\rm spr}^{\rm CMB}}
\sim
\frac{2\times10^{-25}}{Q_{\rm eff}},
\]
which is much smaller than one throughout the regime
\[
Q_{\rm eff}\gg10^{-20}.
\]

We have thus verified that an individual environmental interaction is
microscopic on the observational scale,
\[
v\tau_{\rm int}\ll\sigma_{\rm CMB},
\]
and that ordinary spreading is negligible even over the
high-probability return interval,
\[
\frac{T_\eta^{\rm CMB}}{T_{\rm spr}^{\rm CMB}}\ll1.
\]
In addition,
\[
\delta a_{\rm CMB}
\ll
\sigma_{\rm CMB},
\qquad
\delta p_{\rm CMB}
\ll
p_{\rm res},
\]
for any \(Q_{\rm eff}\) satisfying the extremely weak lower bound found
above. The Newtonian motion is assumed to vary smoothly on the
environmental time scale, and the force to vary negligibly across the
localized state.

The ordinary part of the full Hamiltonian supplies the Newtonian
conditional drift of the position and momentum coordinates. By the
centered \({\bf RM}\) law, the horizontal environmental increments have
zero conditional mean and produce fluctuations far below the
observational position and momentum resolutions. Independently, the
transverse localization walk returns the state repeatedly to the
localized sector.
Consequently, the positions and momenta recorded at successive entries
into the finite-resolution phase-space sector form a narrow stochastic
sequence centered on the Newtonian trajectory. The motion is therefore
stroboscopically Newtonian even in the limiting environment provided by
the cosmic microwave background.

The cosmic microwave background remains present when ordinary matter and
local thermal radiation are removed. The air calculation is therefore
not needed for the logical conclusion; it demonstrates instead how the
same mechanism operates under familiar terrestrial conditions and
permits comparison with standard Brownian-motion and decoherence
estimates.

\section{Microscopic Tracks in a Bubble Chamber}
\label{sec:microscopic-tracks-in-a-bubble-chamber}

A bubble-chamber track appears to display a microscopic particle moving
continuously along a classical trajectory. In the present framework,
however, the track is not identified with a continuous microscopic path
in \(\mathbb R^3\). It is a sequence of localized records produced by
repeated measurement-like interactions of the particle with the chamber
medium.
This situation lies between the two measurement regimes considered in
this chapter. The measured object is microscopic, and each individual
record is therefore produced by the \({\bf RM}\) dynamics responsible
for microscopic state reduction and the Born rule. At the same time, the
interactions occur frequently enough, and ordinarily perturb the
particle's momentum weakly enough, that the sequence of recorded
positions forms a narrow track centered on a classical trajectory.

A charged particle moving through the chamber repeatedly ionizes atoms
or initiates local bubble formation. Such interactions distinguish the
position of the particle with a finite microscopic localization
resolution. They therefore define a family of detector-resolvable
position classes
\[
\{\varphi_a\},
\]
where \(a\) labels the center of the corresponding localized
record-forming event.
During a record-forming interaction, we use the extended form of
conjecture \({\bf RM}\). The state evolves according to the full linear
Schr\"odinger equation
\[
i\hbar\frac{d\psi}{dt}
=
\bigl(
\widehat h+\widehat h_{\bf RM}(t)
\bigr)\psi,
\]
where \(\widehat h\) generates the ordinary dynamics and
\(\widehat h_{\bf RM}(t)\) represents the unresolved interaction with
the chamber medium.

The random state-space motion enters one of the detector-defined
localized classes. The transition to a particular class carries the corresponding Born
weight determined by the state immediately before the interaction. Once such a localized event has occurred, its amplification
into a visible bubble is an ordinary macroscopic process described in
Section~\ref{sec:from-state-reduction-to-macroscopic-recording}. The
bubble records the position class reached by the microscopic state but
does not introduce an additional collapse law.
Repetition of the process produces a sequence of localized records
\[
a_1,a_2,\ldots,a_N.
\]
The visible track is the macroscopic record of this sequence.

Let \(\ell_{\rm rec}\) denote the typical separation between successive
detector-resolved record-forming events along the visible track, and let
\(u\) be the speed of the charged particle. This scale need not coincide
with the much smaller separation between individual microscopic
ionization events, since several such interactions may contribute to a
single detector-resolved bubble. Representative orders of magnitude are
\[
\ell_{\rm rec}
\sim
10^{-4}\text{--}10^{-3}\ {\rm m},
\qquad
u
\sim
10^7\text{--}10^8\ {\rm m/s}.
\]
The corresponding time between successive localized records is
\[
T_{\rm rec}
\sim
\frac{\ell_{\rm rec}}{u}
\sim
10^{-12}\text{--}10^{-10}\ {\rm s},
\]
to order of magnitude. Thus, although the particle is microscopic, the
chamber repeatedly produces localized position records on a very short
time scale.

Between effective record-forming interactions, the unresolved
record-producing part of \(\widehat h_{\bf RM}(t)\) is absent or
negligible, and the state evolves predominantly under the ordinary
Hamiltonian. During a record-forming event, the two terms act together
through the full Hamiltonian
\[
\widehat h+\widehat h_{\bf RM}(t).
\]
We shall not attempt to factor the corresponding unitary evolution into
ordinary and \({\bf RM}\) parts, and no commutativity of
\(\widehat h\) and \(\widehat h_{\bf RM}(t)\) is assumed. Instead, we
estimate the variation of the ordinary Schr\"odinger drift during the
short interaction window and the accumulated stochastic perturbations of
the particle's position and momentum.

Let the spatial resolution of an individual microscopic
record-forming interaction be
\[
\sigma
\sim
10^{-6}\ {\rm m}.
\]
This is the localization scale of the detector-defined position classes
and need not coincide with the coarser spatial resolution with which
individual bubbles are reconstructed macroscopically.

In the nonrelativistic approximation, the free spreading time of a
localized state of mass \(m\) is
\[
T_{\rm spr}
=
\frac{m\sigma^2}{\hbar}.
\]
For an electron,
\[
T_{\rm spr}^{(e)}
\sim
\frac{(9\times10^{-31})(10^{-12})}{10^{-34}}
\sim
10^{-8}\ {\rm s},
\]
whereas for a proton,
\[
T_{\rm spr}^{(p)}
\sim
\frac{(1.7\times10^{-27})(10^{-12})}{10^{-34}}
\sim
10^{-5}\ {\rm s}.
\]
Consequently,
\[
\frac{T_{\rm rec}}{T_{\rm spr}}
\ll1
\]
for both electrons and heavier charged particles. The ordinary spreading
between successive records is therefore negligible on the microscopic
localization scale \(\sigma\). For relativistic particles, the ordinary
Hamiltonian and spreading estimate must be modified, but the separation
between the short recording time and the dynamical spreading time
remains the relevant condition.

The \({\bf RM}\) interaction is active during the short record-forming
event itself. Let \(\ell_{\rm int}\) be the microscopic interaction
length. The duration of such an event is then of order
\[
\tau_{\rm int}
\sim
\frac{\ell_{\rm int}}{u}.
\]
Taking
\[
\ell_{\rm int}
\sim
10^{-10}\text{--}10^{-9}\ {\rm m}
\]
gives
\[
\tau_{\rm int}
\sim
10^{-18}\text{--}10^{-16}\ {\rm s}.
\]

During this interaction window, the physical displacement generated by
the ordinary motion is
\[
u\tau_{\rm int}
\sim
\ell_{\rm int}.
\]
Since
\[
\ell_{\rm int}\ll\sigma,
\]
the deterministic displacement during a single record-forming
interaction is negligible at the localization scale. Ordinary spreading
during the same interval is smaller still:
\[
\frac{\tau_{\rm int}}{T_{\rm spr}}
\ll1.
\]

No smallness of the tangent displacement in Fubini--Study units is
required. The tangent position and momentum components represent the
ordinary motion of the particle. What is required is that this motion
vary smoothly during the short record-forming interaction and that
ordinary spreading remain negligible on that time scale. The ordinary
tangent field may then be regarded as approximately constant during the
interaction, even though the environmental motion itself may be strong
in the ambient projective space.

Let
\[
a(\psi)
=
\langle\psi,\widehat x\psi\rangle,
\qquad
p(\psi)
=
\langle\psi,\widehat p\psi\rangle.
\]
Along the full evolution,
\[
\frac{d}{dt}a
=
\frac{i}{\hbar}
\left\langle
[
\widehat h+\widehat h_{\bf RM}(t),
\widehat x
]
\right\rangle
\]
and
\[
\frac{d}{dt}p
=
\frac{i}{\hbar}
\left\langle
[
\widehat h+\widehat h_{\bf RM}(t),
\widehat p
]
\right\rangle.
\]
The ordinary Hamiltonian supplies the deterministic Hamiltonian drift.
After coarse-graining the repeated interactions with the chamber
medium, their systematic part contributes the mean stopping force,
whereas the centered part contributes stochastic fluctuations. 

In the diffusion limit, the induced coordinate motion may therefore be
written schematically as
\[
da_t
=
v_t\,dt+dM_t^a+r_a(t)\,dt
\]
and
\[
dp_t
=
F_{\rm eff}(a_t,p_t)\,dt+dM_t^p+r_p(t)\,dt,
\]
where \(F_{\rm eff}\) includes the external force and the mean stopping
force of the chamber medium, \(M_t^a\) and \(M_t^p\) are centered
stochastic processes induced by the random Hamiltonian, and \(r_a,r_p\)
denote small finite-width and geometric corrections. These coordinate
equations are consequences of the full linear Schr\"odinger evolution
and are not introduced as an independent dynamics.

We next describe the localization coordinate. Let
\[
s
=
\ln\frac{\delta_x}{\sigma},
\]
where \(\delta_x\) is the spatial spread of the microscopic state. The localized sector associated with the record-forming interaction is
\[
s\leq0.
\]
Unlike the macroscopic-body analysis, the state immediately before a
record-forming interaction need not lie on the boundary \(s=0\). We
therefore write
\[
s_0>0
\]
for its initial transverse coordinate relative to the localized sector.
During the interaction, the isotropic \({\bf RM}\) law induces a
centered one-dimensional Gaussian walk in the \(s\)-direction. Because
the localization direction is orthogonal to the horizontal position and
momentum directions, its increment is statistically independent of the
horizontal stochastic increments.

Let \(\tau_{\rm hit}\) denote the first step at which the walk enters the
localized half-line:
\[
\tau_{\rm hit}
=
\inf
\left\{
n\geq1:s_n\leq0
\right\}.
\]
For a centered symmetric walk beginning at \(s_0>0\), the large-\(n\)
survival probability has the form
\[
P_{s_0}(\tau_{\rm hit}>n)
\sim
\frac{C(s_0)}{\sqrt n},
\]
where \(C(s_0)\) depends on the initial distance from the localized
sector and on the scale of the transverse increments. The universal
content is the \(n^{-1/2}\) decay. Thus the state enters the localized sector associated with the
record-forming interaction with probability approaching one,
although the number of \({\bf RM}\) steps required depends on its
initial transverse distance.

For a prescribed failure probability \(\eta\), the corresponding
high-confidence number of steps satisfies
\[
n_\eta
\sim
\frac{C(s_0)^2}{\eta^2}.
\]
If \(\tau_{\bf RM}\) denotes the characteristic duration of one
\({\bf RM}\) step during the record-forming interaction, the associated
localization time is
\[
T_\eta^{\rm loc}
\sim
\frac{C(s_0)^2\tau_{\bf RM}}{\eta^2}.
\]
For a visible chamber record, this time must satisfy
\[
T_\eta^{\rm loc}
\lesssim
\tau_{\rm int}.
\]
This is the condition that the record-forming interaction last long
enough for the microscopic state to reach one of the detector-defined
localized classes with high probability.
Once a localized class has been reached, the subsequent ordinary
evolution begins from the state actually reached in that class. No
projection is inserted into the dynamics. The ionization or bubble
amplifies the resulting position distinction and thereby forms a
physical record.

The deterministic tangent displacement between two successive records
is
\[
\Delta a_{\rm cl}
\sim
uT_{\rm rec}
\sim
\ell_{\rm rec}.
\]
When the localization scale is small compared with the separation of
successive detector-resolved records,
\[
\sigma\ll\ell_{\rm rec},
\]
the successive localized outcomes are well separated, and the centers
of the recorded classes advance by approximately the classical tangent
displacement from one interaction to the next. The observed track is therefore
naturally stroboscopic.

The record-forming interactions also transfer momentum to the particle.
For a proton or heavier charged particle with classical momentum of
order
\[
p_{\rm cl}
\sim
10^{-20}\text{--}10^{-19}\ {\rm kg\,m/s}
\]
and an ionization-scale energy transfer
\[
\Delta E
\sim
10^1\text{--}10^3\ {\rm eV},
\]
the corresponding longitudinal momentum decrement is of order
\[
q_\parallel
\sim
\frac{\Delta E}{u}
\sim
10^{-25}\text{--}10^{-23}\ {\rm kg\,m/s}.
\]
Thus
\[
\frac{q_\parallel}{p_{\rm cl}}
\ll1.
\]
The systematic part of these decrements produces the mean stopping force
already included in \(F_{\rm eff}\). Their fluctuations, together with
the transverse recoil components, perturb the classical momentum only
weakly.

In a homogeneous chamber, the transverse momentum transfers have no
preferred direction, and the fluctuations about the mean longitudinal
loss are centered. Let \(q_{\rm rms}\) denote the characteristic
root-mean-square fluctuating momentum transfer accumulated over one
detector-resolved record interval. If \(N\) such records are formed
along a visible track, the accumulated root-mean-square momentum
fluctuation is of order
\[
\delta p_N
\sim
q_{\rm rms}\sqrt N,
\]
up to correlations and order-one geometric factors.
The requirement that the recorded sequence remain narrowly concentrated
about the effective classical trajectory is
\[
\delta p_N
\ll
p_{\rm cl},
\]
or, more generally,
\[
\delta p_N
\ll
p_{\rm res},
\]
where \(p_{\rm res}\) is the momentum resolution relevant to the track.
Likewise, the accumulated transverse displacement generated by the
stochastic momentum transfers must remain small compared with the
spatial resolution:
\[
\delta a_N
\ll
\sigma.
\]
Under these conditions, the centered environmental fluctuations modify
the recorded trajectory only weakly.

Let \(t_k\) denote the successive record times, and define
\[
a_k
=
\mu_x\bigl(\psi(t_k)\bigr),
\qquad
p_k
=
\mu_p\bigl(\psi(t_k)\bigr).
\]
The state at \(t_k\) is the state actually reached by the full stochastic
unitary evolution. No projection is applied.
Let \((a_{\rm cl}(t),p_{\rm cl}(t))\) denote the effective classical
trajectory generated by the deterministic tangent drift, including the
mean stopping force, with initial data \((a_k,p_k)\) at time \(t_k\).
Between successive record-forming interactions, the localized state
evolves into a narrow packet whose center follows this trajectory to the
accuracy established above. The next \({\bf RM}\) measurement produces
a detector-defined localized outcome with the corresponding Born
weight. Provided that ordinary spreading, stochastic recoil, and
variation of the effective force across the packet remain small, the
distribution of outcomes at a given next record time remains narrowly
concentrated about the propagated classical center. Consequently,
\[
\mathbb E\bigl(
a_{k+1}\mid a_k,p_k,t_{k+1}
\bigr)
=
a_{\rm cl}(t_{k+1})+o(\sigma)
\]
and
\[
\mathbb E\bigl(
p_{k+1}\mid a_k,p_k,t_{k+1}
\bigr)
=
p_{\rm cl}(t_{k+1})+o(p_{\rm res}).
\]

The microscopic and macroscopic aspects of the process must therefore
be distinguished. Each record is a microscopic position measurement:
the \({\bf RM}\) walk drives the state into one detector-defined
localized class with the corresponding Born weight. The visible
track, however, is a macroscopic sequence of such records. The effective
ordinary dynamics supplies the classical tangent drift between them,
while the centered environmental fluctuations produce only small
perturbations of position and momentum.

For a typical visible track, the momentum transferred in each
microscopic ionization is only a small fraction of the particle
momentum. Over a sufficiently short track segment, the accumulated
multiple-scattering and energy-loss fluctuations therefore remain small
compared with the relevant momentum resolution and with the microscopic
localization scale. The conditions
\[
\delta p_N\ll p_{\rm res},
\qquad
\delta a_N\ll\sigma
\]
are thus normally satisfied on the resolved segments of tracks produced
by sufficiently energetic charged particles.
They need not hold uniformly for every particle or along an entire
track. Near the end of its range, a particle loses momentum more rapidly
and undergoes larger deflections. Electrons and positrons may also suffer
substantial radiative losses and produce visibly spiraling or irregular
tracks. Such cases lie outside, or at the boundary of, the narrow-track
approximation used here.

The relevant conditions for a smooth track segment are therefore
\[
\frac{T_{\rm rec}}{T_{\rm spr}}
\ll1,
\qquad
\ell_{\rm int}\ll\sigma,
\qquad
T_\eta^{\rm loc}
\lesssim
\tau_{\rm int},
\]
together with
\[
\delta p_N
\ll
p_{\rm res},
\qquad
\delta a_N
\ll
\sigma,
\]
and the requirement that the deterministic motion and effective force
vary smoothly on the scale of a record-forming interaction.
For the energetic charged particles that produce ordinary reconstructible
bubble-chamber tracks, these conditions are physically typical over the
observed smooth portions of the track.
Under these conditions, the successive recorded positions and momenta
form a narrow stochastic sequence centered on the effective classical
trajectory.

Thus a bubble-chamber track does not establish that the microscopic
particle possessed a continuous trajectory in ordinary space between
the recorded events. What is observed is a succession of localized
outcomes whose centers lie close to a classical trajectory. In this
sense, the track is the microscopic counterpart of stroboscopic
Newtonian motion: each point is produced by a Born-rule measurement,
while the ordered sequence of points is classical to the accuracy of the
chamber.

\section{Laboratory Position Measurement of a Microscopic Particle}
\label{sec:laboratory-position-measurement-of-a-microscopic-particle}

We now consider a laboratory position measurement of a microscopic
particle produced by weak, frequent photon scattering. Suppose that a
sufficiently massive microscopic particle is prepared in a normalized
state \(\psi_0\) and illuminated by radiation. The scattered photons are
detected by a macroscopic apparatus, and their collective record
distinguishes the position of the particle with a finite resolution
\(\sigma\).

The measuring apparatus therefore distinguishes a family of
finite-resolution position classes
\[
\{g_{a,\sigma}\},
\]
rather than individual rays of projective Hilbert space. A recorded
outcome centered at \(a\) means that the state has entered the
equivalence class of states whose position expectation value is \(a\)
and whose position spread does not exceed the detector resolution
\(\sigma\).

We restrict attention to the weak repeated-interaction regime. Here
``weak'' means that a single effective scattering event produces only a
small change of the particle state and transfers momentum that is small
relative to the tolerated mechanical disturbance. No single photon is
assumed to localize the particle at the final resolution \(\sigma\).
Instead, each effective scattering event carries partial position
information, and many such events cumulatively drive the state into one
of the detector-defined localized classes.

The single-event spatial resolution, which depends on the photon
wavelength, scattering angle, optical geometry, and numerical aperture,
may therefore be much coarser than \(\sigma\). This is consistent with
the weakness of an individual event. The final resolution \(\sigma\) is
the scale at which the accumulated photon record distinguishes the
localized alternatives.

The illuminating radiation may be prepared as a localized wave packet
or as a coherent field containing a large mean number of photons. The
entire incident packet is not regarded as one step of the state-space
walk. Rather, each effective scattering event that transfers information
between the radiation field and the particle produces one step. Photons
that pass through the interaction region without an effective
interaction do not contribute such steps.

Before and during the measurement, we use the extended form of
conjecture \({\bf RM}\). The particle evolves under the full Hamiltonian
\[
\widehat h_{\rm tot}(t)
=
\widehat h+\widehat h_{\bf RM}(t),
\qquad
\widehat h
=
-\frac{\hbar^2}{2m}\Delta+\widehat V,
\]
and hence satisfies
\[
i\hbar\frac{d\psi}{dt}
=
\widehat h_{\rm tot}(t)\psi.
\]
The photons, detector, and surrounding apparatus are not included
explicitly as additional dynamical variables. Their unresolved effective
action on the particle state is represented by the random Hermitian
Hamiltonian \(\widehat h_{\bf RM}(t)\). For every realization of this
Hamiltonian, the evolution is linear and unitary.

The \({\bf RM}\) walk is homogeneous and isotropic in projective state
space. In particular, its restriction to the localized position
manifold \(M_3^\sigma\) gives a homogeneous and rotationally symmetric
spatial random walk. After passage to the Brownian limit, its covariance
has the form
\[
\mathbb E(da\,da^T)
=
2D_a I_3\,dt.
\]
In the Fubini--Study coordinates
\[
d\widetilde a
=
\frac{da}{2\sigma},
\]
this becomes
\[
\mathbb E(d\widetilde a\,d\widetilde a^T)
=
\frac{D_a}{2\sigma^2}I_3\,dt.
\]
Thus changing the spatial scale changes only the diffusion coefficient
and preserves isotropy. By Brownian scaling, the spatial step size or
time parametrization may be rescaled to the Fubini--Study scale
associated with the detector resolution \(\sigma\).

This spatial isotropy is the ingredient needed for the normal-to-Born
relation. The transition weight associated with a localized state
\(g_{a,\sigma}\) is determined by
\[
\bigl|
(g_{a,\sigma},\psi_0)
\bigr|^2.
\]
For an actual finite-resolution detector, however, the outcomes are
spatial cells rather than continuously labeled individual Gaussian
states. If the detector distinguishes a cell \(I\), the corresponding
Born probability is
\[
P_{\psi_0}(I)
=
\int_I
|\psi_0(x)|^2\,dx,
\]
to the accuracy determined by the detector resolution. Each individual
measurement follows one stochastic unitary path and produces one
localized outcome.

We next consider the mechanical disturbance caused by the photons. A
photon of wavelength \(\lambda\) has momentum
\[
p_\gamma
=
\frac{h}{\lambda}
=
\frac{2\pi\hbar}{\lambda}.
\]
The magnitude of the momentum transferred to the particle in one
scattering event is
\[
q_\gamma
=
\left|
p_{\gamma,\mathrm{in}}
-
p_{\gamma,\mathrm{out}}
\right|,
\]
where \(p_{\gamma,\mathrm{in}}\) and \(p_{\gamma,\mathrm{out}}\) are the
incoming and outgoing photon momentum vectors.

In an approximately elastic event,
\[
q_\gamma
\approx
2p_\gamma\sin\frac{\theta}{2},
\]
where \(\theta\) is the scattering angle. Small-angle scattering may
therefore transfer only a small fraction of the photon momentum,
whereas large-angle scattering produces recoil of order \(p_\gamma\).

The weak mechanical-disturbance condition is
\[
q_\gamma\ll p_{\rm tol},
\]
where \(p_{\rm tol}\) is the largest momentum disturbance that does not
appreciably alter the prepared particle state for the purpose of the
measurement. Generic strong, large-angle scattering is outside the
regime considered here.

The physical recoil \(q_\gamma\) and the size of the state-space step
are related but should not be identified. For two Gaussian states with
the same center and momenta differing by \(q_\gamma\), the overlap is
\[
\left|
\left(
g_{a,\sigma}e^{\frac{ipx}{\hbar}},
g_{a,\sigma}e^{i(p+q_\gamma)x/\hbar}
\right)
\right|
=
\exp\left[
-\frac{\sigma^2q_\gamma^2}{2\hbar^2}
\right].
\]
The corresponding Fubini--Study distance is
\[
\delta_{\gamma,p}
=
\arccos
\exp\left[
-\frac{\sigma^2q_\gamma^2}{2\hbar^2}
\right].
\]
In the weak projective regime,
\[
\frac{\sigma q_\gamma}{\hbar}\ll1,
\]
this reduces to
\[
\delta_{\gamma,p}
\sim
\frac{\sigma q_\gamma}{\hbar}.
\]
The full photon-induced step may also contain position, localization, and
other normal components. The mechanical condition
\(q_\gamma\ll p_{\rm tol}\) and the projective small-step condition are
therefore distinct.

Let \(E_s\) be the unit normal direction corresponding locally to a
change in the position spread. Write the localization component of the
\(j\)-th photon-induced step as
\[
\xi_jE_s,
\]
and define
\[
\kappa_s^2
=
\mathbb E(\xi_j^2).
\]
In the weak repeated-interaction regime,
\[
\kappa_s\ll1,
\]
but many effective scattering events occur during the record-forming
interval. Their cumulative effect produces the approximately Brownian
normal motion that drives the state into the detector-defined localized
set.

Let
\[
\mathcal C
=
\bigcup_\alpha\mathcal C_\alpha
\]
be the union of the localized classes distinguished by the detector, and
let
\[
d_0
=
\rho_{\rm FS}(\psi_0,\mathcal C)
\]
be the initial Fubini--Study distance from the state to this set. Near a
nearest regular point of \(\mathcal C\), introduce a local normal
coordinate \(r\), with
\[
r_0=d_0,
\qquad
\mathcal C:\ r\leq0.
\]
The normal random walk may then be written locally as
\[
r_n
=
d_0+\sum_{j=1}^n\xi_j,
\]
where the increments are approximately independent and identically
distributed, with a continuous symmetric law satisfying
\[
\mathbb E(\xi_j)=0,
\qquad
\mathbb E(\xi_j^2)=\kappa_s^2.
\]

Let
\[
\tau_{\rm det}
=
\inf
\left\{
n\geq1:r_n\leq0
\right\}
\]
be the first effective scattering event at which the state enters the
localized set. Since the initial state need not lie on its boundary, the
large-\(n\) survival probability has the form
\[
P_{d_0}(\tau_{\rm det}>n)
\sim
\frac{C(d_0)}{\sqrt n},
\]
where \(C(d_0)\) depends on the initial normal distance and on the scale
and distribution of the increments. The universal content is the
\(n^{-1/2}\) first-passage behavior.

A prescribed failure probability \(\eta\) therefore corresponds to
approximately
\[
n_\eta
\sim
\frac{C(d_0)^2}{\eta^2}
\]
effective scattering events. If \(\Gamma_\gamma\) is the rate of
effective events that produce state-space increments, the associated
high-confidence localization time is
\[
T_\eta
\sim
\frac{C(d_0)^2}
{\eta^2\Gamma_\gamma}.
\]

When the increments are small compared with \(d_0\) and many events
occur before localization, the normal motion may be approximated by
Brownian motion. Its survival probability is then
\[
P_{d_0}(\tau_{\rm det}>n)
\approx
\operatorname{erf}
\left(
\frac{d_0}
{\sqrt{2n}\,\kappa_s}
\right),
\]
and hence
\[
n_\eta
\approx
\frac{d_0^2}
{
2\kappa_s^2
\bigl(\operatorname{erf}^{-1}\eta\bigr)^2
}.
\]
The corresponding localization time is
\[
T_\eta
\approx
\frac{d_0^2}
{
2\kappa_s^2
\bigl(\operatorname{erf}^{-1}\eta\bigr)^2
\Gamma_\gamma
}.
\]
The ordinary Hamiltonian remains active throughout the measurement and
need not be separated from \(\widehat h_{\bf RM}(t)\). During the
localization interval \(T_\eta\), what is required is not that the
tangent deterministic displacement be small in Fubini--Study units, but
that the ordinary motion remain smooth on the measurement time scale and
that spreading not destroy the localization being produced by the
measurement.

For a nonrelativistic particle, the characteristic spreading time at
resolution \(\sigma\) is
\[
T_{\rm spr}
=
\frac{m\sigma^2}{\hbar}.
\]
A sufficient condition for negligible ordinary spreading during the
measurement is therefore
\[
\frac{T_\eta}{T_{\rm spr}}
\ll1,
\]
together with the requirement that the deterministic force vary
negligibly across the localized state during the measurement interval.

The cumulative mechanical disturbance must be controlled separately. If
\(n_\eta\) effective scattering events occur and their fluctuating recoil
components are centered and approximately independent, the accumulated
root-mean-square recoil is of order
\[
\delta p_{\rm meas}
\sim
q_{\rm rms}\sqrt{n_\eta},
\]
where \(q_{\rm rms}\) denotes the characteristic fluctuating component
of one recoil. Any nonzero mean recoil must be added separately to the
deterministic momentum change. The weak repeated-interaction regime
requires
\[
\delta p_{\rm meas}\ll p_{\rm tol},
\]
together with the requirement that any systematic accumulated recoil
also remain below \(p_{\rm tol}\).

The simplified first-passage description is therefore valid when
\[
\kappa_s\ll1,
\qquad
n_\eta\gg1,
\qquad
\frac{T_\eta}{T_{\rm spr}}\ll1,
\qquad
\delta p_{\rm meas}\ll p_{\rm tol},
\]
and when the photon-induced increments are approximately stationary and
independent over the relevant measurement interval. Under these
conditions, many weak effective scattering events generate an
approximately Brownian \({\bf RM}\) motion, while ordinary spreading
and accumulated mechanical disturbance remain negligible at the
measurement resolution.

To indicate the physical scale of the conditions, consider a
\(^{87}{\rm Rb}\) atom confined in a one-dimensional box of length
\[
L=10^{-5}\ {\rm m}
\]
and measured with position resolution
\[
\sigma=5\times10^{-7}\ {\rm m}.
\]
Suppose that effective weak scattering events occur at the representative
rate
\[
\Gamma_{\bf RM}\sim2\times10^4\ {\rm s^{-1}}.
\]
Such rates are typical of fluorescence measurements of trapped atoms,
although the precise value depends strongly on the atomic transition,
detuning, optical collection efficiency, and cooling mechanism.

If the atom is prepared in an exact energy eigenstate of the box, its
ordinary Schr\"odinger evolution changes only its overall phase:
\[
\psi(t)
=
e^{-iE_1t/\hbar}\psi(0).
\]
Its ray in projective Hilbert space is therefore stationary before the
measurement interaction is included. The value of \(E_1/\hbar\) does
not measure projective motion.
For a preparation that is not an exact energy eigenstate, the relevant
ordinary projective time scale is instead determined by the energy
spread. A sufficient condition for little ordinary projective evolution
over the localization interval is
\[
\frac{\Delta E\,T_\eta}{\hbar}\ll1,
\]
where \(\Delta E\) is the energy uncertainty of the prepared state.

The localization time cannot be inferred from the scattering rate alone,
because it also depends on the elementary normal step scale
\(\kappa_s\) and on the initial projective distance \(d_0\). In the
Brownian approximation, the median localization time is
\[
T_{1/2}
\approx
\frac{2.2\,d_0^2}
{\kappa_s^2\Gamma_{\bf RM}},
\]
whereas a failure probability \(\eta\) gives
\[
T_\eta
\approx
\frac{d_0^2}
{
2\kappa_s^2
\bigl(\operatorname{erf}^{-1}\eta\bigr)^2
\Gamma_{\bf RM}
}.
\]
For example, taking
\[
d_0=0.3,
\qquad
\kappa_s=0.05,
\qquad
\Gamma_{\bf RM}=2\times10^4\ {\rm s^{-1}},
\]
gives a median localization time of order
\[
T_{1/2}\sim4\times10^{-3}\ {\rm s}.
\]
A \(95\%\)-confidence time is much longer, of order
\[
T_{0.05}\sim5\times10^{-1}\ {\rm s},
\]
because the one-dimensional first-passage distribution has a long
\(N^{-1/2}\) survival tail.

These values are representative rather than predictions of conjecture
\({\bf RM}\). A quantitative prediction requires a microscopic
calculation or experimental calibration of \(\kappa_s\). They show,
however, that for plausible weak-scattering rates and step sizes,
localization on millisecond-to-subsecond time scales is compatible with
the condition that the ordinary center-of-mass evolution remain small.
The accumulated recoil condition must be checked separately for the
chosen optical geometry and trapping or cooling scheme.

We now apply these considerations to two preparations of a heavy
microscopic particle. In the first example, a heavy atom is placed in a
known energy eigenstate of a box. In the second, a heavy neutral atom is
prepared in a coherent superposition of two spatially separated packets.
The examples illustrate, respectively, position measurement of a
spatially extended single wave function and measurement of a
superposition of two approximately localized alternatives.

\subsection{An Atom Prepared in a Box Eigenstate}
\label{subsec:an-atom-prepared-in-a-box-eigenstate}

Consider a neutral atom confined in a one-dimensional box of length
\(L\). We assume that its internal state is fixed and that only its
center-of-mass motion is relevant. If \(m_A\) is the mass of the atom,
the normalized center-of-mass energy eigenstates are
\[
\psi_n(x)
=
\sqrt{\frac{2}{L}}\,
\sin\frac{n\pi x}{L},
\qquad
0<x<L,
\]
with energies
\[
E_n
=
\frac{n^2\pi^2\hbar^2}{2m_A L^2}.
\]
The preparation is therefore unambiguous: before measurement, the atom
is in the known state \(\psi_n\), whose position density is
\[
|\psi_n(x)|^2
=
\frac{2}{L}
\sin^2\frac{n\pi x}{L}.
\]

If the box remains present during the measurement, the ordinary
Schr\"odinger evolution of the initial state is
\[
e^{-\frac{i\widehat h t}{/\hbar}}\psi_n
=
e^{-\frac{iE_nt}{\hbar}}\psi_n.
\]
It therefore changes only the overall phase and leaves the initial
projective state fixed. Before the measurement-induced random motion
begins, the ordinary Hamiltonian produces no projectively observable
change of the prepared state.

During the measurement, however, the state satisfies the full linear
Schr\"odinger equation
\[
i\hbar\frac{d\psi}{dt}
=
\bigl(
\widehat h+\widehat h_{\bf RM}(t)
\bigr)\psi.
\]
Once the \({\bf RM}\) motion has carried the state away from the exact
eigenstate \(\psi_n\), the ordinary Hamiltonian may generate
projectively nontrivial motion. At a state \(\psi\), its
Fubini--Study speed is
\[
v_{\widehat h}(\psi)
=
\frac{\Delta E(\psi)}{\hbar},
\]
where
\[
\Delta E(\psi)^2
=
\langle\widehat h^2\rangle_\psi
-
\langle\widehat h\rangle_\psi^2.
\]
Let \(\Delta E_{\rm eff}\) denote a characteristic upper bound on this
energy uncertainty over the states explored during the record-forming
interval. The projective displacement attributable to the ordinary
Hamiltonian over a measurement time \(T\) is then bounded in order of
magnitude by
\[
\varepsilon_A(T)
\lesssim
\frac{\Delta E_{\rm eff}T}{\hbar}.
\]
Thus the ordinary Hamiltonian contribution to the projective motion is
negligible during the record-forming process when
\[
\frac{\Delta E_{\rm eff}T}{\hbar}
\ll1.
\]
This condition does not require a factorization of the ordinary and
measurement-induced unitary evolutions.

Let
\[
\mathcal C
=
\bigcup_\alpha\mathcal C_\alpha
\]
be the union of the localized classes distinguished by the detector.
The detector resolves the box into intervals \(I_\alpha\) of width of
order \(\sigma\), with \(\mathcal C_\alpha\) denoting the localized
classes whose centers lie in \(I_\alpha\). By the normal-to-Born
relation, the probability of the corresponding outcome is
\[
P(\mathcal C_\alpha\mid\psi_n)
=
\int_{I_\alpha}
|\psi_n(x)|^2\,dx
=
\int_{I_\alpha}
\frac{2}{L}
\sin^2\frac{n\pi x}{L}\,dx.
\]
Each individual run follows one stochastic unitary path and produces one
localized outcome.

Let
\[
d_0
=
\rho_{\rm FS}(\psi_n,\mathcal C)
\]
be the initial distance from the prepared eigenstate to the
detector-defined localized set. For a
finite-resolution detector,
\[
0\leq d_0<\frac{\pi}{2}.
\]
Near a nearest regular point of \(\mathcal C\), introduce a local normal
coordinate \(r\), with
\[
r_0=d_0,
\qquad
\mathcal C:\ r\leq0.
\]
The normal component of the elementary \({\bf RM}\) motion is then
represented locally by
\[
r_N
=
d_0+\sum_{j=1}^N\xi_j,
\]
where the increments have a continuous symmetric distribution satisfying
\[
\mathbb E(\xi_j)=0,
\qquad
\mathbb E(\xi_j^2)=\kappa_s^2.
\]

Let
\[
\tau_{\rm det}
=
\inf
\left\{
N\geq1:r_N\leq0
\right\}
\]
be the first step at which the state enters one of the detector-defined
localized classes. Since the walk begins at a positive distance
\(d_0\), its large-\(N\) survival probability has the form
\[
P_{d_0}(\tau_{\rm det}>N)
\sim
\frac{C(d_0)}{\sqrt N},
\]
where \(C(d_0)\) depends on the initial distance and on the scale and
distribution of the elementary normal increments.

For a prescribed failure probability \(\eta\), the corresponding
high-confidence number of elementary \({\bf RM}\) steps is therefore
\[
N_\eta
\sim
\frac{C(d_0)^2}{\eta^2}.
\]
If \(\Gamma_{\bf RM}\) denotes the rate of elementary \({\bf RM}\)
steps during the measurement interaction, the corresponding measurement
time is
\[
T_\eta
\sim
\frac{C(d_0)^2}
{\eta^2\Gamma_{\bf RM}}.
\]

When the elementary normal increments are small compared with \(d_0\)
and many steps occur before localization, the normal motion may be
approximated by Brownian motion. In that regime,
\[
P_{d_0}(\tau_{\rm det}>N)
\approx
\operatorname{erf}
\left(
\frac{d_0}
{\sqrt{2N}\,\kappa_s}
\right),
\]
and
\[
N_\eta
\approx
\frac{d_0^2}
{
2\kappa_s^2
\bigl(\operatorname{erf}^{-1}\eta\bigr)^2
}.
\]
The corresponding measurement time is
\[
T_\eta
\approx
\frac{d_0^2}
{
2\kappa_s^2
\bigl(\operatorname{erf}^{-1}\eta\bigr)^2
\Gamma_{\bf RM}
}.
\]

A quantitative estimate of the localization time requires the initial
projective distance \(d_0\), the elementary normal step scale
\(\kappa_s\), and the effective step rate \(\Gamma_{\bf RM}\). For a
specified measurement arrangement, these quantities can be estimated
from the physical parameters of the apparatus. The prepared state and
the detector resolution determine
\[
d_0
=
\rho_{\rm FS}(\psi_n,\mathcal C),
\]
whereas the photon wavelength, scattering geometry, coupling strength,
and detection efficiency determine the information acquired in one
effective event and hence the scale \(\kappa_s\). The effective
scattering rate determines \(\Gamma_{\bf RM}\).

In the Brownian regime, these quantities give
\[
T_\eta
\approx
\frac{d_0^2}
{
2\kappa_s^2
\bigl(\operatorname{erf}^{-1}\eta\bigr)^2
\Gamma_{\bf RM}
}.
\]
When the elementary steps are not small, the corresponding localization
time must instead be obtained from the full discrete first-passage law.

A concrete experimental setup therefore permits one to check whether
localization occurs during the record-forming interval and to estimate
its characteristic time. It also permits one to verify that the
ordinary projective evolution and the accumulated mechanical recoil
remain negligible at the resolution of the measurement. These estimates
are apparatus-dependent and are not universal constants of conjecture
\({\bf RM}\).

A sufficient condition for using the fixed initial state \(\psi_n\) in
the first-passage and outcome-weight calculation throughout the
high-confidence localization interval is
\[
\frac{\Delta E_{\rm eff}T_\eta}{\hbar}
\ll1.
\]
Using the general first-passage estimate, this becomes
\[
\frac{
\Delta E_{\rm eff}C(d_0)^2
}{
\hbar\eta^2\Gamma_{\bf RM}
}
\ll1.
\]
In the Brownian regime, it becomes
\[
\frac{
\Delta E_{\rm eff}d_0^2
}{
2\hbar\kappa_s^2
\bigl(\operatorname{erf}^{-1}\eta\bigr)^2
\Gamma_{\bf RM}
}
\ll1.
\]

When this condition holds, the contribution of the ordinary Hamiltonian
to the projective motion during localization is negligible, and the
outcome weights are accurately determined by the prepared eigenstate:
\[
P(\mathcal C_\alpha\mid\psi_n)
=
\int_{I_\alpha}
\frac{2}{L}
\sin^2\frac{n\pi x}{L}\,dx.
\]
When the condition does not hold, the state still evolves linearly and
unitarily under
\[
\widehat h+\widehat h_{\bf RM}(t),
\]
but the first-passage problem must be evaluated with the ordinary
Schr\"odinger drift included. In that case the outcome probabilities
are determined by the full stochastic evolution and cannot in general
be obtained from the fixed initial-state weights alone.

In the regime in which the ordinary Hamiltonian contribution remains
negligible during localization, repetition of the preparation and
measurement yields the position distribution of the atomic
center-of-mass eigenstate, while each individual run produces one
detector-defined localized outcome. The outcome weights follow from the
normal-to-Born relation, and the localization time is determined by the
initial distance to the detector-defined set together with the strength
and rate of the elementary measurement-induced steps.

\subsection{A Heavy Atom Prepared in a Two-Packet Superposition}
\label{subsec:a-heavy-atom-prepared-in-a-two-packet-superposition}

Consider next a heavy neutral atom initially confined in a symmetric
double-well potential. By coherently splitting a single localized packet,
one may prepare a state of the form
\[
\psi_0
=
\frac{
g_{a,\sigma_0}
+
g_{b,\sigma_0}
}{
\sqrt{
2\bigl(
1+\operatorname{Re}
(g_{a,\sigma_0},g_{b,\sigma_0})
\bigr)
}
}.
\]
When the two packets are well separated,
\[
(g_{a,\sigma_0},g_{b,\sigma_0})\simeq0,
\]
and hence
\[
\psi_0
\simeq
\frac{
g_{a,\sigma_0}
+
g_{b,\sigma_0}
}{\sqrt2}.
\]

The double well is used only to prepare the state and is switched off
immediately before the photon measurement begins. During the
measurement, the atom therefore satisfies the full linear Schr\"odinger
equation
\[
i\hbar\frac{d\psi}{dt}
=
\bigl(
\widehat h+\widehat h_{\bf RM}(t)
\bigr)\psi,
\]
where \(\widehat h\) is the ordinary free Hamiltonian and
\(\widehat h_{\bf RM}(t)\) represents the unresolved
photon--atom--detector interaction. For every realization of the random
Hamiltonian, the evolution is linear and unitary. No factorization of
the corresponding unitary evolution is assumed, and no commutativity of
\(\widehat h\) and \(\widehat h_{\bf RM}(t)\) is required.

Assume that each packet lies well within the corresponding localized
class distinguished by the detector. In the idealization in which the
two alternatives are represented by the packet rays themselves, their
Born weights are
\[
P_a
=
\left|
(g_{a,\sigma_0},\psi_0)
\right|^2
\simeq\frac12,
\qquad
P_b
=
\left|
(g_{b,\sigma_0},\psi_0)
\right|^2
\simeq\frac12.
\]
Thus each individual run produces one of the two localized outcomes,
with equal probability in the symmetric case.
More generally, for
\[
\psi_0
=
\alpha g_{a,\sigma_0}
+
\beta g_{b,\sigma_0},
\qquad
|\alpha|^2+|\beta|^2=1,
\]
the corresponding Born weights are approximately
\[
P_a\simeq|\alpha|^2,
\qquad
P_b\simeq|\beta|^2,
\]
provided that the packets are well separated and lie within distinct
detector-resolved classes.

Let
\[
\mathcal C
=
\mathcal C_a\cup\mathcal C_b
\]
be the union of the two detector-defined localized classes. In the
symmetric case, the initial state has the same Fubini--Study distance
from the two alternatives:
\[
d_a
=
\rho_{\rm FS}(\psi_0,\mathcal C_a)
\simeq
\frac{\pi}{4},
\qquad
d_b
=
\rho_{\rm FS}(\psi_0,\mathcal C_b)
\simeq
\frac{\pi}{4}.
\]
Here the equality is exact in the idealization in which each class is
represented by the corresponding packet ray. For finite-resolution
classes, the actual distance to the class may be smaller.

Because the symmetric state has two equally near localized
alternatives, the distance to
\(\mathcal C_a\cup\mathcal C_b\) need not define a smooth normal
coordinate at the initial state. The localization process is therefore
more properly viewed as a first-passage problem with two competing
channels. When the ordinary evolution is negligible over the localization
interval, the Born weights of the prepared state determine the relative
probabilities of the two channels, whereas the first-passage dynamics
determines the time required to reach their union.
For an order-of-magnitude estimate of this time, one may consider the
normal motion toward either branch. Let \(r\) be a local normal
coordinate associated with one of the two localized classes. In the
symmetric case,
\[
r_0
\simeq
\frac{\pi}{4}.
\]
The corresponding elementary normal motion is represented locally by
\[
r_N
=
\frac{\pi}{4}
+
\sum_{j=1}^N\xi_j,
\]
where the increments have an approximately continuous symmetric
distribution satisfying
\[
\mathbb E(\xi_j)=0,
\qquad
\mathbb E(\xi_j^2)=\kappa_s^2.
\]

Let
\[
\tau_{\rm det}
=
\inf
\left\{
N\geq1:
\psi_N\in\mathcal C_a\cup\mathcal C_b
\right\}
\]
be the first step at which the state reaches either localized class. The
large-\(N\) survival probability has the characteristic first-passage
form
\[
P(\tau_{\rm det}>N)
\sim
\frac{C}{\sqrt N},
\]
where \(C\) depends on the initial geometry and on the scale and
distribution of the elementary increments.

For a prescribed failure probability \(\eta\), the corresponding
high-confidence number of steps is therefore of order
\[
N_\eta
\sim
\frac{C^2}{\eta^2}.
\]
If \(\Gamma_{\bf RM}\) denotes the effective rate of elementary
measurement-induced steps, the associated localization time is
\[
T_\eta
\sim
\frac{C^2}
{\eta^2\Gamma_{\bf RM}}.
\]

When the elementary increments are small compared with \(\pi/4\), and
many steps occur before localization, each local normal channel may be
approximated by Brownian motion. Using the common initial distance
\[
d_0\simeq\frac{\pi}{4},
\]
one obtains the characteristic estimate
\[
T_\eta
\approx
\frac{(\pi/4)^2}
{
2\kappa_s^2
\bigl(\operatorname{erf}^{-1}\eta\bigr)^2
\Gamma_{\bf RM}
}.
\]
Because the full problem has two competing absorbing alternatives, this
formula should be understood as an order-of-magnitude estimate of the
time required to reach their union, rather than as an exact reduction of
the complete projective motion to a single normal coordinate.

For a specified experimental arrangement, the quantities entering this
estimate can be determined from the measurement parameters. The photon
wavelength, scattering geometry, coupling strength, and detection
efficiency determine the information acquired in one effective event and
hence the normal step scale \(\kappa_s\). The effective scattering rate
determines \(\Gamma_{\bf RM}\), while the prepared state and the
finite-resolution detector classes determine the initial distances
\(d_a\) and \(d_b\).
A concrete apparatus therefore permits an estimate of the localization
time and a check that localization occurs during the record-forming
interval. When the elementary steps are not small, the full discrete
two-channel first-passage law must be used instead of the Brownian
approximation.

We must compare the resulting localization time with the ordinary free
evolution of the atom. Take, for illustration,
\[
m_A\sim3\times10^{-25}\ {\rm kg},
\qquad
\sigma_0\sim5\times10^{-7}\ {\rm m}.
\]
The free spreading time of either packet is
\[
T_{\rm spr}
=
\frac{m_A\sigma_0^2}{\hbar}
\sim
7\times10^{-4}\ {\rm s}.
\]

If the residual mean velocity of either packet is \(v\), its physical
translation during the localization interval is of order
\[
vT_\eta.
\]
For the prepared two-packet geometry to remain effectively unchanged
during the measurement, it is sufficient that
\[
vT_\eta\ll\sigma_0,
\]
or, more generally, that the translation be small compared with the
spatial scales on which the detector distinguishes the two alternatives.

The free spreading of each packet is negligible when
\[
\frac{T_\eta}{T_{\rm spr}}
\ll1.
\]
These are physical smallness conditions on the change of the prepared
packet geometry. No requirement that the ordinary tangent displacement
be small in Fubini--Study units is imposed.
These conditions control the ordinary evolution directly and do not
require the ordinary and measurement-induced Hamiltonians to commute or
their unitary evolutions to be factored. If they fail, the underlying
dynamics remains unchanged, but the two-channel first-passage problem
must be evaluated with the ordinary Schr\"odinger evolution included.
The final outcome probabilities are then determined by the full
stochastic evolution and need not coincide with the Born weights of the
initial state \(\psi_0\).

When
\[
vT_\eta\ll\sigma_0,
\qquad
\frac{T_\eta}{T_{\rm spr}}\ll1,
\]
and any residual force produces only a negligible deformation of the
two-packet geometry during the measurement, the prepared state remains
effectively unchanged before localization occurs. The outcome weights
may then be evaluated from \(\psi_0\):
\[
P_a\simeq|\alpha|^2,
\qquad
P_b\simeq|\beta|^2.
\]
Each individual run nevertheless reaches only one detector-defined
localized class.

The outcome probabilities and the localization time describe distinct
aspects of the measurement. When the ordinary evolution is negligible
over the localization interval, the Born weights of the prepared state
determine the relative probabilities of the two localized outcomes,
whereas the first-passage dynamics determines how long it takes the
state to reach
\[
\mathcal C_a\cup\mathcal C_b.
\]
For a specified photon--atom measurement arrangement, the elementary
step scale and rate provide an apparatus-dependent prediction for the
time-dependent probability that localization has occurred.

\chapter{Two-Particle States and Measurement}
\label{chap:two-particle-measurement}

Chapters~\ref{chap:measurement-rm-walk} and~\ref{chap:macro-micro-measurement} treated the state reduction of a microscopic
particle and the stabilization of macroscopic motion. We now consider a
composite system. No new dynamical postulate is required. The same
\({\bf RM}\) conjecture is applied to the projective state space of the
tensor-product Hilbert space.

The many-particle localized manifolds and their Newtonian tangent
dynamics were constructed in Chapters~\ref{chap:classical-submanifolds} and~\ref{chap:newtonian-dynamics-from-schrodinger-dynamics}. We shall not repeat that
construction here. Our present concern is the effect of measurement on a two-particle
state, first when both particles are microscopic and then when one of
the particles is replaced by a macroscopic measuring device.

\section{The \({\bf RM}\) Walk on Tensor-Product State Space}
\label{sec:the-rm-walk-on-tensor-product-state-space}

Let
\[
\mathcal H
=
\mathcal H_1\otimes\mathcal H_2
\]
be the Hilbert space of two distinguishable particles, whose individual
Hilbert spaces are \(\mathcal H_1\) and \(\mathcal H_2\), respectively.
Let
\[
\Psi\in\mathcal H,
\qquad
\|\Psi\|=1,
\]
be their joint state. During a measurement or unresolved interaction with the environment,
the stochastic component of conjecture \({\bf RM}\) is applied to the
composite state itself. An elementary \({\bf RM}\) step has the form
\[
\Psi_{k+1}
=
\exp\left(
-\frac{i}{\hbar}\widehat H_k\Delta t
\right)\Psi_k,
\]
where the operators
\[
\widehat H_1,\widehat H_2,\ldots
\]
are independent random Hermitian operators drawn from the GUE on the
relevant finite-dimensional approximation to the tensor-product Hilbert
space. When ordinary dynamics is also present, this stochastic component
is combined with the ordinary Hamiltonian as in the extended form of
conjecture \({\bf RM}\) introduced in the preceding chapter.

Note that there is no separate \({\bf RM}\) walk for each particle. A
single random Hamiltonian acts on the joint state in the tensor-product
Hilbert space. Consequently, the walk takes place in the projective state space
\[
\mathbb P(\mathcal H_1\otimes\mathcal H_2),
\]
and may pass through product states, superpositions of product states,
and entangled states.
The unitary invariance of the GUE implies, exactly as in the one-particle
case, that the resulting walk is homogeneous and isotropic with respect
to the Fubini--Study metric of the composite projective space. The
distribution of a projective increment is therefore independent of the
initial state and has no preferred tangent direction. This statement
holds both at product states and at entangled states.

To relate the composite walk to ordinary Brownian motion, consider the
localized product state
\[
g_{a,\sigma_1}\otimes g_{b,\sigma_2},
\]
where \(g_{a,\sigma_1}\) and \(g_{b,\sigma_2}\) are localized states of
the first and second particles, respectively.
For two such states,
\[
\begin{split}
&
\left|
\left(
g_{a,\sigma_1}\otimes g_{b,\sigma_2},
g_{c,\sigma_1}\otimes g_{d,\sigma_2}
\right)
\right|^2
\\
&\hspace{2cm}
=
\exp\left[
-\frac{|a-c|^2}{4\sigma_1^2}
-\frac{|b-d|^2}{4\sigma_2^2}
\right].
\end{split}
\]
Hence the induced Fubini--Study line element on the two-particle
configuration manifold is
\[
ds^2
=
\frac{|da|^2}{4\sigma_1^2}
+
\frac{|db|^2}{4\sigma_2^2}.
\]
Introducing the normalized coordinates
\[
u=\frac{a}{2\sigma_1},
\qquad
v=\frac{b}{2\sigma_2},
\]
we obtain
\[
ds^2=|du|^2+|dv|^2.
\]

The tangential restriction of the isotropic \({\bf RM}\) walk to this
product manifold is therefore ordinary isotropic Brownian motion in the
joint coordinates \((u,v)\), with \(u\) and \(v\) describing the first
and second particles, respectively. In particular, over a short time interval,
\[
\mathbb E(du)=0,
\qquad
\mathbb E(dv)=0,
\]
\[
\mathbb E(du\,du^{T})=2D\,dt\,I,
\qquad
\mathbb E(dv\,dv^{T})=2D\,dt\,I,
\qquad
\mathbb E(du\,dv^{T})=0.
\]
Equivalently,
\[
\mathbb E[da\,da^{T}]
=
8\sigma_1^2D\,dt\,I,
\qquad
\mathbb E[db\,db^{T}]
=
8\sigma_2^2D\,dt\,I,
\]
and the cross-covariance vanishes.
The independence of the two coordinate increments applies only to the
tangential Brownian motion on the classical product manifold. It does not imply that an arbitrary
two-particle quantum state factorizes. Entanglement is a property of the
state \(\Psi\), whereas the covariance relations above describe the
restriction of the isotropic ambient walk to the localized product
coordinates.

The Brownian calibration on the localized product manifold fixes the
distance dependence of the transition law. There the Gaussian
transition law is related by the normal-to-Born correspondence to
\[
\cos^2\rho_{\rm FS}.
\]
Since the ambient \({\bf RM}\) walk is homogeneous and isotropic, its
transition law depends only on Fubini--Study distance, and the same
relation therefore extends to arbitrary states in
\[
\mathbb P(\mathcal H_1\otimes\mathcal H_2).
\]
Accordingly, for normalized two-particle states \(\Psi\) and \(\Phi\),
\[
W(\Psi,\Phi)
=
\cos^2\rho_{\rm FS}(\Psi,\Phi)
=
|(\Phi,\Psi)|^2.
\]
Thus the Born rule retains the same geometric form in the tensor-product
state space. As before, the finite-dimensional \({\bf RM}\) construction gives a
stochastic unitary realization of the dynamics, while in infinite
dimension its Gaussian generator law extends exactly as a cylindrical
Gaussian law on self-adjoint Hilbert--Schmidt probes. The controlled
large-\(N\) analysis provides a dynamical route consistent with the same
Born weights.

\section{Measurement of One Particle in a Two-Particle State}
\label{sec:measurement-of-one-particle-in-a-two-particle-state}

Let
\[
\Psi\in\mathcal H_1\otimes\mathcal H_2
\]
be a normalized state of two distinguishable particles. Suppose that the
position of the first particle is measured, while the second particle is
not observed. The measuring device distinguishes a collection of
finite-resolution cells
\[
I_1,I_2,\ldots
\]
in the position coordinate of the first particle.
The outcome associated with \(I_j\) records the position of the first
particle only. It therefore corresponds to a detector-defined
equivalence class of composite states for which the first particle is
localized in \(I_j\), while the state of the second particle is not
resolved.

In the idealized sharp-cell description, the corresponding spectral
sector of the composite Hilbert space is
\[
\mathcal H_{I_j}\otimes\mathcal H_2,
\]
where
\[
\mathcal H_{I_j}
=
\left\{
\psi\in\mathcal H_1:
\operatorname{supp}\psi\subset I_j
\right\}.
\]
The orthogonal projector onto this sector is
\[
P_{I_j}\otimes I.
\]
The relationship between the spectral sector and the finite-resolution equivalence class was discussed earlier. 
The former is used to calculate the Born weight of the latter.

As shown in Section~\ref{sec:the-rm-walk-on-tensor-product-state-space}
and Chapters~\ref{chap:brownian-born} \& \ref{chap:measurement-rm-walk},
the normal-to-Born relation, together with the distance-only dependence
of the homogeneous and isotropic \({\bf RM}\) transition law, fixes the
weights of the detector-defined outcomes to be the corresponding Born
weights. Hence the probability of recording the first particle in the
cell \(I_j\) is
\begin{equation}
\label{P_I_j}
W_j
=
\left\|
(P_{I_j}\otimes I)\Psi
\right\|^2.
\end{equation}
In the position representation,
\[
W_j
=
\int_{I_j}\int_{\mathbb R^3}
|\Psi(x,y)|^2\,dy\,dx.
\]
Thus the probability is obtained by integrating the joint position
density over the recorded cell of the first particle and over all
positions of the unobserved particle.

The projector in~\eqref{P_I_j} is used only to specify the spectral
weight associated with the detector outcome. It is not an additional
term in the physical dynamics. During measurement, the state evolves
linearly and unitarily in the full projective space
\[
\mathbb P(\mathcal H_1\otimes\mathcal H_2),
\]
with the \({\bf RM}\) component providing the stochastic
measurement-induced motion. The recorded outcome is determined by the
detector-defined equivalence class reached by this evolution.
For a product state
\[
\Psi=\psi_1\otimes\psi_2,
\]
the weight becomes
\[
\begin{split}
W_j
&=
\left\|
P_{I_j}\psi_1\otimes\psi_2
\right\|^2
\\
&=
\|P_{I_j}\psi_1\|^2\|\psi_2\|^2
\\
&=
\|P_{I_j}\psi_1\|^2.
\end{split}
\]
The measurement probability therefore depends only on the state of the
first particle.

If both positions are measured, with detector cells \(I_j\) for the
first particle and \(J_k\) for the second, the joint weight is
\[
W_{jk}
=
\left\|
(P_{I_j}\otimes P_{J_k})\Psi
\right\|^2.
\]
For a product state this factorizes:
\[
W_{jk}
=
\|P_{I_j}\psi_1\|^2
\|P_{J_k}\psi_2\|^2
=
W_j^{(1)}W_k^{(2)}.
\]
Thus product states give statistically independent position records.

For an entangled state, the corresponding weights do not generally
factorize. Consider, for example,
\[
\Psi
=
\sqrt{p_L}\,g_L\otimes h_L
+
\sqrt{p_R}\,g_R\otimes h_R,
\qquad
p_L+p_R=1,
\]
where \(g_L\) and \(g_R\) are approximately orthogonal localized states
of the first particle, and \(h_L\) and \(h_R\) are normalized states of
the second particle. Suppose that the detector cells \(I_L\) and \(I_R\)
resolve the supports of \(g_L\) and \(g_R\), respectively. Then
\[
P_{I_L}g_L\simeq g_L,
\qquad
P_{I_L}g_R\simeq0,
\]
and similarly with \(L\) and \(R\) interchanged. It follows that
\[
W_L
=
\left\|
(P_{I_L}\otimes I)\Psi
\right\|^2
\simeq p_L,
\]
and
\[
W_R
=
\left\|
(P_{I_R}\otimes I)\Psi
\right\|^2
\simeq p_R.
\]
These are the probabilities of the two detector-defined outcomes. The \({\bf RM}\) walk acts on the single state
\[
\Psi\in\mathcal H_1\otimes\mathcal H_2
\]
and reaches one of the corresponding detector-defined equivalence
classes in the full projective space.

For the particular two-component state considered here, the
\(L\)-equivalence class is represented, to the accuracy of the detector
resolution, by
\[
\Psi_L
\simeq
g_L\otimes h_L,
\]
while the \(R\)-equivalence class is represented by
\[
\Psi_R
\simeq
g_R\otimes h_R.
\]
The corresponding outcome weights are \(p_L\) and \(p_R\).
There is therefore no separate reduction of the first particle followed
by a subsequent change in the second. There is one state, one
measurement process in the tensor-product state space, and one
detector-defined equivalence class reached by that process.

For either outcome, the state relevant to subsequent measurements of the
second particle is obtained from a representative two-particle state
\(\Psi_j\) in the corresponding equivalence class by tracing over the
first factor:
\[
\rho_{2,j}
=
\operatorname{Tr}_1
|\Psi_j\rangle\langle\Psi_j|.
\]
Here \(\operatorname{Tr}_1\) denotes the partial trace over
\(\mathcal H_1\). More explicitly, if \(\{e_n\}\) is any orthonormal
basis of \(\mathcal H_1\), then
\[
\rho_{2,j}
=
\sum_n
\left(
\langle e_n|\otimes I
\right)
|\Psi_j\rangle\langle\Psi_j|
\left(
|e_n\rangle\otimes I
\right).
\]
Equivalently, \(\rho_{2,j}\) is the unique operator on
\(\mathcal H_2\) satisfying
\[
\operatorname{Tr}_{\mathcal H_2}
\left(
\rho_{2,j}B
\right)
=
\left\langle
\Psi_j,
(I\otimes B)\Psi_j
\right\rangle
\]
for every observable \(B\) acting on the second particle.

For the two equivalence classes above,
\[
\rho_{2,L}
\simeq
|h_L\rangle\langle h_L|,
\qquad
\rho_{2,R}
\simeq
|h_R\rangle\langle h_R|.
\]
Thus, if the two-particle state reaches the \(L\)-equivalence class, a
later measurement of the second particle tests the state \(h_L\); if it
reaches the \(R\)-equivalence class, it tests the state \(h_R\). The
later measurement does not create either state. It probes the
second-particle state associated with the equivalence class already
reached by the system.

The distinction between product and entangled states is therefore
direct. For a product state, the joint outcome weights factorize, and
the one-particle records are statistically independent. For an
entangled state, the \({\bf RM}\) dynamics acts on the single two-particle
state and carries it into one of the detector-defined equivalence
classes associated with the possible records. In the special
two-component example above, these classes are represented approximately
by the product states
\[
g_L\otimes h_L,
\qquad
g_R\otimes h_R.
\]
Only after such a two-particle equivalence class has been reached is it
meaningful to discuss the corresponding one-particle states separately.

In both cases, when the first particle is measured, the outcome weights
for the two-particle state are
\[
W_j
=
\left\|
(P_{I_j}\otimes I)\Psi
\right\|^2,
\]
which is the Born rule on the tensor-product state space.

\section{Identical Particles}
\label{sec:identical-particles}

The preceding discussion assumed that the two particles are
distinguishable. In that case the Hilbert space is the full tensor
product
\[
\mathcal H\otimes\mathcal H,
\]
and the two tensor factors may be associated with the two particles.

For identical particles, the same tensor-product Hilbert space is used
as the ambient space, but only one of its exchange-symmetry subspaces is
physical. Let \(\widehat S\) denote the exchange operator,
\[
\widehat S(\psi\otimes\phi)
=
\phi\otimes\psi.
\]
The symmetric and antisymmetric subspaces are
\[
\operatorname{Sym}\mathcal H
=
\left\{
\Psi\in\mathcal H\otimes\mathcal H:
\widehat S\Psi=\Psi
\right\},
\]
and
\[
\operatorname{Alt}\mathcal H
=
\left\{
\Psi\in\mathcal H\otimes\mathcal H:
\widehat S\Psi=-\Psi
\right\}.
\]
The first is the Hilbert space of symmetric two-particle
states, whereas the second is the Hilbert space of antisymmetric
two-particle states. Thus,
\[
\Psi(x_1,x_2)
=
\Psi(x_2,x_1)
\]
for a symmetric state, while
\[
\Psi(x_1,x_2)
=
-\Psi(x_2,x_1)
\]
for an antisymmetric state. When the spin degrees of freedom are included, the symmetry of the
spatial part combines with that of the spin part to give the required
bosonic or fermionic symmetry of the full state.

Let \(g_a\) and \(g_b\) be normalized localized one-particle position
states centered at \(a\) and \(b\). The corresponding normalized
symmetric and antisymmetric states are
\[
\Phi_{a,b}^{(+)}
=
\frac{
g_a\otimes g_b+g_b\otimes g_a
}{
\sqrt{2\bigl(1+|(g_a,g_b)|^2\bigr)}
},
\]
and
\[
\Phi_{a,b}^{(-)}
=
\frac{
g_a\otimes g_b-g_b\otimes g_a
}{
\sqrt{2\bigl(1-|(g_a,g_b)|^2\bigr)}
}.
\]
The second expression is defined only when \(g_a\) and \(g_b\) are not
the same one-particle state.

The exchange
\[
(a,b)\longleftrightarrow(b,a)
\]
does not produce a new physical configuration. Indeed,
\[
\Phi_{b,a}^{(+)}
=
\Phi_{a,b}^{(+)},
\qquad
\Phi_{b,a}^{(-)}
=
-\Phi_{a,b}^{(-)},
\]
and the minus sign in the antisymmetric case is projectively
irrelevant. Thus the classical configuration associated with two
identical particles is an unordered pair of points. Its configuration
space is
\[
\frac{
\mathbb R^3\times\mathbb R^3
}{
S_2
},
\]
where \(S_2\) acts by exchanging the two coordinates.
For antisymmetric position states, the set corresponding to \(a=b\) is
excluded.

This is the essential difference from distinguishable particles. For
distinguishable particles, the point
\[
(a,b)
\]
is represented by the product state
\[
g_a\otimes g_b.
\]
It means that particle \(1\) is localized near \(a\), while particle
\(2\) is localized near \(b\). The corresponding classical
configuration space is
\[
\mathbb R^3\times\mathbb R^3,
\]
embedded in the tensor-product state space as the manifold of localized
product states
\[
M_3^\sigma\otimes M_3^\sigma=\left\{g_a\otimes g_b:\ a,b\in\mathbb R^3\right\}.
\]

For identical particles, however, the unsymmetrized product
\[
g_a\otimes g_b
\]
is not a physical state. The physical localized states are instead the
symmetric or antisymmetric combinations
\[
\Phi_{a,b}^{(\pm)}.
\]
These states do not belong to the manifold
\[
M_3^\sigma\otimes M_3^\sigma
\]
and therefore cannot be represented by points of the classical
configuration space
\[
\mathbb R^3\times\mathbb R^3
\]
of two individually localized particles. 

When the packets are well separated,
\[
|(g_a,g_b)|\ll1,
\]
their exchange overlap is negligible. Then
\[
\Phi_{a,b}^{(\pm)}
\simeq
\frac{1}{\sqrt2}
\left(
g_a\otimes g_b
\pm
g_b\otimes g_a
\right).
\]
Locally, the corresponding state carries the classical position data of
two particles localized near \(a\) and \(b\), although neither particle
can be assigned a definite position in space.

The ordinary Schr\"odinger dynamics preserves the symmetry sector. If
the two-particle Hamiltonian is
\[
\widehat H
=
\widehat h\otimes I
+
I\otimes\widehat h
+
\widehat V,
\]
and the interaction \(\widehat V\) is invariant under exchange, then
\[
[\widehat H,\widehat S]=0,
\]
where \(\widehat S\) exchanges the two tensor factors. Hence a symmetric
initial state remains symmetric and an antisymmetric initial state
remains antisymmetric.

The same applies to the \({\bf RM}\) dynamics. No modification of the
conjecture is required; the random Hamiltonian acts on the physical
Hilbert space,
\[
\operatorname{Sym}\mathcal H
\qquad\text{or}\qquad
\operatorname{Alt}\mathcal H.
\]
Thus the \({\bf RM}\) walk takes place in
\[
\mathbb P\bigl(\operatorname{Sym}\mathcal H\bigr)
\]
or
\[
\mathbb P\bigl(\operatorname{Alt}\mathcal H\bigr),
\]
with the random operators drawn from the GUE on the corresponding
finite-dimensional coarse-grained space.

A position measurement must likewise be formulated without assigning
identities to the particles. Let \(I\subset\mathbb R^3\) be a detector
region, and let \(P_I\) be the corresponding one-particle position
projector. For distinguishable particles, one may ask whether the first
particle lies in \(I\), using
\[
P_I\otimes I.
\]
For identical particles, this is not an observable question. The
physical question is how many particles are recorded in \(I\).

The corresponding number operator is
\[
\widehat N_I
=
P_I\otimes I
+
I\otimes P_I.
\]
For two particles, its possible eigenvalues are
\[
0,\qquad 1,\qquad 2.
\]
Let \(P_{I,n}\) denote the projector onto the sector in which \(n\)
particles are recorded in \(I\). 
The detector outcomes \(n=0,1,2\) define the corresponding
finite-resolution equivalence classes, whose Born weights are
\[
P(N_I=n)
=
\|P_{I,n}\Psi\|^2.
\]

The corresponding projectors may be written explicitly. Let \(I^c\)
denote the complement of \(I\) in space. The projector onto the sector
with both particles outside \(I\) is
\[
P_{I,0}
=
P_{I^c}\otimes P_{I^c},
\]
the projector onto the sector with both particles inside \(I\) is
\[
P_{I,2}
=
P_I\otimes P_I,
\]
and the projector onto the sector with exactly one particle in \(I\) is
\[
P_{I,1}
=
P_I\otimes P_{I^c}
+
P_{I^c}\otimes P_I.
\]
These operators preserve both the symmetric and antisymmetric
two-particle sectors.

For a symmetric or antisymmetric state localized near two well-separated
points \(a\) and \(b\), with \(a\in I\) and \(b\notin I\), one obtains
\[
P(N_I=1)\simeq1.
\]
The measurement records one particle in \(I\) and one outside it. It does
not determine which of two pre-existing labeled particles entered the
detector, because no such particle labels are present in the physical
state.

Similarly, if a detector resolves two disjoint regions \(I\) and \(J\),
the physical outcome is the occupation configuration
\[
(N_I,N_J).
\]
For a state with one localized packet in each region, the record is
\[
N_I=1,
\qquad
N_J=1.
\]
For a symmetric position state, two particles may be recorded in the
same detector region:
\[
N_I=2.
\]
For an antisymmetric position state, the wave function vanishes when the
two position arguments coincide, although two particles may still be
recorded in the same finite region provided their localized position
states are distinct.

\section{A Particle and a Macroscopic Measuring Device}
\label{sec:a-particle-and-a-macroscopic-measuring-device}

We now include the measuring device explicitly in the system. Let
\[
\mathcal H
=
\mathcal H_{\rm p}\otimes\mathcal H_{\rm D}
\]
be the Hilbert space of the particle--device system, and let
\[
\Phi\in\mathcal H
\]
be its normalized joint state. Although the device is macroscopic, it is
described quantum mechanically in the same tensor-product state space as
the particle. The particle and the device are assumed to be
distinguishable.

This description must agree with the particle-centered
description used in Chapter~\ref{chap:macro-micro-measurement}. There, the measuring device and the
environment were not included explicitly; their effective
action on the particle was represented by the \({\bf RM}\) Hamiltonian
\(
\widehat h_{\bf RM}(t).
\)
Here, the same conjecture \({\bf RM}\) is applied directly to the joint
particle--device state.
During the measurement interaction, the total Hamiltonian has the form
\[
\widehat H_{\rm tot}
=
\widehat H_0+\widehat H_{\bf RM}(t),
\]
where
\[
\widehat H_0
=
\widehat h_{\rm p}\otimes I
+
I\otimes\widehat h_{\rm D}
+
\widehat H_{\rm int}.
\]
The first two terms describe the free dynamics of the particle and the
device, while \(\widehat H_{\rm int}\) accounts for the ordinary
interaction between them. The random term
\(\widehat H_{\bf RM}(t)\) represents the stochastic component of the
interaction among the particle, the device, and the environment.
On successive short time intervals, its values are taken to be
independent random Hermitian operators drawn from the GUE on the
relevant finite-dimensional coarse-grained sector of the
particle--device Hilbert space.

Thus the full dynamics acts on the joint state:
\[
\Phi_{k+1}
=
\exp\left[
-\frac{i}{\hbar}
\left(
\widehat H_0+\widehat H_{{\bf RM},k}
\right)\Delta t
\right]\Phi_k,
\]
and the resulting motion takes place in
\[
\mathbb P(\mathcal H_{\rm p}\otimes\mathcal H_{\rm D}).
\]
There is no separate stochastic evolution for the particle and another
for the device. The \({\bf RM}\) Hamiltonian acts on the single joint
state.

Let
\[
a=(a_1,a_2,a_3)
\]
denote the orthogonal position coordinates of the particle, and let
\[
A=(A_1,\ldots,A_N)
\]
denote the relevant orthogonal macroscopic pointer coordinates of the
device.
Localized product particle--device states may be represented by
\[
g_{a,\sigma_{\rm p}}
\otimes
G_{A,\sigma_{\rm D}},
\]
where \(\sigma_{\rm p}\) and \(\sigma_{\rm D}\) are the corresponding
localization scales. For two such states,
\[
\begin{split}
&
\left|
\left(
g_{a,\sigma_{\rm p}}
\otimes
G_{A,\sigma_{\rm D}},
g_{b,\sigma_{\rm p}}
\otimes
G_{B,\sigma_{\rm D}}
\right)
\right|^2
\\
&\hspace{2cm}
=
\exp\left[
-\frac{|a-b|^2}{4\sigma_{\rm p}^2}
-\frac{|A-B|^2}{4\sigma_{\rm D}^2}
\right].
\end{split}
\]
Hence the induced Fubini--Study line element on the localized product
manifold
\(
M_3^{\sigma_{\rm p}}\otimes M_3^{\sigma_{\rm D}}
\)
is
\[
ds^2
=
\frac{d|a|^2}{4\sigma_{\rm p}^2}
+
\frac{d|A|^2}{4\sigma_{\rm D}^2}.
\]
Introducing the normalized coordinates
\[
u
=
\frac{a}{2\sigma_{\rm p}},
\qquad
v
=
\frac{A}{2\sigma_{\rm D}},
\]
we obtain
\[
ds^2=d|u|^2+d|v|^2.
\]

The tangential restriction of the isotropic \({\bf RM}\) walk to this
product manifold is therefore Brownian motion in the orthonormal
coordinates \(u\) and \(v\). Over a short interval \(dt\),
\[
\mathbb E(du)=0,
\qquad
\mathbb E(dv)=0,
\]
\[
\mathbb E(du\,du^{T})=2D\,dt\,I,
\qquad
\mathbb E(dv\,dv^{T})=2D\,dt\,I,
\qquad
\mathbb E(du\,dv^{T})=0.
\]
In the original coordinates,
\[
\mathbb E(da\,da^{T})
=
8\sigma_{\rm p}^2D\,dt\,I,
\qquad
\mathbb E(dA\,dA^{T})
=
8\sigma_{\rm D}^2D\,dt\,I,
\]
and
\[
\mathbb E(da\,dA^{T})=0.
\]
Thus isotropy refers to the Fubini--Study-normalized coordinates, not to
equal Euclidean displacements of the particle and the macroscopic
device.

As discussed earlier for macroscopic bodies, continued environmental
monitoring repeatedly returns the device state to its localized
manifold, preventing any appreciable accumulation of its Brownian
displacement in ordinary space. In the present geometry, the sharp
localization of the device is represented by the small value of
\(\sigma_{\rm D}\). Indeed, the
\({\bf RM}\) increments \(du\) and \(dv\) have the same distribution,
but
\[
da
=
2\sigma_{\rm p}\,du,
\qquad
dA
=
2\sigma_{\rm D}\,dv.
\]
Consequently, the same Fubini--Study-normalized displacement produces a
much smaller displacement of the macroscopic device in the coordinate
\(A\). During the short interval in which the microscopic reduction
takes place, the device therefore remains essentially at its initial
macroscopic position.

The two localization scales have different physical origins. The scale
\(\sigma_{\rm p}\) is fixed by the resolution of the measuring device:
it determines which particle positions the device can distinguish. By
contrast, once the device itself is included in the quantum system, its
position is monitored by the surrounding environment. This
environmental monitoring sharply localizes the macroscopic device and
therefore determines the small localization scale
\(\sigma_{\rm D}\).

The preceding geometric argument can be expressed directly in the
finite-dimensional approximation used in conjecture \({\bf RM}\).
Suppose that initially
\[
\Phi_0
=
\psi_0\otimes F_0,
\qquad
F_0=G_{A,\sigma_{\rm D}}.
\]
Choose a finite family of localized device states
\[
F_j
=
G_{A_j,\sigma_{\rm D}}
\]
obtained by spatially displacing \(F_0\). Because
\(\sigma_{\rm D}\) is very small, states whose centers differ by even a
small macroscopic displacement are nearly orthogonal:
\[
\left|
(F_0,F_j)
\right|^2
=
\exp\left[
-\frac{(A-A_j)^2}{4\sigma_{\rm D}^2}
\right]
\ll1,
\qquad
j\neq0.
\]
These states may be used as an approximately orthogonal basis of the
finite-dimensional representation of \(\mathcal H_{\rm D}\).

Let
\[
e_1,\ldots,e_m
\]
be an orthonormal basis of the corresponding finite-dimensional
representation of \(\mathcal H_{\rm p}\). After one \({\bf RM}\) step,
write
\[
\Phi_1
=
\sum_{i,j}
c_{ij}\,e_i\otimes F_j.
\]
By Brownian scaling, the time interval may be chosen sufficiently small
that the Fubini--Study distance between \(\Phi_0\) and \(\Phi_1\) is,
with overwhelming probability, arbitrarily small.

Every term containing \(F_j\) with \(j\neq0\) is nearly orthogonal to
the initial state \(\psi_0\otimes F_0\), independently of its particle
factor. Hence, if such terms had nonnegligible total weight, the
Fubini--Study displacement of the joint state could not be small.
It follows that
\[
\sum_i\sum_{j\neq0}|c_{ij}|^2
\ll1.
\]
Consequently,
\[
\Phi_1
=
\psi_1\otimes F_0
+
\eta_1,
\qquad
\|\eta_1\|^2\ll1,
\]
where \(\psi_1\) is a normalized particle state up to a correction of
the same negligible order.

The difference between the two factors is due to their very different
localization scales. Since
\[
\sigma_{\rm D}\ll\sigma_{\rm p},
\]
a physical displacement that makes two device states nearly orthogonal
may correspond to only a small change of the particle state. Thus a
small \({\bf RM}\) step can change the particle factor while any
appreciable spatial displacement of the device factor would make the
total step too large.

For any prescribed failure probability \(\eta\), let \(N_\eta\) be a
number of \({\bf RM}\) steps sufficient for the joint state to reach one of the detector-defined equivalence classes with probability at least
\(1-\eta\). For this fixed high-confidence number of steps,
\(\sigma_{\rm D}\) may be sufficiently small that the probability of
acquiring appreciable weight in displaced device sectors during any of
these steps is negligible. Hence, with probability at least
\(1-\eta\),
\[
\Phi_k
=
\psi_k\otimes F_0+\chi_k,
\qquad
\|\chi_k\|^2\ll1,
\qquad
k=1,\ldots,N_\eta
\] 
throughout the microscopic reduction process.
The \({\bf RM}\) evolution of the
joint state therefore reduces, to the required accuracy, to the
\({\bf RM}\) evolution of its particle factor.

The same conclusion may be stated without referring to a particular
basis. Let \(\Delta_A\) be the interval of pointer positions corresponding to
the initial macroscopic state of the device, and let
\[
\mathcal H_A
\subset
\mathcal H_{\rm D}
\]
be the corresponding coarse-grained device sector. Denote by \(Q_A\)
the orthogonal projector onto \(\mathcal H_A\), and set
\[
Q_A^\perp=I-Q_A.
\]
For a joint state \(\Phi\),
\[
(I\otimes Q_A)\Phi
\]
is its component in which the device remains in the sector
\(\mathcal H_A\), whereas
\[
(I\otimes Q_A^\perp)\Phi
\]
is its component in which the device has left that sector.

The basis argument above gives
\[
\left\|
(I\otimes Q_A^\perp)\Phi_k
\right\|^2
\ll1,
\qquad
k=1,\ldots,N_\eta.
\]
The quantity on the left is the Born weight of the component in which
the device has left its initial macroscopic position sector.
Equivalently,
\[
\Phi_k
=
(I\otimes Q_A)\Phi_k
+
O(\varepsilon),
\qquad
\varepsilon\ll1.
\]
At the level of finite-resolution equivalence classes, this may be
written as
\[
\{\Phi_k\}
\simeq
\{\psi_k\}\otimes\{F_0\}.
\]

This conclusion concerns the \({\bf RM}\) part of the evolution. It
does not imply that the device remains fixed under its ordinary
Hamiltonian. Since the device remains separable from the particle to the accuracy
considered here, its ordinary evolution is governed by the restriction
of the particle--device Hamiltonian \(\widehat H_0\) to the localized
product manifold.
As established earlier, this restriction yields the usual Newtonian
dynamics of the macroscopic device. 
This ordinary Newtonian motion will provide the mechanism of recording
discussed in Section~\ref{sec:from-state-reduction-to-macroscopic-recording}.

The fact that the particle--device state remains effectively separable
also establishes the equivalence between the \({\bf RM}\)-based
dynamics of the joint system and the particle-centered description
used in Chapter~\ref{chap:macro-micro-measurement}. During the microscopic reduction
process, the joint state has, to the required accuracy, the form
\[
\Phi_k
=
\psi_k\otimes F_0
+
O(\varepsilon),
\qquad
\varepsilon\ll1,
\]
where the device factor remains in its initial macroscopic state.

Consider the steps of the joint \(\bf RM\) walk from
\[
\psi_k\otimes F_0
\]
for which the displacement of the device state is negligible. To the
same accuracy, such steps have the form
\[
\psi_k\otimes F_0
\longmapsto
\psi_{k+1}\otimes F_0.
\]
They lie in the projective subspace
\[
\mathbb P\bigl(
\mathcal H_{\rm p}\otimes\operatorname{span}\{F_0\}
\bigr),
\]
which is naturally identified with
\[
\mathbb P(\mathcal H_{\rm p}).
\]

The joint \({\bf RM}\) steps have a Gaussian isotropic distribution in
the tensor-product state space. Their orthogonal components in the
fixed-device subspace and its complement are therefore independent.
Consequently, restricting attention to steps for which the device
displacement is negligible does not change the distribution of the
component in the fixed-device subspace. That component is itself
Gaussian and isotropic.
Under the identification
\[
\psi\otimes F_0
\longleftrightarrow
\psi,
\]
it has precisely the distribution of the usual one-particle
\({\bf RM}\) step. Since successive joint steps are independent, the
successive particle steps are independent as well. Hence
\[
\psi_k\longmapsto\psi_{k+1}
\]
is the same \({\bf RM}\) random walk in
\[
\mathbb P(\mathcal H_{\rm p})
\]
as in the particle-centered description.

The Born probabilities agree directly. If \(P_\alpha\) denotes the
projector onto a particle outcome sector, then
\[
\begin{split}
\left\|
(P_\alpha\otimes I)
(\psi_0\otimes F_0)
\right\|^2
&=
\left\|
P_\alpha\psi_0\otimes F_0
\right\|^2
\\
&=
\|P_\alpha\psi_0\|^2,
\end{split}
\]
because \(F_0\) is normalized. The particle--device and particle-centered
descriptions therefore yield both the same \({\bf RM}\) walk of the
particle and the same Born probabilities. The particle-centered
description is obtained simply by leaving the negligibly changing
device factor implicit.

\section{From State Reduction to Macroscopic Recording}
\label{sec:from-state-reduction-to-macroscopic-recording}

It is important to distinguish the reduction of the microscopic state
from the formation of the macroscopic record. These are two stages of
the same measurement process, but they arise from different components
of the dynamics.
For the particle--device system, the total Hamiltonian is
\[
\widehat H_{\rm tot}
=
\widehat H_0+\widehat H_{\bf RM}(t),
\]
where
\[
\widehat H_0
=
\widehat h_{\rm p}\otimes I
+
I\otimes\widehat h_{\rm D}
+
\widehat H_{\rm int}.
\]
The random term \(\widehat H_{\bf RM}(t)\) produces the isotropic
stochastic component of the motion in
\[
\mathbb P(\mathcal H_{\rm p}\otimes\mathcal H_{\rm D}).
\]
The ordinary Hamiltonian \(\widehat H_0\) produces the Schr\"odinger
evolution of the particle and the device, including their interaction
described by \(\widehat H_{\rm int}\).

Before the microscopic particle has been localized by the measurement,
its state need not lie near the classical phase-space manifold. The
\({\bf RM}\) Hamiltonian acts on the joint particle--device state, but,
as shown in
Section~\ref{sec:a-particle-and-a-macroscopic-measuring-device}, the
device remains in its localized macroscopic equivalence class during
this stage. The resulting motion therefore reduces, to the required
accuracy, to the particle-centered \({\bf RM}\) motion and brings the
particle into one of the detector-defined localized equivalence classes. If the possible outcomes are labeled by \(\alpha\), the particle state
reaches
\(
\{g_{a_\alpha}\}
\)
while the device remains in its initial macroscopic equivalence class
\(
\{\Psi_A\}.
\)
The probability of this outcome is
\[
P_\alpha
=
\left\|
(P_\alpha\otimes I)\Phi_0
\right\|^2.
\]
Here \(\{\Psi_A\}\) is the initial macroscopic equivalence class of the
device. As explained in Section~\ref{sec:a-particle-and-a-macroscopic-measuring-device}, this class remains
stable during the microscopic reduction process.

The arrival of the particle state in
\(\{g_{a_\alpha}\}\) is the reduction of the state. At this point the
particle has a definite position at the resolution of the device. As
shown in Section~\ref{sec:localized-equivalence-classes}, such
localization also determines the linear phase of the state, identifying
it with a member of an equivalence class on the classical phase-space
manifold \(M_{3,3}^{\sigma_{\rm p}}\). The localized particle--device states
therefore form the submanifold
\[
M_{{\rm p},{\rm D}}
=
M_{3,3}^{\sigma_{\rm p}}
\otimes
M_{3,3}^{\sigma_{\rm D}}
\]
of the particle--device Hilbert space.

Let
\[
(a,p)
\]
be the classical position and momentum coordinates of the particle, and
let
\[
(A,P)
\]
be the corresponding coordinates of the pointer degree of freedom. The restriction of the ordinary Schr\"odinger action to this product
manifold is
\[
S
=
\int
\left[
p\cdot\dot a
+
P\cdot\dot A
-
H_{\rm cl}(a,p,A,P)
\right]dt,
\]
where
\[
H_{\rm cl}
=
\frac{|p|^2}{2m}
+
\frac{|P|^2}{2M_{\rm D}}
+
V(a,A).
\]
As shown in Chapter~\ref{chap:newtonian-dynamics-from-schrodinger-dynamics}, this expression is accurate when the localization
scales are small compared with the spatial scale on which \(V\) varies.

The tangent component of the ordinary Schr\"odinger evolution therefore
gives
\[
\dot a
=
\frac{p}{m},
\qquad
\dot p
=
-\nabla_a V(a,A),
\]
and
\[
\dot A
=
\frac{P}{M_{\rm D}},
\qquad
\dot P
=
-\nabla_A V(a,A).
\]
Thus, after localization, the particle and the device interact through
the coupled Newton equations.

This is the recording stage. The localized particle outcome supplies
the initial condition for the ordinary particle--device interaction.
Different values of \(a_\alpha\) produce different forces on the
device and hence different pointer trajectories,
\[
A(t)
=
A_\alpha(t).
\]
After a sufficiently long interaction, these trajectories enter
macroscopically distinguishable device classes,
\[
\{\Psi_A\}
\longrightarrow
\{\Psi_{A_\alpha}\}.
\]
At the level of localized equivalence classes, the recording process is
therefore
\[
\{g_{a_\alpha}\}\otimes\{\Psi_A\}
\quad\longrightarrow\quad
\{g_{a_\alpha}(t)\}\otimes\{\Psi_{A_\alpha}(t)\}.
\]

The \({\bf RM}\) dynamics does not produce the systematic pointer
trajectory. Its role in the microscopic stage is to bring the particle
into one of the detector-defined localized equivalence classes. Once
this has occurred, the tangent component of the ordinary Schr\"odinger
dynamics gives the Newtonian interaction of the localized particle with
the localized macroscopic device. It is this ordinary interaction that
produces the systematic pointer displacement and hence the record.
The recording need not literally be a pointer displacement. The same
structure applies when an atom is ionized, a bubble is formed, a
scintillation event occurs, an electric pulse is amplified, or a stable
mark is produced. In each case, the microscopic state first reaches a
localized detector-defined class. The subsequent process is an ordinary
interaction among localized degrees of freedom and is therefore
described, to the accuracy of the relevant resolution, by classical
dynamics.

After the new device class \(\{\Psi_{A_\alpha}\}\) has formed, continued
environmental monitoring keeps it stable while it evolves according to
Newtonian dynamics. For a properly prepared device, the outcome is
therefore retained as a persistent classical record.
The two mechanisms are complementary. The ordinary Schr\"odinger
dynamics supplies the Newtonian tangent motion of localized systems,
whereas the \({\bf RM}\) term supplies the stochastic state-space motion
that produces localization with Born probabilities and maintains the
localization of the macroscopic device. Both components are present
throughout the measurement. During microscopic reduction, the
\({\bf RM}\) component determines the stochastic approach to a
detector-defined outcome class. After localization, the ordinary
Hamiltonian supplies the systematic Newtonian motion that records the
outcome, while environmental monitoring continues to keep the device
localized.

\chapter{Quantum Correlations and Locality}
\label{chap:correlations-locality}

The double-slit experiment and the EPR experiment are usually treated as
different quantum paradoxes. In the state-space formulation they have the same
geometric origin. In the double-slit experiment, a one-particle state lies away
from the classical space manifold \(M^\sigma_3\). In the EPR experiment, a
two-particle entangled state lies away from the product classical phase-space
manifold
\[
M^\sigma_{1,1}\otimes M^\sigma_{1,1}.
\]
In both cases the apparent paradox comes from trying to describe a state-space
process as if it were already a process on a classical manifold.

\section{Single-Particle Case: The Double-Slit Experiment}
\label{sec:single-particle-case-the-double-slit-experiment}

The double-slit experiment and the formation of the interference pattern
on the target screen were discussed in Section~\ref{sec:detector-class-weights-and-interference}. Our purpose here
is to consider several versions of the experiment and, most importantly,
to give them a simple geometric interpretation.

The source, the slit screen, and the target screen are macroscopic
quantum systems and are described, together with the particle, in a
tensor-product Hilbert space. During the stages in which their
macroscopic states do not change, the relevant states have the form
\[
\psi\otimes F_{\rm src}\otimes F_{\rm sl}\otimes F_{\rm tar}.
\]
Being macroscopic, the three apparatus factors belong to the
corresponding localized equivalence classes and, as discussed in
Section~\ref{sec:a-particle-and-a-macroscopic-measuring-device}, remain
there to the required accuracy.

For two particle states \(\psi\) and \(\phi\), we have
\[
\begin{split}
&
\left|
\left(
\psi\otimes F_{\rm src}\otimes F_{\rm sl}\otimes F_{\rm tar},
\phi\otimes F_{\rm src}\otimes F_{\rm sl}\otimes F_{\rm tar}
\right)
\right|
\\
&\hspace{4cm}
=
|(\psi,\phi)|.
\end{split}
\]
The corresponding Fubini--Study distances between the total states and
the corresponding particle states are therefore equal. Thus, as long as the macroscopic factors remain in their fixed
equivalence classes, the geometry relevant to the experiment may be
described by the motion of the particle state in
\[
\mathbb P(\mathcal H_{\rm p}).
\]

Whenever the particle state is sufficiently close to the classical
phase-space submanifold of \(\mathbb P(\mathcal H_{\rm p})\), its
dynamics is approximately Newtonian. The particle is then observed as a
classical point object moving in \(\mathbb R^3\).

At the source, the particle is prepared in a localized state and is
therefore in a localized equivalence class of \(M_3^\sigma\). As
discussed earlier, its local phase also determines the corresponding
momentum, placing it on \(M_{3,3}^\sigma\). It may consequently be
interpreted as a classical point object moving toward the slit screen. After
passing through the screen with both slits open, its state is
approximately
\[
\psi_{\rm sl}
=
\alpha g_{a,\sigma}
+
\beta g_{b,\sigma},
\qquad
|\alpha|^2+|\beta|^2=1,
\]
where \(a\) and \(b\) are the slit positions. We assume that the slit
separation is sufficiently large compared with the slit width, so that
\[
(g_{a,\sigma},g_{b,\sigma})\simeq0.
\]

For such well-separated states, the nearest points of
\(M_3^\sigma\) are approximately \(g_{a,\sigma}\) and
\(g_{b,\sigma}\). Consequently,
\[
\cos\rho\left(\psi_{\rm sl},M_3^\sigma\right)
\simeq
\max\{|\alpha|,|\beta|\}.
\]
For a symmetric superposition,
\[
|\alpha|=|\beta|=\frac{1}{\sqrt2},
\]
this gives
\[
\rho\left(\psi_{\rm sl},M_3^\sigma\right)
\simeq
\frac{\pi}{4}.
\]
Thus, with both slits open, the interaction with the slit screen
generally moves the state away from the classical position manifold and
gives the system non-particle-like, or wave-like, properties.

The particle state nevertheless remains a single point of projective
state space, and the state-space picture of its interaction with the
slit screen is simple. Before reaching the screen, the path of the
state lies close to the classical submanifold and may be identified
with the usual path in \(\mathbb R^3\). During the interaction with the
screen, the distance between the state and \(M_3^\sigma\) increases.
The state-space path leaves the classical submanifold and passes around
the obstruction represented by the slit screen through the ambient
state space.
If one slit is closed, the outgoing state is approximately either
\(g_{a,\sigma}\) or \(g_{b,\sigma}\). Its distance from
\(M_3^\sigma\) is then nearly zero. The state subsequently moves away
from the fixed-width manifold as it spreads, but remains close to the
classical sector. The state-space and classical-space pictures therefore
approximately coincide.

Suppose next that a which-way measurement is performed at the slits.
The \({\bf RM}\) dynamics then brings the state into one of the two
localized equivalence classes,
\[
\psi_{\rm sl}
\overset{{\bf RM}}{\longrightarrow}
\{g_{a,\sigma}\}
\qquad\text{or}\qquad
\{g_{b,\sigma}\},
\]
with probabilities approximately
\[
|\alpha|^2
\qquad\text{and}\qquad
|\beta|^2.
\]
The state has therefore returned to a localized equivalence class of \(M_3^\sigma\).
To a classical observer, this appears as a particle having passed
through one of the slits. The state-space picture is different: the
state first moved away from the classical submanifold and was then
returned to it by the \({\bf RM}\) dynamics, which gives the Born
probabilities for finding the particle near the upper or lower slit.

Afterward, the localized state evolves according to the ordinary
Schr\"odinger equation. It spreads while moving toward the target
screen, but it is no longer a coherent superposition of the two slit
alternatives.  As long as it remains sufficiently close to the classical
sector, it retains an approximately particle-like description while
moving toward the target screen.
At the target screen, the \({\bf RM}\) measurement brings the state into
one of the localized target-screen classes. Repeating the experiment
many times and averaging over the two possible which-way outcomes gives
the sum of the two approximately normal single-slit distributions, with
no interference term.

If no which-way measurement is performed, the state evolves instead as
\[
\psi(t)
=
\alpha g_a(t)+\beta g_b(t).
\]
As the two packets spread, they overlap by the time they reach the
target screen. The \({\bf RM}\) walk then brings the resulting state
back to the classical sector and to one of the localized target-screen
classes, with Born weights determined by
\[
\begin{split}
\left|
\alpha g_a(x,t)+\beta g_b(x,t)
\right|^2
={}&
|\alpha|^2|g_a(x,t)|^2
+
|\beta|^2|g_b(x,t)|^2
\\
&+
2\operatorname{Re}
\left[
\alpha\overline{\beta}\,
g_a(x,t)\overline{g_b(x,t)}
\right].
\end{split}
\]
On repetition of the experiment, the final term produces the
interference pattern.

The particle--wave distinction is therefore geometric. Near
\(M_3^\sigma\), the state admits a particle-like position description. Away from \(M_3^\sigma\), the
state has no definite classical position and may display wave-like
behavior. A which-way measurement returns the state to the classical
manifold at the slits, whereas without such a measurement the coherent
two-slit state remains away from it until the final detection.

\section{The EPR State and the Product Classical Manifold}
\label{sec:the-epr-state-and-the-product-classical-manifold}

The original EPR setting involves two particles whose position and momentum
data are correlated. This form of the experiment is closer to the double-slit
case than the spin formulation. The alternatives are not spin directions, but
localized regions of the two-particle phase-space manifold. For simplicity we
continue using one spatial dimension. The generalization to three dimensions
is straightforward.

Let
\[
\mathcal H
=
\mathcal H_1\otimes\mathcal H_2
=
L^2(\mathbb R)\otimes L^2(\mathbb R)
\simeq
L^2(\mathbb R^2)
\]
be the two-particle Hilbert space. We write
\[
\varphi_{a,p}(z)
=
g_{a,\sigma}(z)e^{\frac{ipz}{\hbar}}
\]
for a localized one-particle phase-space state. The two-particle classical
phase-space submanifold is represented by product states of the form
\[
\Phi_{a_1,p_1;a_2,p_2}
=
\varphi_{a_1,p_1}\otimes \varphi_{a_2,p_2}.
\]
Equivalently, in the coordinate representation,
\[
\Phi_{a_1,p_1;a_2,p_2}(z_1,z_2)
=
\bigl(\varphi_{a_1,p_1}\otimes \varphi_{a_2,p_2}\bigr)(z_1,z_2)
=
g_{a_1,\sigma}(z_1)e^{\frac{ip_1z_1}{\hbar}}
g_{a_2,\sigma}(z_2)e^{\frac{ip_2z_2}{\hbar}}.
\]
The projective image of these product states will be denoted by
\[
M^\sigma_{1,1}\otimes M^\sigma_{1,1}.
\]
Thus, in this notation,
\[
M^\sigma_{1,1}\otimes M^\sigma_{1,1}
\subset
\mathbb P(\mathcal H_1\otimes\mathcal H_2)
\]
is the product classical phase-space manifold of two particles in one spatial
dimension. A point of this manifold represents a state in which the two particles
have definite classical position and momentum data
\[
(a_1,p_1;a_2,p_2).
\]

The EPR state is not such a point. It is an entangled state. A normalizable
version may be written as
\[
\Psi(z_1,z_2)
=
C
\exp\left[
-\frac{(z_1-z_2-d)^2}{4\Delta^2}
\right]
\exp\left[
-\frac{(z_1+z_2)^2}{16\Lambda^2}
\right],
\]
where
\[
\Delta \ll \sigma,
\qquad
\Lambda \gg \sigma .
\]
Thus the relative position is sharply localized, while the center coordinate
is broad. In momentum representation the same state has sharply localized
total momentum. More precisely,
\[
\delta(z_1-z_2) \sim \Delta,
\qquad
\delta(p_1+p_2) \sim \frac{\hbar}{\Lambda}.
\]
In the ideal EPR limit,
\[
\Delta \to 0,
\qquad
\Lambda \to \infty,
\]
one obtains the formal constraints
\[
z_1-z_2=d,
\qquad
p_1+p_2=0.
\]
Thus a measurement of \(z_1\) determines \(z_2=z_1-d\), and a measurement of
\(p_1\) determines \(p_2=-p_1\).

The two-particle state \(\Psi\) is a single point of the state space
\[
\mathbb P(\mathcal H_1\otimes\mathcal H_2),
\]
not two points in the one-particle state spaces.
This EPR state lies away from the product classical position manifold.
Indeed, the maximal squared overlap of \(\Psi\) with the product classical
phase-space manifold is
\[
\max_{\Phi\in M^\sigma_{1,1}\otimes M^\sigma_{1,1}}
|(\Psi,\Phi)|^2
=
\frac{8\sigma^2\Delta\Lambda}
{(\Delta^2+2\sigma^2)(2\Lambda^2+\sigma^2)}.
\]
The maximum is attained at
\[
a_1=\frac d2,
\qquad
a_2=-\frac d2,
\qquad
p_1=p_2=0.
\]
Therefore
\[
\rho\!\left(
\Psi,
M^\sigma_{1,1}\otimes M^\sigma_{1,1}
\right)
=
\arccos
\left[
\left(
\frac{8\sigma^2\Delta\Lambda}
{(\Delta^2+2\sigma^2)(2\Lambda^2+\sigma^2)}
\right)^{1/2}
\right].
\]
In the ideal EPR limit,
\[
\Delta \to 0,
\qquad
\Lambda \to \infty,
\]
the maximal squared overlap tends to zero:
\[
\max_{\Phi\in M^\sigma_{1,1}\otimes M^\sigma_{1,1}}
|(\Psi,\Phi)|^2
\longrightarrow 0.
\]
Hence
\[
\rho\!\left(
\Psi,
M^\sigma_{1,1}\otimes M^\sigma_{1,1}
\right)
\longrightarrow
\frac{\pi}{2}.
\]
Thus, in the ideal EPR limit, the state approaches the maximal
Fubini--Study distance from the product classical manifold.

\section{Detector Sectors and Outcome Weights}
\label{sec:detector-sectors-and-path-weights}

Let Alice measure the position of particle \(1\). Her apparatus
distinguishes finite-resolution intervals \(I_\alpha\). Let Bob measure
the position of particle \(2\), with intervals \(J_\beta\). Each pair
\((I_\alpha,J_\beta)\) defines a detector equivalence class corresponding
to the joint position record.
In the sharp-bin idealization, the associated spectral sector is
\[
\mathcal H^{zz}_{\alpha\beta}
=
\mathcal H_{I_\alpha}
\otimes
\mathcal H_{J_\beta},
\]
where
\[
\mathcal H_{I_\alpha}
=
\{\psi\in L^2(\mathbb R):\psi(z)=0
\ {\rm for\ a.e.}\ z\notin I_\alpha\},
\]
and similarly for \(\mathcal H_{J_\beta}\).

If \(P^z_{I_\alpha}\) and \(P^z_{J_\beta}\) denote the corresponding
projections, the Born weight of the joint outcome is
\[
W^{zz}_{\alpha\beta}(\Psi)
=
\left\|
\left(
P^z_{I_\alpha}\otimes P^z_{J_\beta}
\right)\Psi
\right\|^2.
\]
Equivalently,
\[
W^{zz}_{\alpha\beta}(\Psi)
=
\cos^2
\rho\left(
[\Psi],
\mathbb P(\mathcal H^{zz}_{\alpha\beta})
\right).
\]
Thus the Born weight of a detector-defined equivalence class is
determined by the distance from \([\Psi]\) to its associated spectral
sector.

Because \(\Psi\) is concentrated near
\[
z_1-z_2=d,
\]
the dominant sectors are those for which
\[
J_\beta \simeq I_\alpha-d.
\]
Thus, if Alice records \(z_1\in I_\alpha\), Bob's conditional position
distribution is concentrated near the translated interval \(I_\alpha-d\). This
is the position part of the EPR correlation.

Similarly, Alice and Bob may instead use momentum-sensitive apparatuses. Let
\(K_\alpha\) and \(L_\beta\) be momentum intervals, and let
\(P^p_{K_\alpha}\), \(P^p_{L_\beta}\) be the corresponding momentum-space
projections. The joint momentum sectors are
\[
\mathcal H^{pp}_{\alpha\beta}
=
\mathcal H_{K_\alpha}
\otimes
\mathcal H_{L_\beta},
\]
and their weights are
\[
W^{pp}_{\alpha\beta}(\Psi)
=
\left\|
\left(
P^p_{K_\alpha}\otimes P^p_{L_\beta}
\right)\Psi
\right\|^2 .
\]
Since the EPR state is concentrated near
\[
p_1+p_2=0,
\]
the dominant sectors satisfy
\[
L_\beta \simeq -K_\alpha.
\]
Thus, if Alice records \(p_1\in K_\alpha\), Bob's conditional momentum
distribution is concentrated near \(-K_\alpha\).

When the position measurements are performed, the \({\bf RM}\)
measurement dynamics acts on the single two-particle state, whose ray is
\[
[\Psi]\in
\mathbb P(\mathcal H_1\otimes\mathcal H_2).
\]
The Born weights of the possible joint outcomes are determined by the
Fubini--Study distances from \([\Psi]\) to the corresponding
detector-defined outcome classes.
For position measurements, the outcomes of appreciable weight are those
compatible with
\[
z_1-z_2=d.
\]
Thus
\[
J_\beta\simeq I_\alpha-d.
\]
In the ideal EPR limit, incompatible joint outcomes have zero weight.
For finite \(\Delta\), their weights decrease rapidly as their distance
from the correlation condition increases.

A momentum measurement is ultimately also recorded as a position
measurement. The measuring apparatus converts different momentum
intervals into different localized detector records. Thus the
\({\bf RM}\) reduction still occurs at the level of the final position
record, while the corresponding outcome weights may be expressed in
terms of the momentum projectors of the measured state.
Since the EPR state is concentrated near
\[
p_1+p_2=0,
\]
the momentum outcomes of appreciable weight satisfy
\[
L_\beta\simeq-K_\alpha.
\]
In the ideal EPR limit, incompatible joint momentum outcomes have zero
weight.

\section{Measurement Arrangements and RM Paths}
\label{sec:measurement-arrangements-and-rm-paths}

Notice that the \({\bf RM}\) walk itself is not tied to any particular choice
of settings of Alice's and Bob's devices. In particular, the stochastic motion
in state space is not defined by whether both devices are active or only one
of them is active, nor by whether the devices are arranged to measure position
or momentum. The devices, together with the unresolved environmental degrees
of freedom, provide the noise that is represented by the \({\bf RM}\) random
walk of the state. At the same time, the devices define the detector-defined equivalence classes that
constitute the possible outcomes of the measurement arrangement. The devices do not select a particular \({\bf RM}\) path, just as the
medium surrounding a Brownian particle does not select a particular
Brownian path.

Suppose first that Alice and Bob use position-measuring devices. If the two
devices are not arranged so that Bob's interval is the translate
\(I_\alpha-d\) of Alice's interval, then they cannot both record positions
compatible with the EPR correlation. 
If Alice records the position of particle \(1\) as \(a_1\), this
means that the \({\bf RM}\) walk has brought the two-particle state
into the corresponding localized equivalence class of
\[
M^\sigma_1\otimes M^\sigma_1.
\]
In the ideal-resolution notation this class is represented by
\[
g_{a_1,\sigma}(z_1)
\otimes
g_{a_1-d,\sigma}(z_2).
\]
The correlated outcome sector is therefore characterized by
\[
a_2=a_1-d,
\]
independently of whether Bob is measuring the position of particle \(2\) at
that moment.

Likewise, if Alice measures momentum, her apparatus converts the
momentum outcome into a localized macroscopic record. If that record
corresponds to the value \(p_1\), then the EPR correlation implies
\[
p_2=-p_1.
\]
This statement is independent of whether Bob is measuring momentum at that
moment. Alice's record selects the correlated sector of the joint state reached
by the \({\bf RM}\) walk. Bob's apparatus, when present and properly aligned
with that sector, may record the corresponding value, but it does not receive
this value by a signal sent from Alice's side.

This is similar to the double-slit experiment. Suppose two detectors are
placed near the two slits to measure the particle's position. If one detector
clicks, then the other is guaranteed not to click. In this case the state
becomes concentrated near the clicked detector, independently of whether the
second detector remains on, is turned off, or is displaced. If no which-slit
record is formed, the coherent state continues to the screen, and the
probabilities are computed from its distances to the screen sectors.

A position measurement and a momentum measurement on Alice's side
define different detector-defined outcome classes for the same initial
entangled state \(\Psi\). They do not reveal two simultaneously existing
classical outcomes of the same \({\bf RM}\) path. The corresponding
families of \({\bf RM}\) paths, Fubini--Study distances, and outcome
weights therefore differ, while the underlying homogeneous and isotropic
\({\bf RM}\) law remains the same.

\section{Bell Inequalities and Hidden Classical Data}
\label{sec:bell-inequalities-and-hidden-classical-data}

The EPR paradox arises if one assumes that the observed correlations are
generated by pre-existing local data. A concrete classical version of
this assumption represents each member of the ensemble by a point of the
product classical phase-space manifold
\[
M^\sigma_{1,1}\otimes M^\sigma_{1,1}.
\]
Equivalently, one considers product states
\[
\Phi_{a_1,p_1;a_2,p_2}
=
\varphi_{a_1,p_1}\otimes \varphi_{a_2,p_2},
\qquad
\varphi_{a,p}(z)=g_{a,\sigma}(z)e^{\frac{ipz}{\hbar}},
\]
together with a probability measure
\[
d\mu(a_1,p_1;a_2,p_2)
\]
on the product classical phase space.

The product classical phase-space model above is one realization of the
assumptions entering Bell's theorem. More generally, the hidden variable
\(\lambda\) in the theorem need not consist of classical positions and
momenta; it may belong to an arbitrary space. For two observers, Alice
and Bob, the essential assumption is that, once \(\lambda\) is specified,
Alice's outcome depends only on Alice's setting and Bob's outcome only
on Bob's setting. 
The usual Bell argument also assumes statistical independence of the
hidden-variable distribution from the freely chosen measurement
settings; stochastic local models may be reduced to the same
factorizable structure by enlarging the hidden variable if necessary.

For a Bell-type comparison, let Alice choose between two detector
settings \(\mathcal A,\mathcal A'\) and Bob between
\(\mathcal B,\mathcal B'\). Each measurement is coarse-grained into two
outcomes, labeled \(+1\) and \(-1\). In a deterministic hidden-data
description one therefore has
\[
A_{\mathcal A}(\lambda),
\quad
A_{\mathcal A'}(\lambda),
\quad
B_{\mathcal B}(\lambda),
\quad
B_{\mathcal B'}(\lambda)
\in\{+1,-1\}.
\]
The corresponding correlations are
\[
E(\mathcal A,\mathcal B)
=
\int
A_{\mathcal A}(\lambda)
B_{\mathcal B}(\lambda)
\,d\mu(\lambda),
\]
and similarly for the other three pairs of settings.

For every fixed \(\lambda\), let
\[
\begin{aligned}
S(\lambda)
={}&
A_{\mathcal A}(\lambda)B_{\mathcal B}(\lambda)
+
A_{\mathcal A}(\lambda)B_{\mathcal B'}(\lambda)
\\
&+
A_{\mathcal A'}(\lambda)B_{\mathcal B}(\lambda)
-
A_{\mathcal A'}(\lambda)B_{\mathcal B'}(\lambda).
\end{aligned}
\]
Since
\[
B_{\mathcal B}(\lambda),
\,
B_{\mathcal B'}(\lambda)
\in\{\pm1\},
\]
one of the combinations
\[
B_{\mathcal B}+B_{\mathcal B'}
\qquad\hbox{and}\qquad
B_{\mathcal B}-B_{\mathcal B'}
\]
vanishes, while the other equals \(\pm2\). Hence
\[
|S(\lambda)|=2.
\]
Averaging over \(\lambda\) gives the CHSH inequality
\[
\left|
E(\mathcal A,\mathcal B)
+
E(\mathcal A,\mathcal B')
+
E(\mathcal A',\mathcal B)
-
E(\mathcal A',\mathcal B')
\right|
\leq2.
\]

The essential point of the construction is the existence of a single
setting-independent probability space on which all four possible local
outcomes are defined. The state-space description is different. An
entangled state
\[
[\Psi]
\in
\mathbb P(\mathcal H_1\otimes\mathcal H_2)
\]
is not a point of the product classical manifold and does not represent
pre-existing classical data carried separately by the two particles.
The probabilities of detector outcomes are instead determined by the
position of \([\Psi]\) relative to the detector-defined outcome classes.

For detector arrangements \(\mathcal A\) and \(\mathcal B\), the Born
weights of the joint outcome classes are
\[
W_{\alpha\beta}(\mathcal A,\mathcal B)
=
\left\|
\left(
P^A_\alpha(\mathcal A)
\otimes
P^B_\beta(\mathcal B)
\right)\Psi
\right\|^2,
\qquad
\alpha,\beta=\pm1.
\]
The corresponding correlation is
\[
E(\mathcal A,\mathcal B)
=
\sum_{\alpha,\beta=\pm1}
\alpha\beta\,
W_{\alpha\beta}(\mathcal A,\mathcal B).
\]
Changing the detector settings changes the detector-defined outcome
classes and hence the families of \({\bf RM}\) paths leading to them.
The corresponding Born weights are nevertheless determined by the same
initial entangled state \(\Psi\). For suitable incompatible settings,
the resulting correlations violate the CHSH inequality. They therefore
cannot be represented by a single setting-independent local
hidden-variable model of the form assumed above.

\section{Bell Correlations in State Space}
\label{sec:bell-correlations-in-state-space}

A single stochastic \({\bf RM}\) trajectory evolves in an ambient space
larger than the space of directly observed records, while the detector
settings determine which regions of that space represent the possible
outcomes. We first illustrate this mechanism with an elementary
Brownian model. The example shows how a process that is local in a
higher-dimensional space may display apparently nonlocal correlations
when observed only through lower-dimensional records.

\subsection{A Three-Dimensional Brownian Model}
\label{subsec:three-dimensional-brownian-analogy}

Consider a Brownian particle moving in
\[
\mathbb R^3
\]
along a trajectory
\[
X_t=(x_t,y_t,z_t).
\]
Suppose that Alice has access only to the \(x\)-coordinate of the
particle's recorded position, while Bob has access only to its
\(y\)-coordinate. Alice may
choose between settings \(1\) and \(2\), and Bob between settings
\(I\) and \(II\). For example, with setting \(1\) Alice tests whether
the recorded event has \(x\)-coordinate \(x_1\), assigning the outcomes
\(+1\) and \(-1\) according as it does or does not. With setting \(2\)
she similarly tests whether the \(x\)-coordinate is \(x_2\). Bob's
settings \(I\) and \(II\) analogously test whether the \(y\)-coordinate
is \(y_1\) or \(y_2\), respectively.

For each pair of active settings
\[
(\alpha,\beta)
\in
\{1,2\}\times\{I,II\},
\]
let the detector arrangement define a recording surface or region
\[
\Sigma_{\alpha\beta}\subset\mathbb R^3.
\]
The four settings may correspond to different locations or segments
along the \(z\)-direction, which neither Alice nor Bob observes directly.

For a realized Brownian path \(\omega\), define
\[
\tau_{\alpha\beta}(\omega)
=
\inf
\left\{
t>0:
X_t(\omega)\in\Sigma_{\alpha\beta}
\right\}.
\]
The point
\[
X_{\tau_{\alpha\beta}}(\omega)
\]
determines both records. Its \(x\)-coordinate determines Alice's
outcome, while its \(y\)-coordinate determines Bob's outcome.

In three dimensions, different setting pairs can select different
events on the same Brownian path. For example,
\[
\tau_{1I}(\omega)
\neq
\tau_{1II}(\omega),
\]
and the corresponding points
\[
X_{\tau_{1I}}(\omega),
\qquad
X_{\tau_{1II}}(\omega)
\]
may have different \(x\)-coordinates. Thus it may happen that
\[
A(1,I,\omega)
\neq
A(1,II,\omega),
\]
even though Alice's setting is unchanged. Similarly,
\[
B(1,I,\omega)
\neq
B(2,I,\omega)
\]
may occur with Bob's setting held fixed.
This dependence does not result from any nonlocal change of the
Brownian trajectory. The motion remains local in \(\mathbb R^3\);
changing the detector arrangement changes the point of the ambient
trajectory that becomes the recorded event in the observed
\((x,y)\)-space.

The four outcomes are therefore not, in general, representable as
functions
\[
A_\alpha(\omega),
\qquad
B_\beta(\omega)
\]
defined independently of the remote setting. Instead, the observed
records have the form
\[
A=A(\alpha,\beta,\omega),
\qquad
B=B(\alpha,\beta,\omega).
\]
The Bell product representation need not exist for the projected
description, even though the underlying Brownian motion is local in
\(\mathbb R^3\).

To make the connection with Bell correlations explicit, suppose that
each recording surface is divided into four disjoint regions
\[
\Sigma_{\alpha\beta}^{rs},
\qquad
r,s=\pm1,
\]
where entry into \(\Sigma_{\alpha\beta}^{rs}\) produces the pair of
records
\[
A=r,
\qquad
B=s.
\]
Let
\[
p_{\alpha\beta}^{rs}
=
P
\left\{
X_{\tau_{\alpha\beta}}
\in
\Sigma_{\alpha\beta}^{rs}
\right\}
\]
be the corresponding first-entry probabilities. Then
\[
E_{\alpha\beta}
=
\sum_{r,s=\pm1}
rs\,p_{\alpha\beta}^{rs}.
\]
Because the four detector arrangements select different recording
regions in the ambient space, the four probability distributions
\[
p_{1I}^{rs},
\qquad
p_{1II}^{rs},
\qquad
p_{2I}^{rs},
\qquad
p_{2II}^{rs}
\]
need not be marginals of one joint distribution for four
counterfactual outcomes in the observed \((x,y)\)-space. Their
correlations may therefore violate the CHSH inequality.

For example, the recording regions may be chosen so that
\[
E_{1I}
=
E_{1II}
=
E_{2I}
=
-\frac{1}{\sqrt2},
\qquad
E_{2II}
=
\frac{1}{\sqrt2}.
\]
Then
\[
\left|
E_{1I}
+
E_{1II}
+
E_{2I}
-
E_{2II}
\right|
=
2\sqrt2>2.
\]

This example is not intended as a derivation of quantum correlations
from ordinary Brownian motion. It is a direct illustration of how
Bell-type correlations may arise from a local stochastic process in a
larger ambient space. The Brownian motion is local in
\(\mathbb R^3\), but only its detector-defined records in the
\((x,y)\)-space are observed. Different measurement arrangements select
different recording events of the ambient motion. Consequently,
correlations that appear nonlocal in the observed space may arise from
an entirely local stochastic process in the larger space.
The analogy with the \({\bf RM}\) model is direct. The \({\bf RM}\)
motion is local in projective state space, while observations are
confined to detector-defined equivalence classes associated with
classical records. Bell-type nonlocality may therefore reflect the
geometry of observing a higher-dimensional state-space process through
its classical submanifolds rather than a nonlocal dynamical influence.

\subsection{Two-Spin Correlations}
\label{subsec:two-spin-correlations}

For a single spin-\(\frac12\) particle, the projective spin state space
\[
\mathbb CP^1\simeq S^2
\]
and its relation to the \(2\times2\) GUE were discussed in
Section~\ref{subsec:the-bloch-sphere-and-the-2-2-gue}. The relation
between this two-state representation and the \({\bf RM}\) hitting
process associated with a superposition of two localized states was
discussed in
Section~\ref{sec:the-born-rule-as-a-hitting-probability}. We now pass to
two spin-\(\frac12\) particles.

The full Hilbert space of one nonrelativistic spin-\(\frac12\) particle
is
\[
\mathcal H
=
L^2(\mathbb R^3)\otimes\mathbb C^2.
\]
Spin measurements are ultimately recorded through spatial degrees of
freedom, as in the Stern--Gerlach experiment. Accordingly, the
\({\bf RM}\) walk takes place in the full position--spin state space.
The finite-dimensional spin spaces considered below provide a reduced
description of the spin sector selected by the measurement.

For two particles, the product spin states form
\[
S_A^2\times S_B^2
\simeq
\mathbb CP^1\times\mathbb CP^1,
\]
embedded in the full two-spin projective space
\[
\mathbb P(\mathbb C^2\otimes\mathbb C^2)
=
\mathbb CP^3.
\]
An entangled spin state lies in \(\mathbb CP^3\) but outside the product
submanifold \(S_A^2\times S_B^2\). As in the one-spin case,
\(\mathbb CP^3\) is a finite-dimensional representation of the spin
sector rather than the full position--spin arena of the \({\bf RM}\)
walk.

Let Alice choose a measurement direction \(a\) and Bob a direction
\(b\). Their settings select the two antipodal states
\[
|a,+\rangle,\qquad |a,-\rangle
\]
on Alice's Bloch sphere and
\[
|b,+\rangle,\qquad |b,-\rangle
\]
on Bob's. Equivalently, the corresponding one-particle outcome
projectors are
\[
P^A_r(a)
=
\frac12
\left(
I+r\,\boldsymbol{\sigma}_A\cdot a
\right),
\qquad
P^B_s(b)
=
\frac12
\left(
I+s\,\boldsymbol{\sigma}_B\cdot b
\right),
\]
where
\[
r,s=\pm1.
\]

The four possible joint outcomes are represented in the reduced spin
description by the product states
\[
|a,r\rangle\otimes|b,s\rangle,
\qquad
r,s=\pm1.
\]
At finite detector resolution, they correspond to detector-defined
equivalence classes
\[
\mathcal C_{ab}^{rs}.
\]
Thus each pair of settings \(a,b\) selects a family
\[
\left\{
\mathcal C_{ab}^{rs}
\right\}_{r,s=\pm1}
\]
of four possible joint records.

Consider the singlet state
\[
|\Psi^-\rangle
=
\frac{1}{\sqrt2}
\left(
|+\rangle|-\rangle
-
|-\rangle|+\rangle
\right).
\]
In the reduced spin description this state belongs to
\(\mathbb CP^3\) and lies outside the product-state submanifold. In the full description, a single \({\bf RM}\) trajectory evolves from
the corresponding entangled position--spin state. The spin apparatuses
convert the spin outcomes into spatially localized detector records. For
fixed settings \(a,b\), the joint record specifies which of the four
equivalence classes
\[
\mathcal C_{ab}^{rs}
\]
is reached.

The weight of the sector labelled by \(r,s\) is
\[
W_{rs}(a,b)
=
\left\|
\left(
P^A_r(a)\otimes P^B_s(b)
\right)
\Psi^-
\right\|^2.
\]
By the Born rule established earlier,
\[
W_{rs}(a,b)
=
P(r,s\mid a,b,\Psi^-)
=
\frac14
\left(
1-rs\,a\cdot b
\right).
\]
Hence
\[
E(a,b)
=
\sum_{r,s=\pm1}
rs\,W_{rs}(a,b)
=
-a\cdot b.
\]

Choose four coplanar directions
\[
a_1=0,
\qquad
a_2=\frac{\pi}{2},
\qquad
b_I=\frac{\pi}{4},
\qquad
b_{II}=-\frac{\pi}{4}.
\]
Then
\[
E(a_1,b_I)
=
E(a_1,b_{II})
=
E(a_2,b_I)
=
-\frac{1}{\sqrt2},
\]
while
\[
E(a_2,b_{II})
=
\frac{1}{\sqrt2}.
\]
Therefore
\[
\left|
E(a_1,b_I)
+
E(a_1,b_{II})
+
E(a_2,b_I)
-
E(a_2,b_{II})
\right|
=
2\sqrt2>2.
\]

The structure is the same as in the Brownian model above. The
\({\bf RM}\) motion takes place in the full state space, while different
measurement settings select different detector-defined equivalence
classes as possible records. Thus the four experiments
\[
(a_1,b_I),
\qquad
(a_1,b_{II}),
\qquad
(a_2,b_I),
\qquad
(a_2,b_{II})
\]
do not correspond to one fixed set of four outcome classes. Each pair
of settings selects its own family
\[
\left\{
\mathcal C_{ab}^{rs}
\right\}_{r,s=\pm1}.
\]
There is therefore no single detector-defined decomposition from which
the four counterfactual outcomes
\[
A_1,\qquad A_2,\qquad B_I,\qquad B_{II}
\]
can all be read simultaneously.

Alice's and Bob's records belong to one realization of the two-particle
state. They are not produced by independent \({\bf RM}\) walks assigned
separately to the two particles. A single state moves in
\[
\mathbb P(\mathbb C^2\otimes\mathbb C^2)
\]
in the reduced spin description, and in the full position--spin state
space in the complete description. The detector settings determine
which equivalence classes represent the possible records.
As in the Brownian model, the apparent nonlocality belongs to the
description in terms of the observed records; the underlying
\({\bf RM}\) motion remains local in the ambient state space.

\section{Parameter Independence and Outcome Independence}
\label{sec:parameter-independence-outcome-independence}

The Bell correlations obtained above raise the related questions of
parameter independence, outcome independence, and no-signaling. The
first two are conditional properties and depend on what is included in
the variable \(\lambda\) describing the process, whereas no-signaling is
a statement about the observable marginal probabilities.

Let \(\mathcal A\) and \(\mathcal B\) denote Alice's and Bob's
measurement settings and let \(A\) and \(B\) denote their outcomes.
Parameter independence is the condition
\[
P(A\mid\mathcal A,\mathcal B,\lambda)
=
P(A\mid\mathcal A,\lambda),
\]
and
\[
P(B\mid\mathcal A,\mathcal B,\lambda)
=
P(B\mid\mathcal B,\lambda).
\]
Thus, with \(\lambda\) fixed, the probability distribution of a local
outcome does not depend on the remote setting.

Outcome independence is the condition
\[
P(A\mid B,\mathcal A,\mathcal B,\lambda)
=
P(A\mid\mathcal A,\mathcal B,\lambda),
\]
and
\[
P(B\mid A,\mathcal A,\mathcal B,\lambda)
=
P(B\mid\mathcal A,\mathcal B,\lambda).
\]
Thus, with the settings and \(\lambda\) fixed, knowledge of the remote
outcome provides no additional information about the local outcome.

If both conditions hold, then
\[
P(A,B\mid\mathcal A,\mathcal B,\lambda)
=
P(A\mid\mathcal A,\lambda)
P(B\mid\mathcal B,\lambda),
\]
which is the Bell factorization.

Consider first the two-spin experiment at the level at which
\[
\lambda=\Psi^-,
\]
the singlet state. As shown above,
\[
P(r,s\mid a,b,\Psi^-)
=
\frac14
\left(
1-rs\,a\cdot b
\right).
\]
The marginal probabilities are
\[
P(A=r\mid a,b,\Psi^-)
=
P(B=s\mid a,b,\Psi^-)
=
\frac12.
\]
Alice's marginal is therefore independent of \(b\), and Bob's is
independent of \(a\). Parameter independence holds at this level.

Outcome independence does not. Indeed,
\[
P(B=s\mid A=r,a,b,\Psi^-)
=
\frac12
\left(
1-rs\,a\cdot b
\right),
\]
whereas
\[
P(B=s\mid a,b,\Psi^-)
=
\frac12.
\]
Knowledge of Alice's outcome therefore changes Bob's conditional
probability distribution, and conversely. At the level of the quantum
state, Bell correlations are thus compatible with parameter
independence but not with outcome independence.

The situation looks different when the stochastic realization itself
is included in \(\lambda\). Write
\[
\lambda=(\Psi_0,\omega),
\]
where \(\Psi_0\) is the initial state and \(\omega\) denotes a complete
realization of the setting-independent random data generating the
\({\bf RM}\) motion. For the purpose of comparing detector
arrangements, \(\omega\) specifies a realization of the ambient
stochastic path
\[
\Psi_0,\Psi_1(\omega),\Psi_2(\omega),\ldots.
\]
Different settings do not select a different realization \(\omega\).
They determine different detector-defined equivalence classes and hence
different first-entry events on the ambient path.
We may therefore
write
\[
A=A(a,b,\omega),
\qquad
B=B(a,b,\omega).
\]

At this fully resolved level the conditional outcome probabilities are
either zero or one. Outcome independence therefore holds automatically:
once \(a,b,\omega\) are fixed, knowledge of the other recorded outcome
provides no additional probabilistic information.
If parameter independence also held, the outcomes would satisfy
\[
A(a,b,\omega)=A(a,\omega),
\qquad
B(a,b,\omega)=B(b,\omega).
\]
Together with a setting-independent probability distribution for
\(\omega\), these relations would imply the Bell factorization and
hence the CHSH inequality. Since the \({\bf RM}\) model gives the Born correlations and therefore
violates CHSH, these equalities cannot hold for every realization. For
some settings and realizations one must have
\[
A(a,b,\omega)\neq A(a,b',\omega),
\]
or
\[
B(a,b,\omega)\neq B(a',b,\omega),
\]
or both. Thus parameter independence fails at the level of the
detector-defined records conditioned on a complete stochastic
realization.

The three-dimensional Brownian model gives a simple picture of this
dependence. A complete Brownian realization fixes one path
\[
X_t(\omega)
\]
in the ambient space. Changing Bob's setting while leaving Alice's
setting fixed changes the recording region from, for example,
\[
\Sigma_{ab}
\quad\hbox{to}\quad
\Sigma_{ab'}.
\]
The corresponding recording times
\[
\tau_{ab}(\omega)
\qquad\hbox{and}\qquad
\tau_{ab'}(\omega)
\]
may select different points of the same path and hence different values
of the coordinate observed by Alice. Thus
\[
A(a,b,\omega)\neq A(a,b',\omega)
\]
may occur even though the Brownian path itself has not changed.

The analogous statement in the \({\bf RM}\) model is geometric. For the
same fixed \({\bf RM}\) trajectory, changing Bob's setting from \(b\) to
\(b'\) replaces the detector-defined family
\[
\left\{
\mathcal C_{ab}^{rs}
\right\}_{r,s=\pm1}
\]
by
\[
\left\{
\mathcal C_{ab'}^{rs}
\right\}_{r,s=\pm1}.
\]
The same trajectory may reach these two families at different step
numbers and may then receive different Alice labels. The dependence is therefore a dependence of the observed record on the
detector arrangement, not a nonlocal change of the state-space history
caused by the remote setting.

In actual repetitions of the experiment, one does not have control of a
particular realization \(\omega\). Different runs normally correspond
to different realizations of the \({\bf RM}\) walk. The Born
probabilities are obtained from the ensemble of these realizations. For
example,
\[
P(A=r\mid a,b)
=
\int
P(A=r\mid a,b,\omega)\,d\mu(\omega).
\]
The Born rule gives
\[
P(A=r\mid a,b)
=
\frac12,
\]
independently of \(b\), and similarly
\[
P(B=s\mid a,b)
=
\frac12
\]
independently of \(a\).
This is the no-signaling property. It is an ensemble statement and must
be distinguished from parameter independence conditioned on a fully
specified realization. The latter may fail even though the observable
marginal probabilities remain independent of the remote setting.

More generally, no-signaling follows directly from the completeness of
the remote detector decomposition. For any state \(\Psi\),
\[
\begin{aligned}
\sum_\beta
\left\|
\left(
P^A_\alpha(\mathcal A)
\otimes
P^B_\beta(\mathcal B)
\right)\Psi
\right\|^2
&=
\left\|
\left(
P^A_\alpha(\mathcal A)
\otimes I
\right)\Psi
\right\|^2,
\end{aligned}
\]
because
\[
\sum_\beta P^B_\beta(\mathcal B)=I.
\]
The right-hand side is independent of Bob's detector decomposition.
The same argument applies with Alice and Bob interchanged.

The two levels of description may therefore be summarized as follows.
At the quantum-state level,
\[
\lambda=\Psi^-,
\]
parameter independence holds while outcome independence fails. At the
fully resolved \({\bf RM}\) level,
\[
\lambda=(\Psi_0,\omega),
\]
the outcome for each specified detector arrangement is fixed and
outcome independence holds, while Bell violation requires parameter
independence to fail for at least some realizations.
There is no contradiction between these statements. Parameter
independence and outcome independence are conditional properties and
therefore depend on what information is included in \(\lambda\). The
observable joint and marginal probabilities are unchanged by this
choice of description.

\section{Locality in State Space and Nonlocality in Classical Space}
\label{sec:locality-state-space-classical-space}

Locality in the present framework refers first to the geometry of the
fundamental state space and to the motion of the state within it. 
Here ``local in state space'' means that the evolution is generated
locally along the state-space trajectory; it is not, by itself, a
statement of relativistic spacetime locality.
A realization of the \({\bf RM}\) dynamics is a continuous sequence of
local Fubini--Study displacements. At each step, the state moves only to
a neighboring point of projective Hilbert space. Equivalently, the
Hamiltonian determines a tangent vector at the current state, just as a
Hamiltonian vector field determines the local motion of a state in
classical phase space. 
In this differential-geometric sense, the
state-space dynamics is local.

This notion of locality is different from the locality assumption in
Bell's theorem. Bell factorization concerns the dependence of classical
records on spatially separated detector settings. State-space locality
concerns the motion of the state in its projective arena. Correlations
that are nonlocal in the Bell sense when expressed through classical
records may therefore arise from a trajectory that remains local in the
ambient state space.
Our goal in this section is to clarify the interplay between
locality in state space and locality in classical space.

\subsection*{The Brownian Model}

The three-dimensional Brownian model introduced above gives a simple
analogue of this distinction. Let
\[
X_t=(x_t,y_t,z_t)
\]
be a Brownian path in \(\mathbb R^3\). The motion is continuous and
local in the ambient three-dimensional space. Suppose, however, that
Alice observes only the \(x\)-coordinate of a recorded event and Bob
only the \(y\)-coordinate, while the detector settings determine a
recording region
\[
\Sigma_{ab}\subset\mathbb R^3.
\]
The two observed values are then coordinates of a single point selected
on one three-dimensional trajectory.

From the reduced description in terms of Alice's and Bob's records, one
sees only a stochastic process in the \((x,y)\)-plane. The resulting
correlations need not admit a local hidden-variable description based
solely on this observed two-dimensional process. The apparent
nonlocality arises because the observed values are only two coordinates
of an event selected on a local trajectory in the larger space. In the
full three-dimensional description there is only one continuous
Brownian path.

\subsection*{The Position--Momentum EPR Experiment}

The relation between locality in state space and apparent nonlocality in
classical space becomes especially transparent in the one-dimensional
position--momentum EPR experiment. In the present
framework, the classical phase space of a particle in one spatial
dimension is represented by the localized submanifold
\[
M^\sigma_{1,1}
\]
of projective state space. A position measurement distinguishes
finite-resolution equivalence classes according to their position
coordinate. 
A momentum measurement uses a different measurement arrangement. The
apparatus correlates different momentum values with different
spatially localized detector records. Thus the final
localization is again a position localization, but the detector record
identifies the momentum interval that produced it. This is analogous to
the different recording regions selected by different detector
arrangements in the Brownian model in \(\mathbb R^3\).

Thus ``measure position'' and ``measure momentum'' correspond to
different measurement arrangements and therefore to different families
of detector-defined outcome classes. A single \({\bf RM}\) trajectory
is local in the ambient projective state space, while the actual
arrangement determines which localized records count as possible
outcomes.
This gives a geometric meaning to complementarity. Position and
momentum are not simultaneously existing classical records waiting to
be revealed. Different measurement arrangements associate different
classical records with the same initial quantum state.

For an EPR state, the same idea applies in the two-particle state
space. The joint state is one point of
\[
\mathbb P(\mathcal H_1\otimes\mathcal H_2),
\]
and the \({\bf RM}\) motion is one local trajectory there. When a
measurement produces one detector-defined outcome, the correlations
contained in the entangled state determine the corresponding
conditional state of the other particle. If a second compatible
measurement is performed, its record displays the corresponding
correlation. No signal between the particles is required.

For instance, suppose that the initial entangled state describes the
position correlation
\[
z_1-z_2=d.
\]
If Alice records
\[
a_1,
\]
the corresponding conditional position of the second particle is
centered at
\[
a_2=a_1-d.
\]
If Bob also measures position, the joint records therefore satisfy this
correlation.

The situation is analogous to the Brownian model in \(\mathbb R^3\).
The detector arrangement selects the recording region of the ambient
Brownian path, while the recorded event supplies both its observed
\(x\)- and \(y\)-coordinates. The correlation between these coordinates
does not require communication between them. Likewise, in the
\({\bf RM}\) description, the two spatial records arise from one
two-particle state and one local trajectory in the ambient state space,
rather than from an influence transmitted from one particle to the
other.

\subsection*{The Spin EPR Experiment}

The spin EPR experiment gives a finite-dimensional representation of the
same geometric distinction. The entangled spin state is a single point
of the composite projective spin space, while the measurement settings
select the joint outcome sectors relevant to the recorded result.

For a measurement along the \(z\)-axis, the two perfectly
anticorrelated product states
\[
|+\rangle_1\otimes|-\rangle_2,
\qquad
|-\rangle_1\otimes|+\rangle_2
\]
span a two-dimensional subspace
\[
\mathcal K
=
\operatorname{span}
\left\{
|+\rangle_1\otimes|-\rangle_2,\,
|-\rangle_1\otimes|+\rangle_2
\right\}
\simeq\mathbb C^2.
\]
The singlet state belongs to this subspace:
\[
|\Psi^-\rangle
=
\frac{1}{\sqrt2}
\left(
|+\rangle_1\otimes|-\rangle_2
-
|-\rangle_1\otimes|+\rangle_2
\right).
\]
Its projective space is therefore
\[
\mathbb P(\mathcal K)
\simeq
\mathbb CP^1
\simeq
S^2.
\]

For this fixed preparation and measurement setting, the two possible
joint outcomes are
\[
|+\rangle_1\otimes|-\rangle_2
\qquad\text{and}\qquad
|-\rangle_1\otimes|+\rangle_2,
\]
represented by the two poles of this copy of \(S^2\). The Born rule
assigns zero weight to the other two product outcomes. Thus, although
the actual \({\bf RM}\) walk takes place in the full position--spin
projective state space, the relevant two-outcome geometry may be
pictured on this Bloch sphere, with the singlet represented by a point
on the sphere and the two possible joint records by its poles.
Selection of either pole fixes both records simultaneously: if the spin
of one particle is up, then the spin of the other is down, and vice
versa. Thus Alice's record fixes the corresponding value recorded by
Bob, and conversely. No communication between the two particles is
involved; both values are components of a single joint outcome in the
composite state space.

For general measurement directions \(a,b\), the same principle applies
in the larger reduced two-spin space
\[
\mathbb CP^3
=
\mathbb P(\mathbb C^2\otimes\mathbb C^2).
\]
The settings select the joint outcome sectors
\[
\left\{
\mathcal C_{ab}^{rs}
\right\}_{r,s=\pm1}.
\]
Changing the settings changes these detector-defined sectors, not the
local character of the underlying state-space motion.

In this sense, locality and nonlocality refer to different geometric
levels of description. The \({\bf RM}\) dynamics is local in the
Fubini--Study geometry of projective state space, while the induced
relations among spatially separated classical records may be nonlocal
in the Bell sense. Spatial separation in the classical description does
not imply a corresponding separation of the underlying state into
independently evolving parts in projective Hilbert space.

\chapter{Quantum Paradoxes and the Arrow of Time}
\label{chap:paradoxes-arrow-time}

The double-slit experiment and EPR correlations were discussed in the preceding
chapter. In both cases the apparent paradox came from forcing a state-space
process into the language of classical space. Once the dynamics is formulated in state space, the apparent contradiction no longer arises. The state need not lie on the classical
submanifold before a record is formed. When a record is formed, the
\({\bf RM}\) dynamics brings the state into one of the detector-defined
localized sectors.

In this chapter we apply the same idea to the remaining standard paradoxes.
The issues now concern the selection of the recorded alternatives, the problem
of tails, macroscopic superpositions, observers, repeated measurements, and the
arrow of time. No new postulate is added. The same elements are used throughout:
the projective state space, the classical submanifolds, finite-resolution
equivalence classes, and the \({\bf RM}\) walk.

\section{The Reality of a Superposition}
\label{sec:the-reality-of-a-superposition}

In the present framework, a pure quantum state is represented by a
point in projective state space,
\[
[\psi]\in\mathbb P(\mathcal H),
\]
and its physical evolution is represented by a path of this point. In
the absence of measurement, the path is generated by the ordinary
Schr\"odinger Hamiltonian. During measurement, the evolution includes
the random Hamiltonians of \({\bf RM}\). In either case, for each
realized Hamiltonian history, there is one state and one state-space
trajectory.

Suppose that
\[
\psi=\alpha\psi_1+\beta\psi_2.
\]
This expression does not mean that two physical states, \(\psi_1\) and
\(\psi_2\), coexist. It defines a single ray \([\psi]\), generally
different from both \([\psi_1]\) and \([\psi_2]\). The decomposition
depends on the states chosen to represent the possible alternatives,
whereas the ray \([\psi]\) itself is the physical state. Mathematically,
\(\alpha\psi_1\) and \(\beta\psi_2\) are components of that state
relative to the chosen decomposition.

This distinction becomes especially clear when \(\psi_1\) and
\(\psi_2\) belong to different localized equivalence classes. A
superposition of such states normally lies away from the classical
configuration-space or phase-space submanifold. It therefore does not
represent several classical configurations realized simultaneously.
It represents one genuinely quantum state to which separate classical
positions, momenta, or trajectories cannot be assigned.
For example, a state of the form
\[
\psi=\alpha g_a+\beta g_b,
\]
with \(g_a\) and \(g_b\) localized near distinct positions, is one point
of projective state space. Unless one coefficient is negligible, this
point does not lie in either localized class \(\{g_a\}\) or
\(\{g_b\}\), and in general it lies away from the classical
configuration-space manifold. The statement that the particle is
simultaneously at \(a\) and \(b\) results from assigning classical
position values to a state for which position is not classically
defined.

The Fubini--Study distance from the classical submanifold provides a
geometric measure of this departure from classicality. When the state
lies on, or sufficiently close to, the relevant classical submanifold,
its classical coordinates are defined at the resolution under
consideration. When it lies away from that submanifold, those
coordinates no longer provide a complete description of the state.
The state remains physically real, but it is not a classical
configuration.

Measurement does not reveal which one of several pre-existing
classical alternatives was already realized. The measuring arrangement
defines the detector sectors in which a record can be formed. Under
\({\bf RM}\), the single state follows a stochastic unitary path, and
the measurement probabilities are the weights of the paths whose
endpoints lie in the corresponding detector-defined equivalence
classes,
\[
[\psi]\longrightarrow \mathcal C_j .
\]
Only when the state reaches such a class does the corresponding
observable acquire a definite recorded value. There is one realized
path and one resulting outcome, not a branching into several
simultaneously realized classical worlds.

A superposition is therefore neither a collection of coexisting
classical realities nor merely a statement of ignorance. It is a
genuine physical state represented by a point outside the classical
sector appropriate to the observable under consideration. The physical status of such state-space points is already indicated by
the fact that, when the state is constrained to the classical
phase-space submanifold
\(
\widetilde M^\sigma_{3,3},
\)
its evolution describes the usual Newtonian motion of a body. More
generally, the coefficients and relative phases of a superposition
determine observable interference effects and transition probabilities,
and are therefore physically significant. What the superposition does
not possess are the definite classical properties associated with the
individual terms of a chosen decomposition.

The apparent paradox of superposition arises when the classical
meaning of the component states is transferred to their linear
combination. In the geometry of state space, no such transfer is
justified. Classical properties belong to the corresponding classical
submanifolds and equivalence classes. A state away from these
submanifolds is fully real as a quantum state, but it does not represent
a definite classical configuration until the dynamics brings it into
the appropriate outcome class.

\section{The Preferred Basis Problem}
\label{sec:the-preferred-basis-problem}

The preferred-basis problem asks why measurements yield outcomes in a
particular basis rather than in arbitrary superpositions. In the
present framework, during measurement the state undergoes a homogeneous
and isotropic random walk in projective state space. The \({\bf RM}\)
walk has no preferred basis of its own. It assigns weights to the
possible outcome sectors according to their Fubini--Study distances
from the initial state, but it does not determine which sectors
represent measurement outcomes.

The outcome sectors are selected by the measuring arrangement. A
measuring device determines which sectors of state space are
distinguished by the experiment. For a position measurement on a
particle, these sectors are the localized equivalence classes
\[
\{g_a\},
\]
or, in phase space,
\[
\{g_a\}e^{\frac{ipx}{\hbar}}.
\]
A state that reaches one such class has a definite recorded position at
the resolution of the apparatus. For a system of particles whose
positions are measured, the outcome sectors are formed from tensor
products of the corresponding one-particle classes.

Position has a special operational role because macroscopic records are
ultimately records of positions in space. Other observables are
measured by correlating their possible values with position sectors. A
Stern--Gerlach device correlates spin alternatives with spatially
separated detector regions. A spectrometer correlates momentum values
with positions on a screen. The apparatus therefore maps the accessible
values of the measured observable to detector-defined position sectors.

Thus the preferred basis is not selected abstractly in Hilbert space.
It is fixed by the measuring arrangement, which defines the
observable-related outcome sectors, together with the finite-resolution
equivalence relation imposed by the detector. The \({\bf RM}\) walk supplies the stochastic state-space motion. The measuring
arrangement defines the sectors in which a measurement record can be formed.
For that arrangement, the relevant \({\bf RM}\) paths are those whose endpoints
lie in one of the detector-defined sectors. The relative weights of these
paths give the corresponding Born probabilities. The measured observable
acquires a definite recorded value when the state reaches one of the
detector-defined sectors and a macroscopic record is formed.

\section{The Problem of Tails}
\label{sec:the-problem-of-tails}

In the present formalism, a point of classical space is not identified
with a single ray in Hilbert space. It is an equivalence class of
localized states. In one dimension, the class associated with the
position \(c\) is
\[
\{g_c\}
=
\left\{
[\psi]:
\mu_z(\psi)=c,\quad
\delta_z(\psi)\leq \sigma
\right\},
\]
where
\[
\mu_z(\psi)
=
(\psi,\widehat z\psi),
\qquad
\delta_z^2(\psi)
=
\bigl(
\psi,
(\widehat z-\mu_z(\psi))^2\psi
\bigr).
\]
Thus the same point \(c\) is represented by many different states,
including states of noncompact support. The state has reached the
recorded position \(c\) when it belongs to the class \(\{g_c\}\), or
equivalently when
\[
\rho(\psi,\{g_c\})=0.
\]
Exact vanishing of the wave function outside a finite region is not
required.

There is a simple classical analogue. The normal probability density
\[
p(x)
=
\frac{1}{\sqrt{2\pi}\,\delta}
\exp\left[
-\frac{(x-c)^2}{2\delta^2}
\right]
\]
is nonzero for every \(x\in\mathbb R\). Nevertheless, a classical
Brownian particle described by such a distribution may be localized
near \(c\) to any prescribed finite resolution when \(\delta\) is
sufficiently small. The nonvanishing tails of the probability
distribution do not prevent us from assigning a localized classical
position at that resolution.

The same principle applies to quantum measurement. A real detector has
finite resolution and does not distinguish all rays in Hilbert space.
It distinguishes equivalence classes of states that agree within the
resolution of the apparatus. The parameter \(\sigma\) in the definition
above is determined by this resolution. Since detectors provide our means of assigning positions to physical
systems, the present formalism identifies the corresponding equivalence
classes with points of classical space.

The ``problem of tails'' arises only if collapse is required to produce
an exact state of compact support. A Gaussian wave function, and more
generally a wave function with noncompact support, does not vanish
exactly outside any finite region. Under the compact-support criterion,
such a state would therefore never be fully localized.
That criterion is unnecessary here. The tails are not removed by a
nonunitary operation; they are part of the detector-indistinguishable
state belonging to the same equivalence class. A classical record is a
finite-resolution equivalence class, not an exact ray. The apparent
problem comes from replacing the physical question
\[
\hbox{Can the detector distinguish these states?}
\]
by the mathematical question
\[
\hbox{Does the state vanish exactly outside a region?}
\]
With classical points identified with finite-resolution equivalence
classes, the tails present no problem.

\section{Schr\"odinger's Cat}
\label{sec:schrodinger-s-cat}

Schr\"odinger's cat is usually described by the state
\[
\alpha\,
\varphi_{\rm undecayed}\otimes\Psi_{\rm alive}
+
\beta\,
\varphi_{\rm decayed}\otimes\Psi_{\rm dead},
\]
which appears to represent a superposition of two macroscopically distinct
states of the cat, \(\Psi_{\rm alive}\) and \(\Psi_{\rm dead}\), entangled with
the corresponding states of the decaying atom,
\(\varphi_{\rm undecayed}\) and \(\varphi_{\rm decayed}\). The paradox is that such a state is allowed
by a purely formal application of linear Schr\"odinger evolution,
although no cat is observed as simultaneously alive and dead.

In the present framework, the cat is a macroscopic many-body system
continually monitored by its environment. To the required accuracy, its
state remains in the classical sector formed by tensor products of
localized states belonging to copies of the manifold
\[
\widetilde M_{3,3}^{\sigma_{\rm cat}}.
\]
As in the particle--device system considered earlier, the state of the
microscopic trigger and the cat remains effectively separable during the
reduction stage:
\[
\Phi_k
=
\varphi_k\otimes\Psi_0
+
O(\varepsilon),
\qquad
\varepsilon\ll1 .
\]
Here \(k=0,1,2,\ldots\), and
\[
\Phi_0=\varphi_0\otimes\Psi_0
\]
represents the initial state of the system. 

The \({\bf RM}\) dynamics therefore acts effectively on the state of the
microscopic trigger, while the macroscopic state of the cat remains in the
classical phase-space manifold and evolves classically there.
The trigger state is brought into one of the two outcome classes,
\[
\varphi_0
\longrightarrow
\varphi_{\rm undecayed}
\qquad\text{or}\qquad
\varphi_{\rm decayed},
\]
with the corresponding Born probabilities. Once this reduction has occurred,
the ordinary interaction between the trigger and the macroscopic apparatus
initiates the amplification process. The ordinary dynamics of the localized trigger--cat system then
produces one of two classical trajectories:
\[
\varphi_{\rm undecayed}\otimes\Psi_0
\longrightarrow
\varphi_{\rm undecayed}\otimes\Psi_{\rm alive},
\]
or
\[
\varphi_{\rm decayed}\otimes\Psi_0
\longrightarrow
\varphi_{\rm decayed}\otimes\Psi_{\rm dead}.
\]
Continued environmental monitoring keeps the resulting macroscopic state stable
within the corresponding classical equivalence class.

Thus the dynamics does not produce a persistent state of the form
\[
\alpha\,
\varphi_{\rm undecayed}\otimes\Psi_{\rm alive}
+
\beta\,
\varphi_{\rm decayed}\otimes\Psi_{\rm dead}.
\]
During the microscopic reduction stage, the system remains effectively
in a product state to the accuracy relevant to the macroscopic
description. The microscopic state may be a superposition before
reduction, but the macroscopic cat does not follow two classical
trajectories simultaneously. The \({\bf RM}\)
dynamics first produces one microscopic outcome, and the subsequent ordinary
macroscopic dynamics records that outcome as either an alive cat or a dead cat.

\section{Wigner's Friend}
\label{sec:wigner-s-friend}

Wigner's friend sharpens the same issue: the apparent conflict between
the linearity of quantum mechanics and the definiteness of macroscopic
records. The friend performs a measurement and records a definite
outcome. Wigner, outside the laboratory, is then imagined to assign a
superposition to the combined system consisting of the measured object,
the apparatus, and the friend. Wigner's later observation is supposed
to collapse this superposition into one of several possible outcomes.
Taken as physical descriptions of the same macroscopic situation, the
two accounts are incompatible.

What is missing from this analysis is that the friend and the apparatus
are macroscopic systems continually interacting with their environment.
Under \({\bf RM}\), the measured system, the apparatus, and the friend
remain effectively separable during the reduction and recording
processes. If the friend records the outcome \(c\), the relevant state
of the particle--apparatus--friend system belongs to a class of the
form
\[
\{g_c\}\otimes\mathcal D_c\otimes\mathcal F_c,
\]
where \(\mathcal D_c\) represents the corresponding device record and
\(\mathcal F_c\) the corresponding macroscopic state of the friend.

Wigner may not know which class has been reached and may therefore
assign probabilities to several possible records. This lack of
information, however, is not a physical superposition of mutually
incompatible macroscopic records. If Wigner later interacts with the
laboratory, his own macroscopic state becomes correlated with the
record already formed. Because Wigner himself is macroscopic, the resulting state belongs to a class of
the form
\[
\{g_c\}\otimes\mathcal D_c\otimes\mathcal F_c\otimes\mathcal W_c .
\]
Wigner therefore records the same outcome. No observer-dependent
collapse is required.

The paradox assumes that a laboratory containing a macroscopic
observer can remain an isolated coherent quantum system. This does not
occur in the present framework. Environmental interaction keeps the
macroscopic components in definite classical equivalence classes, and
once the microscopic system reaches a detector-defined outcome class,
ordinary macroscopic dynamics produces a stable record. Later observers
interact with that already existing record; they do not create it by
observation.

\section{The Quantum Zeno Effect}
\label{sec:the-quantum-zeno-effect}

The quantum Zeno effect is usually described as the inhibition of
evolution by sufficiently frequent measurements. In the language of the
projection postulate, repeated projections onto the same eigenspace keep
the state from leaving that subspace.

In the present framework, no independent projection postulate is
introduced. What is usually called projection is produced dynamically
by the \({\bf RM}\) evolution itself. During a measurement, the state
undergoes a stochastic sequence of unitary Schr\"odinger steps,
\[
\psi_{k+1}
=
\exp\left[
-\frac{i}{\hbar}
\left(
\widehat h+\widehat h_{{\bf RM},k}
\right)\Delta t
\right]\psi_k,
\]
where the random Hamiltonians
\(\widehat h_{{\bf RM},k}\) are independently drawn from the GUE on the
relevant finite-dimensional coarse-grained sector. Each individual
step is unitary. State reduction consists in the resulting path
reaching a detector-defined equivalence class with the corresponding
Born probability. Thus projection is not an
additional nonunitary operation appended to the dynamics; it is the
effective description of a unitary \({\bf RM}\) path reaching an
outcome sector.

Let \(\mathcal S\) denote the equivalence class, or outcome sector,
associated with a recorded result. Immediately after measurement, the
state belongs to \(\mathcal S\) at the resolution of the detector. If the same measurement is repeated after a sufficiently short
interval, the ordinary Schr\"odinger evolution has had little time to
carry the state away from \(\mathcal S\). The Born weight of the same
outcome sector is therefore close to one, and the next \({\bf RM}\)
measurement returns the state to \(\mathcal S\), or to a
detector-indistinguishable nearby class, with high probability.
As the time between successive measurements decreases, the state is
repeatedly returned to the same or a nearby sector before it can move
appreciably away from the corresponding region of state space. In this
limit, the probability of leaving a small neighborhood of
\(\mathcal S\) becomes arbitrarily small. The inhibition of evolution is
therefore produced by repeated unitary measurement dynamics rather than
by repeated applications of an independent projection rule.

The recording of the result introduces no additional dynamical effect.
It confirms that the state has reached the corresponding equivalence
class.  After
each return to \(\mathcal S\), or to a nearby class, the subsequent
evolution begins again from that class. Frequent returns therefore
stabilize the state-space path near the recorded sector.
For a position measurement, the relevant sectors are the localized
classes
\(
\{g_c\}.
\)
If the state has reached \(\{g_c\}\), then sufficiently frequent
repetition of the same measurement keeps returning it to this class or
to a nearby class. Its position therefore remains localized to the
resolution of the detector, apart from any ordinary motion generated by
the system Hamiltonian.

As demonstrated in Chapter~\ref{chap:macro-micro-measurement}, the same
mechanism operates continuously for a macroscopic body monitored by its
environment. Environmental interactions repeatedly return the state to
the localized classical sector. When significant excursions in the
normal directions are thereby suppressed, the effective evolution near
the classical phase-space manifold is approximately Newtonian. The body
need not remain at one spatial point: it may follow a Newtonian
trajectory while repeated \({\bf RM}\) returns prevent its state from
making a significant nonclassical excursion away from the localized
sector.
It follows that, in the present framework, the Zeno effect is another
consequence of the same unitary measurement dynamics that produces
state reduction and Born probabilities.

\section{Irreversibility from Random Matrix Stabilization}
\label{sec:irreversibility-from-random-matrix-stabilization}

The \({\bf RM}\) dynamics is unitary for every realized sequence of
Hamiltonians. If the complete sequence of random Hamiltonians is known and each
Hamiltonian can be reconstructed, the corresponding evolution can be
inverted exactly. The irreversibility considered here is therefore not
a violation of microscopic unitarity.

There are three sources of effective irreversibility.
First, the walk takes place in a high-dimensional projective state
space. Once the realized Hamiltonian history is not retained, an
independently generated, statistically equivalent history is exceedingly
unlikely to return the state to a prescribed small neighborhood of its
initial ray. In a finite-dimensional approximation \(\mathbb C^N\), the
probability that a uniformly distributed normalized state \(\psi\) has
fidelity
\[
F(\psi,\psi_0)=|(\psi,\psi_0)|^2
\]
greater than \(1-\varepsilon\) with a fixed normalized state
\(\psi_0\), where \(0<\varepsilon<1\), is
\[
P\bigl(F>1-\varepsilon\bigr)=\varepsilon^{N-1}.
\]
For large \(N\), this probability is negligible, and for every fixed
\(0<\varepsilon<1\),
\[
\lim_{N\to\infty}P\bigl(F>1-\varepsilon\bigr)=0.
\]

Second, physical time reversal of Schr\"odinger dynamics is not obtained
simply by replacing \(t\) with \(-t\). It is implemented by an
antiunitary operator. For a spinless system in the position
representation, the time-reversal operator is complex conjugation,
\[
\mathcal T\psi=\psi^*.
\]
For a time-dependent Hamiltonian,
\[
H(t)\longmapsto
\mathcal T H(-t)\mathcal T^{-1}
=
H^*(-t).
\]
To see how this acts on a realized \({\bf RM}\) history, consider one
step
\[
\psi_k
=
U_k\psi_{k-1},
\qquad
U_k
=
\exp\left(
-\frac{i}{\hbar}H_k\Delta t
\right).
\]
Complex conjugation gives
\[
\mathcal T U_k\mathcal T^{-1}
=
\exp\left(
+\frac{i}{\hbar}H_k^*\Delta t
\right).
\]
Hence the time-reversed state retraces this step according to
\[
\psi_{k-1}^*
=
\exp\left(
-\frac{i}{\hbar}H_k^*\Delta t
\right)\psi_k^*.
\]
Thus, if the forward \({\bf RM}\) history is
\[
H_1,H_2,\ldots,H_n,
\]
physical time reversal requires the complex-conjugate Hamiltonians to be applied in reverse order:
\[
H_n^*,H_{n-1}^*,\ldots,H_1^*.
\]

The same reversal may be expressed in terms of the propagators. The
forward evolution is
\[
\psi_n
=
U_nU_{n-1}\cdots U_1\psi_0,
\]
and exact retracing of the unconjugated states requires
\[
\left(
U_nU_{n-1}\cdots U_1
\right)^{-1}
=
U_1^{-1}U_2^{-1}\cdots U_n^{-1}.
\]
The two descriptions express the same reversal: the antiunitary
transformation gives the time-reversed states and Hamiltonians, while
the inverse propagator gives the corresponding reversed sequence of
steps.

This is the point relevant to the Loschmidt argument. In classical
mechanics, reversing all momenta produces the time-reversed trajectory
of a deterministic Hamiltonian system. In the \({\bf RM}\) dynamics,
however, the Hamiltonian history contains a particular realization of
the random matrices. Applying the time-reversal transformation to the final state does not
reconstruct the realized random Hamiltonian sequence in reverse order. Although the GUE law itself is invariant
under time reversal, a typical realized GUE Hamiltonian is not invariant
under complex conjugation. Retaining the same realized Hamiltonian
history therefore does not cause the state to retrace its path. The
Loschmidt reversal does not undo a generic realized measurement history.

Third, measurement records are finite-resolution equivalence classes.
A record preserves neither the exact ray nor the exact \({\bf RM}\)
trajectory that led to it. It retains only the information accessible
to the detector. Many microscopically distinct rays and many distinct
state-space trajectories correspond to the same recorded class.
Passing from the exact state to the recorded equivalence class therefore
discards information. There is consequently no physical basis for
assuming that the sequence of Hamiltonians required for exact reversal
can be reconstructed artificially. The microscopic information needed
to reconstruct the trajectory cannot be recovered from the record
alone.

None of these mechanisms requires a fundamental nonunitary law.
Each realized step remains unitary. Nevertheless, dynamics is not trajectory-wise time-reversal invariant: a realized \({\bf RM}\) 
evolution is not reversed by simply reversing time while keeping the same stochastic history.
High-dimensional non-recurrence and the loss of
microscopic information in finite-resolution records further make such
reversal physically unavailable.

\section{The Arrow of Time}
\label{sec:the-arrow-of-time}

The arrow of time is not introduced in the present framework as an
additional law. It emerges from the stochastic state-space dynamics and
from the formation and ordering of classical records. An abstract
evolution parameter is still required in the dynamical equations, but
observed time, or clock time, appears at the macroscopic level through
sequences of recorded events.

Classical Newtonian dynamics and Brownian motion were derived in
Chapter~\ref{chap:macro-micro-measurement} from Schr\"odinger dynamics
together with conjecture \({\bf RM}\). This points to the dynamics in
state space, together with the geometry of the classical submanifolds,
as a possible source of the observed time asymmetry.
The deterministic Schr\"odinger evolution, however, is time-reversible
and does not by itself select a direction of time. The von Neumann
entropy of each realized pure state remains constant. Likewise,
deterministic Hamiltonian evolution in classical mechanics is
time-reversible and preserves phase-space volume. This points more
specifically to the \({\bf RM}\) dynamics as a possible origin of the
arrow of time.

Yet the ambient \({\bf RM}\) walk also has no preferred direction. A
state may become more localized or more spread out with equal
probability whenever the corresponding final states lie at the same
Fubini--Study distance from the initial state. This is analogous to a
Brownian particle being equally likely to move in opposite directions.
The motion of an individual state in projective state space therefore
does not by itself define an arrow of time. The asymmetry must arise
from the way this motion is represented on, or constrained by, the
classical submanifolds.

To see this distinction, consider the two-particle state
\[
\Psi
=
\alpha\,g_a^1\otimes g_b^2
+
\beta\,g_c^1\otimes g_d^2.
\]
Under the \({\bf RM}\) walk, it may evolve to
\[
\Psi'
=
\alpha\,g_{a'}^1\otimes g_{b'}^2
+
\beta\,g_{c'}^1\otimes g_{d'}^2,
\]
with
\[
|a'-b'|<|a-b|,
\qquad
|c'-d'|<|c-d|.
\]
A transition to a state in which these separations increase has the
same probability whenever the two final states lie at the same
Fubini--Study distance from the initial state. Thus the state may move
farther from its initial ray in the Fubini--Study metric while the
separations appearing in a chosen classical decomposition decrease.
For many particles, the state may similarly become a superposition of
several nearly localized clusters. Such a state is a well-defined point
of projective state space, even though it does not represent a single
classical configuration in \(\mathbb R^3\).

The situation changes when the state is constrained to, or repeatedly
returned to, the classical configuration-space manifold. A localized
product state of an \(n\)-particle system then represents \(n\)
particles in ordinary space, and the restriction of the state-space
diffusion to this manifold induces Brownian motion of the corresponding
classical coordinates. Although each elementary displacement is locally
unbiased, an initially localized probability distribution spreads over
an increasingly large region of the available configuration space.
Configurations in which all particles remain confined to a small
region constitute only a restricted part of the configurations reached
by the diffusion.

For example, if two Brownian particles move independently and
\(X_i(t)\) and \(X_j(t)\) have diffusion coefficients \(D_i\) and
\(D_j\), with no relative drift, then
\[
\mathbb E|X_i(t)-X_j(t)|^2
=
\mathbb E|X_i(0)-X_j(0)|^2
+
6(D_i+D_j)t.
\]
Thus a localized probability distribution on classical configuration
space spreads under the induced Brownian dynamics. Starting from such a localized distribution, this spreading provides a
stochastic mechanism for the increase of coarse-grained classical
entropy and reproduces the essential statistical mechanism underlying
irreversibility.

This is the first mechanism contributing to the arrow of time in the
present framework. It arises from the geometry of the embedding of
classical space into state space together with conjecture \({\bf RM}\).
The usual statistical explanation of entropy increase therefore appears
here as the restriction of a more fundamental stochastic process in
state space. As in the classical statistical argument, the direction of
the resulting entropy increase refers to evolution away from an
initially localized, low-entropy distribution.

The state-space origin of this process also changes the usual recurrence
and reversibility arguments. These considerations do not provide an
independent direction of time, but they help explain why the stochastic
evolution just described is not generically reversed.  As discussed in the preceding
section, Poincar\'e recurrence would now require recurrence in a
high-dimensional, or infinite-dimensional, state space. For any fixed
time interval, however long, the probability of return to the same
microscopic state tends to zero as the dimension increases.
The Loschmidt argument likewise does not undo an individual realized
\({\bf RM}\) history. A realized trajectory is generated by a
particular sequence of random Hamiltonians. As shown in the previous section, applying the time-reversal
transformation to the final state while retaining the realized
Hamiltonian history does not make the state retrace its path.
Exact retracing requires the corresponding inverse unitaries to be applied in
reverse order and therefore requires reconstruction of the complete
microscopic Hamiltonian history.

There is, finally, a more directly observable origin of the arrow of
time. Classical space in this framework is not a space of exact points,
but a space of finite-resolution equivalence classes. Observations in
classical space are therefore records. A record is formed when the
state reaches a detector-defined equivalence class of a classical
submanifold. The record does not create the outcome; it confirms that
the state has reached the corresponding class.
Once formed, the recorded class supplies the current classical state
from which the next stage of effective evolution proceeds. The past is
therefore represented by an ordered sequence of recorded equivalence
classes,
\[
\mathcal C_1,\mathcal C_2,\ldots,\mathcal C_n,
\]
whereas the future is represented by the weights of possible
\({\bf RM}\) paths from the last recorded class to later classes. The
past is fixed by records; the future remains probabilistic.

This may be compared with coarse graining in classical statistical
mechanics. There too, the microscopic details of the process are not
actually observed, and the description is given in terms of
finite-resolution macroscopic data. In the classical theory, however,
this limitation is regarded as practical rather than fundamental. The
underlying trajectory is assumed to possess definite position and
momentum at every instant, and, in principle, increasingly precise
instruments could reveal increasingly fine details of that trajectory.
The situation is different here. The finite-resolution equivalence
classes are themselves the points of classical space rather than
regions imposed on an already given Euclidean space. Increasing the
precision of an instrument therefore does not uncover an underlying
exact state within a class; it replaces one family of
equivalence classes by a finer one. The exact state in projective
Hilbert space may still exist and evolve continuously, but its detailed
state-space trajectory is not itself a classical observable and cannot
in general be reconstructed from classical observations.

There is a second difference. In classical mechanics no separate
notion of recording is needed to define the history: the particle is
assumed to possess a position and momentum at every instant along its
trajectory, whether or not these quantities are observed. In the present framework, by contrast, a classical event is defined by
the state being recorded in a finite-resolution equivalence class on
the classical submanifold. It is through such events that classical
records are formed and clocks register time. The continuous state-space evolution may
therefore be regarded as occurring with respect to the abstract
evolution parameter, while observed time is associated with the ordered
sequence of recorded events on the classical submanifold.

We thus obtain an irreversible classical description from a unitary,
homogeneous, and isotropic process in state space. None of the
ingredients responsible for the arrow of time is fundamentally
classical. The stochastic spreading of classical configurations, the
failure of macroscopic records to retain the microscopic information
required to reverse a realized \({\bf RM}\) history, and the ordered
accumulation of finite-resolution records all arise from the underlying
state-space dynamics. The information needed to reconstruct the exact
microscopic past is not contained in the macroscopic records. This loss
of recoverable microscopic information, together with the statistical
spreading of classical configurations, gives the effective classical
evolution its direction in time.

\chapter{Comparison with Existing Approaches}
\label{chap:comparison}

The framework developed in this book overlaps with several existing
approaches to measurement and the quantum-to-classical transition. It
also differs from them in a basic way: the proposed framework is centered
on state space rather than on classical space. The purpose of this chapter is
to clarify this relationship. We begin by listing the main
distinguishing features of the framework and then compare them with
the corresponding elements of several standard approaches.

First, a quantum state is a point of projective Hilbert space,
\(
\mathbb P(\mathcal H),
\)
equipped with its Fubini--Study metric, which serves as the arena for all
physical processes in the framework. Physical evolution is
represented by a path of the state in this space. The classical
configuration space and phase space of a particle are embedded
submanifolds of \(\mathbb P(\mathcal H)\):
\[
\widetilde M^\sigma_3\simeq \mathbb R^3,
\qquad
\widetilde M^\sigma_{3,3}
\simeq
\mathbb R^3\times\mathbb R^3.
\]
After the appropriate rescaling, the Fubini--Study metric induced on
these submanifolds becomes the Euclidean metric of classical
configuration space and classical phase space. When Schr\"odinger
dynamics is restricted to the phase-space submanifold, it gives
Newtonian dynamics. The same construction extends to systems of
particles. Their classical configuration and phase spaces are
represented by product submanifolds in the projective tensor-product
state space, and the restriction of Schr\"odinger dynamics to these
submanifolds gives the usual Newtonian dynamics of the particle system.

Second, the stochastic component of measurement and environmental
monitoring is represented by unitary motion in projective state space:
\[
\psi
\longmapsto
\exp\left(
-\frac{i}{\hbar}\widehat h_{\bf RM}dt
\right)\psi,
\qquad
\widehat h_{\bf RM}\in{\rm GUE}.
\]
The stochastic dynamics is related to ordinary microscopic
system--environment dynamics in two steps. A spatially local
Caldeira--Leggett-type interaction generates, in the localized sector,
a homogeneous family of Brownian motions. The common unitary lift of
this Brownian family gives random Hermitian generators with the GUE law,
apart from an overall scale and a projectively irrelevant scalar term.
Thus the Brownian motion on the classical manifold and the
random-matrix motion in projective state space are two levels of the
same stochastic construction. The former gives normal measurement
errors, while the latter gives the corresponding Born probabilities.

Third, measurement outcomes are finite-resolution sectors, not exact
rays. A detector does not resolve all points of Hilbert space. It
assigns the state to equivalence classes such as
\[
\{g_a\}
=
\left\{
[\psi]:
\mu_x(\psi)=a,\quad
\delta_x(\psi)\leq\sigma
\right\}.
\]
Each such class defines a point of the classical-space manifold
\(\widetilde M_3^\sigma\). Both the microscopic system and the
macroscopic apparatus are quantum systems. The apparatus is classical
only because environmental interactions repeatedly stabilize its state
near a classical submanifold. Measurement is then the state-space
dynamics by which a joint state is recorded in one of the
detector-defined sectors. Classical reality is the recorded part of
this state-space dynamics.

\section{Decoherence}
\label{sec:decoherence}

In decoherence theory, a system \(S\) coupled to an environment \(E\)
is described by a joint density operator on
\[
\mathcal H_S\otimes\mathcal H_E.
\]
The joint state evolves unitarily,
\[
\rho_{SE}(t)
=
U(t)\rho_{SE}(0)U(t)^\dagger,
\]
while the reduced state of the system is obtained by tracing over the
environment:
\[
\rho_S(t)
=
\operatorname{Tr}_E\rho_{SE}(t).
\]
For suitable system--environment interactions, the reduced density
matrix becomes approximately diagonal in a pointer basis,
\[
\rho_S(t)
\simeq
\sum_\alpha p_\alpha
|\alpha\rangle\langle\alpha|.
\]
The off-diagonal terms are suppressed, and interference between
macroscopically distinct alternatives becomes practically inaccessible.

The present framework is compatible with this mechanism but addresses
a different question. Decoherence produces an improper mixture in the
reduced density matrix; by itself, it does not specify which outcome is
realized in an individual run. In the present framework, the state
follows a stochastic unitary path in projective state space,
\[
\psi(t+dt)
=
\exp\!\left(
-\frac{i}{\hbar}\widehat h_{\bf RM}(t)\,dt
\right)\psi(t),
\]
and the outcome is the detector-defined equivalence class reached by
this path. The object carrying the outcome information is therefore not
the reduced density matrix alone, but the realized state-space
trajectory together with the equivalence-class structure determined by
the detector resolution.

At the level of physical scales, the two descriptions are connected
more directly. In collisional decoherence and related open-system
models, environmental scattering determines momentum-transfer and
diffusion scales. In the present framework, a spatially local
Caldeira--Leggett-type system--environment interaction produces the
Brownian family on the localized classical manifold. The diffusion
coefficient of this family then fixes the scale of the common
random-matrix lift.
For example, a collision rate \(\Gamma\) with characteristic momentum
transfer \(q\) produces a momentum-diffusion scale of order
\[
D_p\sim \Gamma q^2.
\]
The corresponding position fluctuations depend on the dynamical regime.
In a dissipative medium they may be described by the long-time Brownian
diffusion coefficient, whereas in an underdamped regime they are
obtained by integrating the stochastic momentum increments.

Decoherence theory and \({\bf RM}\) therefore arise from the same
underlying class of system--environment interactions but describe
different aspects of the resulting dynamics. Decoherence describes the
suppression of interference in the reduced density matrix. The present
construction follows the associated stochastic motion of the state in
projective state space and supplies a mechanism by which one
detector-defined equivalence class is reached with the corresponding
Born probability.

\section{Open-System Dynamics and Microscopic Bath Models}
\label{sec:open-system-dynamics-and-microscopic-bath-models}

Microscopic bath models begin with a Hamiltonian for a system coupled
to many environmental degrees of freedom. After tracing out the bath,
one obtains an effective master equation for the reduced density
matrix. In a Markovian regime this often has the form
\[
\frac{d\rho}{dt}
=
-\frac{i}{\hbar}[H,\rho]
+
{\mathcal D}(\rho),
\]
where \({\cal D}\) is a dissipative or decohering term.

A representative example is the Caldeira--Leggett model, in which a
distinguished coordinate \(x\) is coupled linearly to a bath of
oscillators:
\[
H
=
\frac{p^2}{2M}
+
V(x)
+
\sum_j
\left(
\frac{p_j^2}{2m_j}
+
\frac12 m_j\omega_j^2 q_j^2
\right)
-
x\sum_j c_jq_j
+
x^2\sum_j
\frac{c_j^2}{2m_j\omega_j^2}.
\]
Under standard approximations one obtains a master equation containing,
among other terms, a double commutator
\[
[x,[x,\rho]],
\]
which suppresses spatial coherences.

The relation of this microscopic bath dynamics to \({\bf RM}\) can be
made more explicit. The random-matrix dynamics is not a competing
microscopic Hamiltonian. Rather, it is the state-space lift of the
effective Brownian motion generated by the microscopic
system--environment interaction.
For a spatially local Caldeira--Leggett-type coupling, the reduced
dynamics of sufficiently localized states generates a Brownian family
on the classical submanifold. Schematically,
\[
\text{system+bath Schr\"odinger dynamics}
\longrightarrow
\text{Brownian family on }
\widetilde M^\sigma_{3,3}.
\]
The complete Brownian family, taken over the localized initial states,
must then be represented by one common stochastic unitary acting on the
ambient Hilbert space. Reconstruction of the corresponding Hermitian
generator gives, under the homogeneity and independence assumptions
developed earlier, the Gaussian Unitary Ensemble law:
\[
\text{Brownian family on }
\widetilde M^\sigma_{3,3}
\longrightarrow
\text{common GUE lift}
\longrightarrow
{\bf RM}.
\]
Thus the full relation is
\[
\text{system+bath Schr\"odinger dynamics}
\longrightarrow
\text{Brownian family on } \widetilde M^\sigma_{3,3}
\]
\[
\longrightarrow
\text{GUE-generated RM dynamics}.
\]

The distinction from the usual reduced-density-matrix treatment is
therefore not that the two theories begin from unrelated stochastic
assumptions. They begin from the same microscopic system--environment
interaction but retain different information. The master-equation
description follows the reduced density operator after environmental
degrees of freedom are traced out. The present construction instead
uses the Brownian family in the localized sector to reconstruct a
stochastic unitary motion of the state in projective Hilbert space.

The remaining question is one of scope rather than of principle:
how generally realistic measuring devices and environments can be
reduced to the local bath structure required by this derivation.

\section{Continuous Measurement and Quantum Trajectories}
\label{sec:continuous-measurement-and-quantum-trajectories}

Continuous-measurement theory describes monitored quantum systems by
stochastic master equations or stochastic Schr\"odinger equations. For
a continuous measurement of an observable \(A\), a typical conditioned
stochastic master equation has the form
\[
d\rho
=
-\frac{i}{\hbar}[H,\rho]\,dt
-
k[A,[A,\rho]]\,dt
+
\sqrt{2k}
\left(
A\rho+\rho A-2\langle A\rangle\rho
\right)dW_t.
\]
Here \(dW_t\) is a Wiener increment and \(k\) is the measurement
strength. Such equations successfully describe monitored systems,
measurement records, feedback, and quantum trajectories.

The present framework has a similar conditioning structure, but a
different geometric and dynamical origin. In continuous-measurement
theory, the observable is specified in advance, and the stochastic
equation is constructed to condition the state on the corresponding
measurement record. In the present framework, the measured classical
variables arise from localized submanifolds and detector-defined
equivalence classes. The stochastic motion is instead generated by
short random unitary steps,
\[
\psi_{k+1}
=
\exp\left(
-\frac{i}{\hbar}
\widehat h_{{\bf RM},k}\Delta t
\right)\psi_k,
\qquad
\widehat h_{{\bf RM},k}\in{\rm GUE},
\]
with independent Hamiltonians at successive steps. For each realized
Hamiltonian history, the evolution is unitary.
In the present construction these random unitary increments need not be
taken as a primitive measurement noise. A local microscopic
system--environment interaction generates the Brownian family on the
localized manifold, and the common stochastic lift of that family gives
the effective GUE-generated evolution.

Thus the Wiener noise appearing explicitly in continuous-measurement
equations and the random unitary motion used here arise at different
levels of description. In the present framework the Brownian noise is
first obtained in the localized classical sector, and its common lift
becomes isotropic random unitary motion in projective state space. The outcome sectors
are not exact eigenspaces of a preassigned operator. They are
finite-resolution sectors defined by the apparatus. The relevant
measurement probabilities are the weights of the \({\bf RM}\) paths
whose endpoints lie in these sectors. This is why the same stochastic
process can give both classical Brownian errors on
\(\widetilde M^\sigma_3\) and Born-rule probabilities in the full
projective state space, without introducing a fundamental nonlinear or
nonunitary law.

\section{Ehrenfest and Semiclassical Limits}
\label{sec:ehrenfest-and-semiclassical-limits}

The Ehrenfest theorem gives
\[
\frac{d}{dt}\langle \widehat x\rangle
=
\frac{1}{m}\langle \widehat p\rangle,
\qquad
\frac{d}{dt}\langle \widehat p\rangle
=
-\langle \nabla V(\widehat x)\rangle.
\]
If the state remains sufficiently localized, or if the potential is of
a form for which the higher-order corrections vanish or are negligible,
then
\[
\langle \nabla V(\widehat x)\rangle
\simeq
\nabla V(\langle \widehat x\rangle),
\]
and the expectation values approximately satisfy Newton's equations.
This is an important consistency result, but it does not by itself
explain classicality. In particular, it does not explain why an
initially localized state remains localized, why a single outcome is
realized, or why the outcome statistics obey the Born rule.

The Ehrenfest theorem is compatible with the present framework, but it
is not the source of the explanation. The explanation is both geometric and dynamical. The Newtonian dynamics
of a particle is obtained by restricting the Schr\"odinger action to the
localized phase-space submanifold
\(
\widetilde M^\sigma_{3,3}.
\)
For the localized states and potentials considered earlier, or within
the corresponding localization approximation, the restricted action
takes the classical form
\[
S
=
\int
\left[
p\dot a-h(a,p,t)
\right]dt,
\qquad
h(a,p,t)
=
\frac{p^2}{2m}+V(a,t).
\]
Variation yields
\[
\dot a=\frac{p}{m},
\qquad
\dot p=-\nabla V(a).
\]
Equivalently, the tangent component of the Schr\"odinger vector field
along \(\widetilde M^\sigma_{3,3}\) is the Newtonian vector field. In
this restricted dynamics, the commutator form of the Heisenberg
equations reduces to the corresponding Poisson-bracket form of
classical Hamiltonian dynamics.

The additional \({\bf RM}\) stabilization, generated effectively by
system--environment interaction through the Brownian-lift mechanism,
provides the dynamical origin of the restriction by repeatedly returning
the state to a small neighborhood of the localized phase-space
submanifold and suppressing its normal evolution. The resulting effective motion is therefore
approximately Newtonian. Unlike a conventional semiclassical
approximation, the present construction also explains dynamically why
the state remains in the classical sector.

\section{Objective-Collapse Models}
\label{sec:objective-collapse-models}

Objective-collapse models modify the Schr\"odinger equation by adding
stochastic nonlinear terms. In continuous spontaneous localization
models, for example, the state vector satisfies an equation of the
general form
\[
d|\psi_t\rangle
=
\left[
-\frac{i}{\hbar}H\,dt
+
\sqrt{\lambda}\int d^3x\,
\bigl(N(x)-\langle N(x)\rangle_t\bigr)dW_t(x)
-
\frac{\lambda}{2}\int d^3x\,
\bigl(N(x)-\langle N(x)\rangle_t\bigr)^2dt
\right]
|\psi_t\rangle.
\]
Here \(\lambda\) is a collapse rate, \(N(x)\) is a smeared
mass-density operator, and \(dW_t(x)\) is a classical noise field. The
nonlinear terms drive localization and produce single outcomes, but the
dynamics is no longer strictly linear and unitary.

The present framework also produces single outcomes, but by a different
mechanism. For each realization of \(\widehat h_{\bf RM}\), the
evolution is unitary. The resulting random walk is homogeneous and
isotropic in projective state space. Reduction occurs when the
state-space path reaches a detector-defined equivalence class and this
event is recorded. It is not a fundamental nonunitary transformation of
the state vector.

This distinction also changes the interpretation of the parameters.
Objective-collapse models typically introduce universal constants
governing spontaneous localization and may predict effects such as
spontaneous heating or radiation. In the present framework, there is no
universal collapse noise acting independently of measurement or
environmental interaction. The relevant stochastic scales are inherited
from the microscopic system--environment coupling. In the local bath
construction, the environment first determines the Brownian diffusion
coefficient on the localized manifold, and this coefficient fixes the
scale of the corresponding GUE lift. The stochastic parameters are
therefore environmental and dynamical rather than new universal
constants.

Objective-collapse models and the present framework therefore share the
aim of describing outcome formation as a physical process, but they
differ in their dynamical laws. Collapse models introduce a fundamental
stochastic nonunitary modification of Schr\"odinger evolution. The
present framework preserves unitary evolution for every realized
Hamiltonian history. The stochastic dynamics is generated effectively
by system--environment interaction through the Brownian-lift mechanism,
while finite-resolution equivalence classes define the definite
measurement records.

\section{Center-of-Mass Classicalization}
\label{sec:center-of-mass-classicalization}

Another route to classical behavior appeals to the concentration of the
quantum center-of-mass coordinate of a large composite system. For
\(N\) particles with total mass \(M\), the center-of-mass operator is
\[
\widehat R
=
\frac{1}{M}
\sum_{i=1}^N m_i\widehat x_i.
\]
In a many-particle quantum state, its expectation value is
\[
\langle \widehat R\rangle
=
\frac{1}{M}
\sum_{i=1}^N m_i\langle \widehat x_i\rangle,
\]
while its quantum uncertainty is determined by
\[
(\Delta R)^2
=
\langle \widehat R^2\rangle
-
\langle \widehat R\rangle^2.
\]
More explicitly,
\[
(\Delta R)^2
=
\frac{1}{M^2}
\sum_{i,j}
m_i m_j\,
\operatorname{Cov}(\widehat x_i,\widehat x_j).
\]
If correlations among the constituents are sufficiently weak, the
off-diagonal covariance terms are negligible. For equal masses and
comparable single-particle position uncertainties, this gives
\[
\Delta R
\sim
\frac{\Delta x}{\sqrt N}.
\]
Thus the quantum probability distribution of the center-of-mass
coordinate may become very sharply concentrated about
\(\langle\widehat R\rangle\) for a macroscopic body.

This argument is compatible with the present framework, but it
addresses a different question. Center-of-mass concentration is a
statement about the statistical behavior of one collective coordinate.
By itself, it does not derive Newtonian dynamics from Schr\"odinger
dynamics and does not supply a mechanism for suppressing motion away
from the classical submanifold. Nor does it explain state reduction,
outcome uniqueness, or the Born probabilities of measurement outcomes.

Center-of-mass concentration is nevertheless useful. It supports the
physical plausibility of narrow, smooth, and often approximately
Gaussian representative states for macroscopic objects and helps
explain why localized collective coordinates arise naturally in the
large-\(N\) regime. The present framework goes further by explaining
why localized equivalence classes form a classical phase space, why
Schr\"odinger evolution restricted to that space is Newtonian, and why environmentally generated \({\bf RM}\) stabilization keeps
the actual state-space trajectory near the classical sector.

\section{Bohmian Mechanics}
\label{sec:bohmian-mechanics}

Bohmian mechanics supplements the wave function
\[
\Psi(q_1,\ldots,q_N,t)
\]
with an actual configuration of particle positions,
\[
Q_1(t),\ldots,Q_N(t).
\]
These positions evolve according to the guiding equation
\[
\dot Q_k(t)
=
\frac{\hbar}{m_k}
\operatorname{Im}
\left.
\frac{\nabla_k\Psi(q_1,\ldots,q_N,t)}
{\Psi(q_1,\ldots,q_N,t)}
\right|_{q_j=Q_j(t)}.
\]
Bohmian mechanics therefore assigns definite trajectories in
configuration space to the particles at all times.

The present framework introduces no additional particle positions. Its
basic object is the quantum state itself, represented by a point in
projective Hilbert space. A classical position is defined when the
state lies in, or is recorded as belonging to, a localized equivalence
class on the classical submanifold. Away from this sector, the state
need not possess a definite classical position.

The two approaches therefore adopt different ontologies. Bohmian
mechanics combines the wave function with an actual configuration in
ordinary space and introduces a guiding law for that configuration. The
present framework takes projective state space as the primary arena and
obtains ordinary configuration space as a distinguished localized
submanifold. A classical trajectory is not an additional hidden path in
\(\mathbb R^3\), but the effective motion of a state repeatedly
stabilized near the classical sector and represented by a sequence of
localized records.

The difference also affects the description of nonlocality. In Bohmian
mechanics, the velocity of one particle generally depends on the full
configuration and on the joint wave function, so the guiding dynamics
is explicitly nonlocal in configuration space. In the present
framework, an entangled two-particle system is represented by one point
of
\[
\mathbb P(\mathcal H_1\otimes\mathcal H_2),
\]
and the \({\bf RM}\) walk acts on this joint state as a whole, locally in time and in state space. The
correlated outcomes are determined by the Fubini--Study relations
between the initial state and the detector-defined joint outcome
sectors. They do not arise from a signal transmitted between the two
particles through ordinary space.

\section{Everettian Approaches}
\label{sec:everettian-approaches}

Everettian approaches retain universal unitary evolution and interpret
measurement as branching into several simultaneously realized
alternatives:
\[
\sum_\alpha c_\alpha
|\alpha\rangle\otimes|A_0\rangle
\longrightarrow
\sum_\alpha c_\alpha
|\alpha\rangle\otimes|A_\alpha\rangle.
\]
The different terms are taken to describe distinct branches, each
containing a corresponding apparatus record and observer. No single
branch is selected dynamically; all branches persist, and definite
outcomes are understood relative to the branch occupied by the
observer.

The present framework preserves unitary evolution for every
realized Hamiltonian history without requiring branching, through a
different measurement dynamics. A macroscopic apparatus is continually
stabilized by its environment near a localized classical equivalence
class. In the local bath construction, the same system--environment
interaction generates the Brownian family in the localized sector,
whose common lift gives the effective \({\bf RM}\) evolution.

During the reduction stage, the joint state of the measured microscopic
system and the apparatus remains, to the required accuracy, in the
fixed-apparatus sector,
\[
\Phi_k
=
\psi_k\otimes A_0+O(\varepsilon),
\qquad
\varepsilon\ll1.
\]
The \({\bf RM}\) walk acts on the total composite state. However, because
the apparatus remains stabilized in its initial macroscopic class, the
induced motion within the fixed-apparatus sector is represented, to the
same accuracy, by an effective \({\bf RM}\) walk of the microscopic
factor. The joint state is therefore not driven during this stage into a
superposition of macroscopically distinct apparatus records.

The composite state follows one stochastic unitary path, whose
effective motion in the fixed-apparatus sector brings the microscopic
state to one detector-defined outcome class,
\[
[\psi]
\overset{{\bf RM}}{\longrightarrow}
\mathcal C_\alpha.
\]
Only after this reduction does the ordinary interaction Hamiltonian
produce the corresponding macroscopic record,
\[
\mathcal C_\alpha\otimes\mathcal A_0
\longrightarrow
\mathcal C_\alpha\otimes\mathcal A_\alpha.
\]
Since the microscopic outcome has already been selected, this
amplification follows one classical trajectory. It does not produce a
state of the form
\[
\sum_\alpha
c_\alpha\,
\mathcal C_\alpha\otimes\mathcal A_\alpha.
\]

Thus the terms in a microscopic superposition do not represent
simultaneously realized classical worlds. They are components of one
quantum state, represented by one point in projective state space. For
each realization of the random Hamiltonian history, there is one
composite state-space path, one microscopic outcome class, and one
resulting macroscopic record. The alternatives not reached remain
possible outcomes of the stochastic dynamics, but they are not
additional realized worlds.

The role of probability is correspondingly different. In Everettian
approaches, since all branches persist, one must explain why the branch
weights \(|c_\alpha|^2\) govern the probabilities experienced by
branch-relative observers. In the present framework, the Born weights
are the transition probabilities for the alternative \({\bf RM}\)
paths ending in the corresponding detector-defined classes. They are
determined by the Fubini--Study geometry and by the unique extension of
Brownian transition probabilities from the classical submanifold to
the full projective state space.

The two approaches therefore agree that no fundamental nonunitary term
need be added to the Schr\"odinger equation, but they differ both in
their dynamics and in their ontology. Everettian theory allows the
unitary interaction to generate a superposition of macroscopically
distinct records and interprets all of its branches as realized. In the
present framework, environmental \({\bf RM}\) stabilization prevents
such a macroscopic superposition from arising: one microscopic outcome
is reached first, and ordinary macroscopic dynamics then records that
single outcome.

\section{Consistent Histories}
\label{sec:consistent-histories}

The consistent-histories approach seeks to formulate quantum mechanics
in terms of entire sequences of events or properties rather than
individual measurements. It assigns probabilities to alternative
histories of a closed quantum system without introducing an external
observer or assuming that a measurement occurs at every stage. An
important further objective is to identify coarse-grained families
whose histories exhibit approximately classical behavior and may
therefore represent the quasiclassical trajectories observed in the
macroscopic world.

A history is represented by a sequence of coarse-grained alternatives,
described in the Heisenberg picture by projectors
\[
P_{\alpha_1}(t_1),\ldots,P_{\alpha_n}(t_n).
\]
The corresponding class operator is
\[
C_\alpha
=
P_{\alpha_n}(t_n)\cdots P_{\alpha_1}(t_1),
\qquad
\alpha=(\alpha_1,\ldots,\alpha_n).
\]
For an initial state \(\psi\), the history is assigned the weight
\[
p(\alpha)
=
\|C_\alpha\psi\|^2,
\]
provided interference between distinct histories in the chosen family
is sufficiently suppressed. Under this consistency condition, the
weights obey the ordinary rules of probability, and one may reason
about histories without referring to measurements at each time.

The present framework describes macroscopic histories without
introducing an abstract sequence of projectors. The
state follows one stochastic unitary \({\bf RM}\) trajectory in
projective state space. Whenever this trajectory reaches a
detector-defined equivalence class of the classical sector, a record may
be formed. The macroscopic history is the resulting ordered sequence
\[
\mathcal C_1,\mathcal C_2,\ldots,\mathcal C_n.
\]
These classes are not merely coarse-grained labels assigned to an exact
microscopic history after the fact. They are finite-resolution sectors
that define the classical configuration-space and phase-space manifolds
themselves. State-space trajectories may cross these manifolds, and the
resulting sequence of records constitutes the macroscopic history.

The distinction is dynamical. Consistent histories begins with a family
of coarse-grained alternatives satisfying a consistency condition. The
present framework begins with stochastic unitary dynamics and obtains
the classical history from the recorded returns of the state-space path
to finite-resolution classical sectors. For a macroscopic system, this
history is a stroboscopic sequence of returns to the localized
phase-space manifold, with approximately Newtonian motion between
successive records.

\section{Geometric Quantum Mechanics}
\label{sec:geometric-quantum-mechanics}

Geometric quantum mechanics formulates ordinary quantum theory in terms
of projective Hilbert space
\(
\mathbb P(\mathcal H).
\)
Pure states are points of
\(
\mathbb P(\mathcal H),
\)
the Fubini--Study metric provides the natural Riemannian structure, and
observables generate Hamiltonian vector fields. Schr\"odinger evolution
is a Hamiltonian flow on projective state space.

The present framework uses this geometry but adds two further
structures. First, classical configuration space and classical phase
space are represented by localized submanifolds,
\[
\widetilde M^\sigma_3
\subset
\mathbb P(\mathcal H),
\qquad
\widetilde M^\sigma_{3,3}
\subset
\mathbb P(\mathcal H).
\]
After rescaling, the Fubini--Study metric induced on these submanifolds
becomes the Euclidean metric. The tangent component of the
Schr\"odinger vector field along
\(\widetilde M^\sigma_{3,3}\) is the Newtonian vector field.
In addition, the complete Hamiltonian flow on the localized phase-space
manifold determines its common unitary lift uniquely, whenever such a
lift exists, and supplies the classical motions entering the time-sliced
Feynman construction.
Second, measurement and environmental monitoring are represented by
stochastic unitary \({\bf RM}\) dynamics. A local system--environment
interaction generates Brownian motion on the localized classical
manifold, while the common unitary lift of the resulting Brownian
family gives the GUE-generated stochastic motion in projective state
space. Detector-defined equivalence classes are reached with the
corresponding Born probabilities.

The present construction therefore extends the geometric formulation
by introducing localized classical submanifolds, detector-defined
equivalence classes, and a dynamical bridge from microscopic
system--environment Schr\"odinger evolution to Brownian motion on the
classical manifold and from there to stochastic unitary \({\bf RM}\)
motion in projective state space.

\section{Outlook: From Schr\"odinger Dynamics to the {\bf RM} Regime}

The results of this and the preceding chapters suggest a broader
structure for quantum dynamics in state space. Two kinds of evolution
have appeared. Ordinary Schr\"odinger evolution describes the coherent
propagation of a quantum state between measurements. The stochastic
\({\bf RM}\) evolution developed here describes the state-space motion
associated with measurement and yields the Born probabilities for the
possible detector records. Both are evolutions in the same projective
Hilbert space; what differs is the law governing their generators and
the physical role of the resulting motion.

The relation to classical dynamics makes this structure particularly
transparent. Schr\"odinger evolution, when restricted to the classical
phase-space manifold \(M_{3,3}^\sigma\), yields Hamiltonian, and hence
Newtonian, dynamics. Conversely, as shown in
Section~\ref{sec:unitary-lift-hamiltonian-motion}, the complete family
of motions generated by one fixed classical Hamiltonian determines,
whenever a common unitary lift exists, that lift uniquely up to an
overall phase. For the Hamiltonians obtained by restricting
Schr\"odinger dynamics, existence of the lift is already supplied by
the corresponding quantum Hamiltonian.

The same complete Hamiltonian family also enters the time-sliced
construction of the quantum propagator. At each subdivision time, the
Hamiltonian flow supplies the classical motion beginning from every
possible phase-space point. For sufficiently short intervals, the
classical trajectory connecting successive configurations determines
the corresponding short-time action and hence the phase of the
short-time propagator, together with the appropriate amplitude factor.
Repeated composition and integration over the intermediate
configurations then yields the Schr\"odinger propagator in the
fine-partition limit.

The important point is that no independent dynamics has to be assigned
to the individual paths appearing in this construction. All of the
short classical motions are members of the same family generated by
one Hamiltonian. A motion selected at a later subdivision time may
simply continue a Newtonian trajectory selected earlier, if the
intermediate position and momentum agree with that trajectory. Other
choices of intermediate phase-space data select different members of
the same Hamiltonian family. Thus the time-sliced construction
repeatedly draws upon one complete family of Newtonian motions.

The deterministic construction may therefore be summarized as
\[
\text{complete Hamiltonian family}
\quad\longrightarrow\quad
\text{time-sliced Feynman construction}
\quad\longrightarrow\quad
\text{Schr\"odinger propagation}.
\]
The resulting object is a single evolution in projective Hilbert space,
\[
\psi(t)=U(t)\psi(0).
\]

The stochastic construction developed in this chapter has a parallel
structure. A homogeneous family of Brownian motions on the localized
classical manifold admits a common stochastic unitary lift. Under the
homogeneous component independence established above, the
projectively relevant Hermitian generators of this lift have the GUE
law. Conversely, the resulting \({\bf RM}\) evolution, when constrained
to the localized manifold, reproduces the Brownian family.

Thus one has schematically
\[
\begin{array}{ccc}
\text{Schr\"odinger evolution}
& \dashrightarrow &
\text{\({\bf RM}\)/GUE evolution}
\\[2mm]
\Big\updownarrow
&&
\Big\updownarrow
\\[2mm]
\text{Hamiltonian motion on }M_{3,3}^\sigma
& \dashrightarrow &
\text{Brownian motion on }M_{3,3}^\sigma .
\end{array}
\]

The vertical relations are structurally parallel. On the left, an
entire family of Hamiltonian motions generated by one classical
Hamiltonian is represented by one common unitary evolution. On the
right, an entire family of Brownian motions is represented by one
common stochastic unitary evolution. In neither case is a separate
state-space law assigned to each classical trajectory.

The horizontal relations concern a change of physical regime. The
lower arrow represents the familiar passage from microscopic mechanical
motion to Brownian behavior. A particle undergoing repeated interactions
with a many-particle environment follows Newtonian motion between
interactions. If no interaction changes its phase-space data, the same
Newtonian trajectory continues. When an interaction changes the
momentum, the resulting phase-space point becomes the initial point of
another member of the same Hamiltonian family. Under suitable
statistical and scaling conditions, the resulting stochastic sequence
of mechanical motions approaches Brownian motion.

The upper arrow is the corresponding quantum relation obtained in the
present construction. Ordinary Schr\"odinger dynamics governs the full
microscopic system consisting of the object, apparatus, and
environment. Section~\ref{sec:particle-bath-origin-brownian-family}
showed that, for a spatially local oscillator bath with translation and
rotation symmetry, thermal Gaussian statistics, and an appropriate
Ohmic and coarse-grained limit, the restricted particle--bath dynamics
produces precisely the statistical structure required of the Brownian
family used in the lift. Once that family is obtained, its common
stochastic unitary lift is already determined: the projectively
relevant generators have the GUE distribution, and the resulting
evolution is the \({\bf RM}\) walk.

In this picture the \({\bf RM}\) law need not be an additional
fundamental dynamics.  Under the assumptions of
Section~\ref{sec:particle-bath-origin-brownian-family}, it arises as the
common stochastic lift of Brownian statistics generated by an ordinary
particle--environment dynamics.  What remains open is a fully
microscopic derivation controlling the approximations by which the
many-body Schr\"odinger evolution passes to the local Markovian
Brownian regime, and determining the range of systems and environments
for which those assumptions are valid.

The comparison can now be stated more directly at the level of paths.
The deterministic construction begins with the complete Newtonian
family generated by one Hamiltonian. Time slicing selects, at
successive times, members or continuations of members of this family,
and combines the resulting histories with complex amplitudes determined
by the action. The result is the ordinary Schr\"odinger propagator.

The stochastic construction begins instead with a family of Brownian
trajectories. At a finite microscopic or coarse-grained level, these
trajectories may themselves be viewed as stochastic successions of
Newtonian motions whose initial data are altered by interactions with
the environment. The common unitary lift of the entire Brownian family
produces the \({\bf RM}\) evolution in projective Hilbert space.

The two constructions therefore have the parallel form
\[
\begin{array}{ccc}
\text{family of Newtonian paths}
&\longrightarrow&
\text{Schr\"odinger--Feynman propagation},
\\[3mm]
\text{family of Brownian trajectories}
&\longrightarrow&
\text{\({\bf RM}\) measurement dynamics}.
\end{array}
\]
Both produce evolutions in state space, but their physical roles are
different. The first determines the coherent propagation of quantum
amplitudes before measurement. The second determines the stochastic
state-space evolution associated with measurement and, through the
detector-defined outcome sectors, the probabilities of the possible
records.

This distinction is essential. The \({\bf RM}\) construction does not
replace the quantum state by a classical probability distribution.
For every realization \(\omega\), the evolution remains unitary:
\[
\psi_t(\omega)
=
U_t(\omega)\psi.
\]
Hence, if
\[
\psi
=
\sum_k c_k\psi_k,
\]
then
\[
\psi_t(\omega)
=
\sum_k c_k U_t(\omega)\psi_k.
\]
The stochasticity concerns the unitary path followed in state space,
not a replacement of the quantum state by a statistical mixture.

The Born probabilities arise when these stochastic state-space paths
are related to detector outcomes. The localized outcome sectors act as
recording regions, and the probabilities that the \({\bf RM}\)
trajectory reaches the different sectors reproduce the Born weights.
Thus the roles of the two state-space constructions may be summarized
as
\[
\begin{array}{ccl}
\text{Schr\"odinger--Feynman dynamics}
&:&
\text{propagation of amplitudes},
\\[2mm]
\text{\({\bf RM}\) dynamics}
&:&
\text{measurement dynamics and Born probabilities}.
\end{array}
\]

This suggests that ordinary propagation and measurement need not
require two fundamentally independent laws. They may correspond to two
regimes of one underlying many-body Schr\"odinger dynamics. Away from
measurement, the reduced system exhibits the familiar deterministic
unitary propagation. During sufficiently strong interaction with a
macroscopic apparatus and environment, the same microscopic dynamics
may generate, after coarse graining, the stochastic unitary increments
described by the \({\bf RM}\) walk.

A more complete microscopic derivation could go further.  It could
determine the diffusion coefficient and the coarse-graining scale from
the microscopic properties of the system, apparatus, and environment,
rather than treating them as effective parameters.  It could also
establish quantitatively when the Gaussian, Markovian, and homogeneous
component-independence approximations derived in the local bath model
are valid, and how the strength of the resulting state-space diffusion
depends on the size and complexity of the system.

The stochastic formulation opens further questions at the level of
measurement histories. Recording times, first-passage distributions,
persistence of established records, and correlations between successive
measurements can be studied through the \({\bf RM}\) trajectories
themselves. Of particular interest is the dependence of the diffusion
coefficient on macroscopicity. If this coefficient decreases as the
system becomes increasingly macroscopic, the same framework would
describe the transition from appreciable stochastic motion in state
space to effectively deterministic classical motion.

The picture that emerges is therefore a possible unification of
ordinary quantum propagation and measurement dynamics within state
space. The Schr\"odinger--Feynman construction describes coherent
propagation by combining histories assembled from the complete family
of Newtonian motions. The \({\bf RM}\) construction describes a
stochastic unitary regime obtained from the Brownian family and yields
the Born probabilities of detector records. Both act in the same
projective Hilbert space.

The present work therefore provides a concrete route toward a
microscopic origin of the \({\bf RM}\) regime. For the spatially local
particle--bath model considered above, the restricted Schr\"odinger
dynamics yields, under explicit thermal, symmetry, Ohmic, and
coarse-graining assumptions, the homogeneous Brownian family required
by the lift. The common stochastic lift of that family then has the
GUE form and gives the \({\bf RM}\) walk. The remaining problem is to
place this chain on a fully microscopic footing by deriving and
controlling the required approximations for realistic many-body
systems and measuring environments.

\chapter*{Conclusion}
\label{chap:conclusion}
\addcontentsline{toc}{chapter}{Conclusion}
\markboth{Conclusion}{}

The purpose of this book was to develop a state-space framework in
which classical and quantum dynamics, measurement, definite outcomes,
and their probability laws can be described within a common geometric
and dynamical structure. Starting from the projective Hilbert space of
quantum states,
\[
\mathbb P(\mathcal H),
\]
classical configuration space and classical phase space were identified
with localized submanifolds of this space. After the appropriate
rescaling, the induced Fubini--Study metric becomes the Euclidean
metric, while the Schr\"odinger action restricted to the localized
phase-space manifold becomes the classical action.

Classical mechanics is therefore contained in quantum mechanics as the
dynamics on a distinguished localized sector, which, in the
single-particle case, is the manifold
\[
\widetilde M^\sigma_{3,3}
\subset
\mathbb P(\mathcal H).
\]
Its coordinates are the classical position and momentum variables.
Variation of the restricted action gives Hamilton's equations, while
the tangent component of the Schr\"odinger vector field gives the
corresponding classical Hamiltonian vector field. Newtonian motion is
therefore not introduced as an independent postulate; it is the
restriction of Schr\"odinger dynamics to the classical phase-space
sector.

The converse relation is equally important. The complete family of
Hamiltonian motions generated by one fixed classical Hamiltonian
determines, whenever a common unitary lift exists, that lift uniquely
up to an overall phase. The lift is not assigned separately to
individual classical trajectories. One unitary propagator acts
simultaneously on the localized states corresponding to all initial
conditions. For the classical Hamiltonians obtained by restricting
Schr\"odinger dynamics, existence of this lift is already supplied by
the corresponding quantum Hamiltonian.

The same complete Hamiltonian family also supplies the classical
motions needed in the time-sliced construction of the propagator. At
each subdivision time, the Hamiltonian flow provides the trajectory
beginning from every possible phase-space point. For sufficiently short
intervals, the classical trajectory joining two successive
configurations determines the corresponding short-time action and,
together with the appropriate amplitude factor, the short-time
propagator. Repeated composition and integration over the intermediate
configurations then gives the Schr\"odinger propagator in the
fine-partition limit. A trajectory selected at a later subdivision may
simply continue one selected earlier, or it may begin from different
phase-space data. In either case it belongs to the same complete family
generated by the original Hamiltonian.

Thus the complete Hamiltonian flow plays two complementary roles. It
determines the common unitary lift uniquely, whenever such a lift
exists, and it supplies the family of classical motions from which the
time-sliced Schr\"odinger--Feynman propagator is constructed. The same
construction extends to many-particle systems and, in the presence of
interactions, naturally allows the ambient quantum evolution to leave
the product-state manifold and generate entanglement.

The second ingredient was the introduction of equivalence classes of
states associated with the finite resolution of real measuring
devices. A classical outcome is not an exact ray in Hilbert space. It
is a class of states that cannot be distinguished by a detector at the
given resolution. The points of classical configuration space and,
after the inclusion of momentum phase factors, the points of classical
phase space are identified with such classes. This construction removes
the need to represent physical outcomes by exact position eigenstates
or exactly compactly supported states and thereby resolves the usual
problem of tails within the present framework.

The third ingredient is the random-matrix dynamics \({\bf RM}\),
whose statistical form is derived in this book from the common unitary
lift of a homogeneous Brownian family.
During measurement or sufficiently strong environmental monitoring,
the effective state-space dynamics enters, under the conditions
developed in the particle--bath construction, a stochastic unitary
regime generated by successive random Hermitian increments. For every
realization, the evolution remains unitary and linear. The state follows
one path in projective Hilbert space, and a measurement record is formed
when that path enters one of the detector-defined outcome sectors.

The form of the random generator is not chosen arbitrarily. The
classical system used in the reconstruction is a homogeneous family of
independent identical Brownian particles with centered isotropic
Gaussian coarse-grained steps. Its statistical structure was summarized
by homogeneous component independence. Requiring one common linear
unitary lift of this entire Brownian family determines the covariance
of the projectively relevant Hermitian generator. In a
finite-dimensional approximation the resulting centered generator has
the traceless Gaussian Unitary Ensemble law, apart from a
projectively irrelevant scalar contribution, while its overall scale is
fixed by the Brownian diffusion coefficient.

Conversely, the GUE-generated stochastic unitary evolution reproduces
the same Brownian family when resolved along the localized position and
momentum tangent directions and constrained to the classical manifold.
Thus the relation between the Brownian and random-matrix descriptions
is two-sided: a homogeneous Brownian family determines its common
stochastic unitary lift, and the resulting lift reproduces the original
Brownian motion.

The origin of the Brownian input was also examined.  A spatially local
Caldeira--Leggett-type particle--bath model, restricted to the
localized particle--bath manifold, produces under thermal Gaussian,
translation and rotation symmetry, Ohmic, and coarse-graining
assumptions the Gaussianity, isotropy, spatial independence, and
position--momentum component structure required by homogeneous
component independence.  Under these assumptions the chain
\[
\text{particle--bath Schr\"odinger dynamics}
\longrightarrow
\text{Brownian family}
\longrightarrow
\text{GUE lift}
\longrightarrow
{\bf RM}
\]
is explicit.

When the stochastic motion is constrained to the classical
configuration-space manifold, it gives ordinary Brownian motion and the
normal distribution. In the full projective state space, the same
geometry gives the Born transition law. In the measurement construction
the stochastic state-space trajectories reach the detector-defined
outcome sectors with the Born probabilities. Classical normal
probabilities and quantum Born probabilities therefore arise as two
manifestations of the same state-space geometry.

The deterministic and stochastic constructions can now be compared at
the level of the families from which they are built. A fixed classical
Hamiltonian generates Newtonian motions from all possible initial
conditions. The complete family of these motions supplies the
time-sliced Feynman construction and thereby the ordinary
Schr\"odinger propagator. A Brownian particle, at a microscopic or
coarse-grained level, follows Newtonian motion between interactions. If
its phase-space data are unchanged, the same Newtonian trajectory
continues; if an interaction changes the momentum, the resulting
phase-space point becomes the initial point of another member of the
same Hamiltonian family. In the diffusion limit, the resulting
stochastic family becomes Brownian motion.

The two constructions therefore have the parallel form
\[
\begin{array}{ccc}
\text{family of Newtonian paths}
&\longrightarrow&
\text{Schr\"odinger--Feynman propagation},
\\[3mm]
\text{family of Brownian trajectories}
&\longrightarrow&
\text{\({\bf RM}\) measurement dynamics}.
\end{array}
\]
Both produce evolutions in the same projective Hilbert space, but they
play different physical roles. Schr\"odinger--Feynman dynamics
describes the coherent propagation of quantum amplitudes. The
\({\bf RM}\) dynamics describes the stochastic state-space evolution
associated here with measurement and determines the probabilities of
the possible detector records:
\[
\begin{array}{ccl}
\text{Schr\"odinger--Feynman dynamics}
&:&
\text{propagation of quantum amplitudes},
\\[2mm]
\text{\({\bf RM}\) dynamics}
&:&
\text{measurement dynamics and Born probabilities}.
\end{array}
\]

This distinction does not amount to replacing the quantum state by a
classical probability distribution during measurement. For every
realization \(\omega\), the \({\bf RM}\) evolution remains unitary:
\[
\psi_t(\omega)
=
U_t(\omega)\psi.
\]
Hence, if
\[
\psi
=
\sum_k c_k\psi_k,
\]
then
\[
\psi_t(\omega)
=
\sum_k c_kU_t(\omega)\psi_k.
\]
The stochasticity concerns the unitary path followed in state space,
not a replacement of the quantum state by a statistical mixture. The
Born probabilities arise when these stochastic trajectories are
related to the detector-defined outcome sectors.

The microscopic and macroscopic regimes differ in the relative scales
of the deterministic and stochastic parts of the state-space dynamics.
During a microscopic measurement, the stochastic component may dominate
the short record-forming interval, and the realized state-space path is
recorded in one of the possible outcome sectors with the corresponding
Born probability. For a macroscopic body, frequent environmental
interactions repeatedly stabilize the state near the localized
phase-space sector. At the resolution of observation, the ordinary
Hamiltonian supplies the Newtonian tangent drift, while the stochastic
deviations remain small. The observed trajectory is therefore a
sequence of finite-resolution records concentrated near a Newtonian
path.

The estimates developed in the book show how this regime arises in
concrete situations. For a macroscopic body in air, and even in the
limiting environment provided by the cosmic microwave background,
environmental interactions occur sufficiently frequently to maintain
localization while producing only negligible deviations from the
Newtonian trajectory. The same framework accounts for microscopic
tracks in a bubble chamber and for laboratory position measurements of
microscopic particles.

This picture also fixes the meaning of superposition. A superposition
is one physical state, represented by one point of projective state
space. It is neither a collection of simultaneously realized classical
worlds nor merely a statement of ignorance. When the state lies away
from the classical sector associated with an observable, the
corresponding classical value is not defined. A definite value appears
when the state is recorded in the corresponding detector-defined
outcome sector.

The standard quantum paradoxes then acquire a common geometric origin.
In the double-slit experiment, the state leaves the classical position
manifold and returns to an appropriate outcome sector when a position
record is formed. In an EPR experiment, the entangled pair is one state
in the tensor-product projective space, and the observed correlations
are determined by the geometry of this state relative to the detector
sectors. Schr\"odinger's cat and Wigner's friend arise when microscopic
superposition language is extended directly to macroscopic systems
without accounting for their continual environmental stabilization near
the classical sector. The preferred basis is fixed physically by the
measuring arrangement and by the finite-resolution sectors in which
records can be formed.

The same framework gives an account of irreversibility and the arrow of
time. Although the ambient \({\bf RM}\) law itself has no preferred
temporal direction, its restriction to the classical variables produces
Brownian diffusion. Localized probability distributions therefore tend
to spread through classical configuration space, providing a stochastic
contribution to the increase of coarse-grained entropy.

A second source of effective irreversibility is the finite-resolution
character of classical records. Every realized \({\bf RM}\) history is
unitary and has a mathematical inverse if the complete realized
Hamiltonian history is known. Exact reversal, however, requires
reconstructing that particular sequence and applying the inverse
unitaries in reverse order. A finite-resolution classical record
contains neither this sequence nor the exact microscopic ray by which
it was reached. Moreover, for any fixed observation interval, however
long, the probability of returning to the same microscopic state tends
to zero as the dimension of the relevant state space increases. The
classical arrow of time therefore emerges from Brownian spreading, the
loss of microscopic path information in passing to finite-resolution
records, and the ordering of those records into a macroscopic history.

Several problems remain open.  The principal one is to extend the
particle--bath route obtained here to a fully microscopic derivation
for realistic many-body systems, measuring apparatuses, and
environments.  The local oscillator-bath model shows explicitly how
the centered Gaussian increments and homogeneous component independence
required by the Brownian lift can arise.  What remains is to derive and
control the Markovian, thermal, locality, and coarse-graining
approximations directly from the underlying many-body Schr\"odinger
dynamics and to determine their domain of validity.

Such a microscopic treatment could also determine the diffusion
coefficient and the effective coarse-graining scale from properties of
the system, apparatus, and environment.  It could clarify
quantitatively why stochastic state-space motion may be appreciable for
microscopic systems yet negligible for macroscopic ones, and place the
transition between ordinary Schr\"odinger propagation and the
\({\bf RM}\) measurement regime on a firmer dynamical foundation.

Other problems remain. The construction should be extended
systematically to relativistic quantum mechanics and quantum field
theory. This requires identifying the appropriate localized classical
submanifolds for relativistic particles and fields, deriving their
classical dynamics as restricted quantum dynamics, and determining how
environmental stabilization and stochastic unitary motion operate in
the corresponding state spaces. The infinite-dimensional stochastic
process and its path measure also deserve a more complete mathematical
treatment. The relation between commutators and state-space curvature,
and the physical conclusions that may follow from this geometry,
likewise require further investigation.

The main conclusion is that classical space, classical phase space,
Hamiltonian dynamics, definite measurement outcomes, and Born
probabilities need not be appended to quantum mechanics as unrelated
structures. They can all be represented within projective quantum state
space. Ordinary Schr\"odinger dynamics gives coherent propagation and,
on the localized sector, classical Hamiltonian motion. The complete
Hamiltonian family also provides the classical motions entering the
Feynman construction of the propagator. The stochastic \({\bf RM}\)
regime gives measurement dynamics and, on the same localized sector,
Brownian motion. Both are evolutions of the same quantum state in the
same state-space arena.

The resulting possibility is therefore a particularly economical one:
the \({\bf RM}\) regime need not be a modification of quantum mechanics
at all.  The particle--bath construction developed here shows, under
explicit assumptions, how the Brownian input required by the
\({\bf RM}\) lift can arise from ordinary Schr\"odinger dynamics.
If this route can be extended and placed on a controlled microscopic
footing for realistic measurement environments, coherent propagation,
measurement, Born probabilities, and the emergence of the classical
world would become different dynamical manifestations of one underlying
state-space description.

\appendix

\input{RM_Simulations_Appendix_Revised/appendix_rm_simulations_revised.tex}

\backmatter

\chapter{Guide to the Literature}
\label{app:literature-guide}

The purpose of this appendix is to indicate a small number of standard
references for the mathematical and physical background used throughout
the book. It is not intended as a comprehensive literature review.
Whenever possible, established textbooks and monographs are listed
rather than individual research papers.

The references below concern the standard structures on which the
present construction builds. The specific state-space framework
developed in this book, including the localized classical submanifolds,
the relation between Schr\"odinger and Newtonian dynamics, the unitary
lift of Brownian motion, the random-matrix conjecture \({\bf RM}\), and
their applications to measurement and macroscopic dynamics, is
developed in the text itself.

\section*{General References}

\subsection*{Quantum Mechanics and Quantum Foundations}

A useful general reference for the conceptual and mathematical
foundations of quantum mechanics is A.~Peres,
\emph{Quantum Theory: Concepts and Methods}. It discusses pure and
mixed states, observables, measurement, composite systems,
entanglement, and Bell inequalities. It is particularly relevant to
the material in Chapters~5, 7, 8, and 9.

\subsection*{Geometry of Quantum States}

For the geometry of pure and mixed quantum states, including projective
Hilbert space, the Fubini--Study metric, the Bloch sphere,
tensor-product state spaces, and entanglement, a standard reference is
I.~Bengtsson and K.~\.Zyczkowski,
\emph{Geometry of Quantum States: An Introduction to Quantum
Entanglement}. This provides background for the geometric structures
used especially in Chapters~1, 3, 7, and 8.

The localized classical submanifolds and the finite-resolution
equivalence classes introduced in the present book are not part of
that standard construction. They are specific to the framework
developed here.

\subsection*{Classical Hamiltonian Mechanics}

For Hamiltonian mechanics, symplectic geometry, Hamiltonian vector
fields, and Poisson brackets, a standard reference is V.~I.~Arnold,
\emph{Mathematical Methods of Classical Mechanics}. These structures
underlie the comparison between classical Hamiltonian dynamics and
Schr\"odinger dynamics in Chapters~1--3.

In the present book, classical phase space is realized as a localized
submanifold of projective Hilbert space, and Hamiltonian dynamics is
obtained by restricting the Schr\"odinger dynamics to this submanifold.
These identifications are part of the construction developed here.

\subsection*{Path Integrals}

For the time-sliced construction of the quantum propagator and the role
of the classical action in short-time propagation, see
L.~S.~Schulman,
\emph{Techniques and Applications of Path Integration}. This material
provides background for the discussion in Chapter~2 relating the
complete family of classical Hamiltonian motions to the many-path
representation of the quantum propagator.

The connection made in Chapter~2 between this complete Hamiltonian
family and the common unitary lift is specific to the present
state-space formulation.

\subsection*{Brownian Motion and Stochastic Processes}

For Brownian motion, Wiener processes, diffusion equations, stochastic
differential equations, and first-hitting problems, see B.~\O ksendal,
\emph{Stochastic Differential Equations: An Introduction with
Applications}. These standard stochastic structures are used in
Chapters~4--6 and again in the discussion of irreversibility in
Chapter~9.

The identification of Brownian motion on the localized classical
manifold with the constrained form of a stochastic unitary motion in
projective state space is developed in Chapter~4 and is not part of
the standard theory of Brownian motion.

\subsection*{Random Matrices}

For the Gaussian Unitary Ensemble and the general theory of
unitarily invariant random-matrix ensembles, see M.~L.~Mehta,
\emph{Random Matrices}. This provides the standard background for the
GUE used in Chapters~4--6.

The appearance of the GUE in the present framework is derived from a
different direction. Chapter~4 reconstructs the projectively relevant
GUE law from the homogeneous family of Brownian motions on the
localized classical manifold and the requirement of a common linear
unitary lift. Mehta's text is therefore cited for the standard
properties of the ensemble rather than for this reconstruction.

\subsection*{Open Quantum Systems and Decoherence}

For reduced density operators, environmental decoherence, master
equations, and the general theory of open quantum systems, see
H.-P.~Breuer and F.~Petruccione,
\emph{The Theory of Open Quantum Systems}. This provides background
for the environmental discussion in Chapter~6 and for the comparison
with decoherence and microscopic bath models in Chapter~10.

The \({\bf RM}\) description differs from the usual reduced-state
description in that each realized state-space history remains unitary
and is generated by a realized random Hermitian Hamiltonian. The
comparison is discussed in Chapter~10.

\subsection*{Continuous Quantum Measurement}

For continuous monitoring, stochastic Schr\"odinger equations,
conditional states, quantum trajectories, and measurement records, see
H.~M.~Wiseman and G.~J.~Milburn,
\emph{Quantum Measurement and Control}. This reference is particularly
relevant to the comparison with continuous-measurement theory in
Chapter~10.

The stochastic trajectories used in the present book have a different
origin. The \({\bf RM}\) walk is constructed as a stochastic unitary
motion in projective state space whose random Hermitian generator is
related to the Brownian motion of localized states.

\section*{References for the Approaches Discussed in Chapter~10}

Chapter~10 compares the present framework with several established
approaches to quantum measurement and the quantum--classical
transition. For some of these approaches, the original papers remain
the most natural references. The brief list below is intended only to
identify the formulations being compared; it is not a comprehensive
history of these subjects.

\subsection*{Bell's Theorem}

J.~S.~Bell's paper
\emph{On the Einstein Podolsky Rosen Paradox}
is the foundational reference for Bell's theorem. It shows that the
correlations predicted by quantum mechanics cannot, in general, be
reproduced by a local hidden-variable theory satisfying the
corresponding factorization assumptions. Bell's result is relevant to
the discussion of quantum correlations in Chapter~8 and to the
comparison in Chapter~10.

\subsection*{Objective-Collapse Models}

The original spontaneous-localization proposal of
G.~C.~Ghirardi, A.~Rimini, and T.~Weber is presented in
\emph{Unified Dynamics for Microscopic and Macroscopic Systems}.
It introduces a stochastic modification of quantum dynamics designed
to suppress macroscopic spatial superpositions. It provides the
canonical reference for the GRW model discussed in Chapter~10.

The comparison is particularly useful because both GRW-type models and
the \({\bf RM}\) framework introduce stochasticity into the description
of measurement. In the present construction, however, every realized
evolution remains linear and unitary, and no spontaneous projection or
nonunitary collapse law is added.

\subsection*{Bohmian Mechanics}

The foundational reference for Bohmian mechanics is D.~Bohm's
two-part paper
\emph{A Suggested Interpretation of the Quantum Theory in Terms of
``Hidden'' Variables}. It formulates a deterministic dynamics of
particle configurations guided by the wave function.

The comparison in Chapter~10 emphasizes the different role assigned to
configuration space. In Bohmian mechanics, the particle configuration
is an additional fundamental variable. In the present framework,
classical configuration space appears instead as a localized
submanifold of quantum state space.

\subsection*{Everettian Quantum Mechanics}

The original reference for the relative-state formulation is
H.~Everett's paper
\emph{Relative State Formulation of Quantum Mechanics}. It treats
unitary quantum mechanics as universally valid and interprets the
different components of an entangled measurement state without
introducing a collapse postulate.

Chapter~10 compares this with the present framework, in which one
realized \({\bf RM}\) trajectory produces one finite-resolution
classical record rather than a persistent multiplicity of classical
outcomes.

\subsection*{Consistent Histories}

A foundational reference for the consistent-histories formulation is
R.~B.~Griffiths,
\emph{Consistent Histories and the Interpretation of Quantum
Mechanics}. The approach assigns probabilities to suitable sequences
of quantum alternatives when the required consistency conditions are
satisfied.

The comparison with the present framework is especially natural
because both descriptions involve histories. The objects are,
however, different. A consistent history is a sequence of projectors
satisfying a consistency condition, whereas an \({\bf RM}\) history is
a realized stochastic unitary path in projective state space.

\subsection*{Decoherence}

The standard theory of decoherence is covered broadly by the
open-system reference given above. For the foundational development of
environment-induced suppression of interference, one may also consult
the early work of E.~Joos and H.~D.~Zeh on the emergence of classical
properties through interaction with the environment.

The comparison in Chapter~10 distinguishes suppression of off-diagonal
terms in a reduced density matrix from the formation of a definite
record in the \({\bf RM}\) framework.

\section*{Use of the References}

The references in this appendix are intended to provide background for
standard mathematical and physical structures and to identify the main
approaches discussed in Chapter~10. They are not intended to assign
the constructions developed in this book to the cited literature.

In particular, the following elements belong to the framework developed
in the present text: the representation of classical configuration
space and phase space by localized finite-resolution classes in
projective Hilbert space; the derivation of Newtonian motion from the
tangent restriction of Schr\"odinger dynamics; the corresponding
restriction of the Schr\"odinger action; the reconstruction of a common
unitary lift of the Brownian family and its GUE generator; conjecture
\({\bf RM}\); and the use of the resulting state-space dynamics to
describe measurement, macroscopic stabilization, quantum correlations,
and the formation of classical records.

The bibliography following this appendix gives the complete
bibliographic information for all works listed here.

\cleardoublepage

\end{document}

%% file: RM_Simulations_Appendix_Revised/appendix_rm_simulations_revised.tex
\chapter{Numerical Simulations of the \({\bf RM}\) Dynamics}
\label{app:rm-simulations}

\renewcommand{\thesection}{\thechapter.\Roman{section}}
\renewcommand{\thesubsection}{\thesection.\arabic{subsection}}
\renewcommand{\thefigure}{\thechapter.\arabic{figure}}
\renewcommand{\thetable}{\thechapter.\arabic{table}}
\setcounter{figure}{0}
\setcounter{table}{0}

The purpose of this appendix is to give a detailed numerical account of the
finite-dimensional and discretized tests used to check the stochastic constructions
of the book.  None of the simulations is used as an assumption in the analytical
arguments of the preceding chapters.  Their role is instead to answer, for concrete
finite approximations, three questions that recur throughout the book:
\begin{enumerate}
\item does the proposed random-matrix motion have the projective-space symmetries required by the geometric construction;
\item does its restriction to the localized classical sector reproduce the Newtonian dynamics of classical systems and the stochastic laws required by the measurement arguments; and
\item do the same rules remain compatible with the Born rule, detector records, tensor-product systems, entanglement, the macroscopic limit, and effective irreversibility?
\end{enumerate}

The presentation is organized to follow the logical order of the
present book, and each section below states explicitly which part of the main text is
being checked.

\section{Numerical framework and relation to the main text}
\label{sec:app-numerical-framework}

\subsection{Finite-dimensional projective steps}

For the local projective tests the Hilbert space is replaced by
\[
\mathcal H_N=\mathbb C^N,
\qquad
\mathbb P(\mathcal H_N)=\mathbb{CP}^{N-1}.
\]
A normalized state is denoted by \(\psi\).  The book derives in
Section~\ref{sec:the-gaussian-unitary-lift} that the Brownian-normalized Hermitian
increment amplitude has the GUE form.  Since the overall scale is irrelevant for the
isotropy and homogeneity tests, the simulations use a dimensionless Hermitian matrix
\(X\) sampled from the corresponding GUE law and form the first-order displacement
\[
\delta\psi=-iX\psi.
\]
The phase direction is removed by horizontal projection,
\[
\delta\psi_{\perp}
=
\delta\psi-\psi(\psi,\delta\psi).
\]
The local Fubini--Study norm is therefore, to first order,
\[
\|\delta\psi_{\perp}\|.
\]
Thus the covariance of the real coordinates of \(\delta\psi_{\perp}\) is the numerical
object corresponding to the infinitesimal projective covariance discussed in
Sections~\ref{sec:diffusion-and-dimension} and~\ref{sec:the-gaussian-unitary-lift}.

The notation \(X\) is used here only for the dimensionless local test.  In the notation
of Chapter~\ref{chap:brownian-born}, the dimensional Hermitian action increment is
\(dK_N=H_N\sqrt{dt}\), and \(dK_N/\hbar\) is dimensionless.  No Hamiltonian is being
interpreted as acting for a time \(\sqrt{dt}\).

\subsection{Spatial discretization and projected coordinates}

When the localized manifold is tested directly, \(L^2(\mathbb R)\) is represented on
a finite spatial interval and a uniform grid.  The Gaussian representatives and their
tangent vectors are discretized and normalized with the corresponding discrete
approximation to the \(L^2\) inner product.  A projective increment is then resolved
along the tangent directions of the localized manifold or of the detector-adapted
coordinate surface.  This is the finite numerical counterpart of the tangent
projections used in Chapters~\ref{chap:classical-submanifolds},
\ref{chap:brownian-born}, and~\ref{chap:measurement-rm-walk}.

The simulations use three kinds of numerical checks repeatedly:
\begin{itemize}
\item covariance matrices and variances, to test isotropy and independence;
\item Kolmogorov--Smirnov tests and histogram comparisons, to test Gaussianity and
homogeneity; and
\item Monte Carlo frequencies, to test path weights, hitting probabilities, and
record statistics.
\end{itemize}
Tables and figures are numbered independently of the Roman-numbered appendix
sections.  Thus, for example, Section~A.II contains Figure~A.1 and Table~A.1.

\section{GUE isotropy and homogeneity in projective state space}
\label{sec:app-gue-isotropy}

This section tests the finite-dimensional stochastic symmetry used in
Sections~\ref{sec:diffusion-and-dimension}, \ref{sec:the-gaussian-unitary-lift}, and
\ref{sec:the-random-matrix-walk-on-state-space}.  The question is whether GUE
increments fill the real tangent space of complex projective Hilbert space isotropically
and whether their distribution is independent of the base point.

\subsection{Isotropy at one projective point}

The first calculation uses
\[
N=20,
\qquad
\psi=e_0=(1,0,\ldots,0),
\]
so that
\[
\dim_{\mathbb R}T_{\psi}\mathbb{CP}^{19}=38.
\]
For each of \(50{,}000\) independent matrices, the horizontal increment
\(\delta\psi_\perp\) was resolved into its \(38\) real tangent coordinates and their
covariance matrix was computed.  Isotropy requires this covariance to be a scalar
multiple of the identity.

For GUE, the off-diagonal matrix entries are circular complex Gaussians.  Their real
and imaginary parts therefore supply equal variance to the paired real tangent
directions.  GOE provides an instructive control: at \(e_0\), its off-diagonal entries
are real, so \(-iX\psi\) occupies only one member of each real-imaginary tangent pair.
The expected failure of complex-projective isotropy is therefore visible directly in
the covariance spectrum.

\begin{figure}[htbp]
\centering
\includegraphics[width=0.78\textwidth]{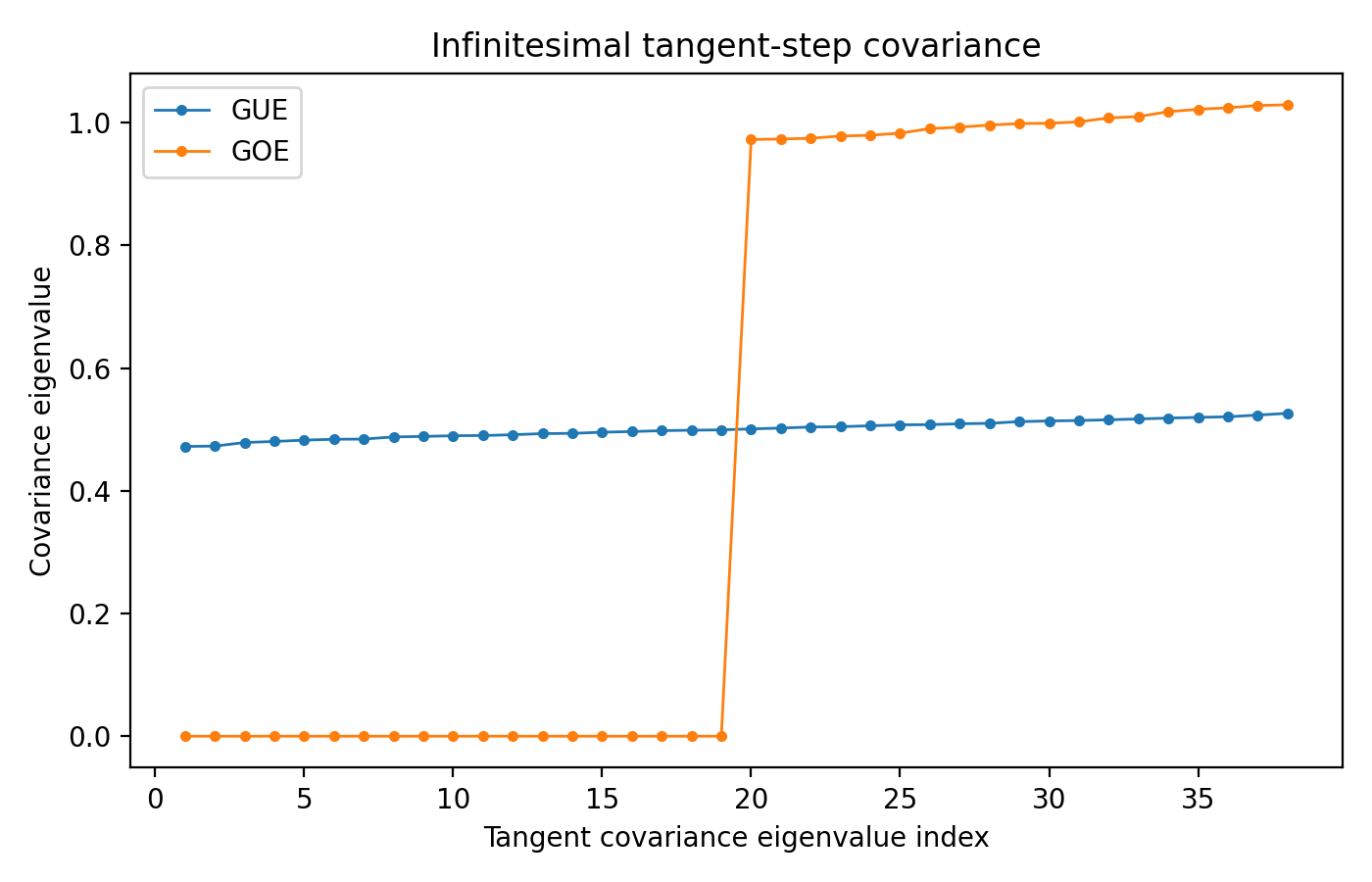}
\caption{Covariance eigenvalues of infinitesimal tangent displacements.  GUE fills
all real tangent directions approximately equally, whereas GOE leaves half of the
real tangent directions with zero variance.}
\label{fig:app-gue-goe-isotropy}
\end{figure}

The numerical result is summarized in Table~\ref{tab:app-gue-goe-isotropy} and
Figure~\ref{fig:app-gue-goe-isotropy}.  GUE has no zero covariance eigenvalues and a
nearly flat spectrum, whereas GOE has exactly \(19\) vanishing eigenvalues.

\begin{table}[htbp]
\centering
\caption{Covariance eigenvalues of infinitesimal projective tangent steps at
\(\psi=e_0\).}
\label{tab:app-gue-goe-isotropy}
\begin{tabular}{lcccc}
\hline
Ensemble & Tangent dim. & Zero eig. & Min eig. & Max eig. \\
\hline
GUE & 38 & 0  & 0.472 & 0.526 \\
GOE & 38 & 19 & 0.000 & 1.029 \\
\hline
\end{tabular}
\end{table}

\subsection{Homogeneity across projective states}

Homogeneity was tested by comparing the distribution of
\(\|\delta\psi_\perp\|^2\) at three normalized states in \(\mathbb C^{20}\): the basis
state \(e_0\), the real equal-amplitude superposition, and a generic complex random
state.  For each state and each ensemble, \(20{,}000\) independent matrices were
sampled.  The distribution at \(e_0\) was used as the reference in two-sample
Kolmogorov--Smirnov tests.

\begin{figure}[htbp]
\centering
\includegraphics[width=0.78\textwidth]{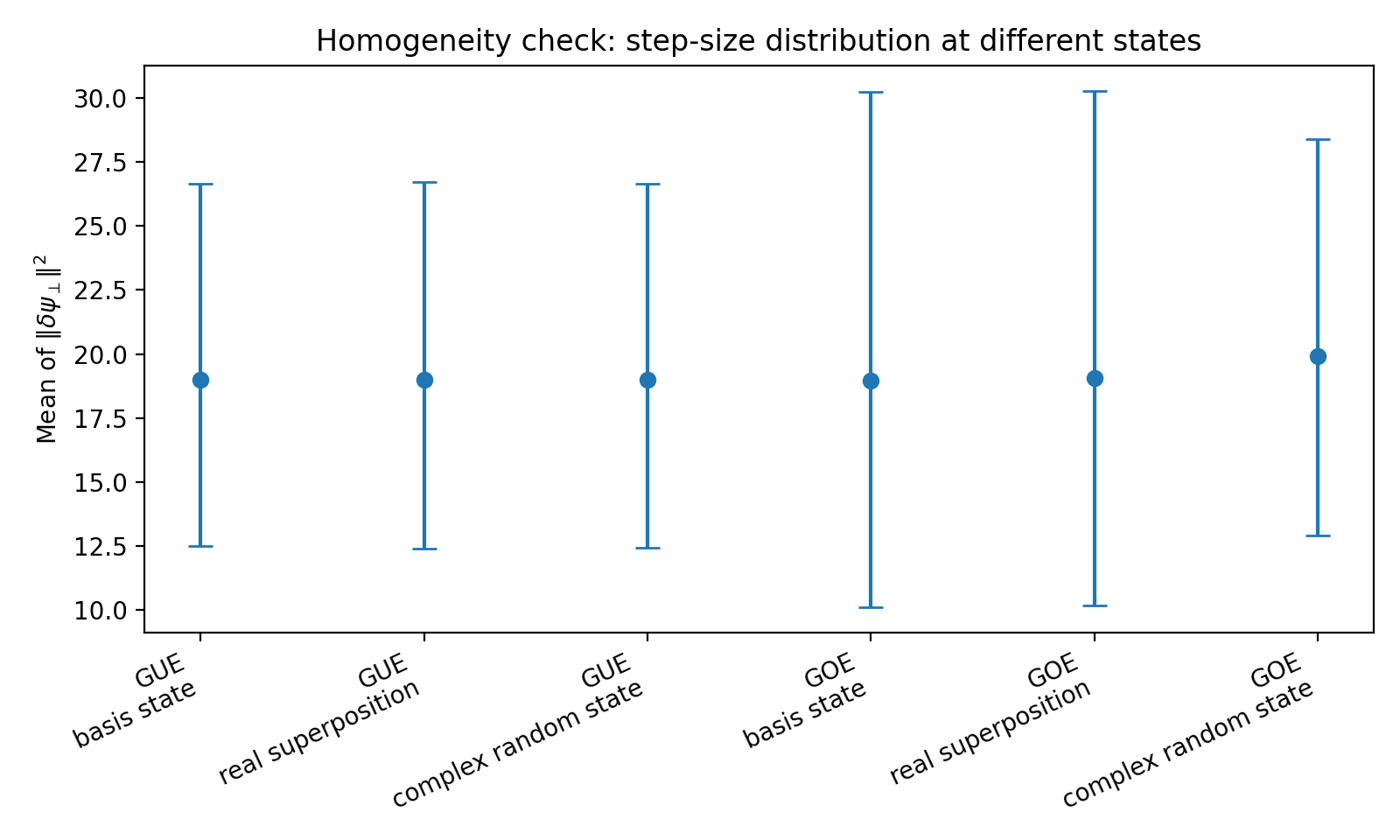}
\caption{Homogeneity check for the projective step-size distribution.  GUE is stable
across the tested states, whereas GOE changes markedly at a complex state.}
\label{fig:app-homogeneity}
\end{figure}

Table~\ref{tab:app-homogeneity-ks} gives the resulting statistics.  The GUE
distributions are statistically indistinguishable at this sample size.  GOE changes
strongly at a genuinely complex state, reflecting its preferred real structure.
Figure~\ref{fig:app-homogeneity} shows the corresponding step-size ranges.

\begin{table}[htbp]
\centering
\caption{Kolmogorov--Smirnov comparison of projective step-size distributions.}
\label{tab:app-homogeneity-ks}
\begin{tabular}{llcc}
\hline
Ensemble & Comparison & KS statistic & \(p\)-value \\
\hline
GUE & \(e_0\) vs. real superposition & 0.0098 & 0.290 \\
GUE & \(e_0\) vs. complex random state & 0.0066 & 0.774 \\
GOE & \(e_0\) vs. real superposition & 0.0139 & 0.0416 \\
GOE & \(e_0\) vs. complex random state & 0.1386 & \(6.46\times10^{-168}\) \\
\hline
\end{tabular}
\end{table}

The two tests therefore verify the finite-dimensional numerical signature assumed in
the subsequent simulations: GUE, but not GOE, produces a homogeneous and isotropic
complex-projective increment law.  This supports the choice of GUE in the unitary
Brownian lift of Section~\ref{sec:the-gaussian-unitary-lift}; it does not by itself prove
the \({\bf RM}\) conjecture of Section~\ref{sec:the-rm-conjecture}.

\section{Brownian restriction to the localized manifold}
\label{sec:app-brownian-restriction}

The next calculations test the central geometric relation of
Sections~\ref{sec:brownian-gaussian-embedding}--\ref{sec:heat-kernel-and-born-weights}:
the same isotropic state-space motion should appear as ordinary Brownian motion when
resolved tangentially on the localized classical manifold.

\subsection{One-dimensional localized Gaussian family}

\begin{figure}[htbp]
\centering
\includegraphics[width=0.76\textwidth]{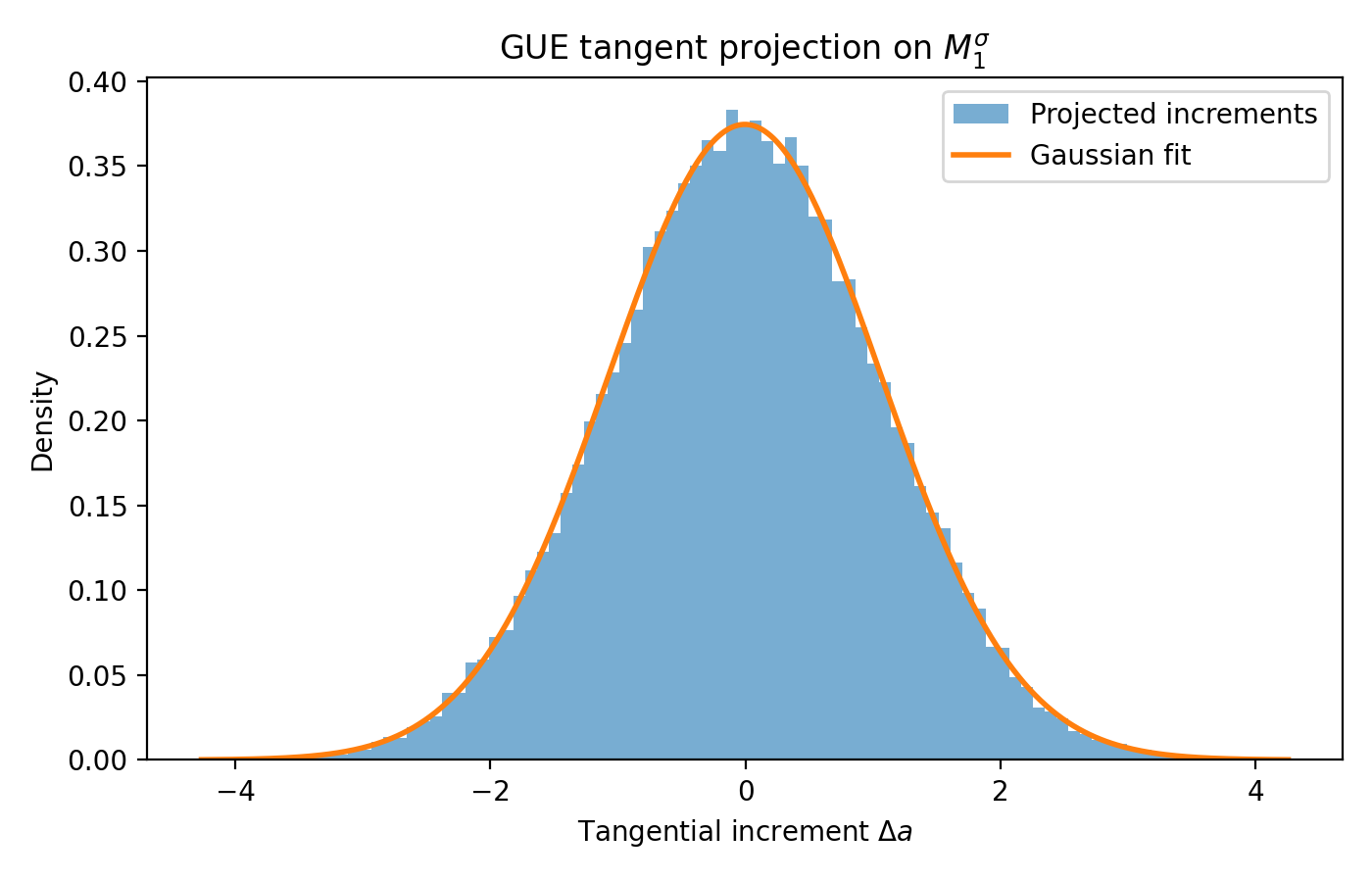}
\caption{Histogram of projected tangential increments at \(a=0\), with the Gaussian
fit used in the normality test.}
\label{fig:app-brownian-increment}
\end{figure}

The numerical model uses
\[
g_{a,\sigma}(x)
=
\left(\frac{1}{2\pi\sigma^2}\right)^{1/4}
\exp\left[-\frac{(x-a)^2}{4\sigma^2}\right],
\qquad
\frac{\partial g_{a,\sigma}}{\partial a}
=
\frac{x-a}{2\sigma^2}g_{a,\sigma}.
\]
The grid is
\[
x\in[-8,8],
\qquad
N=512,
\qquad
\sigma=0.75.
\]
For each GUE-induced horizontal displacement, the induced classical increment is
computed by orthogonal projection,
\[
\Delta a
=
\frac{\operatorname{Re}(\partial_a g_{a,\sigma},\delta\psi_\perp)}
     {\|\partial_a g_{a,\sigma}\|^2}.
\]
Fifty thousand independent increments were generated at each of the centers
\(a=-2,0,2\).  Table~\ref{tab:app-brownian-one-step} shows that the means are
consistent with zero and that the standard deviations agree within sampling error.
The histogram at \(a=0\), shown in Figure~\ref{fig:app-brownian-increment}, is well
fit by a Gaussian.

\begin{table}[htbp]
\centering
\caption{Projected one-step increments on \(M_1^\sigma\).}
\label{tab:app-brownian-one-step}
\begin{tabular}{rccc}
\hline
Center \(a\) & Samples & Mean \(\Delta a\) & Std. dev. \\
\hline
\(-2\) & 50,000 & 0.0052  & 1.0584 \\
0      & 50,000 & \(-0.0048\) & 1.0651 \\
2      & 50,000 & \(-0.0067\) & 1.0625 \\
\hline
\end{tabular}
\end{table}

Normality tests of the standardized increments gave \(p\)-values \(0.767\),
\(0.973\), and \(0.939\) at \(a=-2,0,2\), respectively.  Two-sample tests comparing
\(a=0\) with \(a=-2\) and \(a=2\) gave \(p=0.266\) and \(p=0.937\).  Thus no
statistically significant dependence on the center was observed.

\subsection{Accumulation and Brownian scaling}

\begin{figure}[htbp]
\centering
\includegraphics[width=0.78\textwidth]{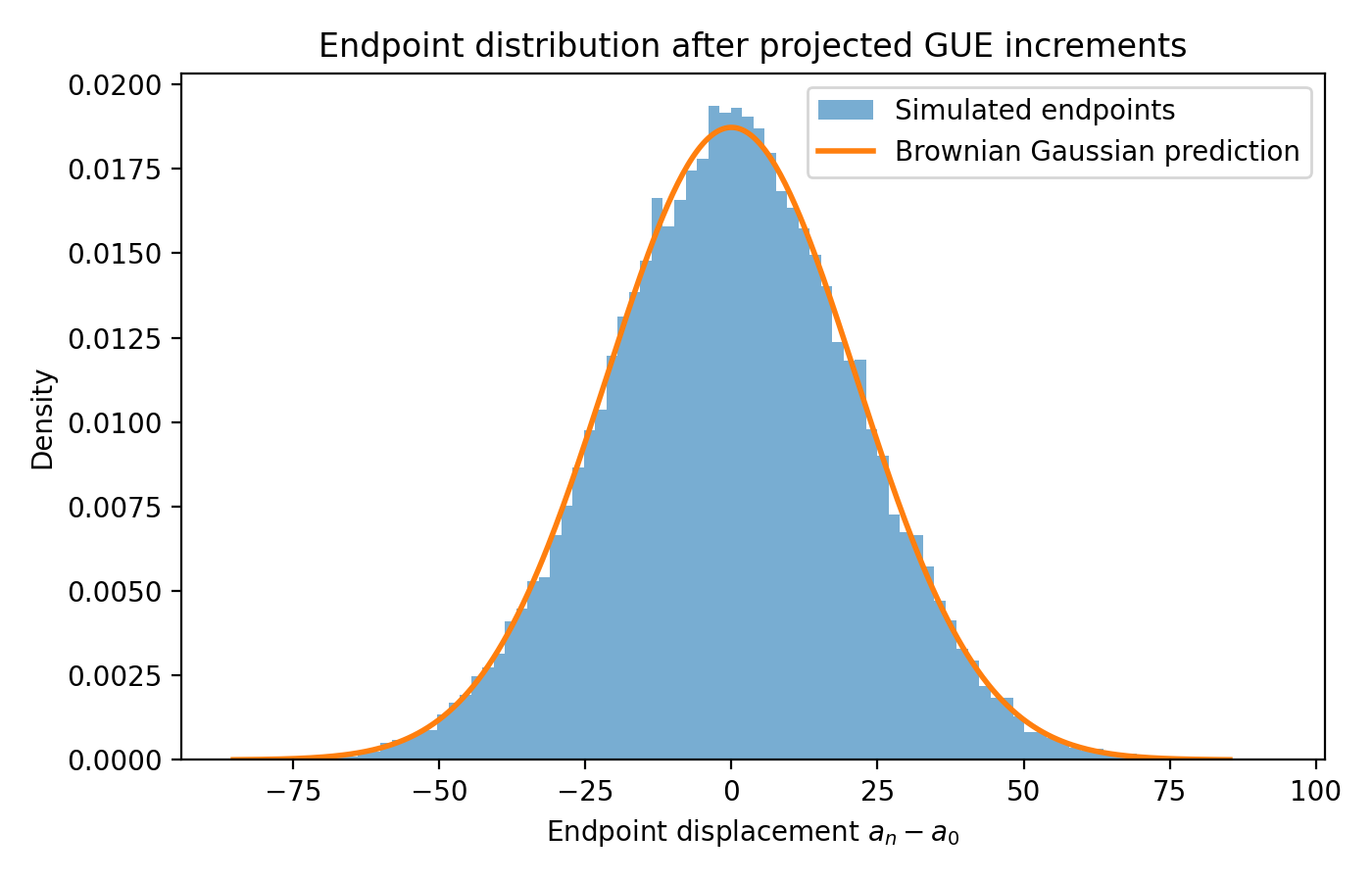}
\caption{Endpoint distribution after \(400\) projected GUE-induced tangent increments
on \(M_1^\sigma\).  The histogram agrees with the Gaussian distribution predicted by
Brownian scaling.}
\label{fig:app-brownian-endpoint}
\end{figure}

To test the diffusion law rather than only the one-step distribution, the measured
standard deviation at \(a=0\),
\[
\sigma_{\Delta a}=1.0651,
\]
was used to generate \(30{,}000\) walks of \(400\) independent steps.  Brownian
scaling predicts
\[
\operatorname{Std}(a_{400}-a_0)
=
\sqrt{400}\,\sigma_{\Delta a}
=
21.3023.
\]
The measured value was \(21.2708\), and the variances were \(452.45\) numerically
and \(453.79\) from the Brownian prediction.  Figure~\ref{fig:app-brownian-endpoint}
shows the full endpoint distribution.

\subsection{Detector-adapted \((\tau,s)\) coordinates}

\begin{figure}[htbp]
\centering
\includegraphics[width=0.68\textwidth]{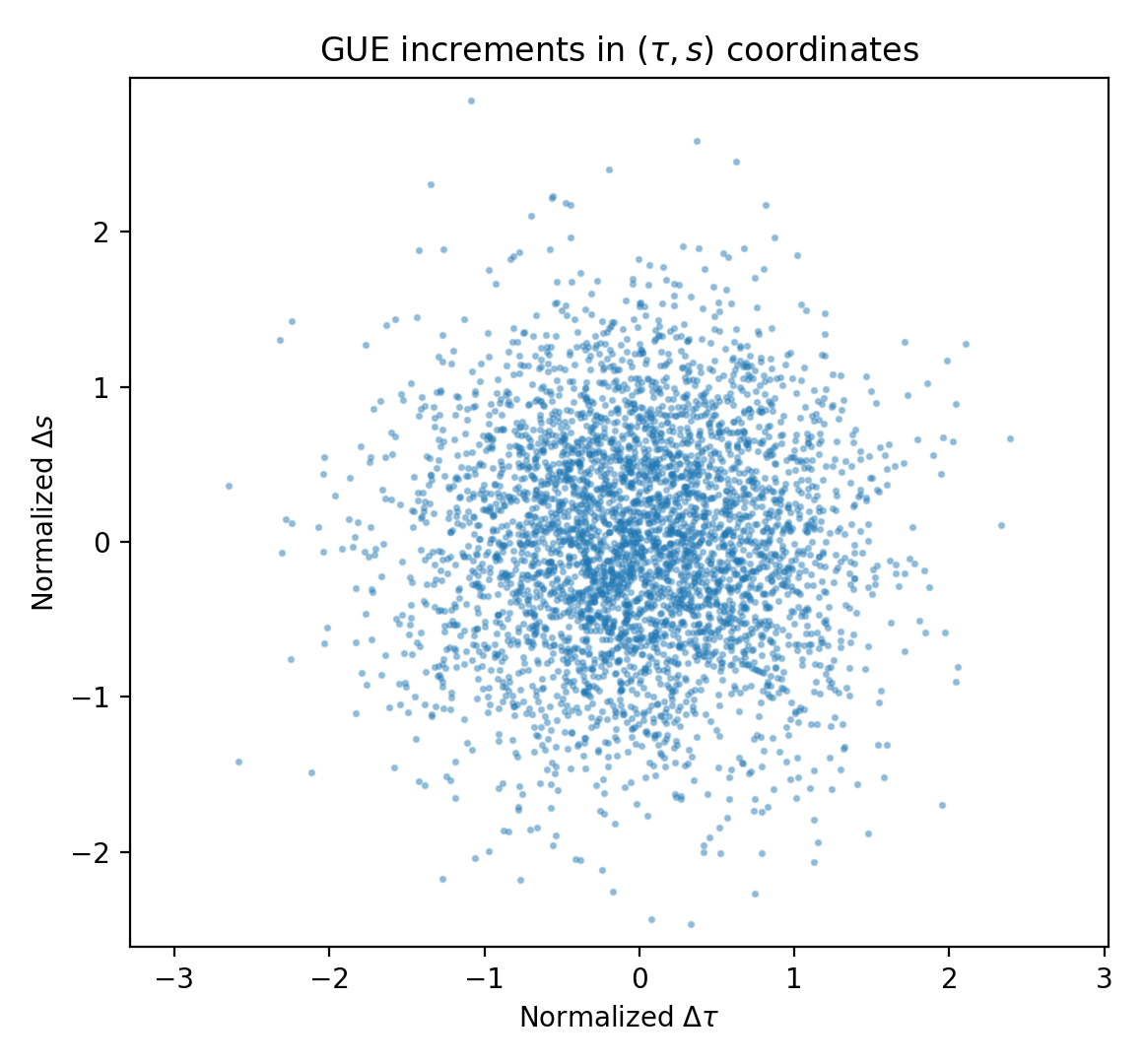}
\caption{Fubini--Study-normalized GUE-induced increments in the \((\tau,s)\)
coordinates.  The nearly circular cloud gives the local numerical signature of
isotropic motion in the orthogonal position and localization directions.}
\label{fig:app-taus-scatter}
\end{figure}

The same restriction was tested in the coordinates used later for detector records,
\[
\tau=\langle x\rangle,
\qquad
s=\ln(\delta_x/\sigma),
\qquad
\delta_x=\sqrt{\langle x^2\rangle-\langle x\rangle^2}.
\]
The localized sector is \(s\le0\).  On the Gaussian surface
\[
\psi_{\tau,s}=g_{\tau,\delta},
\qquad
\delta=\sigma e^s,
\]
the tangent vectors are
\[
\partial_\tau g_{\tau,\delta}
=
\frac{x-\tau}{2\delta^2}g_{\tau,\delta},
\qquad
\partial_s g_{\tau,\delta}
=
\left[-\frac12+\frac{(x-\tau)^2}{2\delta^2}\right]g_{\tau,\delta}.
\]
The calculation used \(N=256\) grid points on \([-10,10]\), with \(\sigma=0.75\),
and \(20{,}000\) independent samples at each of four points.  At \(s=0\), the
numerical metric was
\[
G_{\tau\tau}=0.4444,
\qquad
G_{ss}=0.5000,
\qquad
G_{\tau s}\simeq0.
\]
After multiplication by the square roots of the metric coefficients, the \(\tau\)- and
\(s\)-increments are Gaussian and essentially uncorrelated.  The largest absolute
correlation among the four tested points was about \(6\times10^{-3}\).
Figure~\ref{fig:app-taus-scatter} shows the normalized cloud at \((\tau,s)=(0,0)\).

These tests support the passage used in Chapters~\ref{chap:brownian-born} and
\ref{chap:measurement-rm-walk} from isotropic projective motion to Brownian motion in
the detector-relevant classical coordinates.

\subsection{Position, momentum, and localization increments}
\label{sec:app-aps-increments}

The macroscopic construction of Chapter~\ref{chap:macro-micro-measurement} uses not
only the position and localization directions, but also the momentum direction on the
localized phase-space manifold.  It is therefore useful to check directly that the
three coordinates
\[
(a,p,s)
\]
form mutually orthogonal stochastic directions and that an isotropic GUE-induced
increment gives independent Gaussian components in these coordinates.

We use the one-dimensional Gaussian phase-space surface
\[
\psi_{a,p,s}(x)
=
g_{a,\delta}(x)e^{\frac{ipx}{\hbar}},
\qquad
\delta=\sigma e^s,
\]
with tangent vectors
\[
T_a
=
\frac{x-a}{2\delta^2}\psi_{a,p,s},
\qquad
T_p
=
\frac{i(x-a)}{\hbar}\psi_{a,p,s},
\]
and
\[
T_s
=
\left[
-\frac12+
\frac{(x-a)^2}{2\delta^2}
\right]\psi_{a,p,s}.
\]
The calculation was performed on a uniform grid with
\[
x\in[-10,10],
\qquad
N=256,
\qquad
\sigma=0.75,
\qquad
\hbar=1,
\]
at the point
\[
a=0,
\qquad
p=0.8,
\qquad
s=0,
\]
using \(50{,}000\) independent samples.  At a fixed state, the transverse part of a
GUE-induced infinitesimal displacement is a circular complex Gaussian vector in the
horizontal tangent space.  The simulation samples this equivalent tangent law and
projects each displacement onto \(T_a,T_p,T_s\).

The numerical Gram matrix is
\[
G\simeq
\begin{pmatrix}
0.444444 & -4.6\times10^{-18} & -5.6\times10^{-17}\\
-4.6\times10^{-18} & 0.562500 & 3.7\times10^{-18}\\
-5.6\times10^{-17} & 3.7\times10^{-18} & 0.500000
\end{pmatrix}.
\]
Thus the calculation reproduces the analytical coefficients
\[
G_{aa}=\frac{1}{4\sigma^2},
\qquad
G_{pp}=\frac{\sigma^2}{\hbar^2},
\qquad
G_{ss}=\frac12,
\]
with vanishing mixed terms.  In particular, the ordinary phase-space metric of
Section~\ref{sec:localized-equivalence-classes} is recovered in the \((a,p)\) sector,
while the localization direction is orthogonal to both phase-space directions.

\begin{figure}[htbp]
\centering
\includegraphics[width=0.68\textwidth]{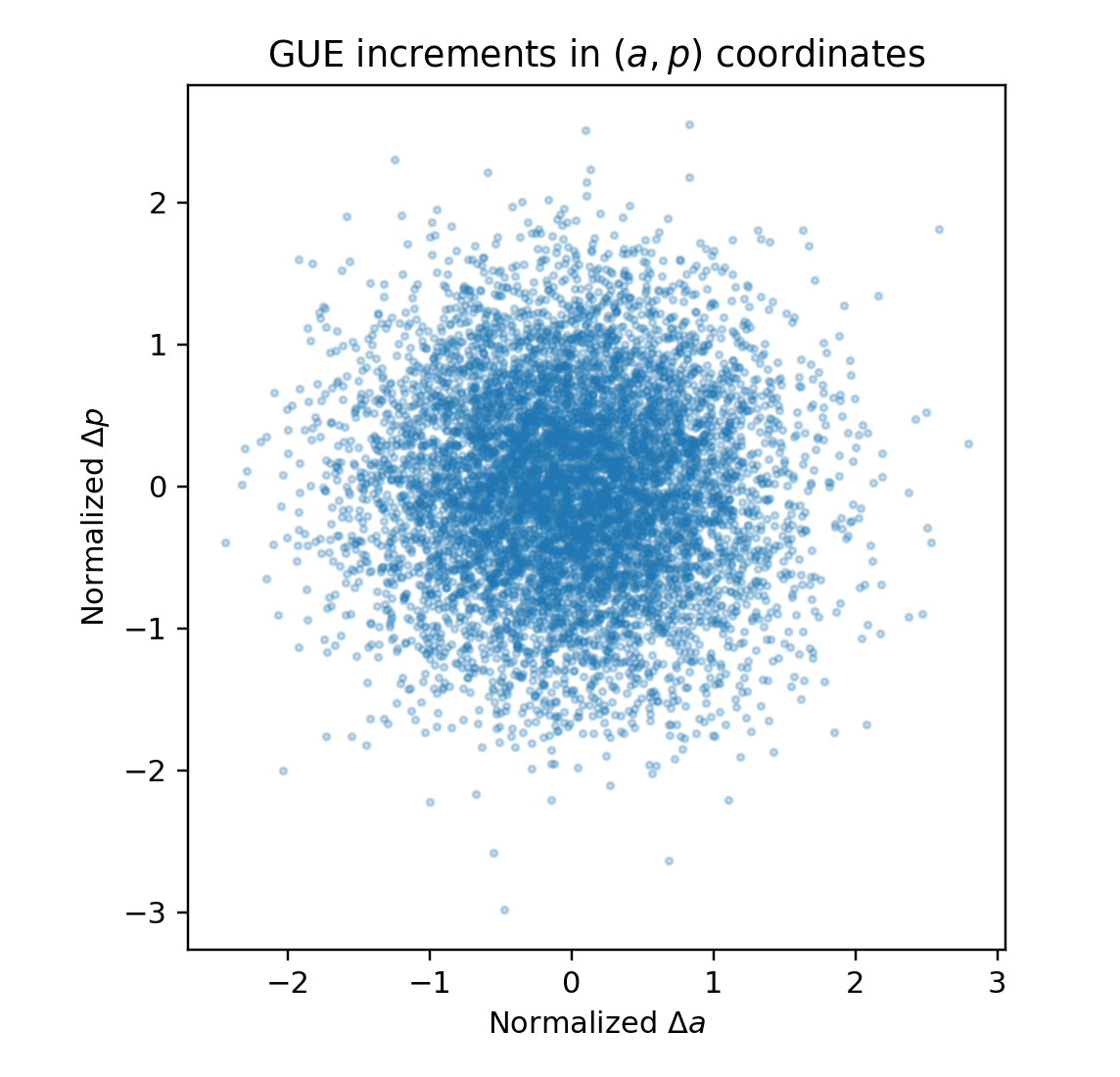}
\caption{Fubini--Study-normalized GUE-induced increments in the position and
momentum coordinates of the localized phase-space manifold.  The circular cloud,
together with Table~\ref{tab:app-aps-covariance}, confirms that the position,
momentum, and localization components are mutually uncorrelated after metric
normalization.}
\label{fig:app-aps-scatter}
\end{figure}

After multiplication by the square roots of the corresponding metric coefficients,
the three projected increments have the common GUE scale.  Dividing out this common
scale gives the covariance matrix shown in Table~\ref{tab:app-aps-covariance}.  The
largest absolute off-diagonal correlation is \(8.6\times10^{-3}\), consistent with
finite sampling.  Figure~\ref{fig:app-aps-scatter} shows the normalized \((a,p)\)
projection; the remaining pairwise projections give the same circular Gaussian
behavior.

\begin{table}[htbp]
\centering
\caption{Standardized covariance of the GUE-induced increments in the
\((a,p,s)\) coordinates.}
\label{tab:app-aps-covariance}
\begin{tabular}{c|rrr}
\hline
 & \(a\) & \(p\) & \(s\) \\
\hline
\(a\) & 1.0000 & -0.0027 & 0.0086 \\
\(p\) & -0.0027 & 1.0000 & 0.0047 \\
\(s\) & 0.0086 & 0.0047 & 1.0000 \\
\hline
\end{tabular}
\end{table}

This additional test supplies the numerical counterpart of the decomposition used in
Section~\ref{sec:newtonian-motion-under-environmental-monitoring}.  For an isotropic
Gaussian projective increment, the mutually orthogonal \(a\)-, \(p\)-, and
\(s\)-directions give independent Gaussian stochastic components.  The deterministic
Schr\"odinger tangent drift may therefore be combined with separate position,
momentum, and localization fluctuations without introducing a preferred direction in
the localized phase-space geometry.

\section{Binary reduction, Born frequencies, and effective absorption}
\label{sec:app-binary-born}

This section tests the bounded two-outcome construction of
Section~\ref{sec:the-born-rule-as-a-hitting-probability}.  It also provides a numerical
illustration of the equivalence-class stabilization discussed in
Section~\ref{sec:the-quantum-zeno-effect}.

\subsection{Unbiased reduction coordinate}

For a two-outcome state
\[
\psi=\alpha|1\rangle+\beta|0\rangle,
\qquad
|\alpha|^2+|\beta|^2=1,
\]
the affine reduction coordinate is
\[
p=|\alpha|^2\in[0,1].
\]
The numerical walk uses a lattice with \(100\) subintervals.  At each step the lattice
index changes by \(\pm1\) with equal probability until it reaches \(0\) or \(100\).
For each of nine values of \(p_0\), \(50{,}000\) independent paths were simulated.
The upper-endpoint frequency is therefore a direct Monte Carlo estimate of the
hitting probability.

\begin{figure}[htbp]
\centering
\includegraphics[width=0.72\textwidth]{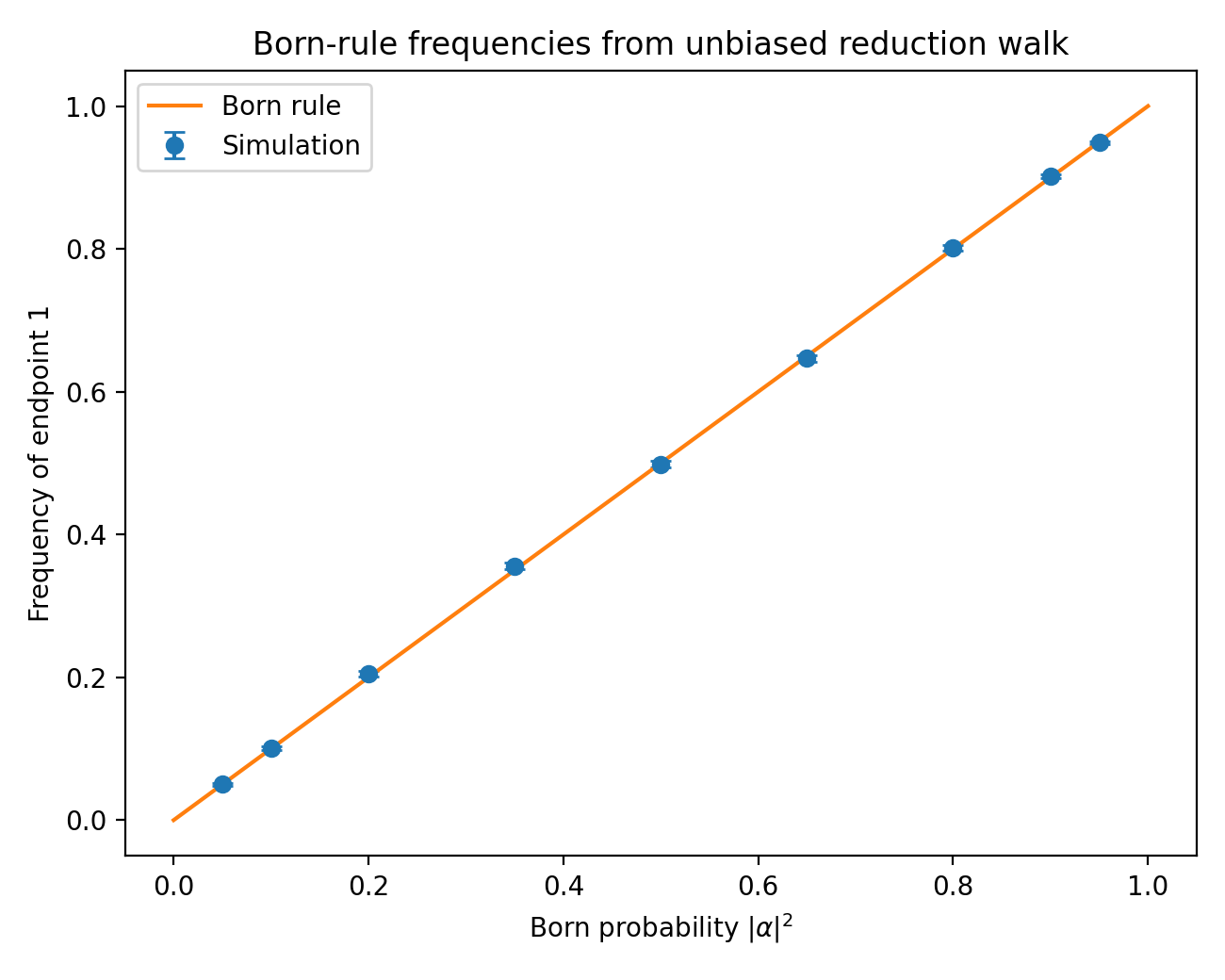}
\caption{Simulated upper-endpoint frequencies for the unbiased reduction-coordinate
walk.  The diagonal is the Born prediction \(P_1=p_0=|\alpha|^2\).}
\label{fig:app-binary-born}
\end{figure}

Table~\ref{tab:app-binary-born} gives the complete result, and
Figure~\ref{fig:app-binary-born} compares the frequencies with the diagonal Born
prediction.  The largest absolute discrepancy in the nine cases is \(0.0060\),
consistent with finite sampling.

\begin{table}[htbp]
\centering
\caption{Born frequencies from the unbiased reduction-coordinate walk.}
\label{tab:app-binary-born}
\begin{tabular}{ccccc}
\hline
\(p_0\) & Paths & Upper frequency & Std. error & Abs. error \\
\hline
0.05 & 50,000 & 0.0501 & 0.0010 & 0.0001 \\
0.10 & 50,000 & 0.1011 & 0.0013 & 0.0011 \\
0.20 & 50,000 & 0.2052 & 0.0018 & 0.0052 \\
0.35 & 50,000 & 0.3560 & 0.0021 & 0.0060 \\
0.50 & 50,000 & 0.4985 & 0.0022 & 0.0015 \\
0.65 & 50,000 & 0.6465 & 0.0021 & 0.0035 \\
0.80 & 50,000 & 0.8014 & 0.0018 & 0.0014 \\
0.90 & 50,000 & 0.9014 & 0.0013 & 0.0014 \\
0.95 & 50,000 & 0.9490 & 0.0010 & 0.0010 \\
\hline
\end{tabular}
\end{table}

\begin{figure}[htbp]
\centering
\includegraphics[width=0.78\textwidth]{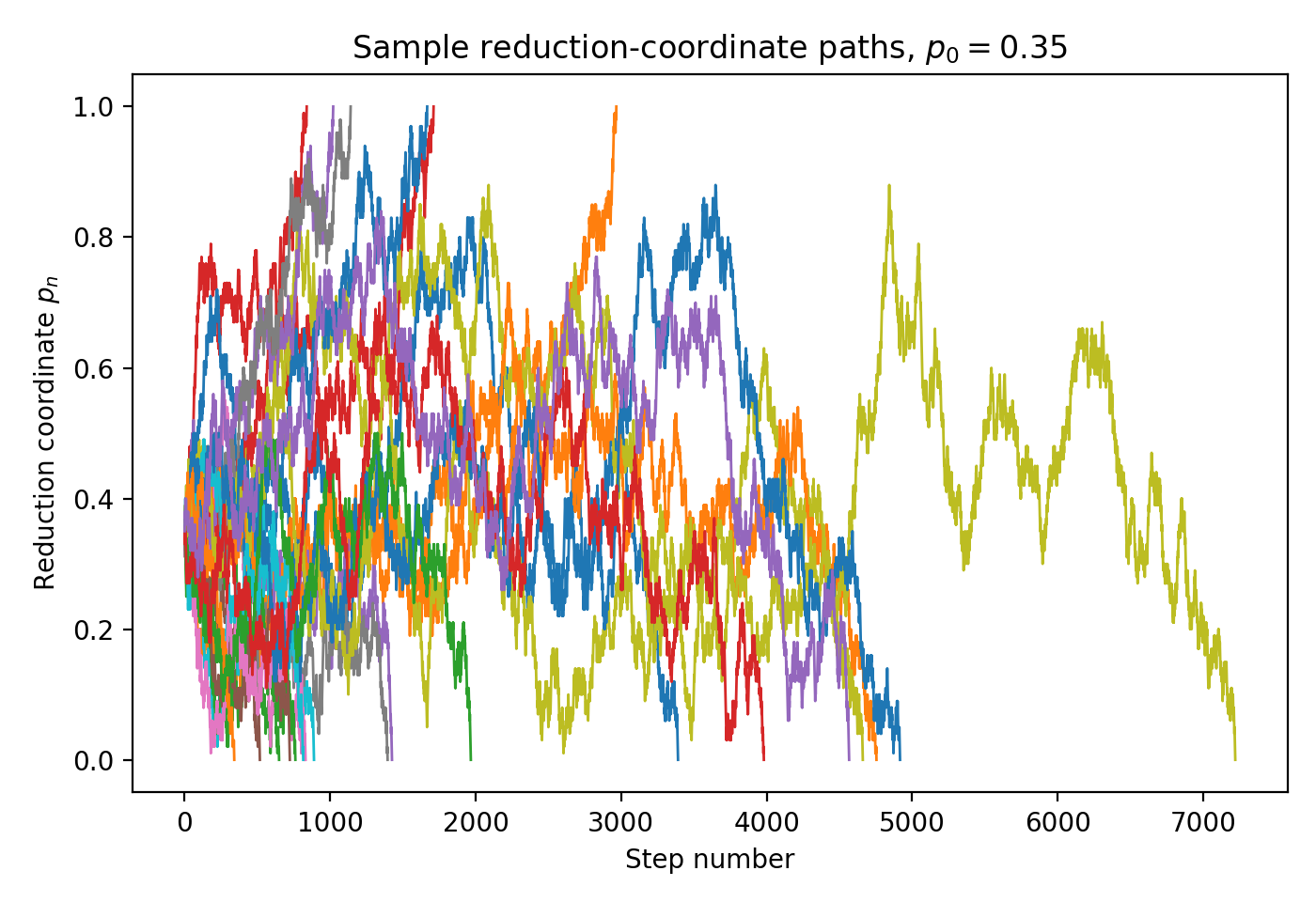}
\caption{Representative reduction-coordinate paths for \(p_0=0.35\).  Each path
fluctuates until it enters one of the two outcome classes.}
\label{fig:app-binary-paths}
\end{figure}

Representative paths for \(p_0=0.35\) are shown in
Figure~\ref{fig:app-binary-paths}.  They emphasize that the frequency law is a property
of an ensemble of unbiased paths; no deterministic bias toward the more probable
endpoint is introduced.

\subsection{Why the detector coordinate appears absorbing}

\begin{figure}[htbp]
\centering
\includegraphics[width=0.74\textwidth]{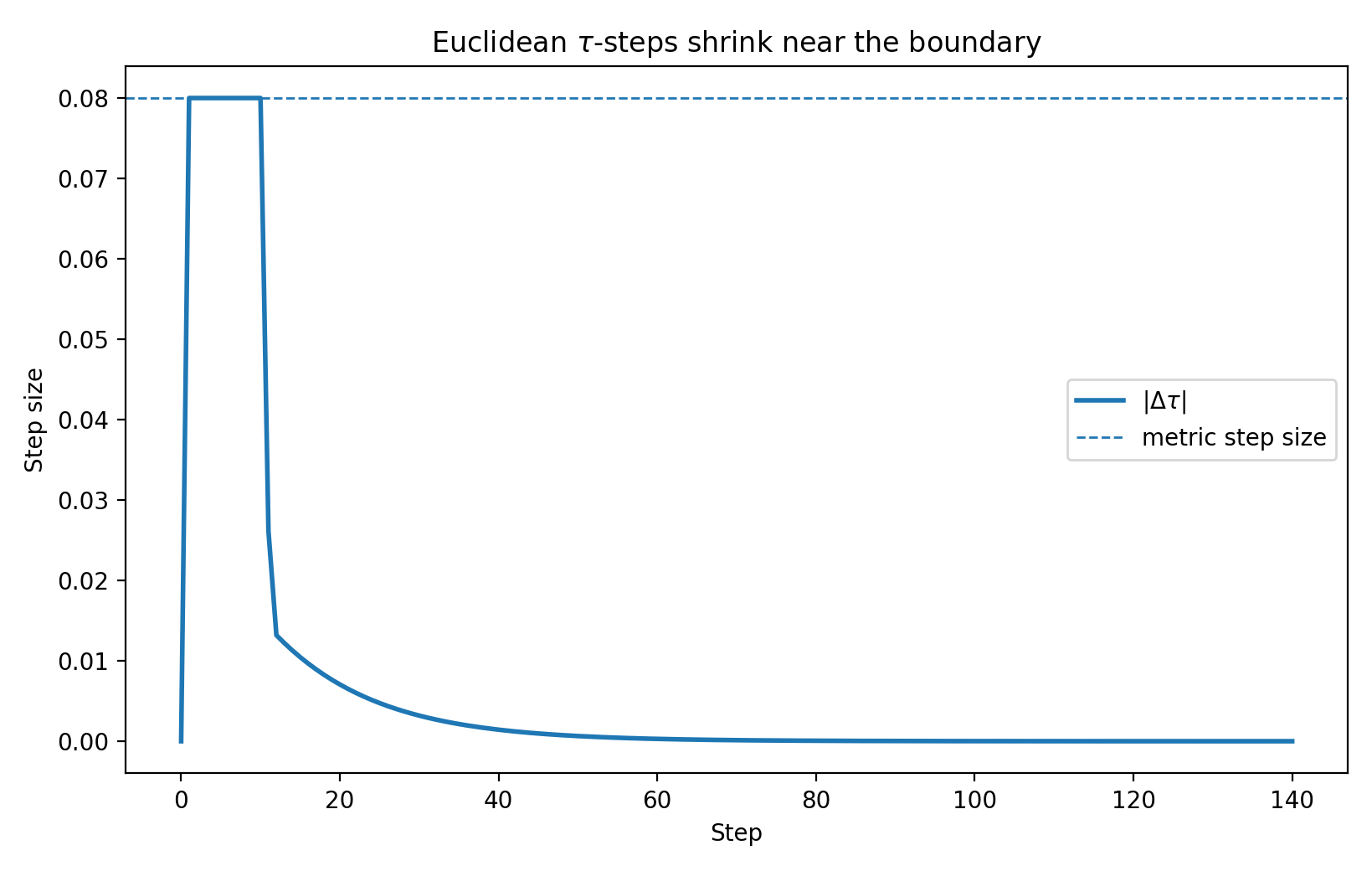}
\caption{Euclidean \(\tau\)-step for equal intrinsic steps of a Gaussian state
constrained to remain inside a detector interval.  The detector coordinate freezes as
the endpoint class is approached.}
\label{fig:app-boundary-step}
\end{figure}

The absorbing-boundary description is a reduced coordinate description of the same
state-space motion.  In a finite interval \([a,b]\), a Gaussian packet constrained to
remain inside the represented detector region must narrow near either endpoint.  If
\(\delta=\sigma e^s\), then near \(b\),
\[
\delta(\tau)\lesssim b-\tau,
\qquad
s(\tau)\sim\ln\frac{b-\tau}{\sigma},
\qquad
\frac{ds}{d\tau}\sim-\frac1{b-\tau}.
\]
For a fixed intrinsic step \(\Delta\ell\) in the \((\tau,s)\) metric,
\[
|\Delta\tau|
\simeq
\frac{\Delta\ell}{\sqrt{1+(ds/d\tau)^2}}
\longrightarrow0
\qquad
(\tau\to b).
\]
Figure~\ref{fig:app-boundary-step} shows this shrinking Euclidean step size.  The
state-space motion need not cease; rather, increasingly more of the motion is resolved
as narrowing of the packet instead of displacement of its detector coordinate.

This gives the numerical content of the equivalence-class form of the Zeno effect used
in Chapter~\ref{chap:paradoxes-arrow-time}: stability means stability of the recorded
finite-resolution class, not freezing of the exact ray.

\section{Finite-resolution double-slit records}
\label{sec:app-double-slit}

The double-slit calculation checks the detector-path rule of
Section~\ref{sec:detector-class-weights-and-interference} and its application in
Section~\ref{sec:single-particle-case-the-double-slit-experiment}.  The final screen is
divided into \(80\) finite-resolution bins \(I_j\) on \([-30,30]\).  For a state
\(\psi_T\) arriving at the screen, the weight of the \(j\)-th record is
\[
W_j(\psi_T)
=
\|P_{I_j}\psi_T\|^2
=
\int_{I_j}|\psi_T(x)|^2\,dx.
\]
The projected \({\bf RM}\) record process is sampled from these calibrated
finite-resolution endpoint weights.

If no slit record has formed, \(\psi_T\) is the coherent two-slit state and the weights
contain the interference cross terms.  If a slit record has already formed, each run
first belongs to the left- or right-slit class and then reaches the final screen in the
corresponding conditional single-slit state.  The final screen rule is unchanged; only
the state reaching the screen is different.

\begin{figure}[htbp]
\centering
\includegraphics[width=0.82\textwidth]{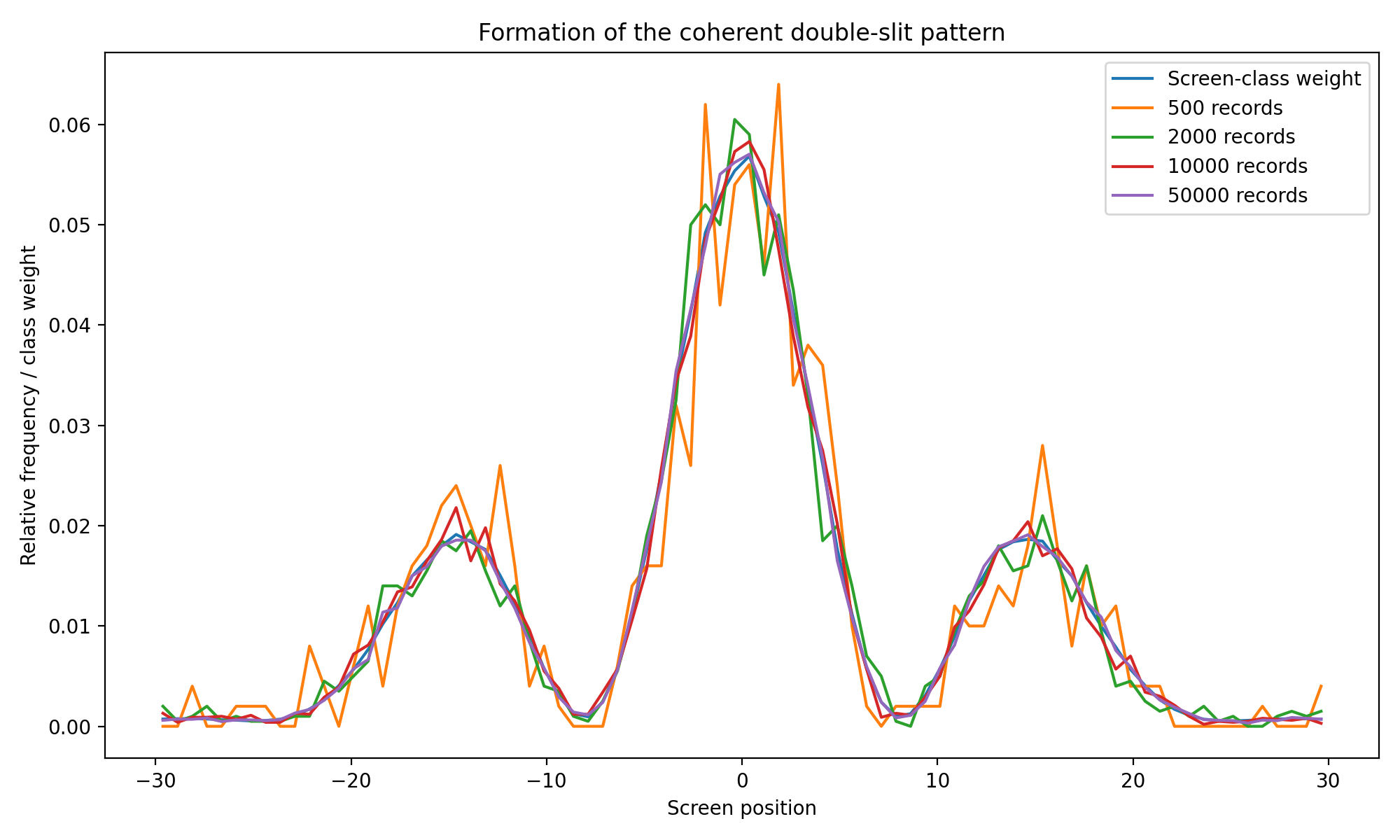}
\caption{Accumulation of coherent two-slit screen records.  The empirical frequencies
converge to the finite-resolution screen-class weights containing the interference
pattern.}
\label{fig:app-double-slit-coherent}
\end{figure}

\begin{figure}[htbp]
\centering
\includegraphics[width=0.82\textwidth]{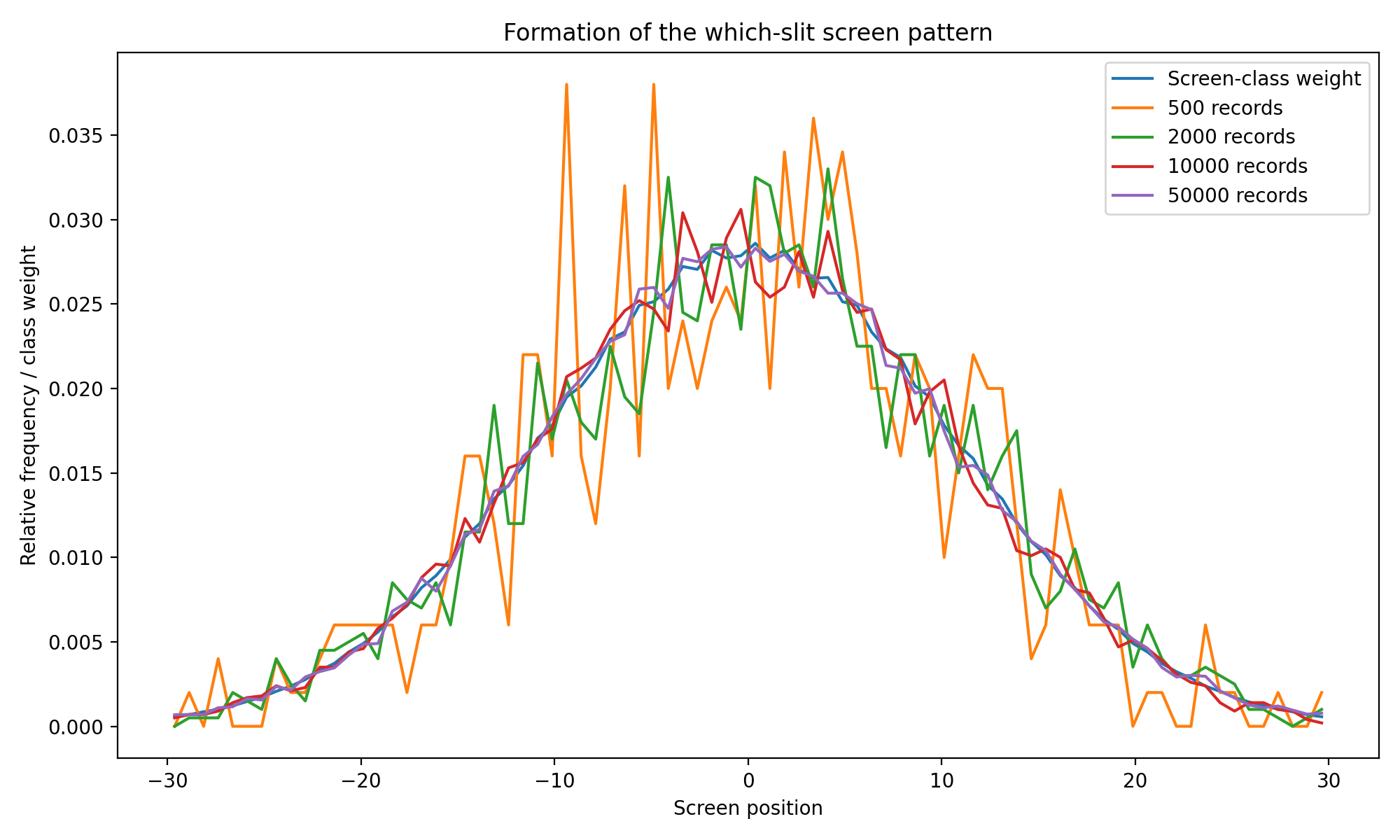}
\caption{Screen records when a which-slit record has formed first.  The accumulated
frequencies converge to the noninterference distribution obtained from the conditional
single-slit alternatives.}
\label{fig:app-double-slit-which}
\end{figure}

Figures~\ref{fig:app-double-slit-coherent} and
\ref{fig:app-double-slit-which} show the accumulation of records after
\(500\), \(2000\), \(10{,}000\), and \(50{,}000\) trials.  The convergence errors are
given in Table~\ref{tab:app-double-slit-errors}.

\begin{table}[htbp]
\centering
\caption{Convergence of cumulative screen-record frequencies to the screen-class
weights.}
\label{tab:app-double-slit-errors}
\begin{tabular}{lrrr}
\hline
Case & Records & \(L^1\)-error & Maximum bin error \\
\hline
Coherent & 500    & 0.2608 & 0.0153 \\
Coherent & 2,000  & 0.1199 & 0.0087 \\
Coherent & 10,000 & 0.0627 & 0.0027 \\
Coherent & 50,000 & 0.0279 & 0.0022 \\
Which-slit & 500    & 0.2822 & 0.0185 \\
Which-slit & 2,000  & 0.1413 & 0.0066 \\
Which-slit & 10,000 & 0.0596 & 0.0032 \\
Which-slit & 50,000 & 0.0275 & 0.0013 \\
\hline
\end{tabular}
\end{table}

The calculation therefore supports the point made in Chapters~\ref{chap:measurement-rm-walk}
and~\ref{chap:correlations-locality}: interference or its absence is determined by the
state presented to the final detector, not by changing the detector-recording rule.

\section{Stroboscopic Newtonian records}
\label{sec:app-newtonian-records}

This section tests the macroscopic regime discussed in
Sections~\ref{sec:components-of-schrodinger-velocity-along-m33-sigma} and
\ref{sec:newtonian-motion-under-environmental-monitoring}.  The deterministic tangent
motion is Newtonian, while the accumulated \({\bf RM}\) contribution is resolved into
a tangential displacement and a localization coordinate.  A classical record is counted
only when the state lies in the localized sector.

\subsection{Conditioned record process}

The numerical model uses a dimensionless harmonic oscillator,
\[
M=1,
\qquad
\omega=1,
\qquad
 a_N(t)=a_0\cos t+p_0\sin t,
\]
with \(a_0=1\), \(p_0=0.35\), and candidate recording times
\[
t_k=k\Delta t,
\qquad
\Delta t=0.05,
\qquad
0\le t\le20.
\]
The detector resolution is \(\sigma=0.10\).  The tangential stochastic displacement is
\[
\eta_k\sim N(0,D_a\Delta t),
\qquad
D_a=0.002,
\]
so that
\[
\sqrt{D_a\Delta t}=0.01,
\qquad
\frac{D_a\Delta t}{\sigma^2}=0.01.
\]
The localization coordinate is represented by
\[
s_{k+1}=s_k+\xi_k,
\qquad
\xi_k\sim N(0,2D_s\Delta t),
\qquad
D_s=0.25.
\]
A record is retained only for \(s<0\); after a record the next interval begins from
\(s_{\rm rec}=-0.5\).  Five thousand independent trajectories were generated.

\begin{figure}[htbp]
\centering
\includegraphics[width=0.82\textwidth]{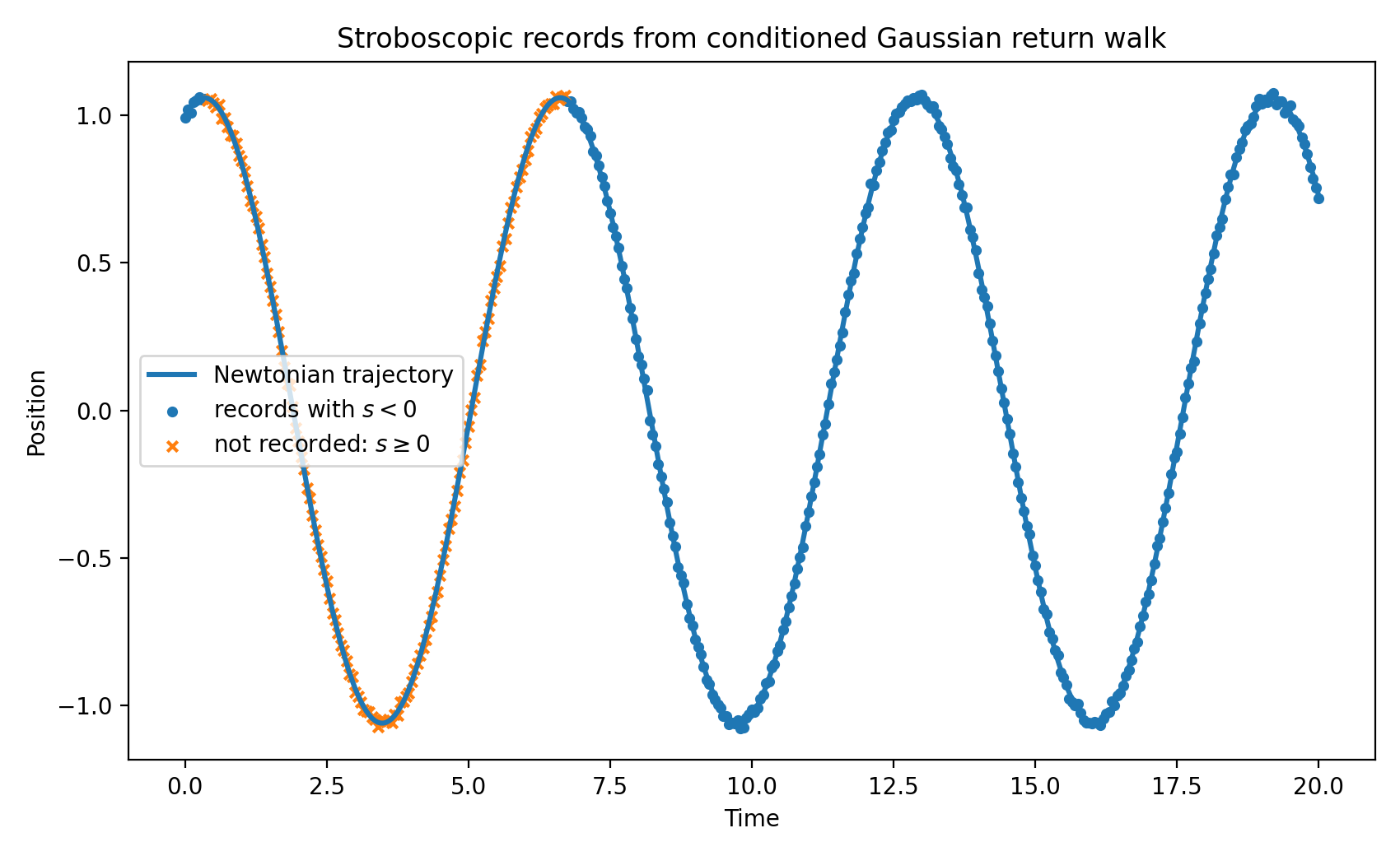}
\caption{Conditioned stroboscopic position records.  The solid curve is the Newtonian
trajectory; only candidate times with \(s<0\) are counted as classical records.}
\label{fig:app-newtonian-records}
\end{figure}

Figure~\ref{fig:app-newtonian-records} shows one representative record sequence.  The
overall fraction of candidate points in the localized sector was \(0.9849\).  On the
recorded subset the RMS deviation from the Newtonian trajectory was \(0.010004\), in
agreement with the imposed scale \(\sqrt{D_a\Delta t}=0.01\).

\begin{table}[htbp]
\centering
\caption{Main parameters and record statistics for the stroboscopic Newtonian test.}
\label{tab:app-newtonian-parameters}
\begin{tabular}{lc}
\hline
Quantity & Value \\
\hline
Candidate interval \(\Delta t\) & 0.05 \\
Number of candidate times & 401 \\
Number of trajectories & 5000 \\
Resolution \(\sigma\) & 0.10 \\
Position diffusion \(D_a\) & 0.002 \\
\(\sqrt{D_a\Delta t}\) & 0.0100 \\
\(D_a\Delta t/\sigma^2\) & 0.01 \\
Localization diffusion \(D_s\) & 0.25 \\
Renewal value \(s_{\rm rec}\) & \(-0.5\) \\
Fraction with \(s<0\) & 0.9849 \\
RMS deviation on recorded points & 0.010004 \\
\hline
\end{tabular}
\end{table}

\subsection{Parameter sweep}

\begin{figure}[htbp]
\centering
\includegraphics[width=0.78\textwidth]{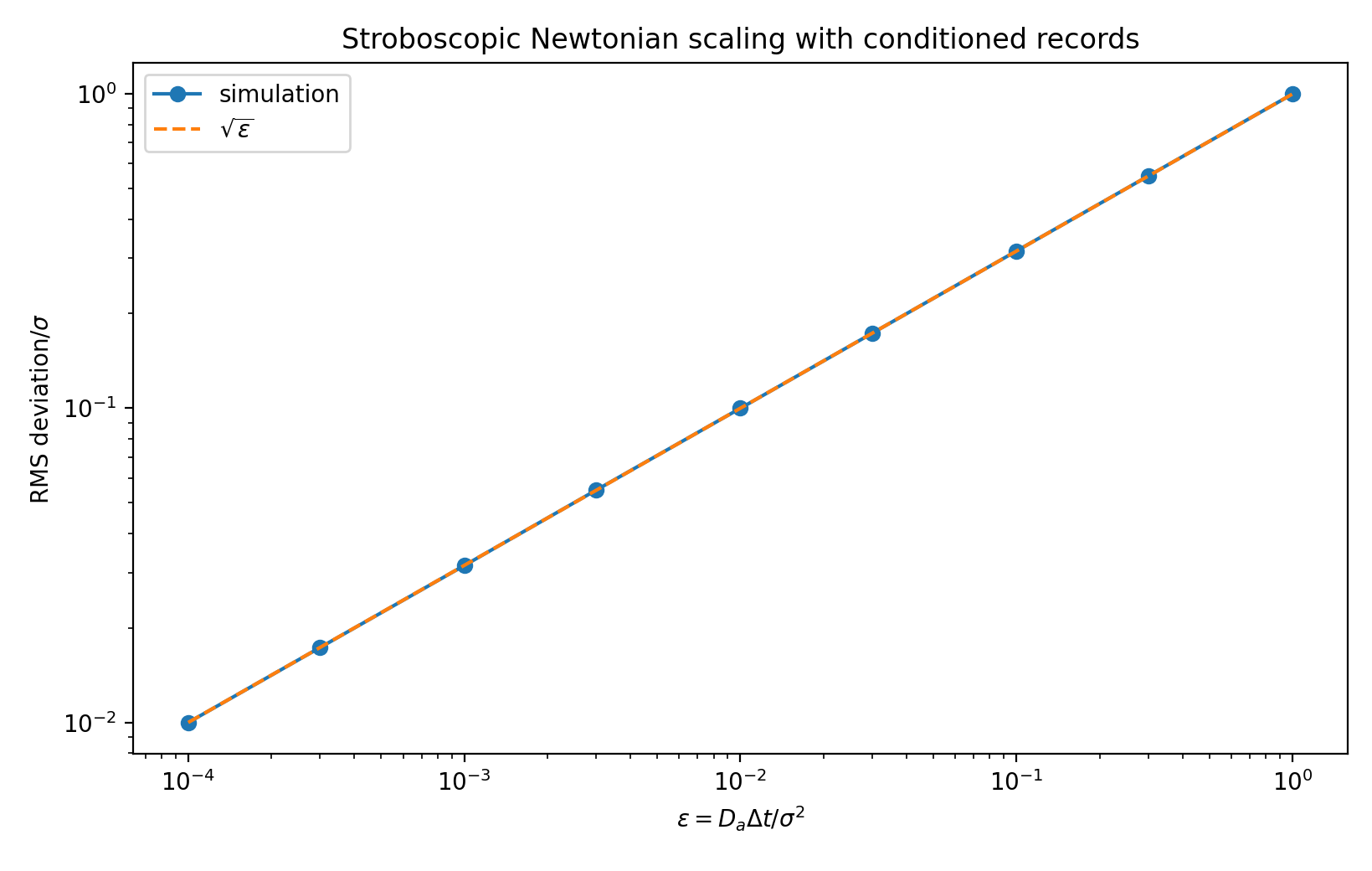}
\caption{Normalized RMS deviation of localized records from the Newtonian trajectory.
The simulation follows the predicted square-root law over four decades in
\(\varepsilon\).}
\label{fig:app-epsilon-sweep}
\end{figure}

The dimensionless control parameter is
\[
\varepsilon
=
\frac{D_a\Delta t}{\sigma^2}.
\]
Since the recorded tangential displacement has variance \(D_a\Delta t\), the predicted
scaling is
\[
\frac{\mathrm{RMS}}{\sigma}\simeq\sqrt{\varepsilon}.
\]
A sweep over four decades in \(\varepsilon\), using \(2500\) trajectories per value,
gives the values in Table~\ref{tab:app-epsilon-sweep} and the curve in
Figure~\ref{fig:app-epsilon-sweep}.  The localized-record fraction remained about
\(0.983\), as expected because \(\varepsilon\) controls the tangential spread rather
than the independent return process in \(s\).

\begin{table}[htbp]
\centering
\caption{Scaling of recorded-position deviations with
\(\varepsilon=D_a\Delta t/\sigma^2\).}
\label{tab:app-epsilon-sweep}
\begin{tabular}{ccc}
\hline
\(\varepsilon\) & Simulated RMS/\(\sigma\) & \(\sqrt{\varepsilon}\) \\
\hline
\(10^{-4}\) & 0.009995 & 0.010000 \\
\(10^{-3}\) & 0.031584 & 0.031623 \\
\(10^{-2}\) & 0.099952 & 0.100000 \\
\(10^{-1}\) & 0.316389 & 0.316228 \\
1 & 0.999496 & 1.000000 \\
\hline
\end{tabular}
\end{table}

The numerical result supports the macroscopic criterion of Chapter~\ref{chap:macro-micro-measurement}:
Newtonian motion is a sequence of finite-resolution localized records when the
stochastic tangential displacement between returns is small on the detector scale and
returns to the localized sector are sufficiently frequent.

\section{Isotropy and homogeneity in tensor-product state space}
\label{sec:app-tensor-isotropy}

Before applying the Brownian calibration to two-particle and particle-device systems,
the same projective symmetry must hold in the full tensor-product Hilbert space.  This
is the numerical premise required by Section~\ref{sec:the-rm-walk-on-tensor-product-state-space}:
the \({\bf RM}\) law cannot be special to product states on the classical product
manifold if it is to act on arbitrary superpositions and entangled states.

The test uses
\[
\mathcal H=\mathbb C^{10}\otimes\mathbb C^{10},
\qquad
\dim_{\mathbb C}\mathcal H=100,
\]
and four representative states: a product basis state, a product superposition, a
two-branch entangled state, and a random entangled state.  Twenty thousand independent
GUE increments were sampled at each state.

\begin{figure}[htbp]
\centering
\includegraphics[width=0.78\textwidth]{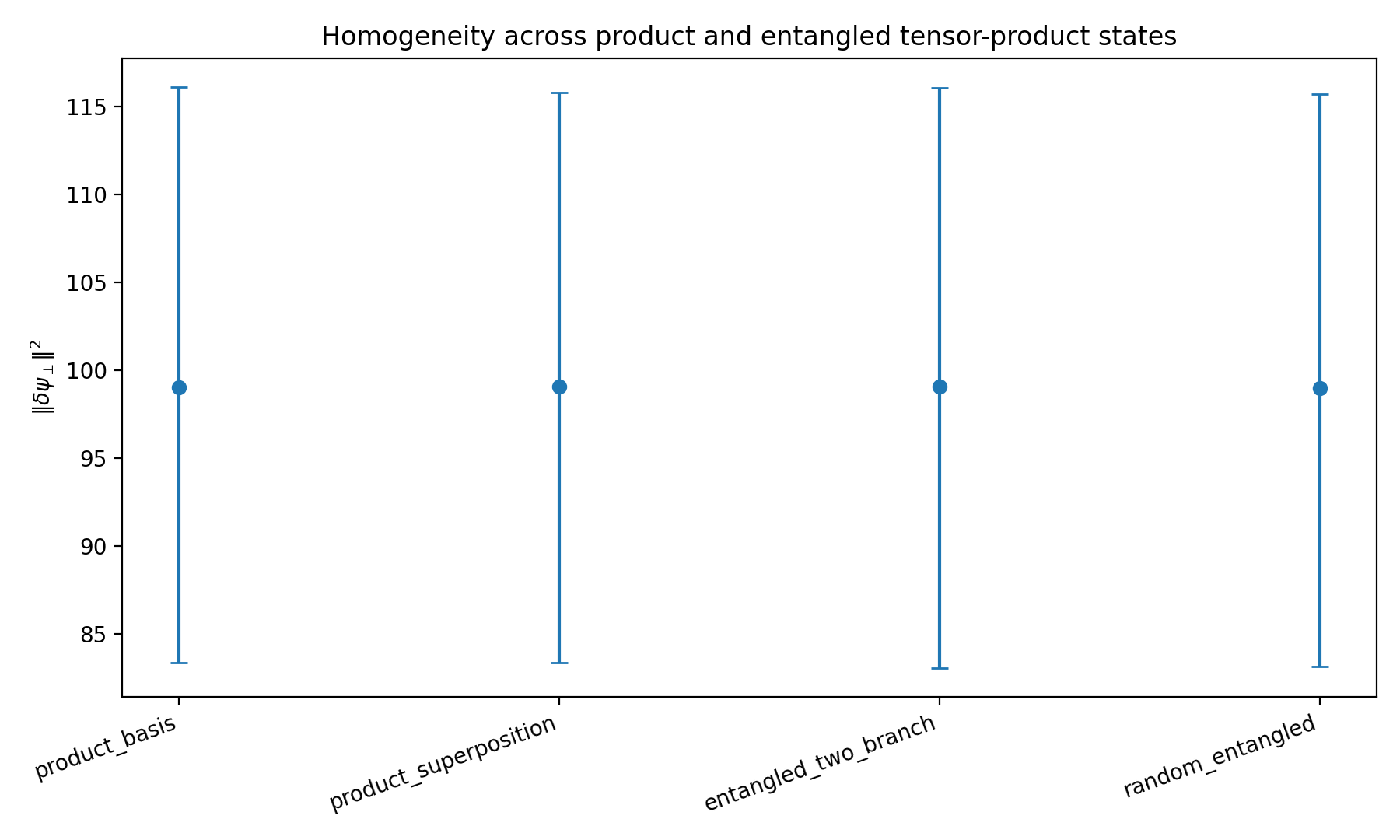}
\caption{Projective step-size distributions at product and entangled states in the
tensor-product Hilbert space.}
\label{fig:app-tensor-homogeneity}
\end{figure}

Homogeneity was tested by the distribution of \(\|\delta\psi_\perp\|^2\).  The mean
values, shown in Table~\ref{tab:app-tensor-homogeneity}, agree at the \(10^{-3}\) relative
level, and two-sample KS tests against the product basis state gave \(p\)-values
\(0.494\), \(0.533\), and \(0.797\).  Figure~\ref{fig:app-tensor-homogeneity} shows the
corresponding quantile ranges.

\begin{table}[htbp]
\centering
\caption{Homogeneity of the GUE projective step size in
\(\mathbb C^{10}\otimes\mathbb C^{10}\).}
\label{tab:app-tensor-homogeneity}
\begin{tabular}{lccc}
\hline
State & Mean \(\|\delta\psi_\perp\|^2\) & 5\% quantile & 95\% quantile \\
\hline
Product basis & 99.054 & 83.385 & 116.109 \\
Product superposition & 99.094 & 83.380 & 115.780 \\
Two-branch entangled & 99.083 & 83.089 & 116.052 \\
Random entangled & 98.978 & 83.166 & 115.696 \\
\hline
\end{tabular}
\end{table}

\begin{figure}[htbp]
\centering
\includegraphics[width=0.80\textwidth]{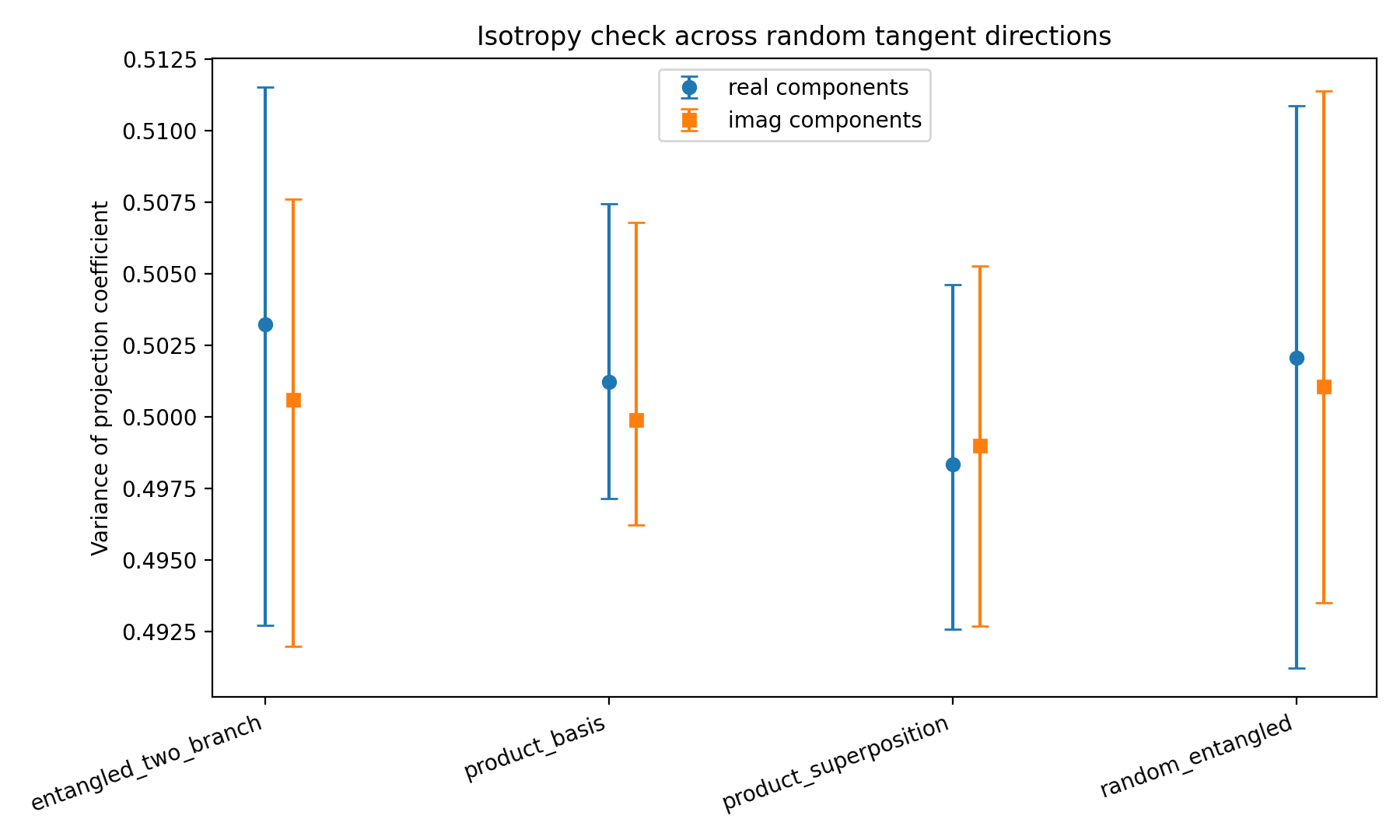}
\caption{Isotropy test in tensor-product state space.  Projection variances are
independent, within sampling fluctuations, of both the tangent direction and whether
the base state is product or entangled.}
\label{fig:app-tensor-isotropy}
\end{figure}

Isotropy was tested separately by projecting the increment onto randomly chosen tangent
directions.  The real and imaginary component variances are close to \(1/2\) for all
four base states, as summarized in Table~\ref{tab:app-tensor-isotropy};
Figure~\ref{fig:app-tensor-isotropy} shows the directional data.

\begin{table}[htbp]
\centering
\caption{Mean projection variance over random tangent directions.}
\label{tab:app-tensor-isotropy}
\begin{tabular}{lcc}
\hline
State & Real-part variance & Imaginary-part variance \\
\hline
Product basis & 0.5012 & 0.4999 \\
Product superposition & 0.4983 & 0.4990 \\
Two-branch entangled & 0.5032 & 0.5006 \\
Random entangled & 0.5021 & 0.5011 \\
\hline
\end{tabular}
\end{table}

This test is logically prior to the two-particle Born calculation below.  It supports the
extension made in Chapter~\ref{chap:two-particle-measurement} from the Brownian
calibration on the product classical manifold to arbitrary joint states in the ambient
tensor-product projective space.

\section{Tensor-product Brownian restriction and the device scaling limit}
\label{sec:app-tensor-brownian}

The product manifold constructed in Section~\ref{sec:systems-of-particles} has a product
Fubini--Study metric.  The present calculation checks the corresponding stochastic
restriction and the metric scaling used later for a device coordinate.

\begin{figure}[htbp]
\centering
\includegraphics[width=0.66\textwidth]{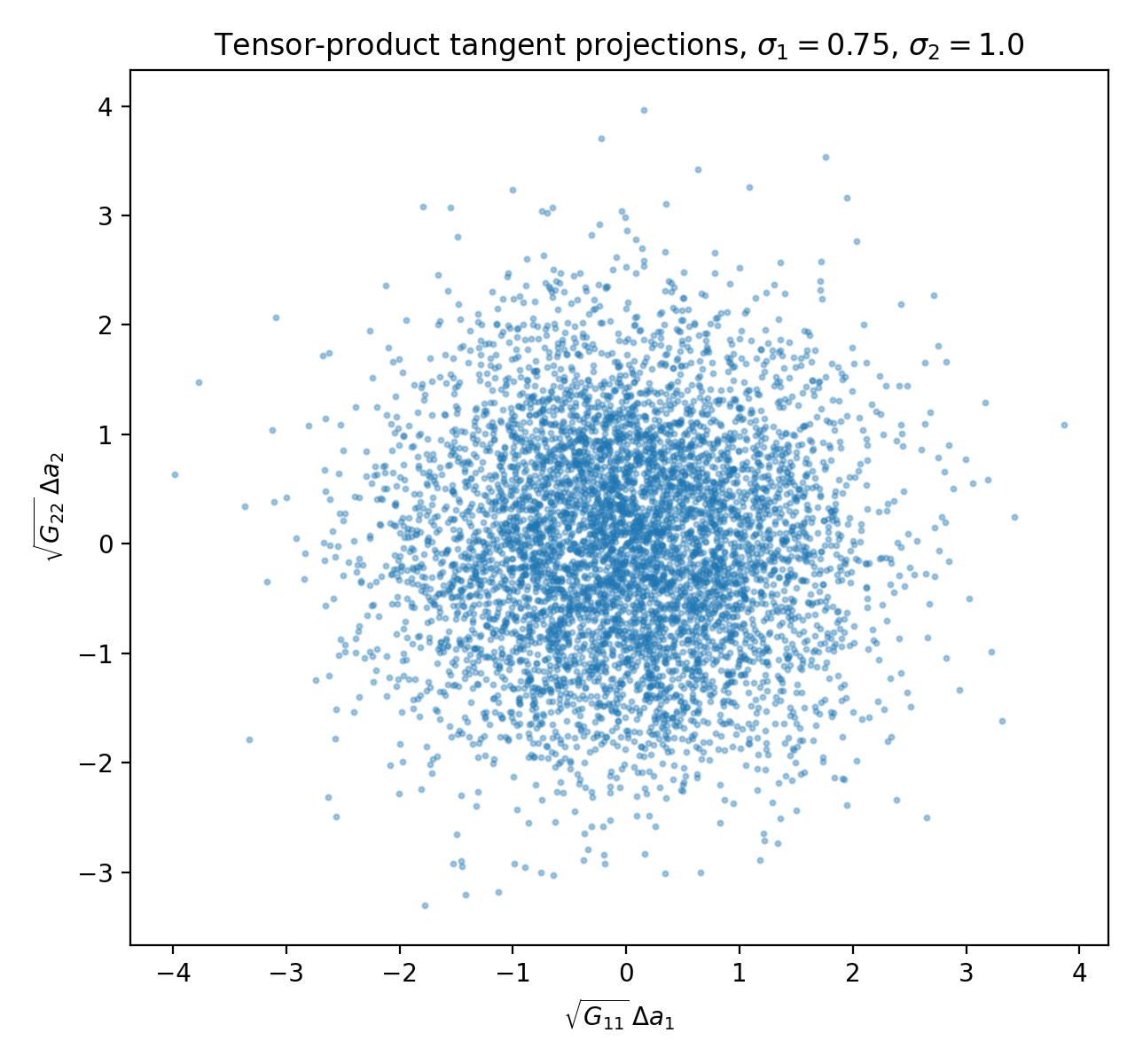}
\caption{Fubini--Study-normalized increments on the product Gaussian manifold.  The
two projected position coordinates have independent Gaussian increments.}
\label{fig:app-tensor-brownian}
\end{figure}

For the product Gaussian state
\[
\Psi_{a,b}(x,y)=g_{a,\sigma_1}(x)g_{b,\sigma_2}(y),
\]
the translation tangents are
\[
T_1=\partial_a g_{a,\sigma_1}\otimes g_{b,\sigma_2},
\qquad
T_2=g_{a,\sigma_1}\otimes\partial_b g_{b,\sigma_2},
\]
with metric
\[
G_{11}=\frac1{4\sigma_1^2},
\qquad
G_{22}=\frac1{4\sigma_2^2},
\qquad
G_{12}=0.
\]
The Fubini--Study-normalized coordinate increments are therefore
\[
z_1=\sqrt{G_{11}}\,\Delta a_1,
\qquad
z_2=\sqrt{G_{22}}\,\Delta a_2.
\]
For \(\sigma_1=0.75\), \(\sigma_2=1.0\), and \(50{,}000\) samples, the measured
standard deviations were
\[
\operatorname{Std}(z_1)=1.0001,
\qquad
\operatorname{Std}(z_2)=0.9962,
\]
with correlation \(-0.0012\).  The circular scatter in
Figure~\ref{fig:app-tensor-brownian} therefore confirms independent Brownian
components in the two orthogonal classical coordinates.

\begin{figure}[htbp]
\centering
\includegraphics[width=0.76\textwidth]{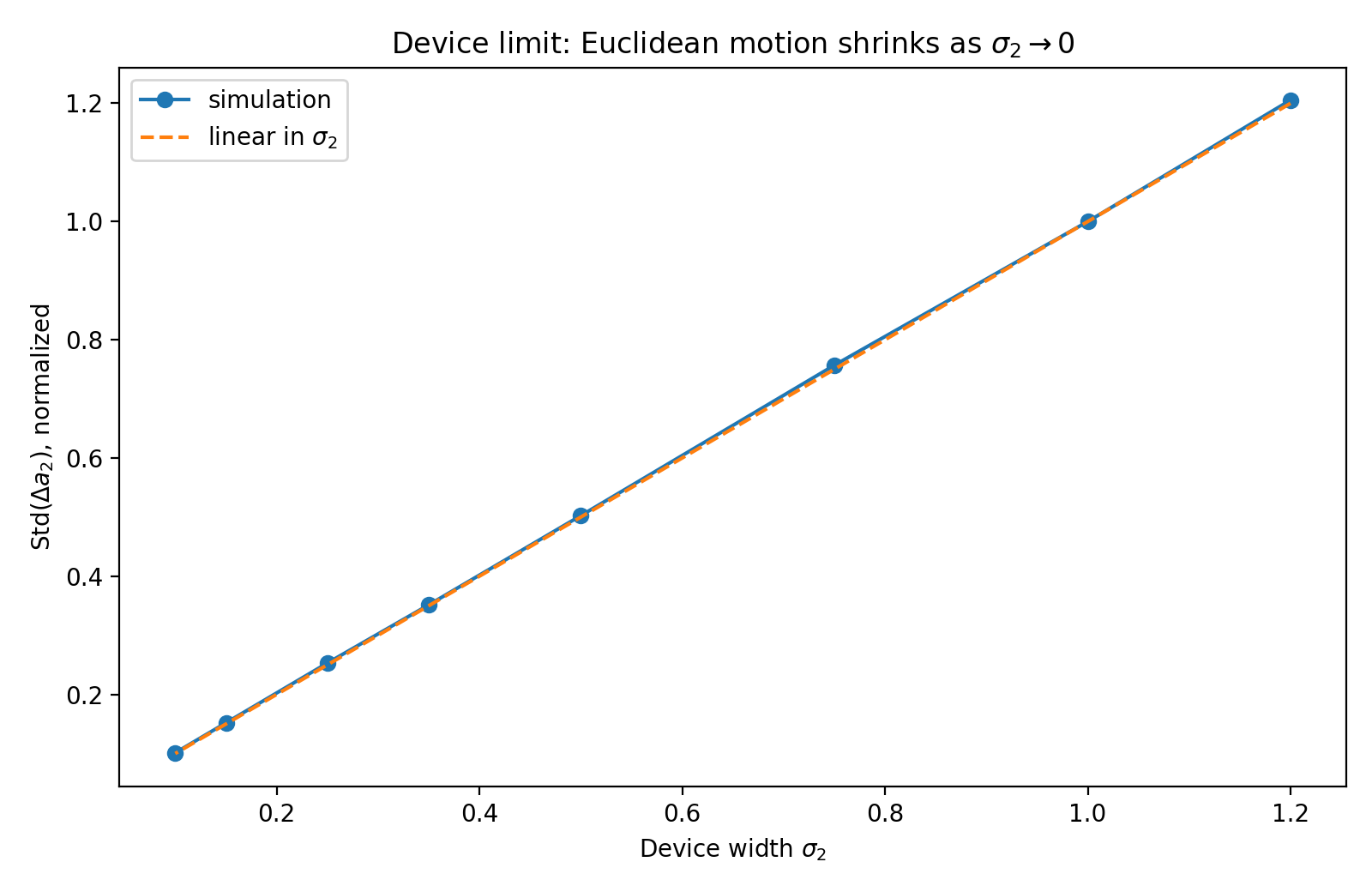}
\caption{Metric origin of the device scaling limit.  The Euclidean standard deviation
of the second projected coordinate is proportional to its localization width
\(\sigma_2\), while the Fubini--Study-normalized step remains of order one.}
\label{fig:app-device-scaling}
\end{figure}

In Euclidean coordinates a unit normalized step satisfies
\[
\operatorname{Std}(\Delta a_i)\propto\sigma_i.
\]
Keeping \(\sigma_1=0.75\) fixed and decreasing \(\sigma_2\) therefore produces the
scaling shown in Figure~\ref{fig:app-device-scaling}.  The normalized standard
deviation follows \(\sigma_2\) to within about one percent over the tested range
\(0.10\le\sigma_2\le1.20\).

This is not yet a complete macroscopic device simulation.  It isolates the geometric
scaling used in Section~\ref{sec:a-particle-and-a-macroscopic-measuring-device}: a
sharply localized pointer coordinate can have a very small Euclidean displacement even
when the ambient projective step has the same normalized stochastic scale as that of
the microscopic particle.

\section{Two-particle Born weights and conditional records}
\label{sec:app-two-particle-born}

The next calculation tests the measurement rule of
Section~\ref{sec:measurement-of-one-particle-in-a-two-particle-state}.  The preceding
tensor-product isotropy test is essential here: it is what allows the Brownian
calibration obtained on the product Gaussian sector to be used for an entangled state
in the full tensor-product projective space.

For a detector cell \(I_j\) in the first coordinate, the detector acts as
\[
P_{I_j}\otimes I,
\]
and the corresponding outcome weight is
\[
W_j
=
\|(P_{I_j}\otimes I)\Psi\|^2
=
\int_{I_j}\int |\Psi(x,y)|^2\,dy\,dx.
\]
After the record \(I_j\), the conditional joint state is
\[
\Psi_j
=
\frac{(P_{I_j}\otimes I)\Psi}
{\|(P_{I_j}\otimes I)\Psi\|}.
\]

\subsection{Product-state factorization}

For a product state \(\Psi(x,y)=\psi_1(x)\psi_2(y)\), the numerical calculation used
Gaussian factors \(g_{-1,0.8}(x)\) and \(g_{1.2,1.1}(y)\) on \([-6,6]^2\), with eight
detector bins in each coordinate.  Direct two-dimensional bin weights were compared
with the products of the one-dimensional marginal weights.  The maximum absolute
difference was
\[
2.78\times10^{-17},
\]
and the total \(L^1\) difference was
\[
8.62\times10^{-17}.
\]
Thus the detector weights reduce numerically to the expected product law when the
state is unentangled.

\subsection{Entangled two-branch state}

\begin{figure}[htbp]
\centering
\includegraphics[width=0.68\textwidth]{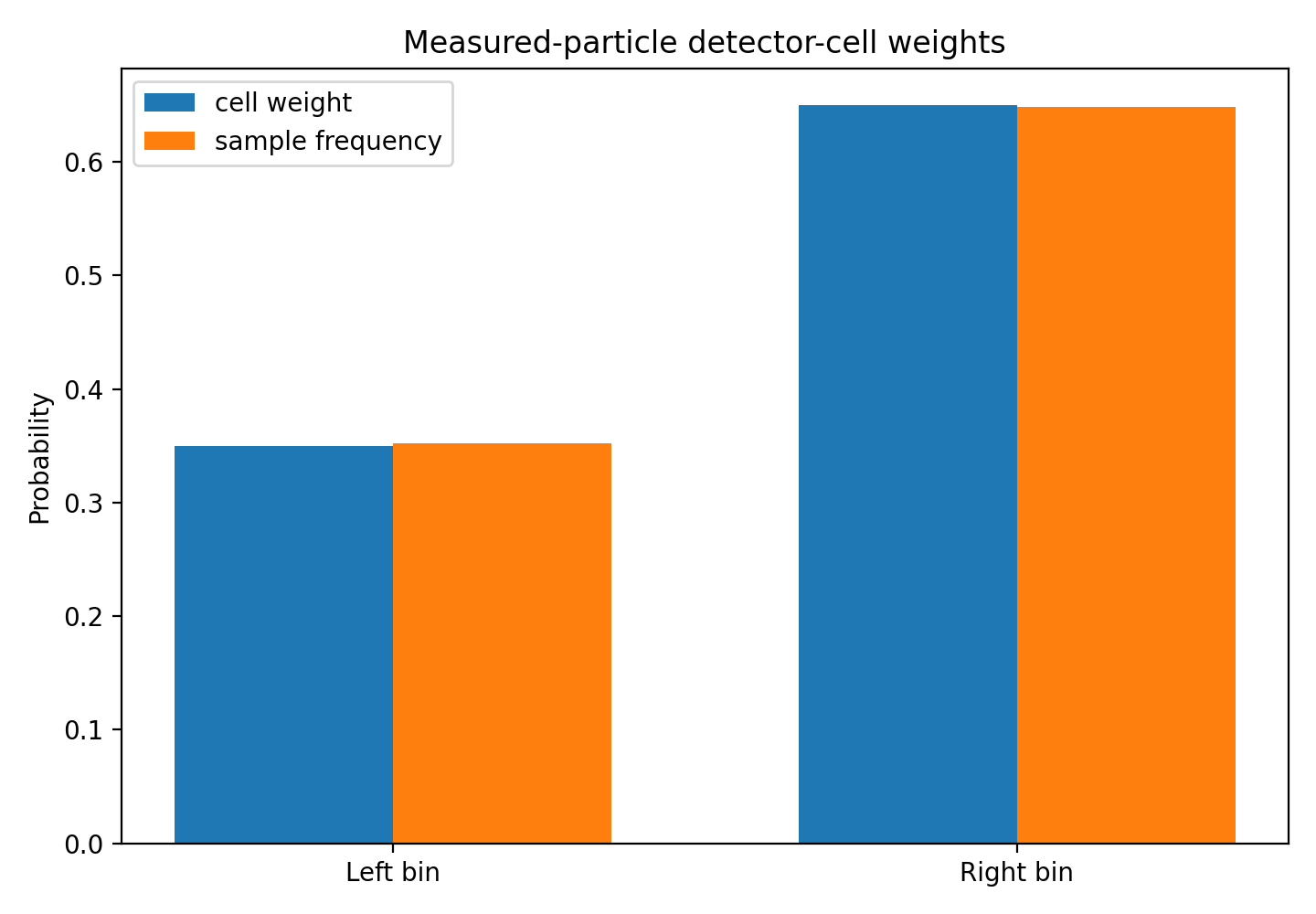}
\caption{Outcome frequencies for the measured particle in the two-branch entangled
state.  The sampled frequencies agree with the detector-class weights \(W_L\) and
\(W_R\).}
\label{fig:app-entangled-frequencies}
\end{figure}

The entangled test uses
\[
\Psi(x,y)
=
\sqrt{p_L}\,g_L(x)h_L(y)
+
\sqrt{p_R}\,g_R(x)h_R(y),
\qquad
p_L=0.35,
\quad
p_R=0.65,
\]
with first-particle packets centered at \(x=-2,2\), partner packets centered at
\(y=-1.5,1.5\), and detector cells
\[
I_L=(-\infty,0),
\qquad
I_R=[0,\infty).
\]
Direct numerical integration gives
\[
W_L=0.350001,
\qquad
W_R=0.649999.
\]
Sampling \(50{,}000\) detector records produced
\[
f_L=0.3519,
\qquad
f_R=0.6481,
\]
as displayed in Figure~\ref{fig:app-entangled-frequencies}.

\begin{figure}[htbp]
\centering
\includegraphics[width=0.80\textwidth]{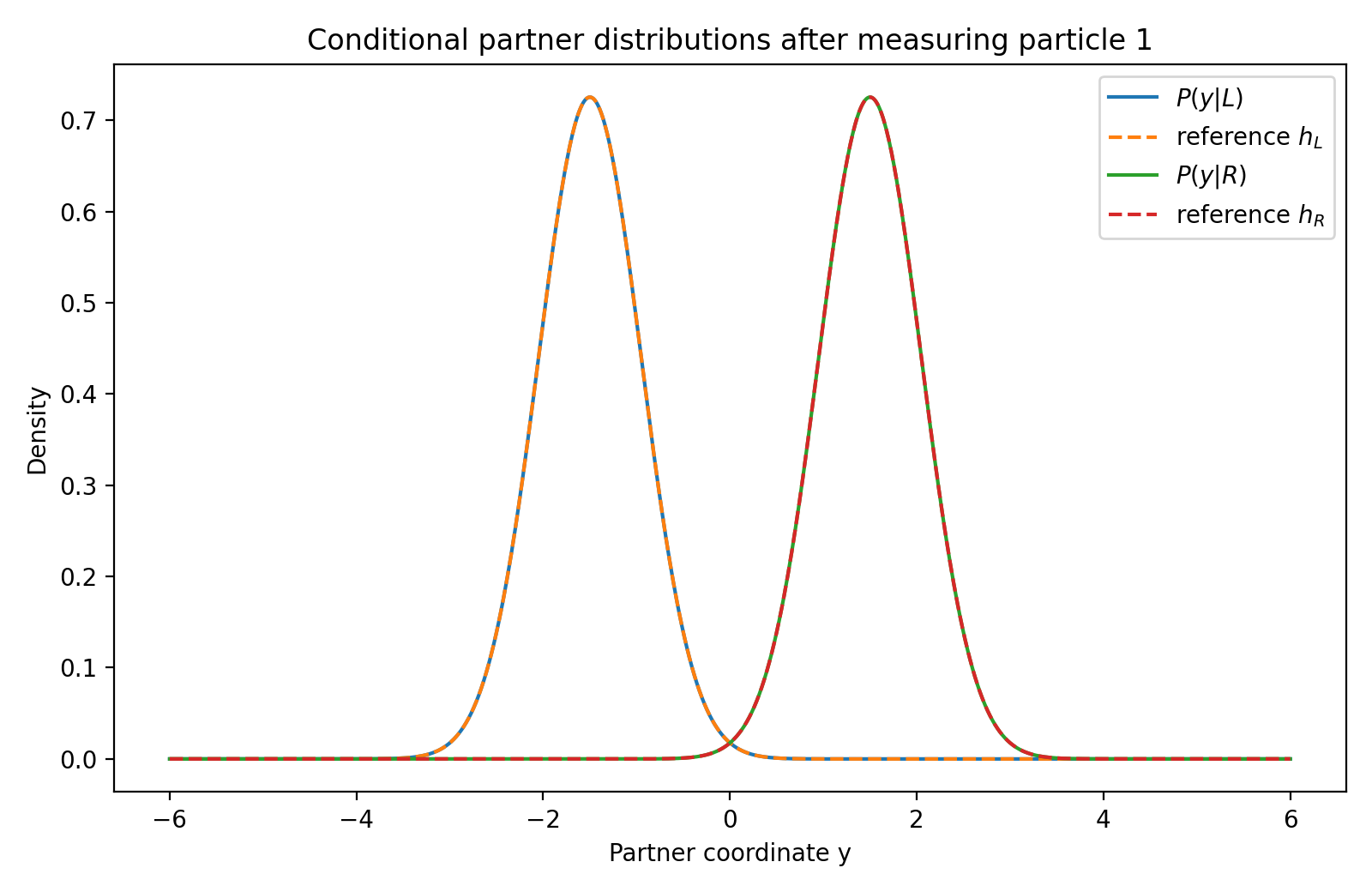}
\caption{Conditional partner distributions after measurement of the first particle.
The numerical conditional distributions coincide, within the quoted finite-grid error,
with the reference packets selected by the left and right records.}
\label{fig:app-conditional-partner}
\end{figure}

For the partner particle, the conditional densities are
\[
\rho_{2|L}(y)
=
\int_{I_L}|\Psi_L(x,y)|^2\,dx,
\qquad
\rho_{2|R}(y)
=
\int_{I_R}|\Psi_R(x,y)|^2\,dx.
\]
The left conditional distribution differs from the reference packet \(h_L\) by
\(1.9\times10^{-5}\) in \(L^1\), and the right distribution differs from \(h_R\) by
\(6.0\times10^{-6}\).  Figure~\ref{fig:app-conditional-partner} presents the comparison
in a simplified grayscale form so that the two conditional alternatives can be read in
the same visual style as the other appendix plots.

The calculation supports the distinction made in Chapters~\ref{chap:two-particle-measurement}
and~\ref{chap:correlations-locality}: the \({\bf RM}\) motion acts on the joint state,
while the detector record is defined in the measured coordinate.  Once that record is
known, the state of the partner is not sampled independently; it is the conditional
state determined by the original entangled state and the recorded detector class.

\section{Particle-device measurement and stability of the record class}
\label{sec:app-particle-device}

This section tests the device limit used in Sections~\ref{sec:a-particle-and-a-macroscopic-measuring-device}
and~\ref{sec:from-state-reduction-to-macroscopic-recording}.  The required statement
is stronger than the one-coordinate scaling of Section~A.VIII: the microscopic
particle must be able to undergo reduction while the macroscopic device remains in
the same finite-resolution record class.

\subsection{Four projected coordinates}

\begin{figure}[htbp]
\centering
\includegraphics[width=0.66\textwidth]{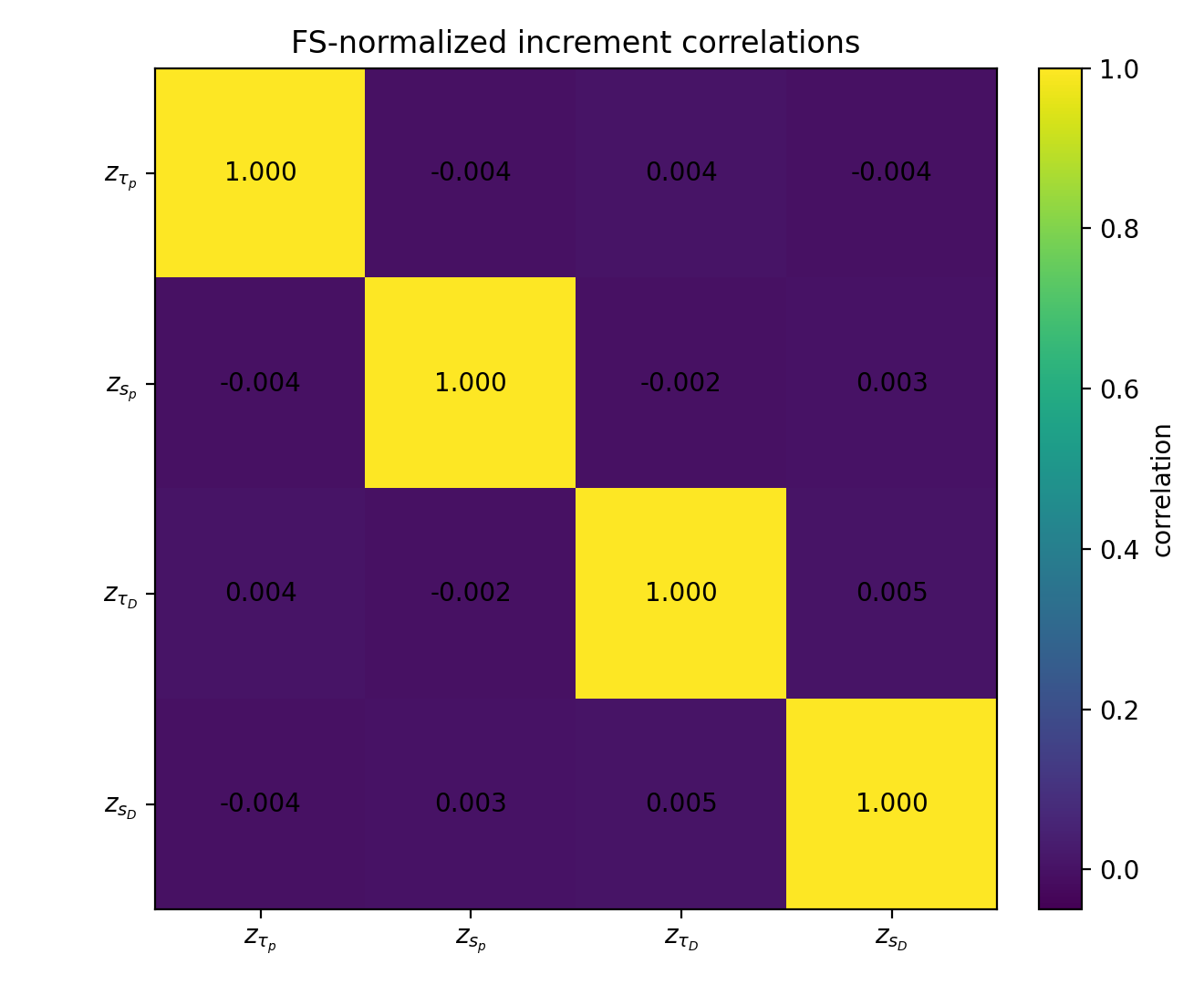}
\caption{Correlation matrix of the normalized particle-device increments
\((z_{\tau_p},z_{s_p},z_{\tau_D},z_{s_D})\).  The small off-diagonal entries confirm
the expected local independence of the four orthogonal coordinates.}
\label{fig:app-four-coordinate}
\end{figure}

Near a product Gaussian particle-device state, the local coordinates are
\[
(\tau_p,s_p;\tau_D,s_D),
\]
where \(\tau\) denotes the position coordinate and
\(s=\ln(\delta/\sigma)\) the localization coordinate.  For the four orthogonal tangent
directions, the metric has diagonal entries
\[
G_{\tau_p\tau_p}=\frac1{4\delta_p^2},
\qquad
G_{s_ps_p}=\frac12,
\qquad
G_{\tau_D\tau_D}=\frac1{4\delta_D^2},
\qquad
G_{s_Ds_D}=\frac12.
\]
With \(\delta_p=0.75\), \(\delta_D=0.10\), and \(80{,}000\) samples, the
Fubini--Study-normalized four-coordinate covariance is essentially diagonal.  The
largest off-diagonal correlation is \(0.0052\).  Figure~\ref{fig:app-four-coordinate}
shows the full correlation matrix.  The measured standard deviations are given in
Table~\ref{tab:app-four-coordinate}.

\begin{table}[htbp]
\centering
\caption{Particle-device projected increments in the four detector coordinates.}
\label{tab:app-four-coordinate}
\begin{tabular}{lcc}
\hline
Increment & Euclidean std. dev. & FS-normalized std. dev. \\
\hline
\(\Delta\tau_p\) & 1.4972 & 0.9981 \\
\(\Delta s_p\)    & 1.4182 & 1.0028 \\
\(\Delta\tau_D\) & 0.2011 & 1.0055 \\
\(\Delta s_D\)    & 1.4172 & 1.0021 \\
\hline
\end{tabular}
\end{table}

The normalized stochastic scale is therefore the same in all four directions, but the
Euclidean device displacement is much smaller because
\[
\Delta\tau_D\sim2\delta_D Z.
\]

\subsection{Finite-resolution device class}

\begin{figure}[htbp]
\centering
\includegraphics[width=0.78\textwidth]{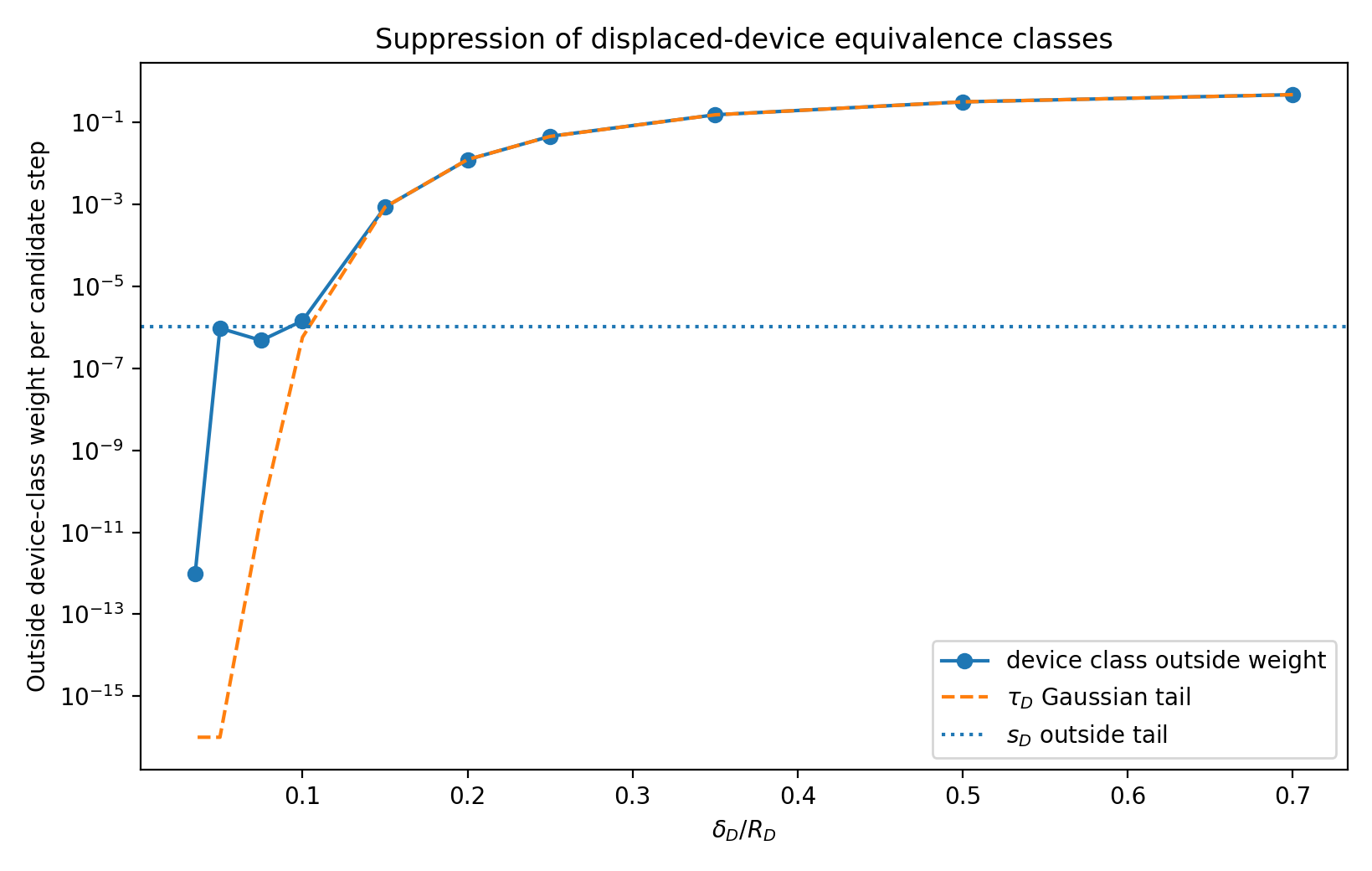}
\caption{Aggregate weight outside the recorded device equivalence class.  The weight
of displaced macroscopic record classes rapidly becomes negligible in the device
limit.}
\label{fig:app-device-outside}
\end{figure}

The recorded device class is defined by
\[
|\tau_D-\tau_{D,0}|\le R_D,
\qquad
s_D<0.
\]
The simulation uses \(R_D=1\), \(s_{\rm rec}=-0.75\), \(D_s=0.25\), and
\(\Delta t=0.05\).  The localization coordinate is sampled as
\[
s_D=s_{\rm rec}+\xi_D,
\qquad
\xi_D\sim N(0,2D_s\Delta t).
\]
Table~\ref{tab:app-device-retention} and Figure~\ref{fig:app-device-outside} show how
the total weight outside the recorded class decreases as \(\delta_D/R_D\) becomes
small.  At \(\delta_D/R_D=0.15\) the outside weight is already
\(8.61\times10^{-4}\), and by \(0.05\) it is about \(7.31\times10^{-7}\).

\begin{table}[htbp]
\centering
\caption{Retention of the finite-resolution device equivalence class.}
\label{tab:app-device-retention}
\begin{tabular}{ccc}
\hline
\(\delta_D/R_D\) & In-class fraction & Outside-class weight \\
\hline
0.70 & 0.5251   & \(4.75\times10^{-1}\) \\
0.50 & 0.6829   & \(3.17\times10^{-1}\) \\
0.35 & 0.8468   & \(1.53\times10^{-1}\) \\
0.25 & 0.9545   & \(4.55\times10^{-2}\) \\
0.20 & 0.9876   & \(1.24\times10^{-2}\) \\
0.15 & 0.9991   & \(8.61\times10^{-4}\) \\
0.10 & 0.999998 & \(1.83\times10^{-6}\) \\
0.075 & 0.999999 & \(1.10\times10^{-6}\) \\
0.050 & 0.999999 & \(7.31\times10^{-7}\) \\
\hline
\end{tabular}
\end{table}

\subsection{Particle reduction while the device class remains stable}

The particle coordinate was then evolved by the same unbiased two-outcome reduction
walk while the device coordinates were sampled simultaneously and tested for class
membership.  To eliminate floating-point ambiguity at the absorbing endpoints, the
particle walk was implemented on the integer lattice
\[
k=0,1,\ldots,20,
\qquad
p=\frac{k}{20},
\]
so that one reduction step changes \(k\) by exactly \(\pm1\).  The endpoints
\(k=0\) and \(k=20\) are therefore represented exactly.  For each initial particle
weight \(p_0\), \(100{,}000\) independent paths were generated.

The device parameters were
\[
\frac{\delta_D}{R_D}=0.05,
\qquad
R_D=1,
\qquad
s_{\rm rec}=-0.75,
\qquad
D_s=0.25,
\qquad
\Delta t=0.05.
\]
At every particle step the device displacement and localization coordinate were
sampled and tested against the same finite-resolution device class used above.
Table~\ref{tab:app-stable-device-born} gives the resulting upper-outcome frequencies
and the measured outside-class weights.  The particle frequencies agree with the Born
values within ordinary Monte Carlo sampling error, while the device outside-class
weight remains of order \(10^{-6}\).

\begin{table}[htbp]
\centering
\caption{Particle Born frequencies with a stable finite-resolution device class.
Each row uses \(100{,}000\) independent reduction paths.}
\label{tab:app-stable-device-born}
\begin{tabular}{cccc}
\hline
Initial weight \(p_0\) & Upper frequency & Std. error & Device outside-class weight \\
\hline
0.10 & 0.10091 & 0.00095 & \(1.39\times10^{-6}\) \\
0.25 & 0.25041 & 0.00137 & \(7.99\times10^{-7}\) \\
0.35 & 0.35019 & 0.00151 & \(1.43\times10^{-6}\) \\
0.50 & 0.49976 & 0.00158 & \(9.97\times10^{-7}\) \\
0.65 & 0.64929 & 0.00151 & \(7.68\times10^{-7}\) \\
0.80 & 0.80100 & 0.00126 & \(7.83\times10^{-7}\) \\
0.90 & 0.90116 & 0.00094 & \(8.33\times10^{-7}\) \\
\hline
\end{tabular}
\end{table}

\begin{figure}[htbp]
\centering
\includegraphics[width=0.72\textwidth]{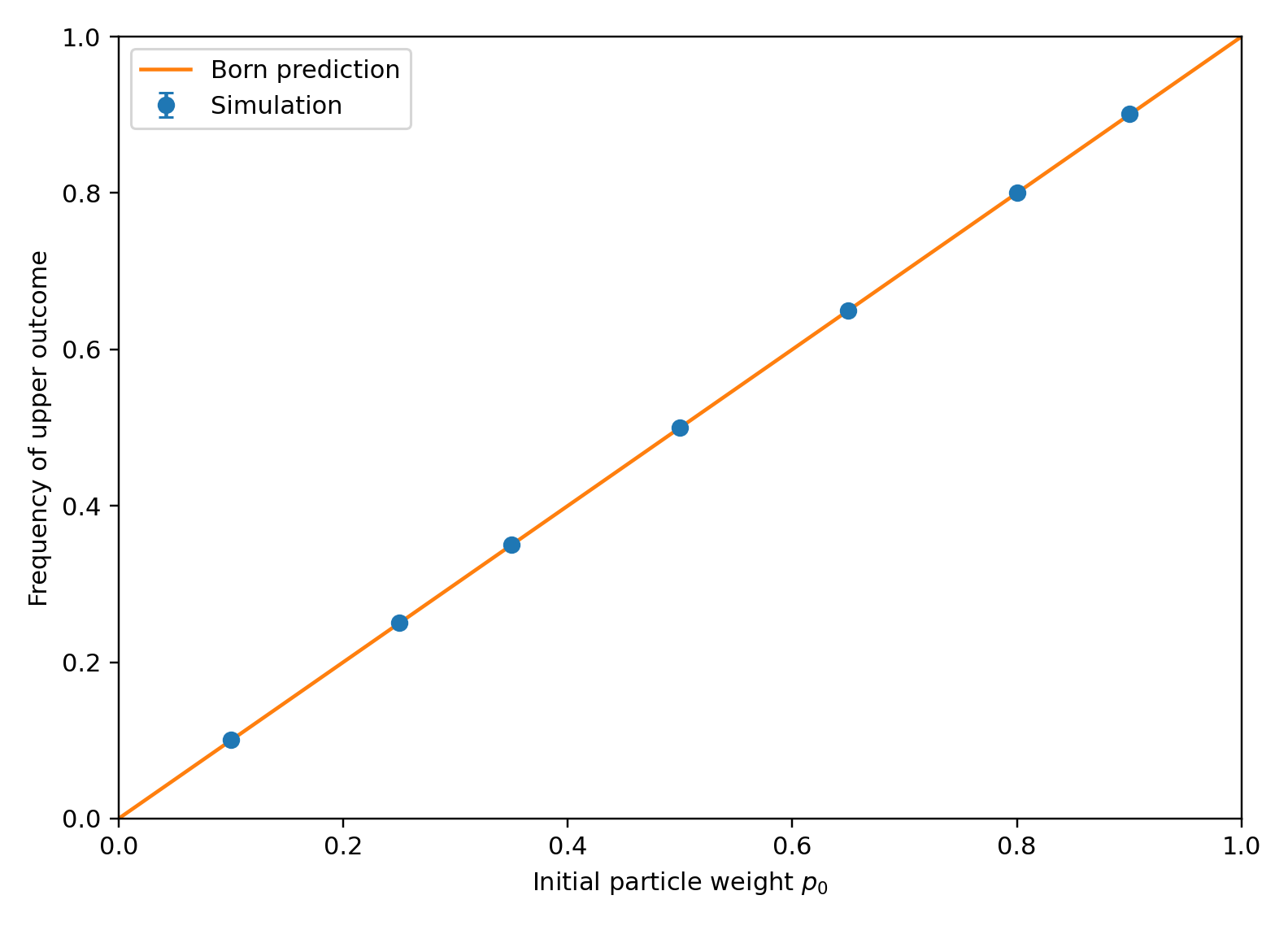}
\caption{Particle outcome frequencies in the stable-device simulation.  Error bars
show the binomial standard errors; the diagonal is the Born prediction.  The device
outside-class weight during the same runs is of order \(10^{-6}\), as listed in
Table~\ref{tab:app-stable-device-born}.}
\label{fig:app-stable-device-born}
\end{figure}

Figure~\ref{fig:app-stable-device-born} compares the measured particle frequencies
with the Born prediction.  The diagonal agreement is obtained while the device remains
inside the same recorded equivalence class with overwhelming probability.

The important numerical conclusion is not that the microscopic device ray is frozen.
It is that the particle retains its reduction statistics while the aggregate weight of
macroscopically displaced device classes is negligible.  This is precisely the
finite-resolution statement required by Chapter~\ref{chap:two-particle-measurement}:
\[
\text{particle reduction with Born weights}
\quad+\quad
\text{stable device equivalence class}.
\]

\section{Effective irreversibility of realized \({\bf RM}\) histories}
\label{sec:app-irreversibility}

The final numerical tests correspond to Sections~\ref{sec:irreversibility-from-random-matrix-stabilization}
and~\ref{sec:the-arrow-of-time}.  The book distinguishes microscopic unitary
invertibility from effective irreversibility after the realized random history and the
microscopic final ray are no longer retained.

\subsection{Exact inverse versus a statistically equivalent reverse history}

The numerical test uses \(N=32\), \(180\) GUE steps, and step strength \(0.22\).  A
single forward history is first generated.  Three reversal procedures are then
compared:
\begin{enumerate}
\item apply the exact inverse unitaries in reverse order;
\item use the complex-conjugated Hamiltonian history associated with antiunitary time
reversal; and
\item discard the realized history and use independently sampled GUE matrices with
the same ensemble law.
\end{enumerate}
Only the first procedure contains the information needed to retrace the particular
realized path.

\begin{figure}[htbp]
\centering
\includegraphics[width=0.82\textwidth]{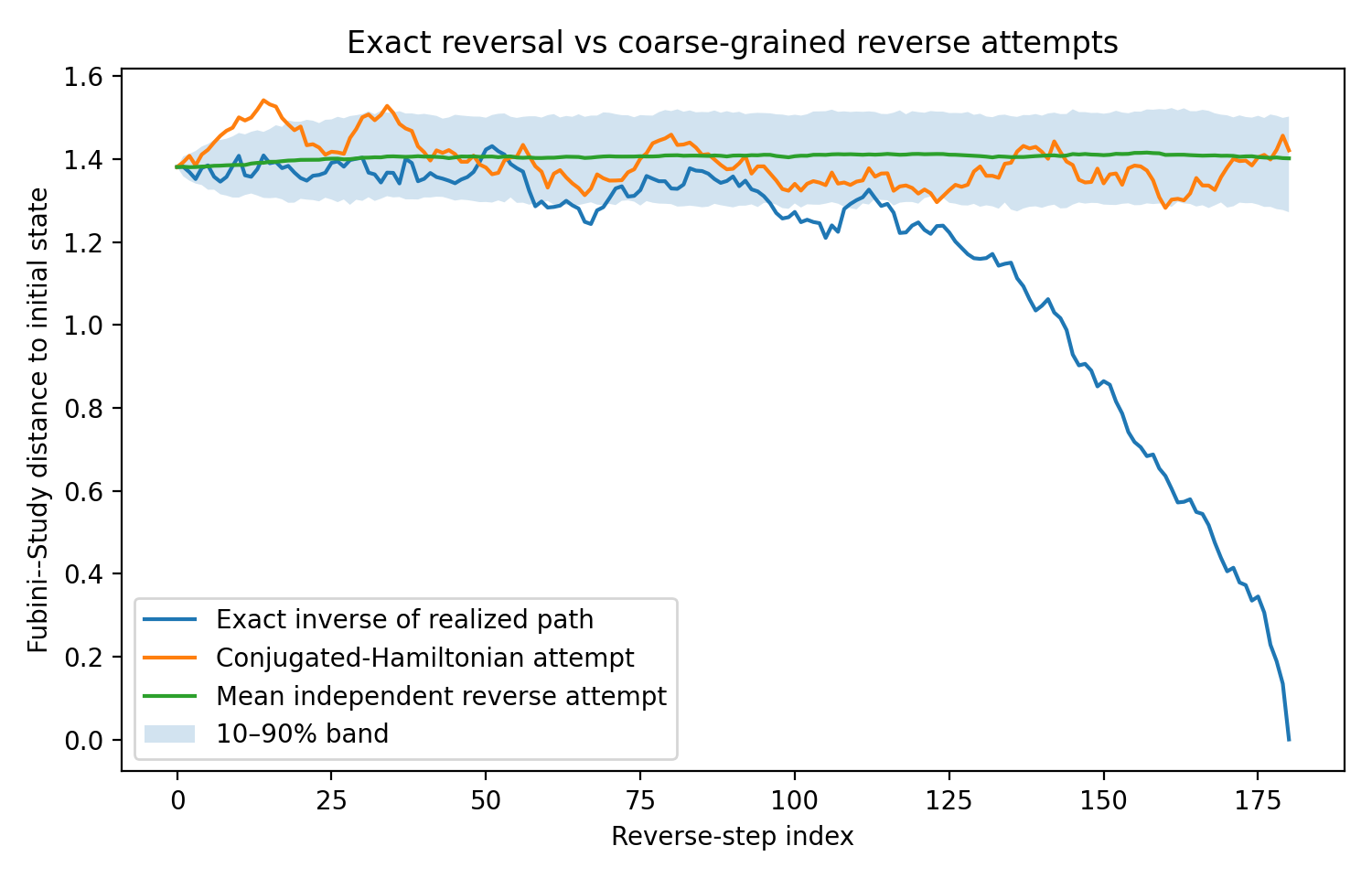}
\caption{Exact inversion of a realized \({\bf RM}\) path compared with reversal
procedures that do not reproduce the realized Hamiltonian history.  Only the exact
inverse returns to the initial ray.}
\label{fig:app-loschmidt}
\end{figure}

Figure~\ref{fig:app-loschmidt} shows the Fubini--Study distance during these reversal
experiments.  The exact inverse returns to zero distance.  The conjugated-history
attempt and the independent reverse histories remain far from the initial ray.

\begin{figure}[htbp]
\centering
\includegraphics[width=0.76\textwidth]{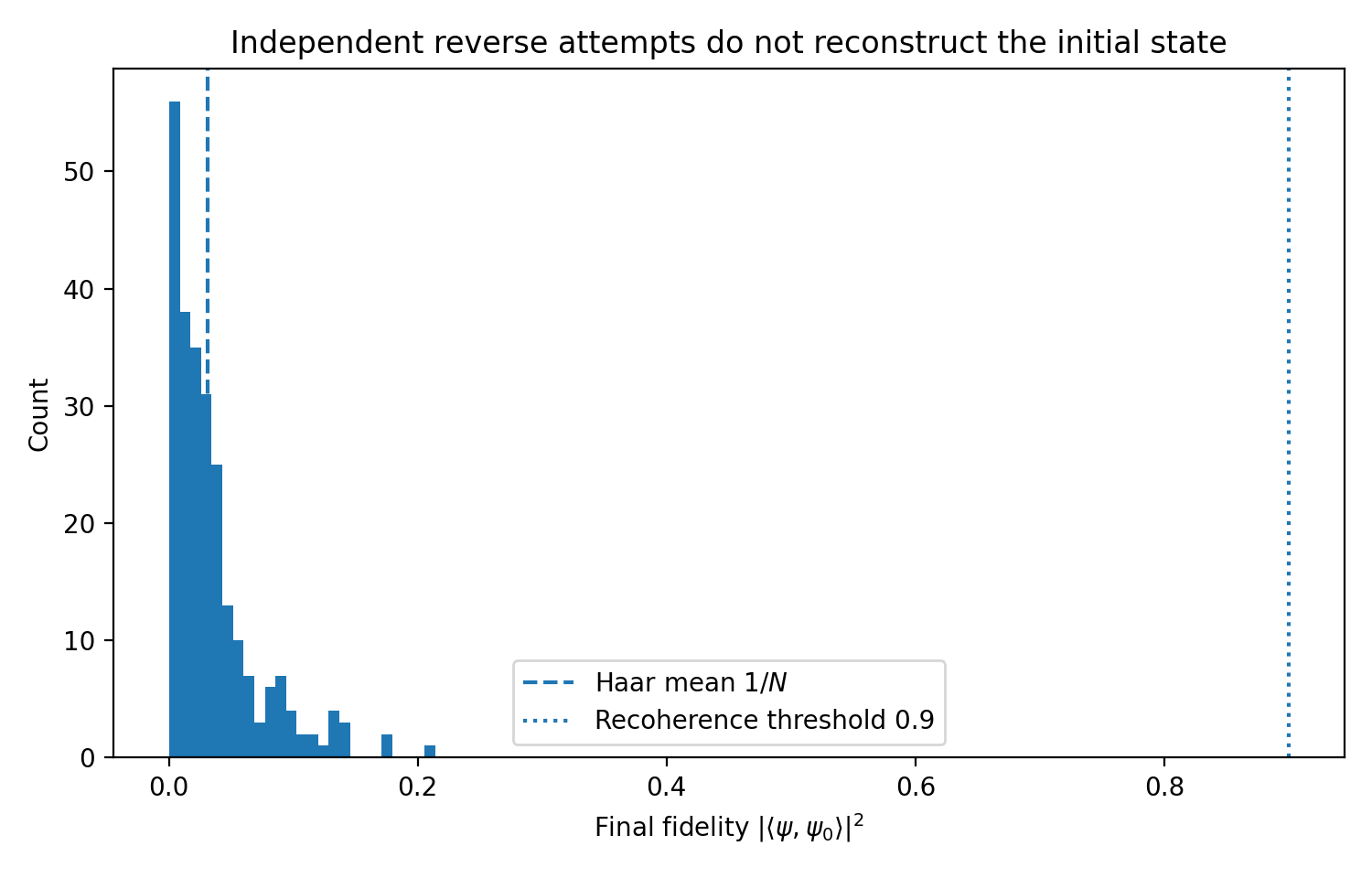}
\caption{Final fidelities with the initial state after reverse attempts using freshly
sampled GUE histories.  The distribution remains near the random-state scale
\(1/N\).}
\label{fig:app-reverse-fidelity}
\end{figure}

For \(250\) independent reverse attempts the mean final fidelity was \(0.035447\),
close to the random-state scale \(1/N=0.03125\).  No attempt reached fidelity larger
than \(0.9\), either at the final time or at an intermediate time.  The numerical
summary is given in Table~\ref{tab:app-irreversibility} and the final-fidelity
distribution in Figure~\ref{fig:app-reverse-fidelity}.

\begin{table}[htbp]
\centering
\caption{Summary of the realized-history reversal simulation.}
\label{tab:app-irreversibility}
\begin{tabular}{lc}
\hline
Quantity & Value \\
\hline
Hilbert-space dimension \(N\) & 32 \\
Number of \({\bf RM}\) steps & 180 \\
Step strength & 0.22 \\
Independent reverse attempts & 250 \\
Final FS distance after exact inverse & 0.000000 \\
Final FS distance after conjugated-history attempt & 1.420865 \\
Mean final fidelity, independent attempts & 0.035447 \\
Median final fidelity & 0.024513 \\
Maximum final fidelity & 0.213899 \\
Random-state scale \(1/N\) & 0.03125 \\
Recoherence events with fidelity \(>0.9\) & 0 \\
\hline
\end{tabular}
\end{table}

The high-dimensional suppression discussed analytically in
Section~\ref{sec:irreversibility-from-random-matrix-stabilization},
\[
P(F>1-\epsilon)=\epsilon^{N-1},
\]
provides the geometric background to this calculation, but the figures retained here
are the actual path-reversal simulations rather than a plot of the analytical formula.

\subsection{Role of finite-resolution records}

The particle-device calculation of Section~A.X supplies the measurement-specific part
of the irreversibility argument.  A detector record keeps neither the exact ray nor the
realized sequence of random Hermitian increments.  It retains only a device
equivalence class.  Since Table~\ref{tab:app-device-retention} shows that the aggregate
weight of macroscopically displaced classes vanishes in the device limit, many
microscopically different histories are represented by the same stable record.  Exact
reversal would therefore require reconstructing information that is absent from the
record description.

The numerical tests thus support the three distinctions made in Chapter~\ref{chap:paradoxes-arrow-time}:
unitarity of every realized path, suppression of accidental return once the realized
history is lost, and additional coarse-graining when only a finite-resolution record is
retained.

\section{What the simulations establish and what they do not}
\label{sec:app-computational-scope}

The calculations in this appendix are finite-dimensional or spatially discretized tests.
Their status should therefore be kept separate from the analytical results of the book.
They establish the following finite-model facts with direct numerical evidence:
\begin{itemize}
\item GUE increments have the isotropy and homogeneity required in complex projective
state space, while GOE does not;
\item the restriction of the GUE-generated projective walk to the localized Gaussian
manifold has Gaussian, translation-invariant increments consistent with the Brownian
family used in the stochastic lift;
\item the detector-adapted $(\tau,s)$ coordinates and the full localized
$(a,p,s)$ coordinates inherit the predicted orthogonal Gaussian increment law;
\item the binary reduction-coordinate walk reproduces the two-outcome Born hitting
probabilities and exhibits effective endpoint stabilization in the detector coordinate;
\item the same detector-class rule produces coherent and which-slit screen records,
depending on the state arriving at the screen;
\item conditioned localized records remain close to Newtonian motion in the regime
$D_a\Delta t/\sigma^2\ll1$;
\item the GUE projective law remains isotropic and homogeneous at product and
entangled states in tensor-product Hilbert space;
\item the product Gaussian restriction yields independent Brownian coordinates, while
the Euclidean motion of a sharply localized device coordinate is suppressed by the
induced metric;
\item two-particle detector weights factor for product states and give the expected
conditional partner state for entangled states;
\item particle reduction is compatible with stability of a finite-resolution device
class in the device limit; and
\item a realized ${\bf RM}$ history is exactly invertible when retained, but a fresh
statistically equivalent GUE history does not reconstruct it.
\end{itemize}

The simulations do \emph{not} construct the infinite-dimensional cylindrical law or
the full path measure discussed in Sections~\ref{sec:the-gaussian-unitary-lift} and
\ref{sec:a-path-measure-from-transition-weights}; those are analytical questions.
Nor do the simulations by themselves provide the microscopic origin of the Brownian
family used in the stochastic lift. That step is supplied separately by the
Caldeira--Leggett-type construction developed in the main text: under the stated
locality, symmetry, thermal, Ohmic, and coarse-graining assumptions, the
system--environment Schr\"odinger dynamics yields the Brownian family on the
localized classical manifold, whose common stochastic lift gives the GUE-generated
${\bf RM}$ regime. What remains open is to derive and control these approximations
for realistic many-body systems and measuring environments, and to determine the
range of validity and universality of the resulting effective dynamics.

The numerical role is therefore narrower but important: the simulations check that the
finite realizations of the stochastic lift and detector dynamics behave in the way
required by the geometric framework, across the one-particle, macroscopic,
tensor-product, entangled, and particle-device regimes used in the main text. They
complement, rather than replace, the analytical construction of the Brownian family
from the microscopic system--environment model.